\documentclass[12pt,document,nofootinbib,superscriptaddress,onecolumn,preprintnumbers,balancelastpage]{article}
\pdfoutput=1
\usepackage{latexsym}
\usepackage{amssymb}
\usepackage{epsfig,amsmath,graphics}
\usepackage{listings}
\usepackage{color}
\usepackage{dcolumn}
\usepackage{slashed}
\usepackage{cancel}
\usepackage{comment,latexsym} 
\usepackage{bm}
\usepackage{verbatim}
\usepackage{tabularx}
\usepackage{dcolumn}
\definecolor{Mahogany}{rgb}{0.62,0.24,0.15}
\definecolor{colorLink}{rgb}{0.6,0,0}
\definecolor{colorCite}{rgb}{0,.6,0}
\definecolor{colorURL}{rgb}{0,0.6,0.0}
\usepackage[colorlinks=true,linktocpage=true,linkcolor=black,citecolor=colorCite,urlcolor=colorURL]{hyperref}
\usepackage{setspace}
\usepackage{xspace}
\usepackage{multirow}
\usepackage{titlesec}
\usepackage[bbgreekl]{mathbbol}
\usepackage{booktabs}
\usepackage{makecell}
\usepackage{array}
\usepackage{authblk}
\usepackage{xfrac}
\usepackage{colortbl,colordvi,xcolor,color}
\usepackage{gensymb}
\usepackage{cite}

\usepackage{cleveref}

\crefname{table}{Table}{Tables}
\crefname{equation}{Eq.}{Eqs.}
\crefname{appendix}{App.}{Apps.}
\crefname{section}{Sec.}{Secs.}
\crefname{figure}{Fig.}{Figs.}

\usepackage[letterpaper, margin=1in]{geometry}

\newcolumntype{L}[1]{>{\raggedright\let\newline\\\arraybackslash\hspace{0pt}}m{#1}}
\newcolumntype{C}[1]{>{\centering\let\newline\\\arraybackslash\hspace{0pt}}m{#1}}
\newcolumntype{R}[1]{>{\raggedleft\let\newline\\\arraybackslash\hspace{0pt}}m{#1}}

\usepackage{sectsty}
\sectionfont{\raggedright}

\expandafter\def\expandafter\normalsize\expandafter{%
    \normalsize
    \setlength\abovedisplayskip{8pt}
    \setlength\belowdisplayskip{8pt}
    \setlength\abovedisplayshortskip{8pt}
    \setlength\belowdisplayshortskip{8pt}
}

\usepackage{feynmp-auto}
\titleformat{\section}{\center\normalfont\fontsize{14}{15}\bfseries}{\thesection.}{1em}{}
\titleformat{\subsubsection}{\center\normalfont\fontsize{12}{15}}{\thesubsubsection.}{1em}{}

\definecolor{colorTC}{rgb}{.2,.7,.2}

\definecolor{colorKF}{rgb}{.5,.1,.8}

\definecolor{colorNC}{rgb}{0,0,1}

\newcolumntype{a}{>{\columncolor{blue}}c}

\def\dd{\text{d}}
\def\be{\begin{equation}}
\def\ee{\end{equation}}

\newcommand{\dfbd}{\mathord{\buildrel{\lower3pt\hbox{$\scriptscriptstyle\leftrightarrow$}}\over {D^\mu} }}

\begin{document}


\title{
\vspace{-42pt} 
\textbf{\large
The Muon Smasher's Guide
}}

\author[1]{\small Hind Al Ali}
\author[2]{Nima Arkani-Hamed}
\author[1]{Ian Banta}
\author[1]{Sean Benevedes}
\author[3]{Dario Buttazzo}
\author[1]{\hspace{60pt} Tianji Cai}
\author[1]{Junyi Cheng}
\author[4]{Timothy Cohen}
\author[1]{Nathaniel Craig}
\author[5]{Majid Ekhterachian}
\author[6]{\hspace{60pt} JiJi Fan}
\author[7]{Matthew Forslund}

\author[8]{Isabel Garcia Garcia}
\author[9]{Samuel Homiller}
\author[10]{Seth Koren}
\author[1]{\hspace{60pt} Giacomo Koszegi}
\author[5,11]{Zhen Liu}
\author[9]{Qianshu Lu}
\author[12]{Kun-Feng Lyu}
\author[13]{Alberto Mariotti}
\author[1]{\hspace{60pt} Amara McCune}
\author[7]{Patrick Meade}
\author[14]{Isobel Ojalvo}
\author[1]{Umut Oktem}
\author[15,16]{Diego Redigolo}
\author[9]{\hspace{60pt} Matthew Reece}
\author[17]{Filippo Sala}
\author[5]{Raman Sundrum}
\author[18]{Dave Sutherland}
\author[16,19]{Andrea Tesi}
\author[1]{\hspace{60pt}  Timothy Trott}
\author[14]{Chris Tully}
\author[10]{Lian-Tao Wang}
\author[1]{Menghang Wang}

\affil[1]{\footnotesize \textit{Department of Physics, University of California, Santa Barbara, CA 93106, USA}}
\affil[2]{\footnotesize \textit{School of Natural Sciences, Institute for Advanced Study, Princeton, NJ, 08540, USA}}
\affil[3]{\footnotesize \textit{INFN, Sezione di Pisa, Largo Bruno Pontecorvo 3, I-56127 Pisa, Italy}}
\affil[4]{\footnotesize \textit{Institute for Fundamental Science, University of Oregon, Eugene, OR 97403, USA}}
\affil[5]{\footnotesize \textit{Maryland Center for Fundamental Physics, University of Maryland, College Park, MD 20742, USA}}
\affil[6]{\footnotesize \textit{Department of Physics, Brown University, Providence, RI 02912, USA}}
\affil[7]{\footnotesize \textit{C. N. Yang Institute for Theoretical Physics, Stony Brook University, 
 Stony Brook, NY 11794, USA}}

\affil[8]{\footnotesize \textit{Kavli Institute for Theoretical Physics, University of California, Santa Barbara, CA 93106, USA}}
\affil[9]{\footnotesize \textit{Department of Physics, Harvard University, Cambridge, MA 02138, USA}}
\affil[10]{\footnotesize \textit{Department of Physics and Enrico Fermi Institute, University of Chicago, Chicago, IL 60637, USA}}
\affil[11]{\footnotesize \textit{School of Physics and Astronomy, University of Minnesota, Minneapolis, MN 55455, USA}}
\affil[12]{\footnotesize \textit{Department of Physics, The Hong Kong University of Science and Technology,\hspace{-60pt} \newline Clear Water Bay, Kowloon, Hong Kong S.A.R., P.R.C}}
\affil[13]{\footnotesize \textit{Theoretische Natuurkunde and IIHE/ELEM, Vrije Universiteit Brussel,\hspace{-70pt} \newline and International Solvay Institutes, Pleinlaan 2, B-1050 Brussels, Belgium}}

\affil[14]{\footnotesize \textit{Princeton University, Princeton, NJ 08540, USA}}
\affil[15]{\footnotesize \textit{ CERN, Theoretical Physics Department, Geneva, Switzerland}}
\affil[16]{\footnotesize \textit{ INFN Sezione di Firenze, Via G. Sansone 1, I-50019 Sesto Fiorentino, Italy}}
\affil[17]{\footnotesize \textit{LPTHE, CNRS \& Sorbonne Universite, 4 Place Jussieu, F-75252 Paris, France}}
\affil[18]{\footnotesize \textit{INFN Sezione di Trieste, via Bonomea 265, 34136 Trieste, Italy}}
\affil[19]{\footnotesize \textit{Department of Physics and Astronomy, University of Florence, Italy}}

\date{}

\begin{spacing}{1.1}
\maketitle
\end{spacing}

\thispagestyle{empty}

\vspace{-17pt}
\begin{abstract}
\begin{spacing}{1.03}\noindent
We lay out a comprehensive physics case for a future high-energy muon collider, exploring a range of collision energies (from $1$ to $100$ TeV) and luminosities. We highlight the advantages of such a collider over proposed alternatives.  We show how one can leverage both the point-like
nature of the muons themselves as well as the cloud of electroweak radiation that surrounds the beam
to blur the dichotomy between energy and precision in the search for new physics. The physics case is buttressed by a range of studies with applications to
electroweak symmetry breaking, dark matter, and the naturalness of the weak scale.
Furthermore, we make sharp connections with complementary experiments that are probing new physics effects using electric dipole moments, flavor violation, and gravitational waves. An extensive appendix provides cross section predictions as a function of the center-of-mass energy for many canonical simplified models.
\end{spacing}
\end{abstract}

\newpage
\begin{spacing}{1.15}
\clearpage

\pagenumbering{arabic} 
\setcounter{page}{2}

\begin{spacing}{1.2}
\newpage
\setcounter{tocdepth}{2}
\tableofcontents
\end{spacing}

\clearpage


\section{Introduction}
The discovery of the Higgs boson \cite{Aad:2012tfa,Chatrchyan:2012ufa} at the Large Hadron Collider (LHC) marks the end of one era and the dawn of another. The origin of mass has been explained, but in answering this question, the Higgs boson poses a host of others: Is this the Higgs of the Standard Model? Is it the only Higgs, or one of many? Why is electroweak symmetry broken in the first place, and what sets the scale? How, if at all, is the origin of mass connected to the patterns of flavor, the nature of dark matter, or the abundance of matter over antimatter?

These questions make the call to explore shorter distances and higher energies as vibrant and clear as it has ever been. Although the path forward is devoid of guaranteed discoveries, the journey thus far has always been 
more a matter of serendipity than inevitability. We build colliders not to confirm what we already know, but to explore what we do not. In the wake of the Higgs boson's discovery, the question is not whether to build another collider, but which collider to build. 

Over the course of the last decade, consensus has largely coalesced around linear or circular $e^+ e^-$ colliders \cite{Bambade:2019fyw, Charles:2018vfv, Abada:2019zxq, CEPCStudyGroup:2018ghi} and circular $pp$ colliders \cite{Benedikt:2018csr, CEPC-SPPCStudyGroup:2015csa}, both of which constitute natural extensions of past and present machines. The strengths of these two approaches are largely complementary, with the precision of $e^+ e^-$ machines and the power of $pp$ machines paving distinct paths toward the exploration of physics at shorter distances. Loosely speaking, the strength of the former is to reveal the fingerprints that new physics has left on the properties of the Higgs and other electroweak states, while the latter are positioned to produce the new physics directly. This has given rise to a familiar dichotomy between energy and precision as contrasting approaches to search for new physics. 

Enter the muon. The potential advantages of high-energy muon colliders have long been recognized \cite{Budker:1969cd, Parkhomchuk:1983ua, Neuffer:1983jr, Neuffer:1986dg, Barger:1995hr, Barger:1996jm, Ankenbrandt:1999cta}. As a fundamental particle, the muon's full energy is available in a collision, with far cleaner final states relative to those produced by the dissociation of a composite particle like the proton. Its considerable mass suppresses the synchrotron radiation that effectively limits the energies of circular $e^+ e^-$ colliders, making both high energies and high luminosities achievable with a relatively small footprint. This raises the prospect that a muon collider could exceed the direct energy reach of the LHC, while achieving unprecedented precision measurements of Standard Model processes.  The muon allows us to leverage the benefits of both energy and precision in a unified future collider program. 

These advantages come at a cost: the colliding particles are no longer stable. The short lifetime of the muon imposes a series of technical challenges that must be overcome before such a collider can be realized. But progress towards this end has hastened considerably in recent years, spearheaded by the US Muon Accelerator Program (MAP) \cite{Delahaye:2013jla, Delahaye:2015yxa, Ryne:2015xua, Long:2020wfp}, the Muon Ionization Cooling Experiment (MICE) \cite{Mohayai:2018rxn, Blackmore:2018mfr, Bogomilov:2019kfj}, and the Low Emittance Muon Accelerator (LEMMA) concept \cite{Antonelli:2015nla}. Developments on the accelerator side have catalyzed experimental and theoretical activity, reflected by input to the European Particle Physics Strategy Update \cite{Delahaye:2019omf} and the proliferation of studies outlining aspects of the theory case for muon colliders at various energies. Recent contributions include studies of the electroweak boson PDF of the muon \cite{Han:2020uid}; the production of new scalars \cite{Eichten:2013ckl,Chakrabarty:2014pja, Buttazzo:2018qqp,Bandyopadhyay:2020otm, Han:2021udl, Liu:2021jyc} and diverse other states \cite{Costantini:2020stv} in vector boson fusion (VBF); the discovery potential for minimal dark matter \cite{Han:2020uak, Capdevilla:2021fmj}; the measurement of the Higgs self-couplings \cite{Chiesa:2020awd} and couplings to electroweak bosons \cite{Han:2020pif}; the sensitivity to new physics encoded in irrelevant operators \cite{DiLuzio:2018jwd,Buttazzo:2020uzc}; and the coverage of potential BSM explanations for hints from the complementary experiments yielding the muon $g-2$ \cite{Capdevilla:2020qel, Buttazzo:2020eyl, Capdevilla:2021rwo,Chen:2021rnl,Yin:2020afe}, $B$ meson~\cite{Huang:2021nkl}, and $K$ meson~\cite{Huang:2021biu} anomalies.

In this paper, we present an aspirational theory case underlining the physics potential of a high-energy muon collider. 
We aim to identify energy and luminosity goals that would position such a collider as a natural successor to the LHC. Our approach synthesizes some of the qualitative lessons of earlier studies (e.g. \cite{Buttazzo:2018qqp, Costantini:2020stv, Han:2020uak}), identifies entirely new physics objectives, and explores complementarity with forthcoming experiments across various frontiers. We summarize qualitative features of the most important production modes and characterize the electroweak gauge boson content of the initial state for both Standard Model and beyond-the-Standard Model final states; present a range of case studies demonstrating the muon collider's potential to shed light on electroweak symmetry breaking, dark matter, and the naturalness of the weak scale; and sharpen connections to complementary experiments probing new physics through electric dipole moments (EDMs), flavor violation, and gravitational waves. Although our primary focus is on high-energy muon colliders, it bears emphasizing that many of the same physics considerations are applicable to other high-energy lepton colliders, and aspects of this work are relevant to the physics case for potential long-term upgrades of the ILC. 

In the interest of identifying an optimal collider to succeed the LHC, we consider a variety of center-of-mass energies between 1 and 100 TeV, including energy benchmarks associated with various existing proposals. Wherever possible, we remain agnostic about the integrated luminosity attained at a given center-of-mass energy, preferring instead to determine the amount of integrated luminosity required to discover or constrain a particular point in parameter space. We will also provide forecasts, which will make reference to two luminosity scalings. The first, an ``optimistic'' scaling $\mathcal{L}_{\rm int}^{\rm opt}$, assumes integrated luminosity growing with $s$ in order to compensate for the $1/s$ falloff in many interaction cross sections. The second, a more conservative scaling $\mathcal{L}_{\rm int}^{\rm con}$, follows the optimistic scaling up to $\sqrt{s} = 10 \text{ TeV}$, after which it remains flat at 10 ab$^{-1}$ for all subsequent energies. These energy and luminosity benchmarks are enumerated in \cref{tab:energylumi}.

\begin{table}[t]
\renewcommand{\arraystretch}{1.5}
\setlength{\arrayrulewidth}{.3mm}
\setlength{\tabcolsep}{0.8 em}
\begin{center}
\begin{tabular}{c|c|c|c|c|c|c|c|c} 
$\sqrt{s}$ [TeV] & 1 & 3 & 6 & 10 & 14 & 30 & 50 & 100 \\ \hline \hline
$\mathcal{L}_{\rm int}^{\rm opt}$ [ab$^{-1}$] & 0.2 & 1 & 4 & 10 & 20 & 90 & 250 & 1000 \\ \hline
$\mathcal{L}_{\rm int}^{\rm con}$ [ab$^{-1}$] & 0.2 & 1 & 4 & 10 & 10 & 10 & 10 & 10 
\end{tabular}
\end{center}
\caption{Energy and luminosity benchmarks considered in this work.}
\label{tab:energylumi}
\end{table}%

For the most part, the studies presented here involve rate measurements and an accounting of simple irreducible backgrounds. More detailed projections are necessarily subject to a host of experimental considerations, and future developments in accelerator, detector, and theory studies for a muon collider are closely intertwined. Muons at rest have a relatively short lifetime of 2.2 $\mu$s and, while the push to high momentum beams can extend the lab frame lifetime up to the order of seconds, the exponential decay of the muon produces an intense source of collinear off-momentum electrons.
The electrons then interact with the beamline components, producing electromagnetic showers that result in a high flux of
low-energy photons and soft neutrons; these are the primary source of background for a muon collider detector. 
The process of bending and ultimately focusing the beams to generate a high luminosity collision rate directs these off-momentum
backgrounds into collimators upstream and also very close to the interaction point.
The exact share of these backgrounds depends strongly on the machine lattice and the interaction point configuration.  In all cases, the incorporation of shielding cones close to the interaction point has been identified as significant means of mitigating the effects of beam-induced background inside the detector~\cite{Bartosik_2020}. 

Detector studies performed in \cite{mokhov2012detector} show that the current approach to handle high detector backgrounds appears
adequate to preserve the physics capabilities. These studies need to be extended and updated to incorporate recent breakthroughs
in technology and higher center of mass energies. For backgrounds incident on the detector elements, the
primary tool for separating collider events from beam-induced background (BIB) is the new generation of precision timing detectors, which leverage the large investment of effort going into HL-LHC upgrades \cite{CMS_MTD,ATLAS_HGTD}. In the context of the present work, we take this as an encouraging indication that beam-induced backgrounds and associated reconstruction issues can be addressed, but emphasize that all of the studies herein represent estimates in need of detailed experimental study. 

As this study will articulate, there is abundant motivation to build a future muon collider.  The technological challenges do not appear to be insurmountable, and provide a wealth of opportunity to develop new experimental techniques.  The way the muon collider blurs the line between energy and precision opens the door to novel analysis approaches, while motivating new higher order calculations.  And such a machine could readily furnish answers to many of the fundamental questions in particle physics.

The muon-smasher's guide is organized as follows: In \cref{sec:compare}, we sketch many of the main qualitative features of collisions at high-energy muon colliders, with an eye towards their advantages over $pp$ colliders and the interplay between various production modes. In \cref{sec:PDF}, we turn to the physics of the initial state at high-energy muon colliders, characterizing the electroweak gauge boson content of high-energy muons and developing a pragmatic approach to capturing the most important effects. The broad physics case is developed in \cref{sec:physics}, focusing mainly on the central themes (electroweak symmetry breaking, dark matter, and naturalness) highlighted by the discovery of the Higgs. In \cref{sec:comp}, we explore the complementarity of a muon collider with other experiments operating on compatible timelines, with a particular focus on electric dipole moments, flavor violation, and gravitational waves. We summarize the central lessons of the study in \cref{sec:conc}, underlining the energies and luminosities that would position a muon collider to address the questions posed by the Higgs discovery. We reserve a compendium of cross sections and the details of various analyses for \cref{app:simplified}.


\section{Muons vs.~Protons}
\label{sec:compare}
High-energy muon colliders enjoy a host of advantages relative to their proton-proton counterparts, owing in part to the coexistence of scattering processes carrying nearly all of the collider energy (muon annihilation) with those carrying a smaller fraction (vector boson fusion). Before studying the physics potential of muon colliders in specific scenarios, we begin with a general exploration of the properties of muon annihilation and vector boson fusion, with an eye towards the comparison with proton-proton colliders.

\subsection{Muon annihilation}

The canonical class of scattering processes, familiar from lower-energy lepton colliders, is $\mu^+\mu^-$ annihilation. Well above the $Z$ pole, the cross section falls off as $1/s$.
On the one hand, this implies that the rate for producing both SM and BSM final states in $\mu^+ \mu^-$ annihilation falls rapidly with collider energy, although it is worth emphasizing that particles with electroweak quantum numbers still enjoy attobarn-level cross sections at energies as high as $\sqrt{s} = 100$~TeV. On the other hand, these cross sections are relatively insensitive to the mass of the final state particles unless $\sqrt{s}$ is close to the production threshold.  Clearly, muon colliders have considerable discovery potential as long as the final state is sufficiently distinctive.

\begin{figure}[h] 
   \centering
 \includegraphics[trim = 20 0 20 0, clip, width = 1.0\textwidth]{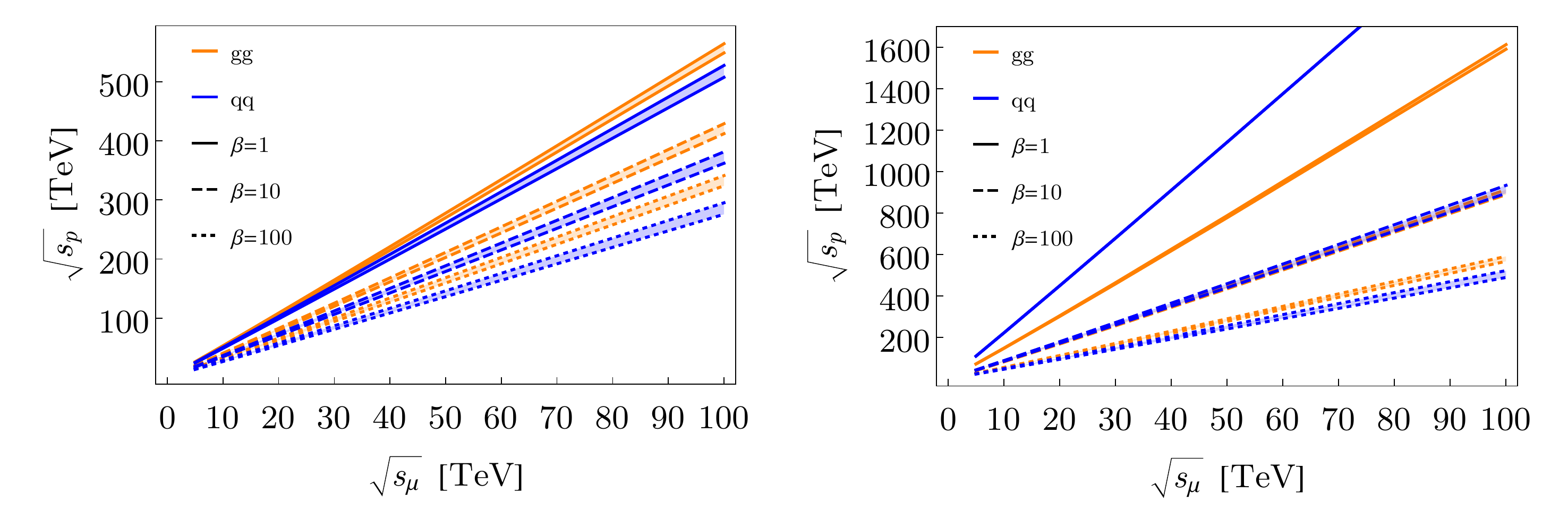} 
   \caption{The c.m.~energy $\sqrt{s_p}$ in TeV at a proton-proton collider versus $\sqrt{s_\mu}$ in TeV at a muon collider, which yield equivalent cross sections. Curves correspond to production via a $gg$ (orange) or $q \bar q$ (blue) initial state at the proton-proton collider, while production at the muon collider is determined by $\mu^+ \mu^-$.  The partonic cross sections are related by $\beta \equiv [\hat \sigma]_p / [\hat \sigma]_\mu$. The bands correspond to two different choices of proton PDF sets, \texttt{NNPDF3.0 LO} (as in \cite{Costantini:2020stv}) and \texttt{CT18NNLO}. The left (right) panel is for $2 \rightarrow 1$ ($2 \rightarrow 2$) scattering. }
   \label{fig:equivxsec}
\end{figure}

To quantitatively compare muon and proton colliders, we can compute the center-of-mass (c.m.) energies at which these two machines have equivalent cross sections~\cite{Delahaye:2019omf,Costantini:2020stv}. As we emphasize here, one of the great qualitative advantages of a muon collider over a $pp$ collider is that the former generates interactions across all values of the partonic c.m.~energy $\sqrt{\hat{s}}$, whereas the latter is dominated by $\sqrt{\hat{s}} \ll \sqrt{s}$ due to compositeness of the proton.  In this section, we estimate that it can require between $\mathcal{O}(1 \text{ - } 10)$ times more energy at a proton collider to achieve the equivalent production rate at a muon machine, see \cref{fig:equivxsec}.  Our focus is on annihilation processes, which are dominated by $x \sim 1$ at a muon collider, where $x$ is the momentum fraction carried by the muon. This is in contrast to  VBF-induced processes when the electroweak bosons radiated in the initial state become relevant, which typically have $x \ll 1$; we discuss qualitative features of VBF in this section, and defer a detailed study to~\cref{sec:PDF}.  The discussion in this section largely reprises the arguments given in~\cite{Costantini:2020stv}.

To make a concrete comparison, we work in terms of generalized parton luminosities.  We assume that the inclusive cross section for the final state $F$ (with unspecified remnants $X$) arising from collisions of (possibly composite) particles $A$ and $B$ takes the form
\begin{align}
\sigma(AB \rightarrow F + X) = \int_{\tau_0}^1 \dd \tau \sum_{ij} \frac{\dd \mathcal{L}_{ij}}{\dd \tau} \,\hat \sigma(ij \rightarrow F)\,,
\end{align}
where hats denote partonic quantities, $\tau = \hat s / s$ in terms of the collider c.m.~energy $\sqrt{s}$ of the collider and partonic energy $\sqrt{\hat s}$, $\tau_0$ is the production threshold, and the parton luminosity is given by
\begin{align}
\frac{\dd \mathcal{L}_{ij}}{\dd \tau} (\tau,\mu_f ) = \frac{1}{1 + \delta_{ij}} \int_\tau^1 \frac{\dd x}{x} \big[f_i(x,\mu_f) f_j(\tau/x,\mu_f) + (i \leftrightarrow j) \big]\,.
\end{align}
Here the $f_i(x,\mu_f)$ are the parton distribution functions (PDFs) for parton $i$ carrying a fraction $x$ of the longitudinal momentum, at factorization scale $\mu_f$, which we take to be $\mu_f = \sqrt{\hat s}/2$ when making~\cref{fig:equivxsec}.

First, we assume that the process results from a $\bm{2 \to 1}$  \textbf{collision}, i.e., $AB \rightarrow Y$ for a final state $Y$ with mass $M = \sqrt{\hat s}$.  In this case, the cross section $\sigma_p$ at a proton-proton collider whose c.m.~energy is $\sqrt{s_p}$ takes the form
\begin{align}
\sigma_p(2 \rightarrow 1) = \int_{\tau_0}^1 \dd \tau \sum_{ij} \frac{\dd \mathcal{L}_{ij}}{\dd \tau} [\hat \sigma_{ij}]_p \,\delta \bigg( \tau - \frac{M^2}{s_p} \bigg)\,.
\end{align}
At a muon collider whose c.m.~energy is $\sqrt{s_\mu}$, the analogous production is dominated at threshold with $s_\mu = \hat s = M^2$ (the $\delta$-function from the phase space measure is absorbed by the narrow width); the cross section may be simply approximated by the partonic one, $\sigma_\mu = [\hat \sigma]_\mu$. We then solve for relation between $s_p$ and $s_\mu$ that yields equivalent cross sections: 
\begin{align}
\sigma_p = \sigma_\mu \qquad \Longrightarrow \qquad \frac{[\hat \sigma]_p}{[\hat \sigma]_\mu} \, \sum_{ij} \frac{\dd \mathcal{L}_{ij}}{\dd \tau} \bigg( \frac{s_\mu}{s_p},\frac{\sqrt{s_\mu}}{2}\, \bigg) \simeq 1 \, ,
\end{align}
where we are making the simplifying assumption that the partonic cross section is universal, $[\hat \sigma_{ij}]_p \simeq [\hat \sigma]_p$. This equation can be solved numerically for $s_p$ in terms of $s_\mu$ for different assumptions about the relation between partonic cross sections:
\begin{align}
\beta \equiv \frac{ [\hat \sigma]_p}{ [\hat \sigma]_\mu}\,.
\end{align}
For example, $\beta \simeq 10$ is reasonable for a situation where a state is produced via QCD (electroweak) processes at the proton (muon) collider.

For $\bm{2 \rightarrow 2}$ \textbf{collisions}, we assume the muon collider is optimized so that $\sqrt{s_\mu}$ is slightly above threshold, while at a proton-proton collider we take $[\hat \sigma]_p \propto 1/\hat s$, which is appropriate far above threshold. Then for the proton-proton case we can write
\begin{align}
\sigma_p(2 \rightarrow 2)= \frac{1}{s_p} \int_{\tau_0}^1 \dd \tau\, \frac{1}{\tau} \sum_{ij} \frac{\dd \mathcal{L}_{ij}}{\dd \tau} [\hat \sigma_{ij}  \hat s]_p\,,
\end{align}
while for the muon collider we have $\sigma_\mu = [\hat \sigma \hat s]_\mu / s_\mu$. In this case, 
\begin{align}
\sigma_p = \sigma_\mu \qquad \Longrightarrow \qquad \frac{s_\mu}{s_p} \frac{[\hat \sigma \hat s]_p}{[\hat \sigma \hat s]_\mu} \int_{s_\mu/s_p}^1 \dd \tau \, \frac{1}{\tau} \frac{\dd \mathcal{L}_{ij}}{\dd \tau} \bigg(\tau,\frac{\sqrt{s_\mu}}{2}\, \bigg) \simeq 1 \,,
\end{align}
which can again be solved numerically for $s_p$ given various assumptions about the ratio of partonic cross sections.

The results of this exercise are shown in Fig.~\ref{fig:equivxsec}. For both $2 \rightarrow 1$ and $2 \rightarrow 2$ processes, the equivalent energy relationship is essentially linear, with only modest dependence on the choice of proton PDFs. A muon collider enjoys considerable advantages in $2 \rightarrow 1$ production, reaching the cross section of a 100 TeV $pp$ collider by $\sqrt{s_\mu} \sim 20$ TeV assuming comparable partonic cross sections. The advantage is even sharper in $2 \rightarrow 2$ production, where a muon collider reaches the cross section of a 100 TeV $pp$ collider between $\sqrt{s_\mu} \sim 5-7$ TeV for comparable partonic cross sections, depending on whether the process is $q \bar q$- or $gg$-initiated. Even allowing for an enhancement of $\beta = 10$ at a $pp$ collider (accounting for the difference between QCD and electroweak partonic cross sections), the $2 \rightarrow 2$ cross section of a 100 TeV $pp$ collider is emulated by a muon collider operating at $\sqrt{s_\mu} \sim 12$ TeV.

Of course, this comparison is necessarily favorable to muon colliders in the sense that it assumes the mass scale of new physics lies just below the c.m.~energy of the collider, although radiative return can salvage some of the same conclusions at the cost of further suppressing the partonic cross section. Developing a comprehensive case for muon colliders still requires investigating production cross sections for a variety of new physics scenarios across a range of c.m.~energies that go well above the production threshold.  That is the motivation for the model dependent studies provided in Secs.~\ref{sec:physics} and \ref{sec:comp} below.

\subsection{Vector boson fusion}

For collisions well above the production threshold, the virtual electroweak gauge boson content of high-energy muon beams becomes increasingly relevant, akin to the virtual gluon content of high-energy proton beams. VBF becomes a correspondingly important channel for the production of SM and BSM particles alike, with cross sections typically scaling with c.m.~energy as $\sim \log(s)$ far above threshold. The practical aspects of VBF at high-energy muon colliders have recently been studied in detail in \cite{Costantini:2020stv}, while the treatment of VBF in terms of electroweak parton distribution functions (PDFs) has been initiated in \cite{Han:2020uid}. Here we summarize the qualitative features of VBF production, reserving a detailed discussion of electroweak PDFs for \cref{sec:PDF}.

\begin{figure}[h] 
   \centering
\includegraphics[trim = 20 0 20 0,width = 0.95\textwidth]{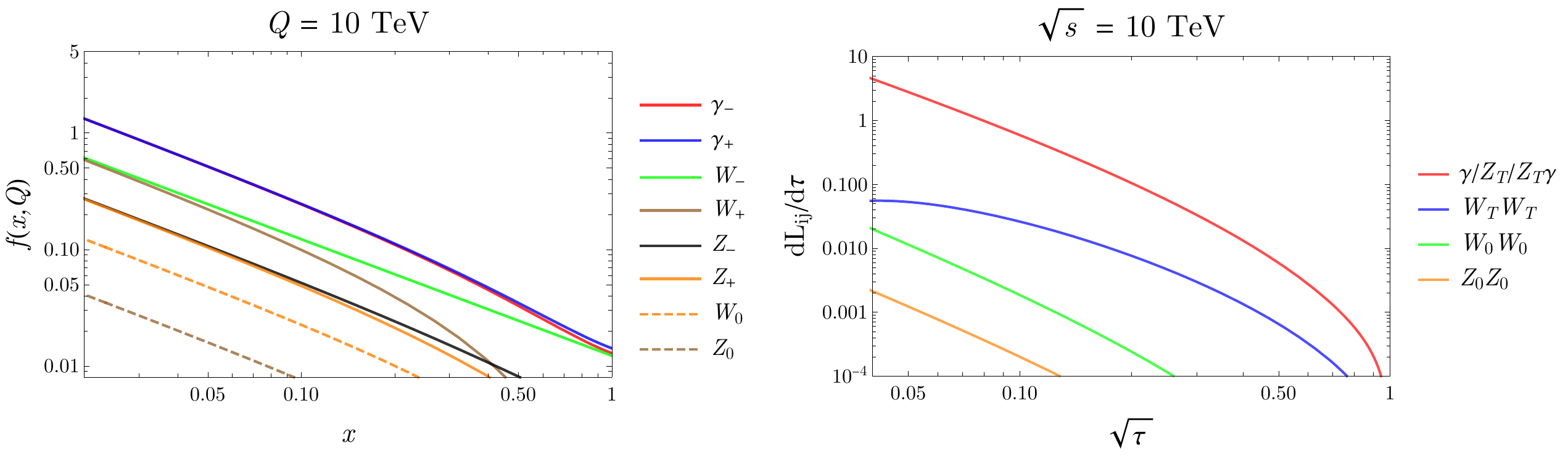}
   \caption{Left: the polarized PDF $f_i(x,Q)$ for electroweak gauge bosons separated by transverse and longitudinal helicities $\pm, 0$ as a function of partonic momentum fraction $x$ at a scale $Q = 10$ TeV. Right: the parton luminosity functions $\dd \mathcal{L}_{ij} / \dd \tau ( \tau, \mu_f)$ as a function of $\sqrt{\tau} = \sqrt{\hat s / s}$ for $Q = \sqrt{\tau s} / 2$ at $\sqrt{s} = 10$ TeV.}
   \label{fig:pdfs}
\end{figure}

For the sake of illustration, representative electroweak PDFs and parton luminosities are shown in \cref{fig:pdfs}. The PDFs $f_i(x,Q)$ are shown as a function of the partonic momentum fraction $x$ at the scale $Q = 10$ TeV for transverse polarizations of the photon and both transverse and longitudinal polarizations of the $W$ and $Z$, while the parton luminosity functions $\dd \mathcal{L}_{ij} / \dd \tau(\tau, \mu_f)$ are shown as a function of $\sqrt{\tau} = \sqrt{\hat s / s}$ with factorization scale $\mu_f = \sqrt{\tau s}/2$ for $\sqrt{s} = 10$ TeV. Details of their derivation and scale dependence are presented in \cref{sec:PDF}.

These distributions illuminate many of the salient features of VBF production, modulo additional dependence on the partonic cross section for the process of interest. All of the electroweak boson PDFs peak at $x \sim 0$, a manifestation of the inherent infrared singularity. Photons constitute by far the largest component away from $x \sim 1$, as their logarithmic enhancement due to soft emission extends all the way down to $m_\mu$, compared to the $W$ and $Z$ who are only non-zero for energies beginning at $\sim m_{W,Z}$. The relative size of $Z$ PDFs relative to $W$ PDFs reflects the familiar suppression of the $Z$ coupling to muons. Longitudinal polarizations of the $W$ and $Z$ are suppressed relative to transverse polarizations due to the former's modest coupling to muons, which is set by the muon mass. 

Although these distributions provide a good qualitative sense of the various contributions to VBF processes, both the convolution of the parton luminosities with the partonic cross section and the imposition of realistic phase space cuts significantly affect the properties of VBF production cross sections. The sizable photon PDF reflects the abundance of soft photons at low scales, which do not contribute significantly to the production of particles with sizable transverse momentum. Even moderate phase space requirements on the final state, such as transverse momentum cuts, reduce the relative logarithmic enhancement enjoyed by the photon. Features of the partonic cross section for a given process can have a significant impact, most notably longitudinal enhancement. As a result, the dominant contribution to VBF production of particles carrying $SU(2)_L$ quantum numbers is often $WW$ fusion, rather than $\gamma \gamma$ fusion, a conclusion borne out in the numerical results of both \cite{Costantini:2020stv} and this study. For partonic processes with enhanced contributions from longitudinal polarizations, such as $W^+ W^- \rightarrow t \bar t$ or $W^+ W^- \rightarrow h$, this enhancement is often sufficient to overcome the relative suppression of the PDFs; this will be illustrated in a number of examples in \cref{sec:EWSB}. As such, high-energy muon colliders are as much longitudinal gauge boson colliders as they are transverse gauge boson colliders, subdominant PDFs notwithstanding. 

\subsection{Annihilation vs.~VBF}

Ultimately, there is a rich interplay between annihilation and VBF production of both SM and BSM particles at high-energy muon colliders.
For Standard Model processes well above threshold, the relative scaling as a function of collider energy $\sqrt{s}$ is \cite{Costantini:2020stv}
\begin{align}
\frac{\sigma_{\rm VBF}^{\rm SM}}{\sigma_{\rm ann}^{\rm SM}} \propto \alpha_W^2 \frac{s}{m_V^2} \log^3 \frac{s}{m_V^2}\,,
\end{align}
where the triple logarithmic enhancement is due to a double collinear logarithm from the two electroweak PDFs and a single soft logarithm. The competition between the coupling suppression and energy growth leads to crossovers between Standard Model cross sections for annihilation and VBF production around energies of $\sqrt{s} \sim$ few TeV, with correspondingly higher crossover energies for higher-multiplicity final states. The scaling is analogous for production of BSM particles with a final-state mass scale $m_X$, for which the relative scaling well above threshold is \cite{Costantini:2020stv}
\begin{align}
\frac{\sigma_{\rm VBF}^{\rm BSM}}{\sigma_{\rm ann}^{\rm BSM}} \propto \alpha_W^2 \frac{s}{m_X^2} \log^2 \frac{s}{m_{V}^2} \log \frac{s}{m_X^2}\,,
\label{eq:BSMVBF}
\end{align}
where $m_V$ is the mass scale of an intermediate state in the production process (often an electroweak vector boson, which in any case is assumed to satisfy $m_V \ll \sqrt{s}$). The collision energy $\sqrt{s}$ at which annihilation and VBF cross sections for BSM final states cross over grows with the mass scale of the final state, but ultimately there are always collision energies at which VBF production wins for a fixed mass scale. This lends credence to the notion of high-energy muon colliders as gauge boson colliders, and highlights the importance of analyzing VBF production modes in characterizing the physics reach of these colliders.

\begin{figure}[h] 
   \centering
 \includegraphics[trim = 20 0 20 0,width = 0.95\textwidth]{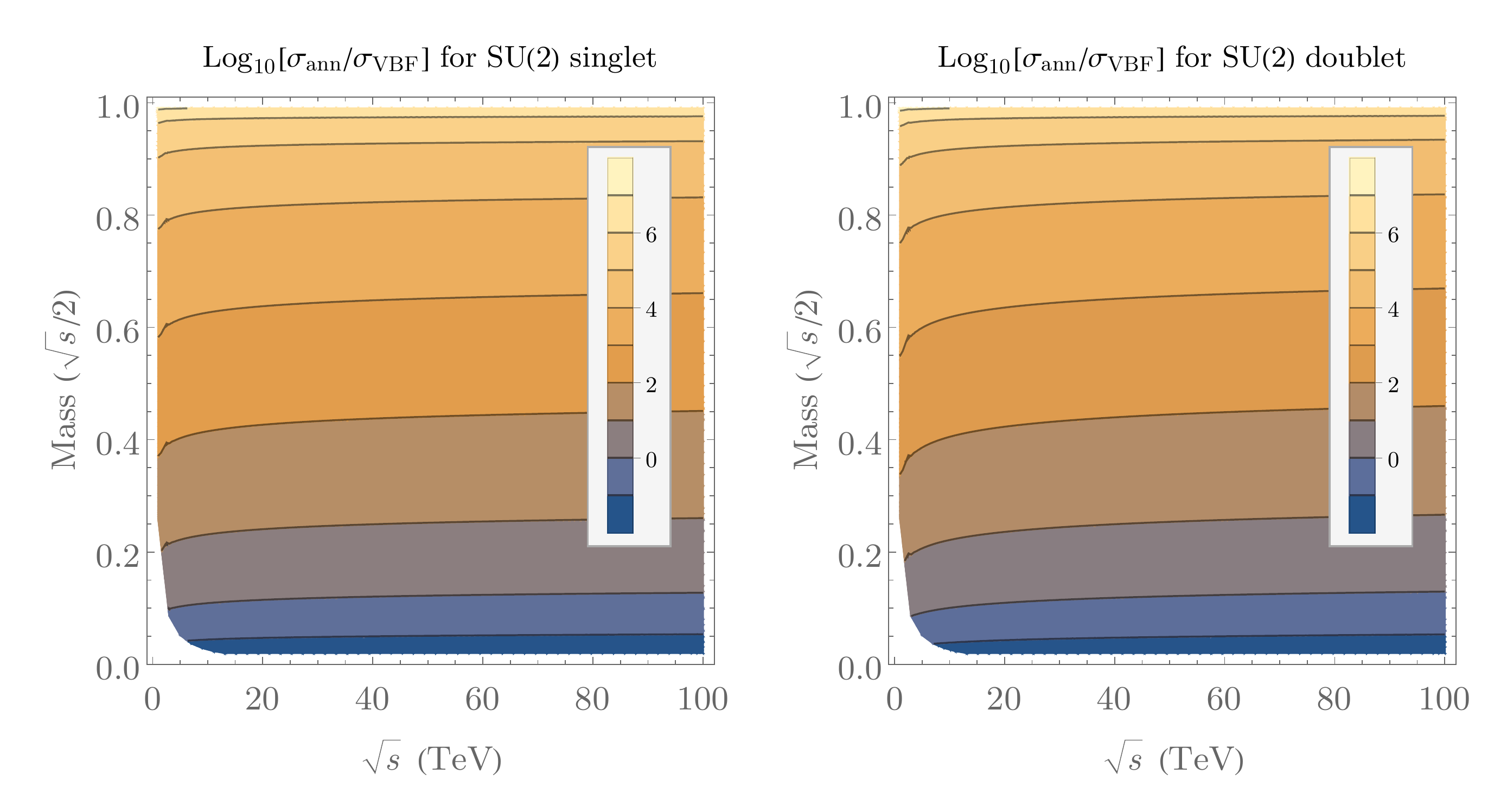} 
   \caption{Left: log ratio of the annihilation cross section $\sigma_{\rm ann}$ and VBF cross section $\sigma_{VBF}$ for a vector-like pair of $SU(2)$ singlet Weyl fermions with hypercharge $\pm 1$ as a function of the collider energy $\sqrt{s}$ and the particle mass relative to threshold. Right: the same ratio for a vector-like pair of $SU(2)$ doublet Weyl fermions with hypercharge $\pm 1/2$.}
   \label{fig:annvvbf}
\end{figure}

Even so, it bears emphasizing that the above scaling assumes muon collisions are occurring well above production threshold. This is likely to be true for most Standard Model processes at a high-energy muon collider, but need not hold for BSM production. Although the underlying approximations break down as $m_X \rightarrow \sqrt{s}$, the naive scaling in~\cref{eq:BSMVBF} indicates that annihilation production once again dominates in this limit. Despite the overall $1/s$ falloff in annihilation cross sections, even a handful of events with sufficiently distinctive final states near threshold may be sufficient for the discovery of new physics.  We caution that projections based on VBF production modes alone would fail to capture these important cases. 

The interplay between annihilation and VBF production is illustrated in \cref{fig:annvvbf}, which shows the ratio of the annihilation cross section $\sigma_{\rm ann}$ and VBF cross section $\sigma_\text{VBF}$ for two representative examples -- a vector-like pair of $SU(2)$ singlet Weyl fermions with hypercharge $\pm 1 $, and a vector-like pair of $SU(2)$ doublet Weyl fermions with hypercharge $\pm 1/2$ -- as a function of the collider energy $\sqrt{s}$ and the particle mass relative to threshold. In both cases, the annihilation cross section is computed analytically while the VBF cross section is computed by convolving partonic cross sections with the corresponding PDFs derived in \cref{sec:PDF}. For both the $SU(2)$ singlet and doublet, the crossover takes place once the fermion mass is above about $10\%$ of $\sqrt{s}/2$. Ultimately the differences in the two cases are modest; although the $WW$ fusion contribution to VBF is much larger for the doublet, the dominant contribution in both cases is ultimately from $\gamma \gamma$ fusion. For sufficiently distinctive final states, this is likely to favor production via annihilation as a discovery mode. Of course, the details depend on the relative sizes of signal and background, another aspect where muon colliders enjoy further advantages over their proton-proton counterparts, as we will now emphasize. 

\subsection{Signal vs.~background}

A final generic advantage of a muon collider over a $pp$ collider that we want to highlight has to do with the comparison of signal and background rates.  Furthermore, this benefit is not restricted to production cross sections for high-mass states. This is not obvious at first glance; for low-mass states, including Standard Model particles, the rate advantage of $pp$ colliders is considerable. For example, at $\sqrt{s} = 14$ TeV, the leading single Higgs production cross section is a factor of $\sim 50$ larger at a $pp$ machine than its $\mu^+ \mu^-$ counterpart. But ultimately, our ability to extract physics from the collider data sensitivity depends on the background rates, and here the advantage is decisively in favor of muon colliders.

\begin{figure}[htbp] 
   \centering
\includegraphics[trim = 20 0 20 0, width = 0.6\textwidth]{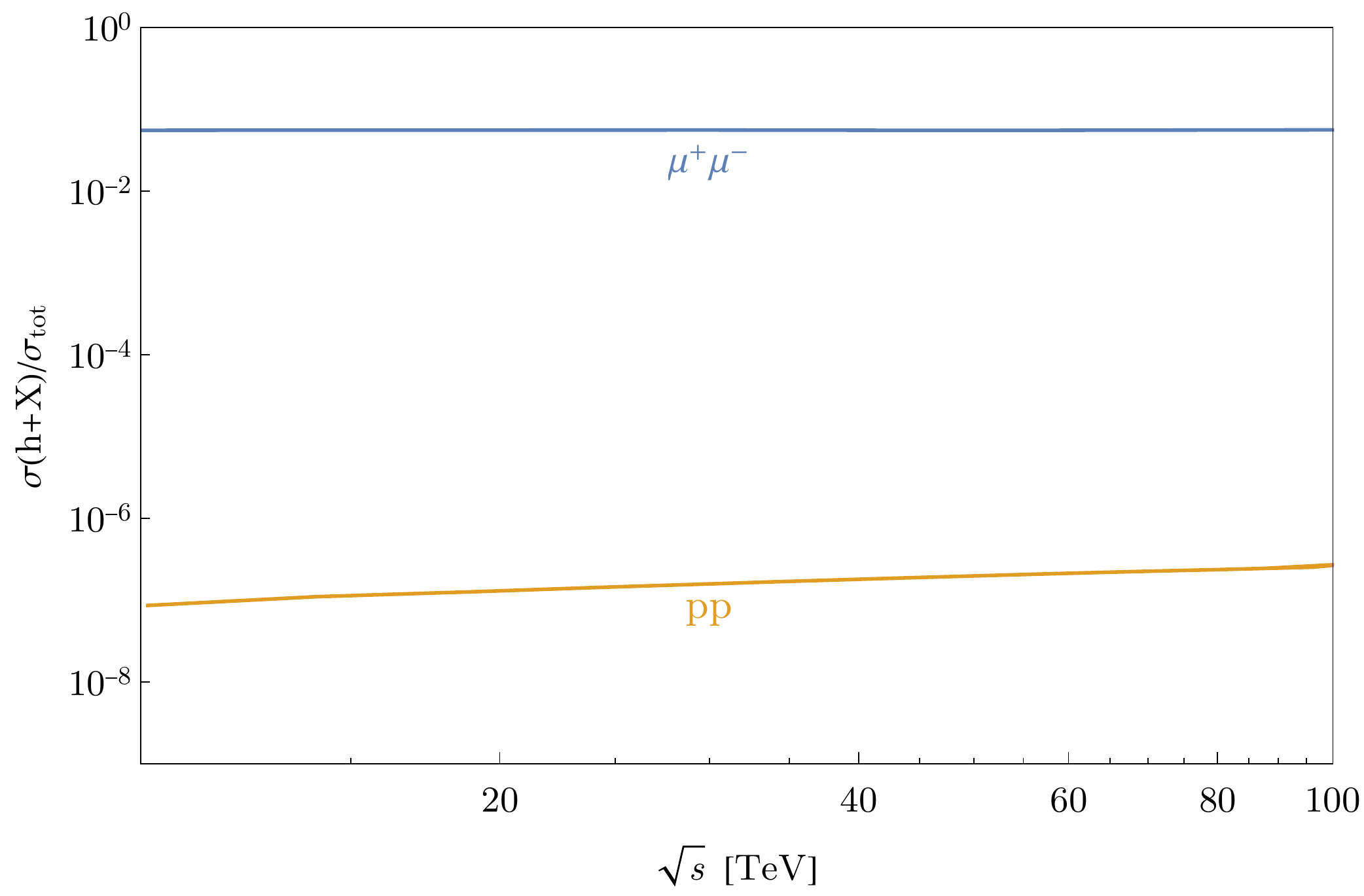} 
   \caption{Higgs production cross section $\sigma(h+X)$ as a fraction of a representative ``total'' cross section $\sigma_{\rm tot}$ for $\mu^+ \mu^-$ and $pp$ colliders. For $\mu^+ \mu^-$ colliders, we compute Higgs production using the LO cross section for $\mu^+ \mu^- \rightarrow h + \nu \bar \nu$, while the ``total'' cross section $\sigma_{\rm tot}$ is taken to be the rate for single electroweak boson production, which is dominated by VBF production of $W,Z,h,\gamma$ at these energies. For $pp$ colliders we take the Higgs production cross section to be the N3LO cross section for $gg \rightarrow h$ \cite{Anastasiou:2016cez} presented in \cite{deFlorian:2016spz}, while the ``total'' cross section $\sigma_{\rm tot}$ is taken to be the $pp \rightarrow b\bar b$ cross section computed by MCFM \cite{Boughezal:2016wmq}.}
   \label{fig:ppcompare}
\end{figure}

 As illustrated in \cref{fig:ppcompare}, the ratio between the single Higgs production cross section and a representative ``total'' cross section at $pp$ and $\mu^+ \mu^-$ colliders operating at the same $\sqrt{s}$ differs by roughly six orders of magnitude and is relatively insensitive to $\sqrt{s}$. Here we have taken the representative ``total'' cross section at a $pp$ collider to be the $pp \rightarrow b \bar b$ cross section, a relevant measure of Standard Model backgrounds; the total or inelastic $pp$ cross sections are orders of magnitude larger. The corresponding ``total'' cross section at a $\mu^+ \mu^-$ collider is taken to be the sum of inclusive single electroweak boson cross sections. Assuming it is possible to achieve comparable integrated luminosities at both experiments, it is clear that a precision Higgs program at a future muon collider provides many opportunities to study the Higgs sector in exquisite detail.   Of course, these statements ultimately rest on details of detector performance and the ability to mitigate the beam induced muon collider backgrounds.  Nonetheless, we see this as a clear sign that the muon collider will be an effective tool to probe both the precision and energy frontiers.


\section{Muon Colliders Are Gauge Boson Colliders}
\label{sec:PDF}
Before getting into the detailed physics case, this section will describe the physics of the initial state at a high energy muon collider.
Naively, the advantage of a lepton collider is that the colliding beams are composed of elementary particles (so that the collision is relatively clean), which are in momentum eigenstates (so that the c.m.~energy for each collision is known).
This can be contrasted against proton colliders, where the beams are composed of composite states, so that the partonic c.m.~energy varies from collision to collision.
To make predictions in this case, one convolves the hard process of interest with universal PDFs.
Additionally, the smashed protons leave a trail of debris in their wake, the so-called underlying event.
As we will argue in this section, making predictions for a muon collider whose beams carry TeVs of energy has aspects in common with both better known types of machines: one must use PDFs, but the collision yields a small number of particles in the initial state that can be modeled reasonably well using perturbation theory.\footnote{Throughout this paper, we treat the muon beams as stable.  Everything we say here is independent of this assumption, as long as our amazing accelerator colleagues can figure out how to provide us with a robust muon beam to play with.}

At the theoretical level, the situation for a muon collider is simplified with respect to a proton collider since perturbative control can be maintained at every step of the calculation.\footnote{Of course, some of these techniques are also relevant for past and proposed electron-position experiments, e.g.,~when predicting VBF initiated processes.
However, the small mass of the electron effectively bounds the maximum energy for circular machines to be near the electroweak scale.}
For example, the boundary conditions for the proton PDFs are set at a scale where QCD is non-perturbative, implying that one must rely on inputs from experiment to numerically determine the proton PDFs.  
All of the complications that stem from this fact are avoided when studying muon PDFs.
The muon colliders we discuss here have energy in the TeV to 100 TeV range, and so the masses of the weak gauge bosons can be treated as a small perturbation, i.e., it is typically reasonable to treat them as massless so that the PDF formalism applies; see~\cref{sec:FiniteMassPDF} for a brief discussion of finite mass effects.
And since the electroweak gauge couplings are relatively small, working with leading order unresummed PDFs provides a reasonable approximation to the resummed result; we will demonstrate the minimal impact of next-to-leading-log corrections in \cref{sec:SubLeadLogs} below.
Interesting complications arise due to electroweak symmetry breaking, but other than treating the mass versus gauge eigenbasis for the electroweak bosons consistently as we do below, these tend to have a small numerical effect on the cross section predictions.
There are additionally subtleties associated with capturing the physics of the longitudinal gauge boson modes, and the interplay with the Goldstone equivalence theorem and unitarity; we will not comment on this further and will simply use the splitting functions in the ``Goldstone Equivalence Gauge'' computed in~\cite{Chen:2016wkt}.
Finally, while it is beyond the scope of this work, we note that one can also include the effects of QCD into the muon PDFs, as was recently described in~\cite{Han:2021kes}.

In the rest of this section, we will first write down the formalism used to solve for the PDFs to leading logarithmic order using leading order splitting functions.   
This will provide us with a framework to explore the accuracy that can be achieved when taking different approximations.
Our goal will be to demonstrate that the leading log (unresummed) PDFs provide a reasonable approximation to the more complete all log order results that result from integrating the DGLAP evolution equations.
Given that the leading log PDFs are easy to understand and can be expressed analytically, we advocate that these are all that are required to make predictions for a future muon collider in the energy range of interest here, unless high precision calculations are needed.

\subsection{From the effective vector approximation to PDFs}
\label{sec:EVA}
The soft and collinear divergences inherent to theories of charged particles coupled to gauge bosons yield physical logarithmic enhancements that can spoil the convergence of perturbation theory.
When considering colliding beams of charged particles, it is important to acknowledge our inherent inability to experimentally distinguish a single state in isolation from one that has emitted a nearly collinear or very soft additional particle.
A framework for addressing this problem was first written down in 1934 by Weizsaecker~\cite{vonWeizsacker:1934nji} and Williams~\cite{Williams:1934ad}; this is what is known as the ``effective photon approximation'' or more generally the ``effective vector approximation'' (EVA):
\begin{equation}
f_\gamma^\text{EVA} (x) \simeq \dfrac{\alpha}{2\pi} P_\gamma (x) \log \dfrac{E^2}{m_\mu^2}\, ,
\label{eq:EPA}
\end{equation}
where the log is the result of soft emissions, and the QED splitting function 
\begin{align}
P_{\gamma}(x) = \frac{1+(1-x)^2}{x} \, 
\label{eq:Pgamma}
\end{align}
can be derived by taking the collinear limit of a tree-level $1\to 2$ process computed using perturbative QED.
Note that in this approximation, $E$ is the beam energy of the colliding charged particle while the emitted photon energy is given by $Q = xE$, where $x$ is the momentum fraction carried by the photon.

Systematically improving this approximation requires developing the relevant DGLAP evolution equations, which allow one to resum the log that appears in \cref{eq:EPA}, yielding the PDFs.
It is perhaps under appreciated that the EVA and PDF approaches already differ at leading log order: the EVA is proportional to $\log E^2$, while the leading log PDF is proportional to $\log Q^2$, where $Q^2$ is an unphysical renormalization scale, whose canonical value $Q^2 \sim (x E)^2$ is typically chosen to minimize higher order logarithms.
With this choice of scale, the unresummed leading log order PDFs are
\begin{equation}
f_\gamma^\text{PDF, LL} (x) \simeq \dfrac{\alpha}{2\pi} P_\gamma (x) \log \dfrac{(x E)^2}{m_\mu^2}\, ,
\label{eq:PDFLLGamma}
\end{equation}
where the splitting function $P_\gamma(x)$ is still given by~\cref{eq:Pgamma}, and we are taking the gauge coupling to be fixed for simplicity.
This is of course the logarithmic behavior one would find when computing at fixed order in perturbation theory, and additionally it follows from solving the DGLAP evolution equations for QED given in~\cref{eq:DGLAP4QED} to leading log order, as it must for self consistency.
As we will see in what follows, simply solving for the leading log approximation given in~\cref{eq:PDFLLGamma} for the full system of partons relevant at a muon collider provides a good approximation to the full solution to the DGLAP equations.
Note that since it is trivial to implement, we do allow the gauge coupling to run (at one-loop order) when computing the leading log PDFs that are used to make some of the cross section predictions below.

\subsection{PDFs with broken electroweak symmetry}
Naively, one might expect that the PDFs for the massive electroweak gauge bosons can be derived by simply using the appropriate splitting functions, and replacing $m_\mu \to m_V$ inside the logarithm.
This is the case for the $W^\pm$ bosons, where the PDFs are given by\footnote{We do note that there are some subtle questions about how to treat the longitudinal components, such that the Goldstone equivalence theorem is respected.
However, this issue only appears at subleading order so we do not have to treat it carefully here; see~\cite{Chen:2016wkt} for a discussion.}
\begin{align}
f_{W^-_T}^\text{EVA} (x) \simeq \dfrac{\alpha_2}{4\pi} P_{W^-_T \leftarrow \mu_L} (x) \log \dfrac{E^2}{m_W^2}\, ,
\end{align}
with splitting functions
\begin{align}
P_{W^-_T \leftarrow \mu_L} (x) = \dfrac{1+(1-x)^2}{x} \, ,
\end{align}
which captures the splitting to $W^-$ summed over both polarizations, assuming that the incoming muon beam has equal left-handed and right-handed helicity.
Note that one must be careful to keep track of the helicity dependence, since the $W^\pm$ couplings are chiral.

The computation of the $Z$-boson PDF is complicated by the fact that it mixes with the photon.
Hence, one must first evolve the photon PDF from the scale $Q^2 = m_\mu^2 \to m_Z^2$, where the evolution equations change.
Noting that the electromagnetic interactions conserve both C and P, we do not need to track the difference in helicities for this step of the calculation.
The DGLAP equations for the photon and muon are
\begin{equation}
\begin{split}
\dfrac{\dd}{\dd \log Q^2} 
\begin{pmatrix}
f_{\gamma}\big(x,Q^2\big)\\[3pt]  f_{\mu}\big(x,Q^2\big)
\end{pmatrix} = \begin{pmatrix}
   \mathcal{P}_{\gamma  \leftarrow \gamma}(x) &\mathcal{P}_{\gamma  \leftarrow \mu}(x) \\[3pt]
  \mathcal{P}_{\mu  \leftarrow \gamma}(x) &   \mathcal{P}_{\mu  \leftarrow \mu}(x)
\end{pmatrix} \otimes  \begin{pmatrix}
f_{\gamma}\big(x,Q^2\big)\\[3pt]  f_{\mu}\big(x,Q^2\big) 
\end{pmatrix}\, ,
\end{split}
\label{eq:DGLAP4QED}
\end{equation}
where the convolution is defined in the standard way:
\begin{equation}
f(x) \otimes g(x) = \int_x^1 \dfrac{\dd z}{z} f\left( z \right) g\left( \dfrac{x}{z} \right)\,.
\end{equation}
The explicit splitting functions are~\cite{Chen:2016wkt}
\begin{subequations}
\begin{align}
\mathcal{P}_{\gamma \leftarrow \mu}(x) &= \dfrac{e^2}{8\pi^2} \dfrac{1 + (1-x)^2}{x} \\[3pt]
\mathcal{P}_{\gamma  \leftarrow \gamma}(x) &= -\dfrac{2}{3} N_\gamma \delta(1-x) \\[3pt]
\mathcal{P}_{\mu  \leftarrow \mu}(x) &= \dfrac{e^2}{8\pi^2} \left(  \dfrac{1 + x^2}{(1-x)_+} + \dfrac{3}{2} \delta(1-x)\right) \, ,
\end{align}
\end{subequations}
where $N_\gamma$ counts the number of ways the photon can annihilate to quark and lepton pairs, and the plus function is defined in the standard way:
\begin{equation}
\int_x^1 \dd z \dfrac{f(z)}{(1-z)_+} = \int_x^1 \dd z \dfrac{f(z) - f(1)}{(1-z)} \, .
\end{equation}
The boundary conditions are
\begin{align}
f_\gamma\big(x, m_\mu^2\big) = 0 \qquad \text{and} \qquad f_\mu\big(x, m_\mu^2\big) = \delta(1-x)\,.
\end{align}

We solve these equations for $f_\gamma$ and $f_\mu$; evaluating them at $Q^2 = m_Z^2$ provides the boundary conditions for the DGLAP evolution equations that include the $Z$ boson.
At this step, we must keep track of the different helicity dependence, so we assign half of the muon PDF to each helicity and then let them evolve independently.
It is also critical to account for the difference between the gauge and mass eigenbases, since the interactions are diagonal in the former while physical processes are computed using the later.
To this end, we need convert from $(\gamma, Z , Z\gamma)$ to $(B, W_3, BW_3)$ using the transformation matrix
\begin{equation}
\begin{pmatrix}
B\\  W_3\\  BW_3\\
\end{pmatrix} = \begin{pmatrix}
\cos^2 \theta_W&  \sin^2\theta_W&  -\cos\theta_W \sin\theta_W \\
\sin^2 \theta_W&  \cos^2\theta_W&  \cos\theta_W \sin\theta_W \\
2\cos\theta_W \sin\theta_W&  -2\cos\theta_W \sin\theta_W&  \cos^2\theta_W - \sin^2\theta_W \\
\end{pmatrix}
\begin{pmatrix}
 \gamma\\ Z\\   Z\gamma
\end{pmatrix}\,,
\label{eq:weakRotationMatrix}
\end{equation}
where $\theta_W$ is the weak mixing angle, and the mixed $Z\gamma$ PDF accounts for possible interference effects among diagrams involving a $Z + \gamma$ initial state. For reference, the parton luminosity for $Z$ and $\gamma$ initial states is derived using the combination
\begin{equation}
\dfrac{\dd L}{\dd \tau} = f_Z(x,Q) \otimes f_\gamma(x,Q) + f_{Z\gamma}(x,Q) \otimes f_{Z\gamma}(x,Q)\,.
\end{equation}

The full DGLAP evolution equations are given in~\cite{Chen:2016wkt}.
To provide an example, the DGLAP evolution equation for $f_{B_-}$ is
%
\begin{equation}
\begin{split}
\dfrac{\dd}{\dd\log Q^2} f_{B_-}\big(x,Q^2\big) &=  \mathcal{P}_{B_- \leftarrow \mu_L}(x) \otimes f_{\mu_L}\big(x,Q^2\big) + \mathcal{P}_{B_- \leftarrow \mu_R}(x) \otimes f_{\mu_R}\big(x,Q^2\big) \\
& + \mathcal{P}_{B_- \leftarrow \nu_L}(x) \otimes f_{\nu_L}\big(x,Q^2\big)  +  \mathcal{P}_{B_- \leftarrow B_-}(x) \otimes f_{B_-}\big(x,Q^2\big) \, ,
\end{split}
\end{equation}
where the splitting functions are given by\footnote{Note that the splitting function $ \mathcal{P}_{B_- \leftarrow B_-}(x)$ is determined by the decay rate of $B_-$. 
This expression includes the decay channel to a pair of massless top quarks, which is no longer a good approximation for small $x$.  
We have checked that this contribution to the $B$ PDF is small, so for our purposes here we will simply use~\cref{eq:BLSplittingFunctions}.
}
\begin{subequations}
\begin{align}
\mathcal{P}_{B_- \leftarrow \mu_L}(x) &= \mathcal{P}_{B_- \leftarrow \nu_L}(x)  = \dfrac{1}{8\pi^2} \left( \dfrac{-g_1}{2} \right)^2 \dfrac{1}{x} \label{eq:PBLMuL}\\[3pt]
\mathcal{P}_{B_- \leftarrow \mu_R}(x)  &= \dfrac{1}{8\pi^2} \left( \dfrac{-g_1}{2} \right)^2 \dfrac{(1-x)^2}{x} \label{eq:PBLMuR} \\[3pt]
\mathcal{P}_{B_- \leftarrow B_-}(x) &= - \dfrac{g_1^2}{8\pi^2} \dfrac{7}{2} \delta(1-x) \, .
\end{align}%
\label{eq:BLSplittingFunctions}
\end{subequations}
%
%

Although we do not write down the DGLAP evolution equation for $f_{B_+}$ explicitly, there is a feature of the splitting functions that is worth noting.
Due to CP invariance, the splitting function  $\mathcal{P}_{B_+\leftarrow\mu_L} = \mathcal{P}_{B_-\leftarrow \mu_R}$, and so we use \cref{eq:PBLMuR} for both.  
Comparing \cref{eq:PBLMuL} with \cref{eq:PBLMuR}, we see that the splitting functions for left-handed muon decaying to left-handed and right-handed gauge boson become equal in the $x\to 0$ limit. 
For contrast, at high $x$ the latter approaches zero. 
This is due to the conservation of angular momentum. 
At high $x$, the probability to reverse the helicity goes to zero, which is one key reasons it is important to use the polarized PDFs for electroweak interactions.

\begin{figure}[t]
\centering
\includegraphics[width=14cm]{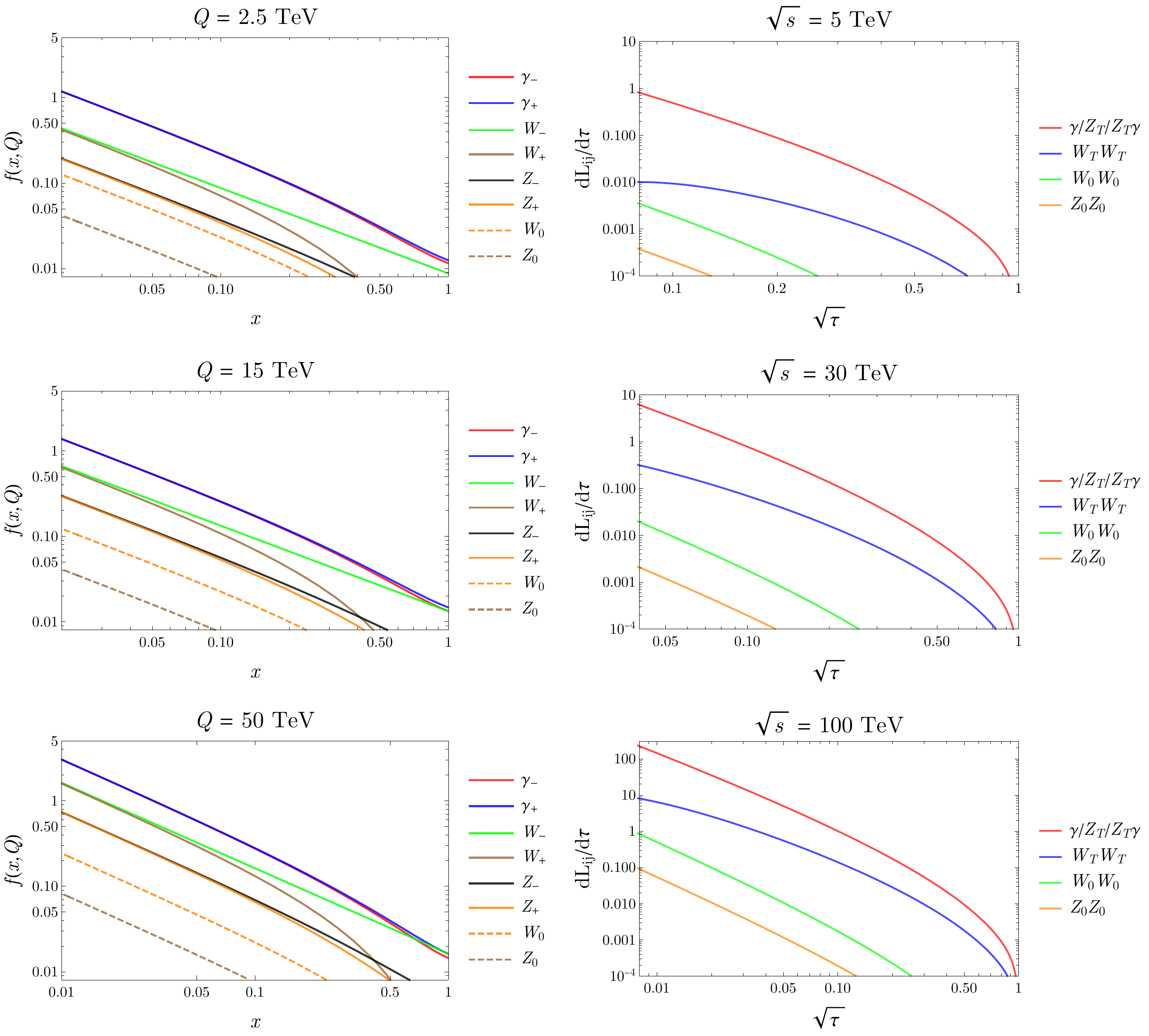}
\caption{Left: The polarized PDF $f_i(x,Q)$ for the electroweak gauge bosons separated by helicity.  Right: The parton luminosity functions $\dd L_{ij}/\dd\tau (\tau, Q = \sqrt{\tau s}/2)$ for gauge bosons separated by helicity.}
\label{fig:LL_PDF_lumi}
\end{figure}

Now that we have set up the detailed formalism, we simply solve the DGLAP evolution equations to leading log order (which does not require performing any convolutions; see~\cref{sec:SubLeadLogs} for more details), including one-loop running gauge couplings. The resulting gauge boson PDFs in the helicity basis are shown in~\cref{fig:LL_PDF_lumi}. For these plots, we align the incoming muon (anti-muon) beam with the positive (negative) $z$-axis, and the positive $z$-axis with positive helicity. 
As we anticipated from the splitting functions, the right-handed gauge boson PDFs approach zero faster than the left-handed PDFs as $x\to 1$. 
We also note an interesting helicity-dependent effect for the photon PDF. 
If we had neglected the impact of rotating between the mass and gauge bases using~\cref{eq:weakRotationMatrix}, then clearly the photon PDF should not show any helicity dependence; see~\cref{eq:DGLAP4QED}.
However, after converting to the gauge basis, the neutral gauge bosons couple to left- and right-handed muons differently. 
This is the origin of the helicity dependence at large $x$ for the photon PDF.
Finally, we note that the PDF for the longitudinal gauge bosons is scale invariant, up to the minor scale dependence from the running coupling. 
Naively, this is simply due to the fact that the longitudinal polarization sum is proportional to $p^2$ which cancels the $p^2$ in the denominator, see~\cite{Dawson:1984gx} for more details.

\subsection{Impact of subleading logs}
\label{sec:SubLeadLogs}
Now that we have computed the leading order PDFs, we will briefly discuss the uncertainty associated with taking the leading log approximation.
Generically, the DGLAP evolution equations can be expressed as
\begin{equation}
\dfrac{\dd}{\dd\log Q^2} \begin{pmatrix}
f_1\\ f_2\\  \vdots\\  f_n
\end{pmatrix} = \begin{pmatrix}
\mathcal{P}_{1\leftarrow 1}& \cdots& \mathcal{P}_{1\leftarrow n}\\
\mathcal{P}_{2 \leftarrow 1}& \cdots& \mathcal{P}_{2 \leftarrow n}\\
\vdots &  & \vdots\\
\mathcal{P}_{n \leftarrow 1}& \cdots& \mathcal{P}_{n \leftarrow n}\\
\end{pmatrix} \otimes \begin{pmatrix}
f_1\\ f_2\\  \vdots\\  f_n
\end{pmatrix}\,,
\label{eq:DGLAPmatrix}
\end{equation}
where we have individual PDFs $f_i$ for the polarization of each particle. 
Typically, to solve this matrix equation to all orders, one diagonalizes this matrix, transforms to Mellin space (where the convolution becomes a product), solves the resulting differential equations, and then transforms and rotates back.

\begin{figure}[t]
\centering
\includegraphics[width=14cm]{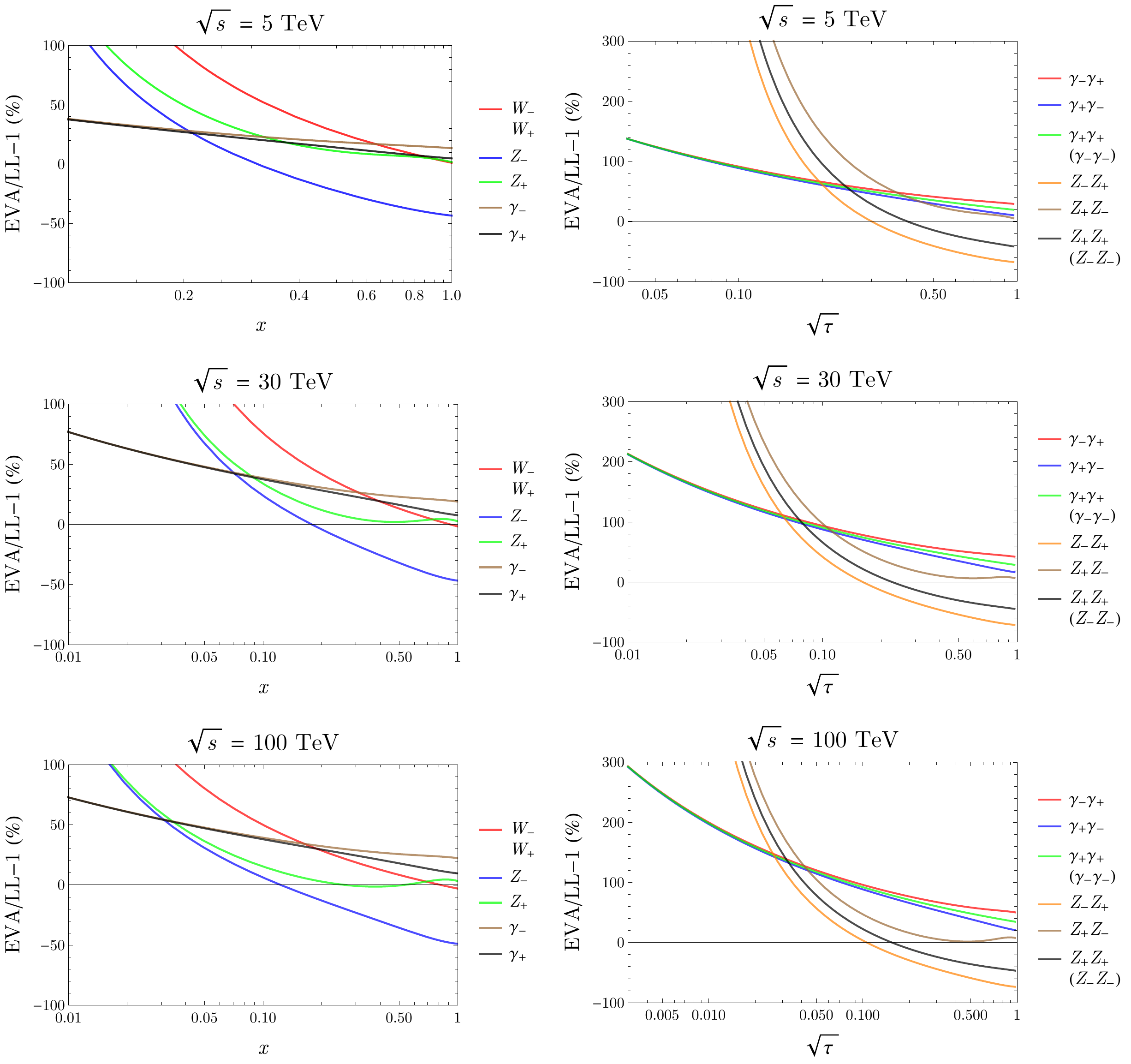}
\caption{The percent difference between the EVA and LL PDFs for the gauge bosons separated by helicity. Left: A comparison of the PDFs taking $Q = x \sqrt{s}/2$ when evaluating the LL PDF.  Right: A comparison of the parton luminosities taking $Q = \sqrt{\tau s}/2$.  }
\label{fig:EVA_LL_NLL}
\end{figure}

As we have already emphasized, the electroweak gauge couplings remain perturbative throughout the range of interest, and so unsurprisingly the unresummed leading log solution provides a reasonable approximation.
In~\cref{fig:EVA_LL_NLL}, we compare the unresummed leading log PDFs against the EVA. 
At low $x$, the EVA deviates significantly from the LL result.
This behavior is easy to understand, since the former is proportional to $\log s/4 m_Z^2$ while the latter is proportional to $\log Q^2/m_Z^2$.   
When $x \to 1$, they approach the same value up to small differences due to the running couplings (we evaluate the gauge couplings at the scale $\sqrt{s}/2$ for the EVA), and the polarization effect for the photon PDFs described above. 
We conclude that what scale appears within the logarithm is an important difference, and the EVA does \emph{not} provide a good approximation.

\begin{figure}[t]
\centering
\includegraphics[width=14cm]{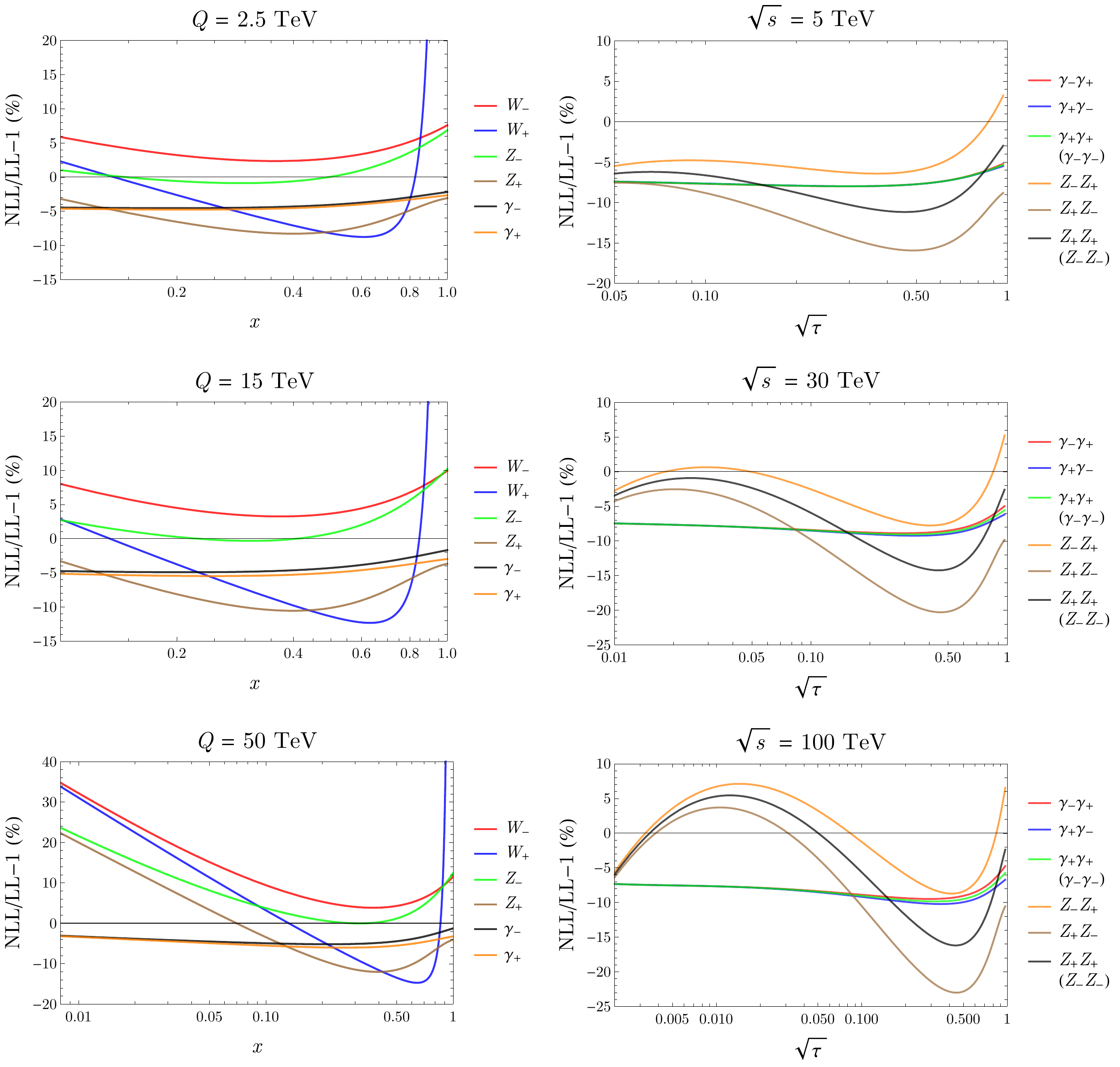}
\caption{The percent difference between the unresummed NLL to LL polarized PDF for different electroweak gauge bosons; see the legend. Left: A comparison of the PDFs for fixed $Q$.  Right: A comparison of the parton luminosities taking $Q = \sqrt{\tau s}/2$.  }
\label{fig:NLL_LL_ratio}
\end{figure}

Next, we turn to exploring the size of next-to-leading-log corrections.
To this end, we will solve~\cref{eq:DGLAPmatrix} iteratively to extract these subleading terms. 
Defining the $n^\text{th}$ logarithmic order as
\begin{equation}
f^{(n)}\big(x,Q^2\big) \sim \big(\alpha \log Q^2\big)^n \, ,
\end{equation}
we can obtain $f^{(n)}\big(x,Q^2\big)$ by inserting $f^{(n-1)}(x)$ into the right-hand-side of the DGLAP equation~\cref{eq:DGLAPmatrix}:
\begin{equation}
\dfrac{\dd}{\dd\log Q^2} f_i^{(n)}\big(x,Q^2\big) = \sum_{j=1}^n \mathcal{P}_{i \leftarrow j}(x)\otimes f_j^{(n-1)}\big(x,Q^2\big) \, .
\end{equation}
At zeroth order, the only non-zero PDF is
\begin{equation}
f^{(0)}_{\mu_L}\big(x,Q^2\big) =  f^{(0)}_{\mu_R}\big(x,Q^2\big) = \delta(1-x) \, ,
\end{equation}
which is simply the statement that the beams would be purely composed of muons in the absence of interactions. 
%

In \cref{fig:NLL_LL_ratio}, we compare the unresummed NLL and LL PDFs and parton luminosity function. The discrepancy for the PDFs between the two levels of approximation are within $\sim 10\%$ ($\sim 40\%$) for $Q = 5 \text{ TeV}$ ($50 \text{ TeV}$),\footnote{This is true except for the $W_+$ PDF.  For this case, the LL pdf goes to zero as $x\to 1$, which is the source of the divergent curve in the plot.  This behavior is simply due to the fact that the probability to emit a single $W_+$ from an on-shell $\mu_L$ is zero.  However, a non-zero contribution appears at NLL in the $x\to 1$ limit, since there can now be multiple emissions.  Note that this issue arrises in a region where the PDF is small, so that this effect has a negligible impact on observables.} and this obviously implies that the parton luminosity is also under control.
We take this as strong evidence that the LL PDFs are sufficient unless one is interested in making precision predictions at a level that is beyond the scope of this work.

\subsection{Finite mass effects}
\label{sec:FiniteMassPDF}
Up until this point, we have treated all the gauge bosons as being massless (up to the fact that the longitudinal mode exists). 
This is a good approximation when the cross section is dominated by partonic collisions whose typical scale is significantly larger than the vector masses. 
However, if the process of interest has non-trivial support at low $\tau$, finite mass effects could become important. 
To understand their numerical impact, we will briefly investigate how mass effects change the calculation of the LL PDF.

The dominant mass effect comes from the propagator. 
Note that in the massless case, the emitted vector boson has virtuality
\begin{equation}
q^2 = - \dfrac{k_T^2}{(1-z)} \, ,
\end{equation}
while for the massive vector boson, the off-shell propagator has virtuality
\begin{equation}
q^2 - m_V^2  =- \dfrac{k_T^2 + (1-z) m_V^2}{1-z} \, ,
\end{equation}
Hence, the splitting function is modified as
\begin{equation}
\mathcal{P}_{V \leftarrow \mu} \rightarrow \dfrac{k_T^4}{\tilde{k}_T^4}   \mathcal{P}_{V \leftarrow \mu}   \quad \quad \text{with}  \quad\quad\tilde{k}_T^2 = k_T^2 + (1-z) m_V^2 \, .
\end{equation}
In order to estimate the impact on the PDFs simply, we neglect the running coupling, so the $Q$ dependence for the massless $W_T$ PDF is 
\begin{equation}
\int_{m_W^2}^{Q^2} \dfrac{\dd k_T^2}{k_T^2} = \log\left( \dfrac{Q^2}{m_W^2} \right)\,,
\end{equation}
while for the massive $W_T$ it is 
\begin{equation}
\int_{m_W^2}^{Q^2} \dfrac{\dd k_T^2}{k_T^2} \dfrac{k_T^4}{(k_T^2 + (1-x) m_W^2)^2} = \log \left( \dfrac{Q^2 + m_W^2(1-x)}{m_W^2(2-x)}\right) + \dfrac{1}{2-x} - \dfrac{Q^2}{Q^2 + m_W^2(1-x)}\,.
\end{equation}
Clearly, the difference is only relevant near threshold with $x \ll 1$.

\begin{figure}[t]
\centering
\includegraphics[width=14cm]{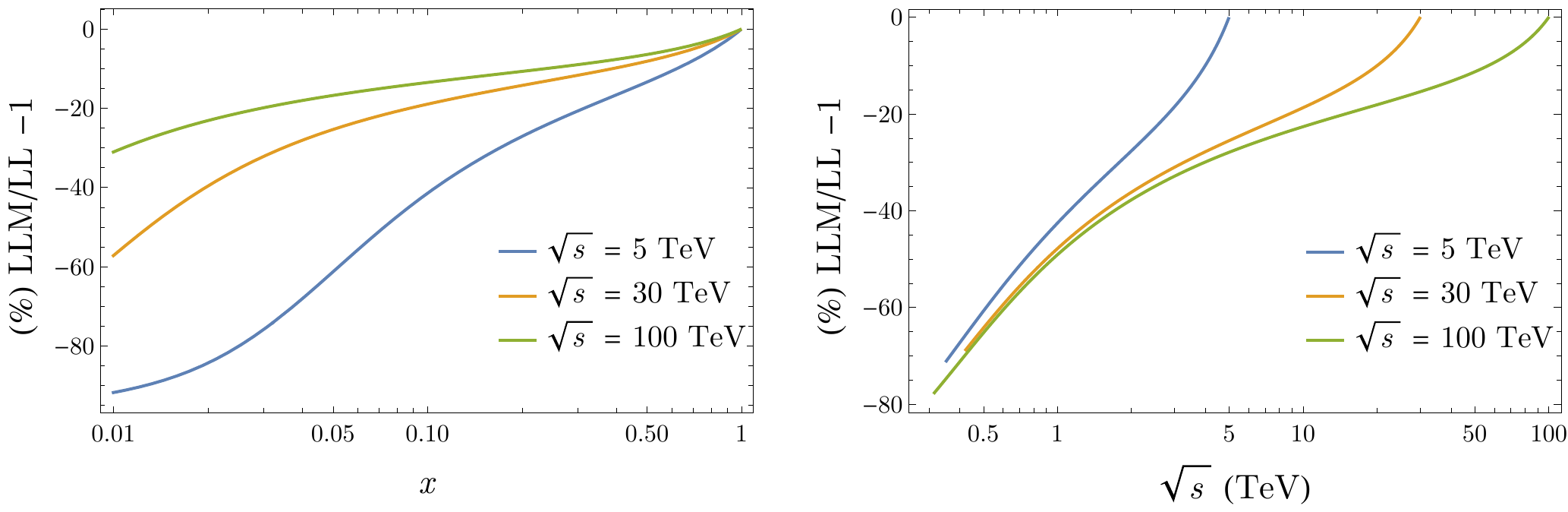}
\caption{This figure compares the massive $W_T$ PDF to massless $W_T$ PDF to LL order.  
Left: A comparison of the PDFs taking $Q = x \sqrt{s}/2$. Right: A comparison of the parton luminosities $W_-^- W^+_+$ taking $Q = \sqrt{\tau s}/2$.  }
\label{fig:finite_mass}
\end{figure}

To get a quantitative sense of the impact of including the finite mass,  \cref{fig:finite_mass} shows the ratio of the massive LL PDF (MLL) to the massless LL PDF.  The left panel shows the ratio of $f(x, Q=x\sqrt{s}/2)$ while the right one displays the ratio of $\dd\mathcal{L}_{ij}/ \dd \tau (\tau, \sqrt{\hat{s}}/2)$ with $\hat{s} = \tau s$. To make the comparison to $m_W$ more explicit, we have converted the $x$-axis of the right panel to $\hat{s}$.
We see that the parton luminosities can be off by as much as $40\%$, even for $\sqrt{\hat{s}}$ above a TeV.  
We conclude that finite mass effects are more important than resumming logs.
We leave the exploration of the impact of these effects for future work.\footnote{We also note that beyond LL order, there are additional important effects that result from maintaining momentum conservation when emitting multiple massive partons.  It would also be interesting to investigate the impact this effect would have on the conclusions comparing NLL to LL PDFs, as we did in \cref{fig:NLL_LL_ratio}.}


\section{Physics}
\label{sec:physics}
We turn next to the physics potential of a high-energy muon collider, focusing on some of the central themes -- electroweak symmetry breaking (EWSB), naturalness, and dark matter (DM) -- that have motivated new physics since the inception of the Standard Model. These considerations provide sharp goalposts for a future collider, indicating energies and luminosities that would enable such a collider to comprehensively explore the underpinnings of the Standard Model.

\subsection{Electroweak symmetry breaking}
\label{sec:EWSB}
The discovery of the Higgs completed the particle content of the SM. However, this discovery has also reinforced the puzzles associated with the Higgs field's role in the SM, generating as much confusion as clarity. The Higgs is the linchpin of the SM, responsible for all of the masses of elementary particles as well as flavor mixings, via EWSB. The majority of the SM parameters associated with the Higgs are not determined by gauge invariance, and their values must be measured. Moreover, the very fact that EWSB occurs via the Higgs is put into the SM by hand, in that we must specify the potential. Before we declare that the SM is complete, we must measure {\em all} of its parameters.

Unfortunately, the path to completing the goal of measuring all the SM parameters is often regarded as requiring two different colliders after the HL-LHC, due to the reliance on two qualitatively different types of observables. The first is to probe the Higgs couplings to other SM particles; we note that the light flavor Yukawa couplings have yet to be measured at all. The second is to explore the Higgs potential itself. To study the couplings of light flavors to the Higgs requires an extremely clean collider environment, which favors lepton colliders, such as the low-energy Higgs factories that have been proposed. Their advantage is clearly illustrated in \cref{fig:ppcompare}, which shows that Higgs production is a relatively large fraction of the total cross section at lepton colliders, once its production via gauge bosons is kinematically allowed. Even at these colliders, there should be sufficient luminosity to probe the Yukawa couplings of the charm quark, while other light flavors pose a significant challenge; the capability to tag and measure light flavors is a subject of ongoing research~\cite{Gao:2016jcm}. However, lepton colliders offer at least the promise of measurements that would be overwhelmingly difficult at hadron colliders, where precision measurement of SM Higgs branching fractions must overcome vast numbers of $u$, $d$, $s$, $c$ and gluon background jets. Furthermore, a future lepton collider running at the Higgs mass pole could measure the $s$-channel resonance production to directly probe the lepton Yukawa coupling, precisely (at a muon collider) or with an upper limit of a few times the SM prediction (at an $e^+e^-$ collider)~\cite{Altmannshofer:2015qra,Greco:2016izi,ValdiviaGarcia:2159683,dEnterria:2017dac}.

While lepton colliders provide a cleaner environment than hadron colliders, their inherent drawback at the energies of the proposed ``Higgs factories" is the small absolute yield of Higgs particles compared to their hadron collider counterparts. For example, the Higgs factories proposed thus far would collect $\mathcal{O}(10^6)$ Higgses, whereas a 100 TeV $pp$ collider would produce $\mathcal{O}(10^{10})$. This is exacerbated when taking into account multi-Higgs production, where a higher c.m.~energy is needed. Only CLIC or the 1 TeV upgrade of an ILC would have sufficient energy for the multi-Higgs production processes to be useful to explore the Higgs potential. Therefore, a common view is that in addition to a Higgs factory, something akin to the FCC-hh is necessary to truly explore the Higgs potential.

A high energy lepton collider, such as a muon collider with $\sqrt{s} \gtrsim \mathcal{O}(10)\,\mathrm{TeV}$, can completely change this narrative by offering both a clean environment for high-precision studies {\em and} the energy needed to produce new final states copiously. We will make this case through examples that follow. A first indication can be gleaned from estimates using the inclusive Higgs cross sections: at an $\mathcal{O}(10)\,\mathrm{TeV}$ muon collider with $\mathcal{O}(10/\mathrm{ab})$ of luminosity, there will be an order of magnitude more Higgs bosons produced as compared to proposed Higgs factories. Additionally, there will be $\mathcal{O}(10^4)$ di-Higgs events, which are completely inaccessible at the low-energy Higgs factories. Although a 100 TeV hadron collider with $\mathcal{O}(10/\mathrm{ab})$ will produce $\mathcal{O}(10^7)$ di-Higgs events, there are severe backgrounds there that grow with collider energy~\cite{Homiller:2018dgu}. Despite the higher yield at a 100 TeV hadron collider such as FCC-hh, the currently best estimated sensitivity to the triple-Higgs coupling of the SM~\cite{Mangano:2020sao} can be matched or exceeded with an $\mathcal{O}(10)\,\mathrm{TeV}$ muon collider~\cite{Han:2020pif}. We perform some new studies in this section to sketch the potential for probing Higgs physics at high energy muon colliders.

Pursuing such a program of future Higgs measurements would not simply complete the SM, it {\em could likely} open the first window to physics beyond the SM. The Higgs boson is unique; it is the only apparently elementary scalar among all the particles observed in the universe. Its distinct properties provide many compelling reasons to investigate it further. The Higgs provides the only source of flavor physics in the SM; the most relevant, invariant portal to other BSM sectors or dark matter; the unitarization of scattering amplitudes in the SM; a window on early universe cosmology via the EW phase transition (EWPT), and potentially EW baryogenesis (EWBG); and, last but not least, the naturalness puzzle. We are strongly motivated to determine whether the Higgs is {\em solely} responsible for EWSB, and whether it is (partially) composite. In this section, we will discuss a muon collider's role in addressing these topic.\footnote{The Higgs potential also lies at the root of deep questions about the stability of the universe, which we will leave for future investigations.} {\em All} of these questions can be attacked by measuring the Higgs's properties with sufficient precision. Many of them benefit from the large Higgs production rate and cleanliness of a high energy muon collider, as well as the dynamical range of c.m.~energy that such a collider achieves by virtue of being a vector boson collider. An apt analogy for the path that started with finding the Higgs and continues by investigating it in sufficient detail is provided by cosmology. While the expansion of the universe was known since Hubble, it was not until many decades later that the right observable was found and measured precisely enough that the accelerated expansion of the universe was conclusively discovered. We are now just beginning to acquire experimental knowledge of the Higgs boson's properties, at a relatively coarse level. We must move toward the new era of precision Higgs physics, which, like precision cosmology, offers the hope of revolutionizing our understanding of the universe.

\subsubsection{Higgs coupling sensitivity estimates from on-shell Higgs processes}

There are many new measurements of Higgs properties that are accessible via higher energies and cleaner environments, and we will explore a sampling of these in the subsequent sections. 
The desire to improve Higgs precision will drive one of the core programs for any future collider. Therefore, it is important to understand how precisely a high energy muon collider could measure Higgs properties on its own, as well as in combination with other colliders. 
The answer to this question depends both on the theoretical framework and experimental details, which leave an enormous range of possibilities that are beyond the scope of this paper to explore. In order to make a first quantitative estimate rather than simply stating that a large number of Higgs particles would be produced, we will make a number of simplifying assumptions. In this section, we focus on the processes in which on-shell Higgs bosons are produced. At a high energy muon collider, off-shell Higgs processes will in some cases offer an even more powerful probe of Higgs properties, a topic to which we will return in \cref{subsec:flavorexotic}.

First, we will adopt the common $\kappa$ fits for Higgs precision~\cite{LHCHiggsCrossSectionWorkingGroup:2012nn,Heinemeyer:2013tqa}. This is not an endorsement of this methodology compared to any other, but a pragmatic choice for the sake of making comparisons, as all future collider proposals have an example of this type of fit (kappa-0 framework)~\cite{deBlas:2019rxi}. The inputs to such a fit are the uncertainties on the cross section measurements in exclusive channels. These depend upon the signal cross section and physics backgrounds, as well as machine backgrounds, detector capabilities, and possible additional theoretical assumptions. The machine backgrounds and detector capabilities are particularly interesting in the context of high-energy muon colliders, as previously discussed. The BIB at muon-colliders serves both as a background to measurements and a driver of detector design. There is no optimized detector design available at all our benchmark c.m.~energies, due to the fact that the BIB depends on the accelerator complex within roughly 25 m on each side of the interaction point. Therefore, we will simply choose our energy and luminosity benchmarks to be 10 TeV and 10/ab, and using the Muon Collider detector card for the {\sc Delphes} fast simulation~\cite{deFavereau:2013fsa}. This choice of ``detector" does not serve as a final word, but allows us to begin exploring how the physics requirements interact with detector design. We do not include BIB, as current full-simulation studies show that it appears to be under control, especially at higher energies~\cite{Bartosik:2019dzq,Bartosik:2020xwr}, and there are potential ways to reduce its effects further. Furthermore, for this toy study we {\em do not} include physics backgrounds.

\begin{table}[ht]
\renewcommand{\arraystretch}{1.3}
\setlength{\arrayrulewidth}{.3mm}
\setlength{\tabcolsep}{0.8 em}
\begin{center}
\begin{tabular}{c|c|c|c|c} 
\multicolumn{5}{c}{10 TeV @ $10\,\textrm{ab}^{-1}$} \\ \hline
Production & Decay & Rate [fb] & $A \cdot \epsilon$ [\%] & $\Delta \sigma / \sigma$ [\%] \\ \hline \hline
\multirow{12}{*}{$W$-fusion} & $bb$ & $490$ & 7.4 & 0.17 \\ \cline{2-5}
& $cc$ & $24$ & 1.4 & 1.7 \\ \cline{2-5}
& $jj$ & $72$ & 37 & 0.19 \\ \cline{2-5}
& $\tau^+ \tau^-$ & $53$ & 6.5 & 0.54 \\ \cline{2-5}
& $WW^*(jj\ell\nu)$ & $53$ & 21 & 0.30 \\ \cline{2-5}
& $WW^*(4j)$ & $86$ & 4.9 & 0.49 \\ \cline{2-5}
& $ZZ^*(4\ell)$ & $0.1$ & 6.6 & 12 \\ \cline{2-5}
& $ZZ^*(jj\ell^+\ell^-)$ & $2.1$ & 8.9 & 2.3 \\ \cline{2-5}
& $ZZ^*(4j)$ & $11$ & 4.6 & 1.4 \\ \cline{2-5}
& $\gamma\gamma$ & 1.9 & 33 & 1.3 \\ \cline{2-5}
& $Z(jj)\gamma$ & $0.9$ & 27 & 2.0 \\ \cline{2-5}
& $\mu^+ \mu^-$ & $0.2$ & 37 & 0.37 \\ \hline
\multirow{2}{*}{$Z$-fusion} & $bb$ & 51 & 8.1 & 0.49 \\ \cline{2-5}
& $WW^*(4j)$ & 8.9 & 6.2 & 1.3 \\ \hline
$W$-fusion $tth$ & $bb$ & $0.06$ & 12 & 12 \\ \hline
\end{tabular}
\end{center}
\caption{
Signal rates and efficiencies for selected Higgs production channels at a $10\,\textrm{TeV}$ muon collider using signal-only selection and the {\sc Delphes} muon collider fast simulation.
}\label{tab:higgsrateprecision}
\end{table}

While our assumptions may seem like too drastic of a simplification, there still is useful sensitivity information despite having made these naively non-conservative estimates. Our signal rates using {\sc Delphes} have a rather small acceptance, given that the detector card limits physics objects to $\vert\eta\vert<2.5$ except for forward muons. There are a number of motivations for the detector card inspired from a hybrid of CLIC and FCC-hh for efficiencies and reconstruction~\cite{delphesTalk}, but these are not optimized for a particular physics target or energy. Additionally, the general acceptance roughly coincides with having BIB-suppressing tungsten nozzles~\cite{Foster:1995ru,Mokhov:2011zzb}, with a $10\degree$ opening angle motivated by 1.5 TeV c.m.~muon collider studies~\cite{Mokhov:2011zzb,Mokhov:2011zzd}. The nozzle opening should be able to be reduced at higher energies since the radiation will be more forward~\cite{Chiesa:2020awd} (and timing should also mitigate BIB effects). Physics backgrounds, of course, potentially matter a great deal more than BIB. However, they are significantly reduced at a lepton collider as shown in \cref{fig:ppcompare}. Detailed sensitivity studies including physics backgrounds have been performed for 3 TeV lepton colliders for CLIC Higgs studies~\cite{Abramowicz:2016zbo}, and thus serve as a proof of principle (or potential floor) for our signal-driven sensitivities. There are a variety of studies that can and should be done in the future, but we hope this serves as a useful starting point by showing the effects of acceptance and efficiency via fast simulation. From the perspective of signal and BIB, we take this to be a conservative starting point.

Table~\ref{tab:higgsrateprecision} shows, for various channels, the results for cross sections, acceptance $\times$ efficiency, and the measurement precision for our 10 TeV muon collider benchmark. The acceptances are based on minimal cuts. As a starting point, we choose these to approximate the one existing full-simulation study that focuses on $h\rightarrow b\bar{b}$~\cite{Bartosik:2020xwr} at 1.5 TeV, which was also extrapolated to 10 TeV. For two-body final states of the Higgs, such as in~\cite{Bartosik:2020xwr}, we make minimal cuts on reconstructed objects such as $p_T>40$ GeV, with the {\sc VLC} jet algorithm \cite{Boronat:2014hva, Boronat:2016tgd} run for an exclusive number of jets with an $R=0.5$ as implemented in {\sc Delphes}, and we use the tight-tagging working point for $b$-tagging. We find that we have good agreement with the very conservative study done in~\cite{Bartosik:2020xwr}, and therefore it serves as another calibration point for extrapolating to the performance in other final states. For all other physics objects, we use the standard parametrization as found in the muon detector card, except for charm-tagging, which is not implemented. For the $c\bar{c}$ final state we apply a flat $20\%$ tagging efficiency for each $c$-jet, as inspired by CLIC~\cite{linssen2012physics}. For greater than two-body final states, we reduce the $p_T$ requirements to 20 GeV and reduce the tagging efficiencies to loose tags for $b$-tagging, such as in $t\bar{t}h$.

We stress again that the performance in the various channels shown in Table~\ref{tab:higgsrateprecision} is in no way optimized and needs further study. However, with these putative sensitivities, we can perform a simple 10-parameter $\kappa$ fit to compare this benchmark to other proposed colliders. Of course, the muon collider can be enhanced with complementary measurements from other colliders as well. Therefore we also perform the fit with the HL-LHC or a 250 GeV $e^+e^-$ collider included. Here we took the CEPC input with full correlation matrix for different channels~\cite{An:2018dwb,CEPCStudyGroup:2018ghi} to represent the 250 GeV $e^+e^-$ collider. The results and discussion on complementarity with other lepton collider Higgs factories would be similar. We present the results of these fits in Table~\ref{tab:higgscouplingfit}.

\begin{table}[t]
\renewcommand{\arraystretch}{1.3}
\setlength{\arrayrulewidth}{.3mm}
\setlength{\tabcolsep}{0.8 em}
\begin{center}
\begin{tabular}{c|ccc}
\multicolumn{4}{c}{Fit Result [\%]} \\ \hline
& \multicolumn{1}{c|}{$10\,\textrm{TeV}$ Muon Collider} & \multicolumn{1}{c|}{with HL-LHC} & with HL-LHC + $250\,\textrm{GeV}$ $e^+e^-$ \\ \hline \hline
$\kappa_W$ & 0.06 & 0.06 & 0.06 \\ \hline
$\kappa_Z$ & 0.23 & 0.22 & 0.10 \\ \hline
$\kappa_g$ & 0.15 & 0.15 & 0.15 \\ \hline
$\kappa_{\gamma}$ & 0.64 & 0.57 & 0.57 \\ \hline
$\kappa_{Z\gamma}$ & 1.0 & 1.0 & 0.97 \\ \hline
$\kappa_c$ & 0.89 & 0.89 & 0.79 \\ \hline
$\kappa_t$ & 6.0 & 2.8 & 2.8 \\ \hline
$\kappa_b$ & 0.16 & 0.16 & 0.15 \\ \hline
$\kappa_{\mu}$ & 2.0 & 1.8 & 1.8 \\ \hline
$\kappa_{\tau}$ & 0.31 & 0.30 & 0.27 \\ \hline
\end{tabular}
\end{center}
\caption{
Results of a 10-parameter fit to the Higgs couplings in the $\kappa$-framework, based on the attainable precision in each on-shell Higgs production and decay channel listed in Table~\ref{tab:higgsrateprecision}. Additionally, we include the effects of adding data sets projected from the HL-LHC and a 250 GeV $e^+e^-$ Higgs factory. One should keep in mind that a muon collider will also strongly constrain Higgs properties via off-shell measurements, which are not included here.
}\label{tab:higgscouplingfit}
\end{table}

It is impossible to directly compare our Higgs sensitivity measurements to other proposed colliders~\cite{deBlas:2019rxi} given the signal-only nature of the results presented here.   However, it is easy to see when looking at Table 3 of~\cite{deBlas:2019rxi} (which served as the input to the European Strategy Report) that the level of signal-only precision reached in our estimates is not orders of magnitude better than proposed Higgs factories. This can be traced back to the rather small $A \cdot \epsilon$ for our ``detector" choice. This motivates exploring detector design to see how much the effective acceptance can be increased while being able to keep BIB in check. Additionally, the cross sections we use are for unpolarized muons, and in fact, the single Higgs VBF cross section for unpolarized muons is not that different at 10 TeV from the polarized cross section for CLIC at 3 TeV. Therefore, it is important to understand the impact and potential for polarization of future high-energy muon colliders. Nevertheless, we emphasize that even in this first simplified study, a 10 TeV muon collider provides similar numbers to other colliders, and there is a great deal of room for improvement. Additionally, given the extra energy, the muon collider can achieve much better precision on Higgs self-interactions~\cite{Han:2020pif} using the same machine. Moreover, the real untapped potential for a high-energy muon collider comes from its ability to make novel measurements of off-shell Higgs couplings. This is a feature that is potentially shared in common with a 100 TeV $pp$ collider, but is relatively unexplored as of yet. In the next section, we explore an example of this type of approach, which could measure the top Yukawa with a precision of $\mathcal{O}(1\%)$. With all of these caveats in mind, a high energy muon collider {\em is} an impressive Higgs factory as well as a discovery machine, and there are numerous interesting avenues for future work related to the Higgs.

\subsubsection{Flavor and exotic couplings}
\label{subsec:flavorexotic}

Flavor physics in the SM {\em only} arises through the Higgs couplings, which determine both the mass pattern of the different generations and the mixings that allow for flavor changing processes. Taking as motivation that flavor is one of the strangest aspects of the SM, there has been a rich history of testing the flavor structure of the SM indirectly using measurements from intensity frontier experiments.  This program has resulted in stringent bounds on flavor changing processes, probing new physics scales that are naively well out of the direct reach of any future energy frontier experiment.  Nevertheless, not all of the SM Yukawas have been measured yet, and large deviations in flavor diagonal Higgs couplings due to BSM physics are possible~\cite{Egana-Ugrinovic:2018znw,Egana-Ugrinovic:2019dqu} as well as smaller flavor-changing BSM Higgs couplings~\cite{Altmannshofer:2016zrn}, depending on the particular flavors involved.  Measuring the SM Yukawas may require more than an $\mathcal{O}(10)\,\mathrm{TeV}$ muon collider. Any channel with a branching fraction similar to $\mathrm{Br}(h\rightarrow \mu^+\mu^-) \sim \mathcal{O}(10^{-4})$ will result in an absolute yield of $10^3$ decays before backgrounds, acceptances, and efficiencies are accounted for. Nevertheless, if detectors are optimized, there is still the possibility to go after first generation couplings directly. If BSM deviations exist, even higher energy muon colliders will only have a greater physics potential.  For example, current LHC data allows an enhancement of $\mathrm{Br}(h\rightarrow d\bar{d})$ by $\mathcal{O}(10^6)$, which is well within the reach of a muon collider~\cite{Egana-Ugrinovic:2019dqu}.   

Beyond just measuring properties of the SM Higgs which should exist, we can use the Higgs as a potential window on unexpected new physics beyond the SM. The $H^\dagger H$ operator is the lowest dimension gauge and Lorentz invariant building block in the SM.  Therefore, if there are new states beyond the SM that are lighter than the Higgs, it is quite likely that the Higgs will have some branching fraction to decay to them. These are known as {\em exotic Higgs decays} and have been a subject of significant recent study~\cite{Curtin:2013fra}.  Given that a muon collider will have a clean environment with the additional benefit that it would produce a larger number of Higgses than the $e^+ e^-$ Higgs factories, it provides an excellent opportunity to investigate exotic Higgs decays further.  Moreover, as at FCC-hh, given that the dynamic range of energies available for the Higgs grows as the lepton collider $\sqrt{s}$ increases, there are a variety of other probes one can employ to test Higgs couplings beyond simply studying branching fractions of on-shell Higgs bosons.

\paragraph{$\textbf{\textit{W}}^\textbf{+}\textbf{\textit{W}}^\textbf{-} \to \textbf{\textit{t}}\bar{\textbf{\textit{t}}}$: a longitudinal scattering case study} \leavevmode\\ 

As an example of the power of having both precision and a dynamic range of energies available to measure Higgs couplings, we consider the classic example of the interplay between Higgs physics and perturbative unitarity.  If the Higgs boson's couplings are not precisely those predicted by the Standard Model, then, in the absence of other new physics, scattering amplitudes of longitudinal gauge bosons will grow with energy and eventually violate perturbative unitarity bounds \cite{Dicus:1992vj, Lee:1977yc, Lee:1977eg, Veltman:1976rt, Chanowitz:1985hj, Appelquist:1987cf}. This allows high-energy colliders to probe new physics operators that involve the Higgs boson  by studying scattering processes with external gauge bosons, rather than Higgses. This approach, very different from that taken at Higgs factories, has been dubbed ``Higgs without Higgs'' \cite{Henning:2018kys}.

A particularly interesting test case for this program at a muon collider is the measurement of the top Yukawa coupling. While an $e^+ e^-$ Higgs factory is especially well-suited to a high-precision measurement of the Higgs coupling to gauge bosons (especially through the Higgsstrahlung process), the top Yukawa coupling will be less constrained. If we assume that the top Yukawa deviates from its Standard Model value by a fraction $\delta_\mathrm{BSM}$, such that 
\begin{align}
y_t \mapsto y_t( 1+ \delta_\mathrm{BSM})\,,
\end{align}
then we expect the 95\% confidence limit on $\delta_\mathrm{BSM}$ after HL-LHC to be $|\delta_\mathrm{BSM}| \lesssim 0.06$ (using Fig.~135 of \cite{Cepeda:2019klc}). This is a much weaker bound than the $\sim 10^{-3}$ precision of the $hZZ$ coupling expected at a Higgs factory. Hence, we focus on the top Yukawa for a first case study of the potential Higgs coupling reach of a muon collider.

The scattering amplitude for top production via longitudinal $W$ bosons when $\delta_\mathrm{BSM} \neq 0$ scales as \cite{Appelquist:1987cf}
\begin{equation}
{\cal M}\big(W^+_L W^-_L \to t {\bar t}\,\big) \simeq -\frac{m_t}{v^2} \delta_\mathrm{BSM}  \sqrt{\hat s}\,, \qquad \text{with}\qquad\sqrt{\hat s} \gg m_t\,. 
\end{equation}
Taking into account only this growing term in the amplitude, we estimate that perturbative unitarity is violated at a scale $\Lambda_\mathrm{BSM} \lesssim \frac{10~\mathrm{TeV}}{\delta_\mathrm{BSM}}$. For small $\delta_\mathrm{BSM}$, this is well above the energy scale of a potential muon collider, so it is theoretically consistent to treat new physics in this sector via the parameter $\delta_\mathrm{BSM}$ without specifying the UV completion. 

\begin{figure}[h]
\centering
\includegraphics[trim = 20 0 20 0, clip, width = 1.0\textwidth]{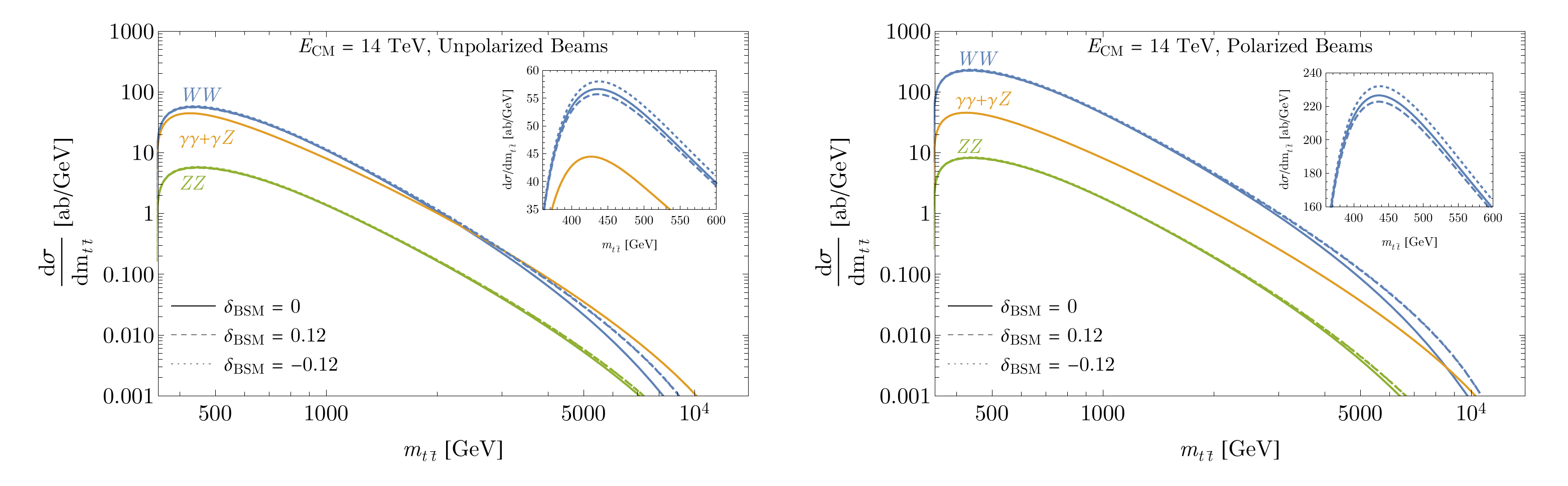}
\caption{Differential cross section for $\mu^+  \mu^- \to t  {\bar t} + X$ from different gauge boson fusion processes at a 14 TeV muon collider, with unpolarized beams (left) or fully polarized (left-handed $\mu^-$ and right-handed $\mu^+$) beams (right). At high energies, a  deviation from the  Standard Model top Yukawa leads to a significant increase in the rates for the $W^+_L W^-_L \to t{\bar t}$ process. At low energies (visible in the insets), it produces either destructive interference ($\delta_\mathrm{BSM} > 0$) or constructive interference ($\delta_\mathrm{BSM} < 0$).
}
\label{fig:topyukawarate}
\end{figure}

\begin{figure}[h]
\centering
\includegraphics[trim = 30 0 30 0, clip, width = 1.0\textwidth]{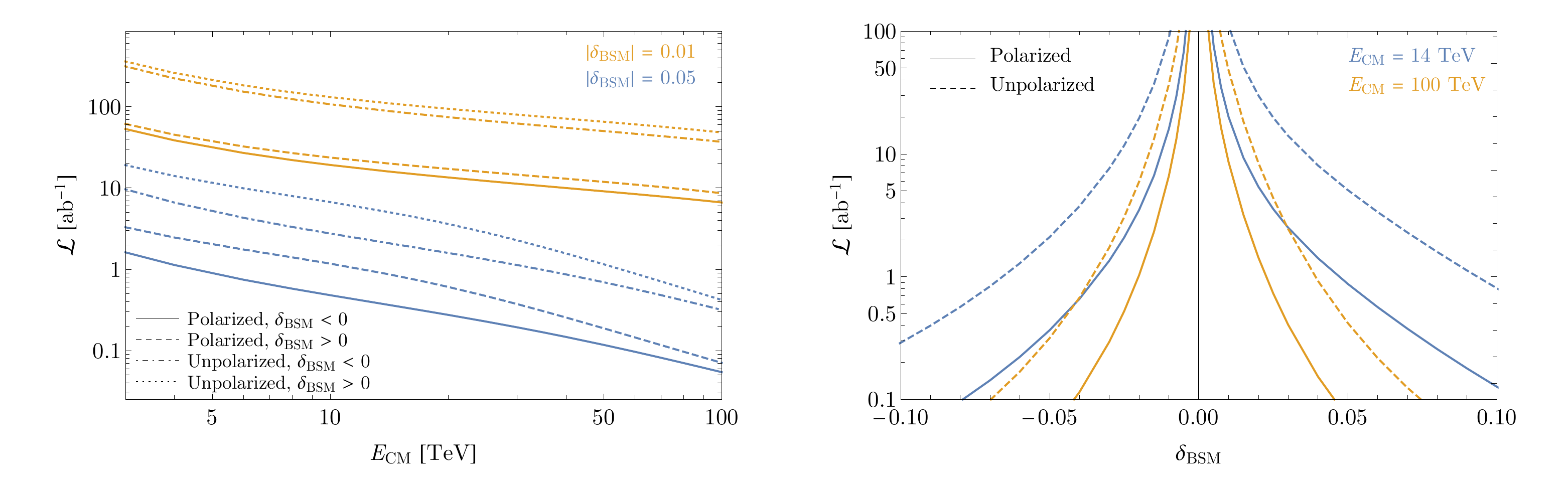}
\caption{Luminosity needed to distinguish a modified top Yukawa coupling parametrized using $\delta_\mathrm{BSM}$ from the Standard Model at 2$\sigma$ confidence, using the differential rate $\dd \sigma/\dd m_{t {\bar t}}^2$ of the process $\mu^+ \mu^- \to t {\bar t} + X$.  Left: The result is shown as a function of the c.m.~energy for various benchmark values of $\delta_\mathrm{BSM}$. Right: The result is shown as a function of $\delta_\mathrm{BSM}$ for $E_\text{CM} = 14$ and $100$ TeV (right).
}
\label{fig:topyukawareach}
\end{figure}

In Fig.~\ref{fig:topyukawarate}, we show  the differential distribution $\dd\sigma/\dd m_{t {\bar t}}$ at a 14 TeV muon collider, both for the case with unpolarized muon beams and the case with a fully left-handed $\mu^-$ and right-handed $\mu^+$. ({\sc FeynArts} \cite{Hahn:2000kx} and {\sc FeynCalc} \cite{Mertig:1990an,Shtabovenko:2016sxi,Shtabovenko:2020gxv} were used to perform these computations.) We see that the $W^+W^-$ initial  state is dominant in the case of polarized beams, increasing the possible sensitivity to  the enhanced $W^+_L W^-_L \to t{\bar t}$ process. In Fig.~\ref{fig:topyukawareach}, we present the $2\sigma$ sensitivity of a muon collider to the parameter $\delta_\mathrm{BSM}$. The sensitivity is computed from the difference in BSM and SM predictions for the differential distribution $\dd\sigma/\dd m_{t {\bar t}}$ integrated over a set of bins.\footnote{We divide the energy range into 20 bins (with smaller bins at lower $m_{t {\bar t}}$, where the cross section is larger) and find the value of $\delta_\mathrm{BSM}$ for which the Poisson log likelihood difference $2 \Delta \log L = 4$, where we compute this difference as  $2 \sum_{i \in \mathrm{bins}} \left(n_i^\mathrm{SM} - n_i^\mathrm{BSM} + n_i^\mathrm{BSM} \log(n_i^\mathrm{BSM}/n_i^\mathrm{SM})\right)$, where  $n_i^\mathrm{(B)SM}$ is the model's predicted mean for bin $i$ and need not  be an integer. This is a rough proxy for the sensitivity that one might obtain by doing pseudo-experiments. We have checked that the result is not very sensitive to the choice of binning.} (This assumes that $t{\bar t}$ events can be detected with high efficiency, an assumption that should be checked with detector simulations in the future.) We find that percent-level deviations in the top Yukawa can be probed with luminosities $\sim 10~\mathrm{ab}^{-1}$ at  a muon collider with polarized beams. The reach with unpolarized beams suffers (requiring roughly an order of magnitude more luminosity to achieve the same sensitivity). Notice that it is easier to probe negative values of $\delta_\mathrm{BSM}$, as these interfere constructively with the Standard Model process at smaller invariant masses $m_{t \bar{t}}$.

Beyond allowing for precision measurements of Higgs couplings to SM particles, the strategy discussed in this section can also be implemented to probe the existence of new degrees of freedom that, although kinematically accessible, are very rarely produced. In  \cref{sec:singlet_invt}, we illustrate this point in the context of one of the most elusive BSM scenarios: a $\mathbb{Z}_2$-symmetric SM-singlet that interacts with the SM only through the Higgs-portal. As we discuss there, a strategy based on exploiting the resulting kinematic features in the differential cross section for the process $\mu^+ \mu^- \rightarrow t \bar t + X$ may be competitive with the traditional missing-mass analysis that is the focus of \cref{sec:singlet_missing}.

\subsubsection{The Higgs potential and the electroweak phase transition}

One of the most intriguing aspects of EWSB is its role in the early universe. Because we can not directly observe the early universe before the time of formation of the CMB other than through gravitational waves, we must make use of particle physics to draw inferences about what occured. Many interesting and yet unmeasured epochs in cosmology are directly intertwined with EWSB. For example, the evolution of neutrinos in the universe and the properties of the cosmic neutrino background depend crucially on the $W$ and $Z$ boson masses. The masses of SM particles arise from EW symmetry breaking, and so may have turned on during the EWPT in a thermal history in which EW symmetry was restored at even earlier times and hotter temperatures. If the EWPT was strongly first order and other sources of CP violation exist -- both of which require new physics beyond the SM -- then EW baryogenesis could explain the matter/antimatter asymmetry in our universe.

Since we can not directly measure the Higgs potential at finite temperature, we are relegated to studying its zero temperature behavior, and possible couplings of the Higgs to other particles.    Unfortunately, we also can not access the Higgs potential away from its minimum at colliders, so we are left to study the shape of the potential locally through its derivatives, i.e., measuring the Higgs self couplings.  This has motivated the intense study of what can be learned from di-Higgs production, since the previous Snowmass process in 2013.  As mentioned earlier, a high energy muon collider has the ability to measure di-Higgs production and thus the triple Higgs coupling with precision similar to or better than a 100 TeV hadron collider.  However, from the perspective of BSM physics, there is rarely {\em just} a shift in the triple Higgs coupling alone.  This is clear from the EFT perspective, where there are multiple operators that can change the di-Higgs production rates. If one wants to focus solely on the $h^6$ operator, this can only be realized in singlet extensions of the SM.   

While singlet extensions of the SM can come in a variety of forms, there is a particularly interesting model that serves as a ``nightmare scenario"~\cite{Curtin:2014jma}, where the singlet is protected from mixing with the SM Higgs by a  $\mathbb{Z}_2$ symmetry and is heavier than $m_h/2$ so that the Higgs can not decay to this state.  By studying this scenario, one can set a worst-case benchmark for how well a strong EWPT can be tested.  Additionally, this  $\mathbb{Z}_2$ singlet model can serve to benchmark the more recent investigations into EW symmetry non-restoration~\cite{Meade:2018saz} and also as a proxy for neutral naturalness \cite{Chacko:2005pe, Craig:2014aea}. 

\begin{figure}[t!] 
   \centering
   \includegraphics[height=2in]{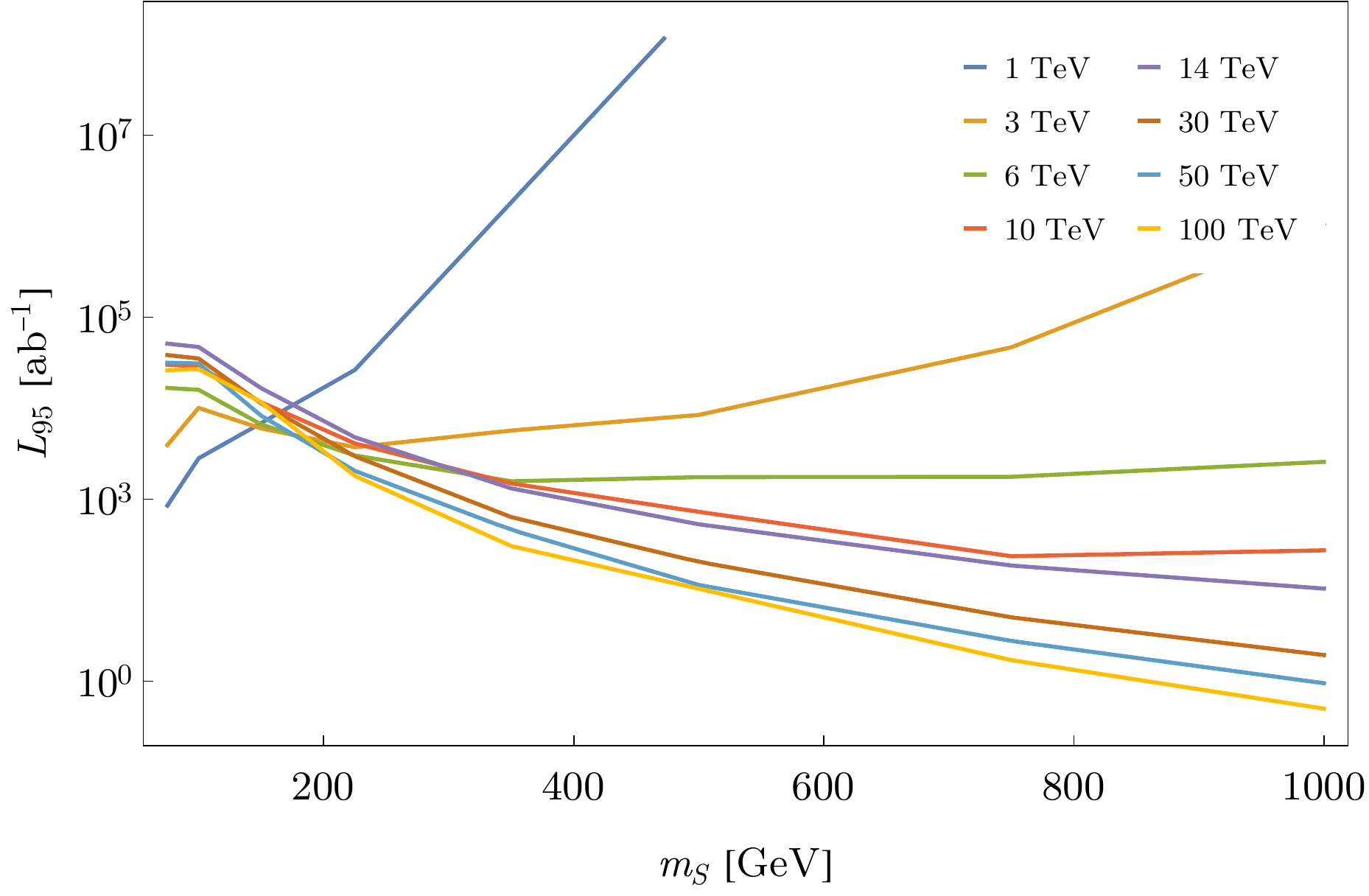} 
      \includegraphics[height=2.1in]{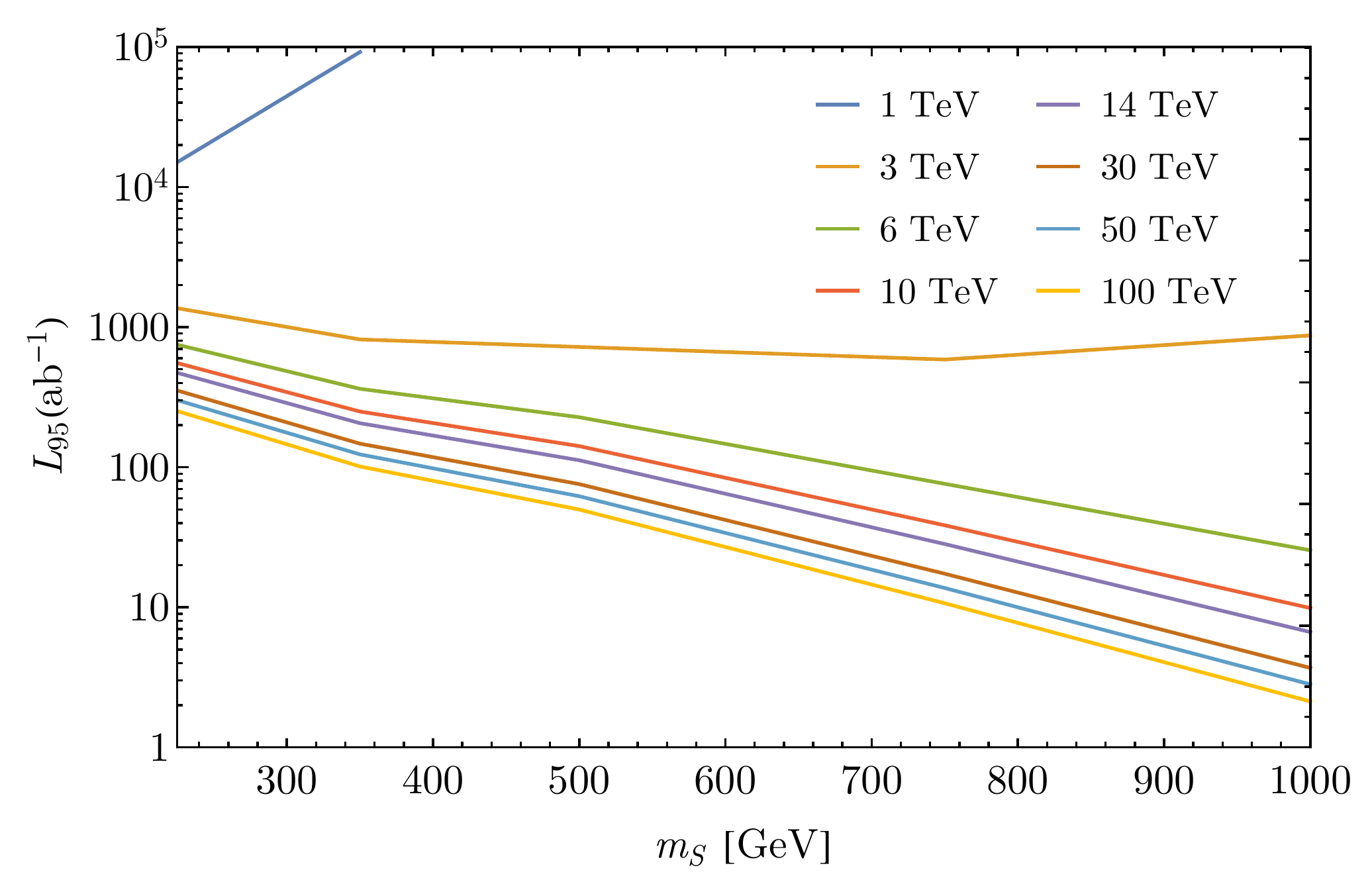} 
   \caption{The integrated luminosity $\mathcal{L}$ in units of ab$^{-1}$ required to exclude a singlet scalar whose mass $m_S$ is due entirely to electroweak symmetry breaking at 95\% CL, for muon colliders operating at various c.m.~energies. Left: Projected exclusion reach from missing energy searches using the naive combination of $SS + \gamma/h/Z$ final states. Right: Projected exclusion reach from the invariant mass distribution of $t \bar t$ pairs produced in VBF.}
   \label{fig:nightmare}
\end{figure}

There are at least two compelling channels in which to probe the ``nightmare scenario'' at a high-energy muon collider, beyond indirect constraints that may be obtained from measurements of the triple Higgs coupling or Higgs couplings to other SM particles. One of these is the natural extension of the strategy pursued at the LHC and proposed for future proton-proton colliders: to use a search for missing energy in conjunction with one or more visible particles produced through ISR or associated production. In \cref{sec:singlet_missing} we present a simplified analysis to assess the prospects of a high-energy muon collider in these final states, combining searches for missing mass in conjunction with an ISR photon, Higgs boson, or $Z$ boson. Alternately, given that missing energy searches are limited by significant backgrounds even at a muon collider, in \cref{sec:singlet_invt} we propose another novel search strategy that leverages the significant contribution of longitudinally polarized $W$ and $Z$ bosons to certain processes. In particular, radiative corrections from a $\mathbb{Z}_2$ singlet scalar with large coupling to the Higgs give rise to a feature as the invariant mass of the final state crosses over the singlet threshold of, e.g., $t \bar t$ pairs produced in vector boson fusion. This feature can be large enough to be distinguished from the Standard Model contribution provided sufficient control over theory systematics in the $t \bar t$ invariant mass distribution, and hence could offer comparable sensitivity to a direct search for the singlet in the missing energy final state. To our knowledge, this is an entirely new way of searching for the ``nightmare scenario,'' and one that leverages the unique strengths of a high-energy muon collider.  The results of the two studies are summarized in \cref{fig:nightmare}, which shows the integrated luminosity required to exclude a singlet scalar obtaining all of its mass from electroweak symmetry breaking for various collider energies.
We conclude that a 10 TeV - 30 TeV muon collider can easily compete with or exceed the reach of a 100 TeV collider for this compelling scenario~\cite{Curtin:2014jma,Craig:2014lda}.

\subsubsection{Additional Higgs bosons}

As a final case study demonstrating the potential for a high-energy muon collider to illuminate the physics of EWSB, we consider the search for additional ``Higgs bosons'' that acquire their Standard Model couplings by mixing with the Higgs. This is exemplified by one of the simplest extensions of the Higgs sector, a real scalar singlet with renormalizable couplings to the SM Higgs.\footnote{The sensitivity of muon colliders to extended Higgs sectors with electroweak doublets was recently studied in \cite{Han:2021udl}.} This encodes a large class of BSM theories which address the stability of the electroweak scale~\cite{Chacko:2005pe,Ellwanger:2009dp}, or relate the baryon asymmetry in the universe today with the EWPT~\cite{Profumo:2007wc,Espinosa:2011ax,Morrissey:2012db,Kotwal:2016tex,Kurup:2017dzf}. More generally, given the diversity of vector bosons and fermions in the SM, it is natural to ask if the scalar sector possesses similar depth.

A scalar singlet SM extension is a very useful benchmark to assess the capabilities of future colliders~\cite{Buttazzo:2015bka,Buttazzo:2018qqp}, since it manifests itself in a two-fold way: indirectly, as modification of the Higgs decay rates, and directly, in single and double production channels. Both these effects are controlled by the same small set of parameters -- notably the singlet mass and its mixing with the Higgs boson -- allowing for an immediate comparison of the direct and indirect reach. As we shall see, the ability of a very high energy lepton collider to discover heavy resonances is crucial to overcoming the limitations of Higgs precision measurements, which are inevitably constrained by systematic uncertainties, and allows the exploration of entirely new territory involving weakly interacting new physics in the 10 TeV range.

The singlet phenomenology is dictated by the following Lagrangian
\be
\mathcal{L}=\mathcal{L}_{\rm SM} + \frac{1}{2}(\partial_\mu S)^2 -V(S) - a_{HS} S |H|^2 -\lambda_{HS} S^2 |H|^2\,,
\ee
where two portal operators between the SM Higgs $H$ and the singlet field $S$ are possible at the renormalizable level. The mass eigenstates are identified via a rotation of an angle $\gamma$,
\be
h = h^0 \cos\gamma  + S \sin\gamma\,,\qquad \text{and}\qquad \phi=S \cos\gamma - h^0 \sin\gamma\,,
\ee
where $h^0$ is the neutral component of the SM Higgs doublet, $h$ is the SM-like state, and $\phi$ the new singlet-like scalar. Since $S$ is a complete singlet, all interactions of $\phi$ with SM states other than the Higgs proceed through this mixing and are controlled by $\sin\gamma$. At the same time, all single Higgs boson couplings to SM fermions and vectors are rescaled by the same factor $\kappa_V = \kappa_f = \cos\gamma$.

The mixing angle can be read directly from the mass matrix 
\be
\gamma \simeq \frac{v(a_{HS} + \lambda_{HS} s)}{V''(s)+ \lambda_{HS} v^2}\,,
\ee
under the assumption of small mixing angle, and where $s$ is the VEV of the singlet. From the general formula above we can distinguish two cases: i) if the singlet gets a VEV $s$ due to its potential and $m_\phi\simeq g_* s$, where $g_*$ is some coupling, then the mixing scales as $\gamma\simeq g_*v/m_\phi$; ii) if the singlet gets a VEV only through its interaction with the SM Higgs then the mixing scales as $\gamma\simeq v a_{HS}/ m_\phi^2$ and can be made arbitrarily small. For instance if we assume $a_{HS}=g_*^2 v$, then $\gamma\simeq (g_*v / m_\phi)^2$  and the mixing decouples with one extra power of the ratio between the EW scale and the singlet mass compared to the case~i). We will be dealing mostly with the first class of models here, but the second scenario is useful to keep in mind. 

\paragraph*{Production modes and decay channels} \leavevmode \\

At a high-energy muon collider, the dominant production mode for the scalar $\phi$ comes from VBF~\cite{Buttazzo:2018qqp}. By exploiting the scattering of equivalent Goldstone bosons we can compute both the single and double production analytically.

\begin{figure}
\includegraphics[width=0.5\textwidth]{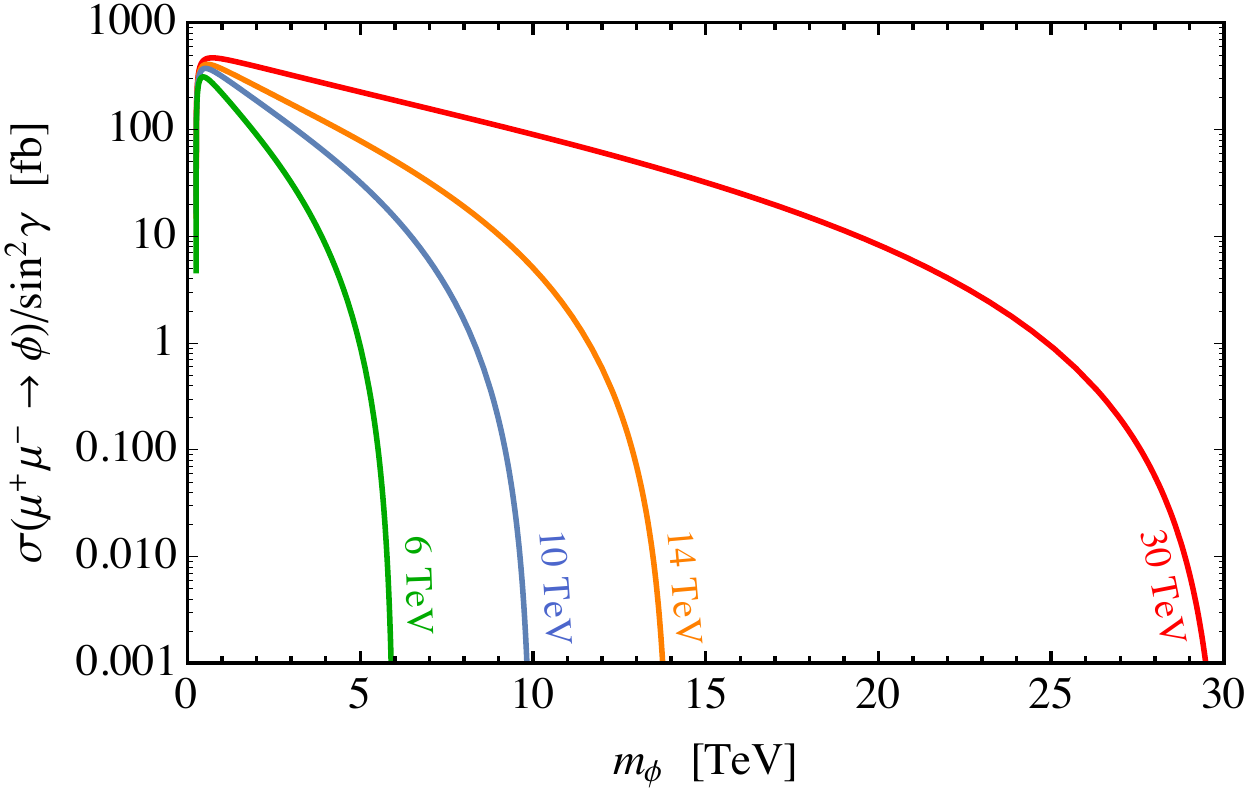}\hfill%
\includegraphics[width=0.48\textwidth]{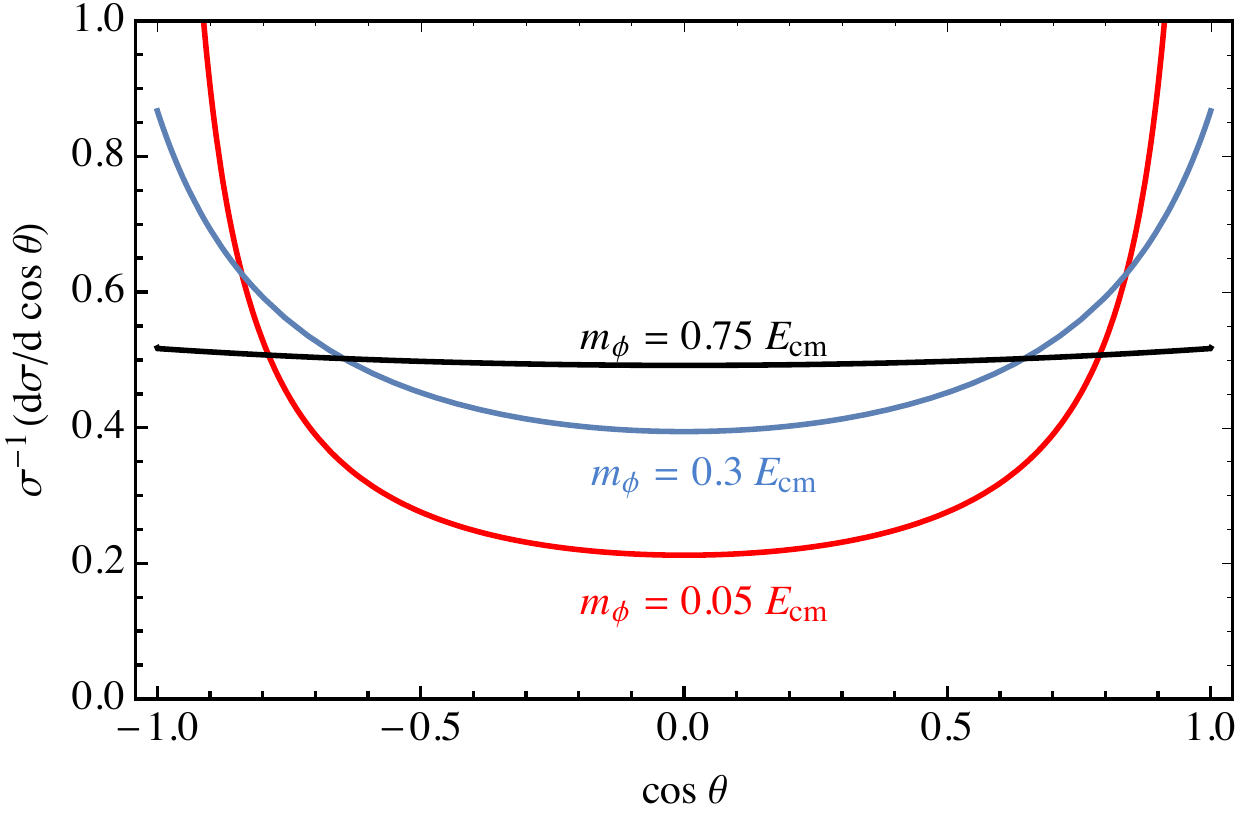}
\caption{Left: Total cross sections for $\ell^+\ell^- \to \phi\nu\bar\nu$ as functions of the mass $m_\phi$, for different collider energies. Right: Differential distributions of the decay products of $\phi\to hh$ in the scattering angle $\theta$, for different values of $m_\phi$; one can see that lighter scalars are more boosted and decay in the forward and backward directions.\label{fig:xsec}}
\end{figure}

Single production proceeds via mixing with the Higgs. Its cross section is proportional to the mixing angle and is only logarithmically sensitive to the mass (at high energies), and it can be simply written as
\be\label{singlet_xsec}
\sigma_{\mu\mu\to \phi}\simeq \sin^2\gamma \frac{g^4}{256\pi^3} \frac{\log (s/m_\phi^2)}{v^2}\,.
\ee
A further dependence on the mass of the singlet-like state is hidden in the mixing $\sin^2\gamma$, as emphasized above. The dependence of the total cross section on the singlet mass is shown in the left panel of Fig.~\ref{fig:xsec}.

Double production, on the contrary, mainly depends on the quartic portal coupling and the mass of the singlet. The total cross section can again be computed analytically exploiting the scattering of equivalent Goldstone bosons, and reads
\be
\sigma_{\mu\mu\to \phi\phi}\simeq \frac{g^4 |\lambda_{HS}|^2}{49152\pi^5} \frac{\log (s/m_\phi^2)
}{m_\phi^2}\,.\label{eq:double}
\ee
The decay channels of the singlet-like scalar are inherited by the mixing with the Higgs boson, and for $m_\phi\gg m_W$ they are related by an approximate $SO(4)$ (custodial) symmetry, which implies
\be
\Gamma(h\to WW)=2\Gamma(\phi\to ZZ)=2\Gamma(\phi\to hh)=\sin^2\gamma\frac{m_\phi^3}{8\pi v^2}\ .
\ee 
The width of the di-top decay channel is  
\begin{equation}
\Gamma(\phi\to t\bar t)=\sin^2\gamma \frac{3 y_t^2 m_\phi}{16\pi}\,,
\end{equation}
and is subleading for all singlet masses.  We note that invisible and/or displaced decays widths could be present if singlet interactions with extra dark sector states are allowed~\cite{Alipour-fard:2018mre}.

\paragraph*{Sensitivity of a muon collider} \leavevmode \\

We now estimate the reach of a muon collider for single scalar production. The main decay channel at a lepton collider is $\phi\to hh \to 4b$, which gives a rather clean signature with a large branching fraction. In the following, we assume ${\rm Br}(\phi\to hh) = 25\%$, as predicted in the $m_\phi \gg m_h$ limit. Other relevant and complementary decay modes are $\phi\to W^+W^-, ZZ$, which are important when the ${\rm Br}(\phi\to hh)$ is small.

We consider the search in the $hh\to 4b$ channel for a resonant $hh$ pair over the SM $\ell^+\ell^- \to hh$ background. The latter is the main source of background, provided the Higgs bosons can be reconstructed with a sufficiently high accuracy. We do not include  other sources of background, which mainly come from $\ell^+\ell^- \to Vh,VV$ (with $V = W,Z$), assuming they can be isolated from the signal, e.g., by including a window cut on the $bb$ invariant masses.
The SM $hh$ background is simulated with {\sc MadGraph} at parton level, without including Higgs decays or detector effects. We instead assume an overall signal selection efficiency of $\epsilon_{hh} = 30\%$, which is consistent with other analyses performed for high-energy lepton colliders~\cite{Buttazzo:2018qqp,deBlas:2018mhx,Roloff:2019crr,Buttazzo:2020uzc}. Furthermore, we impose an acceptance cut $p_T > 20$\,GeV, $\eta > 2$ on the Higgs bosons.
Notice that these cuts are irrelevant for high invariant masses (relevant for heavy singlets), but become important for lower masses, where they cut-off the logarithmic enhancement of the cross section due to the forward singularity of VBF production.

\begin{figure}
\includegraphics[width=0.48\textwidth]{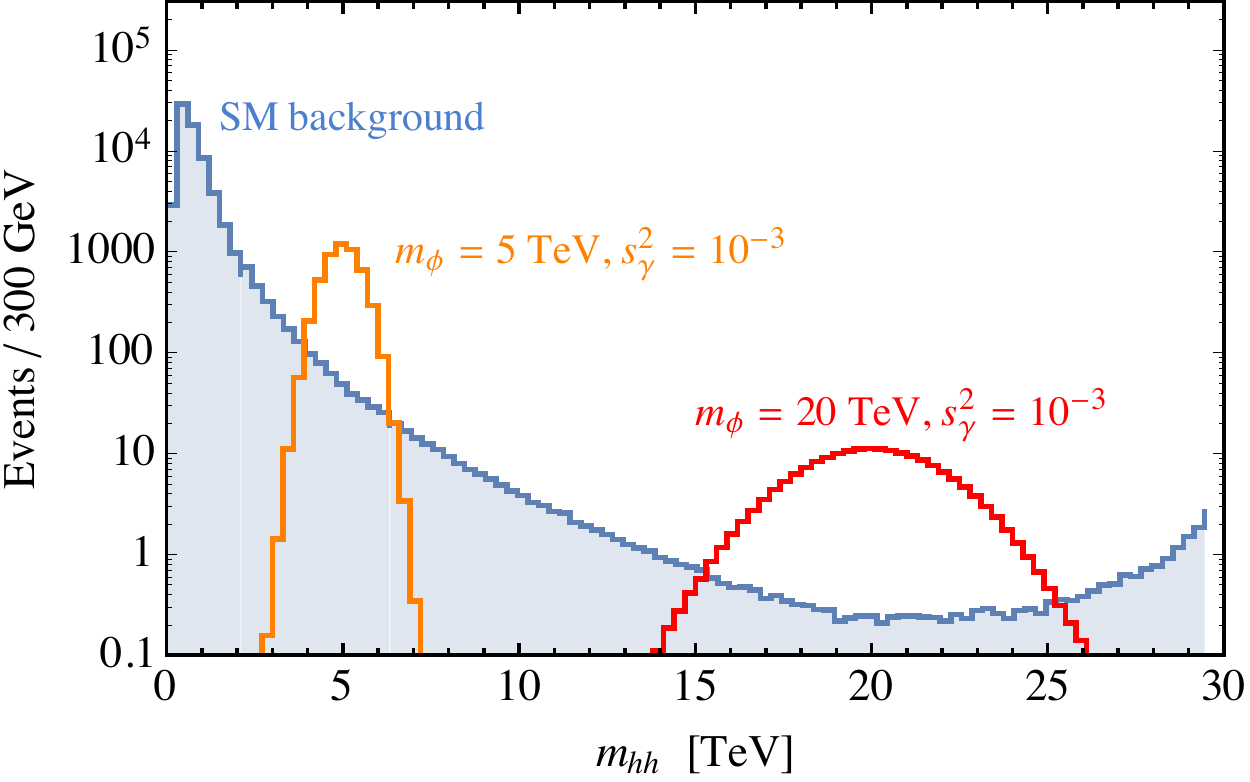}\hfill%
\includegraphics[width=0.48\textwidth]{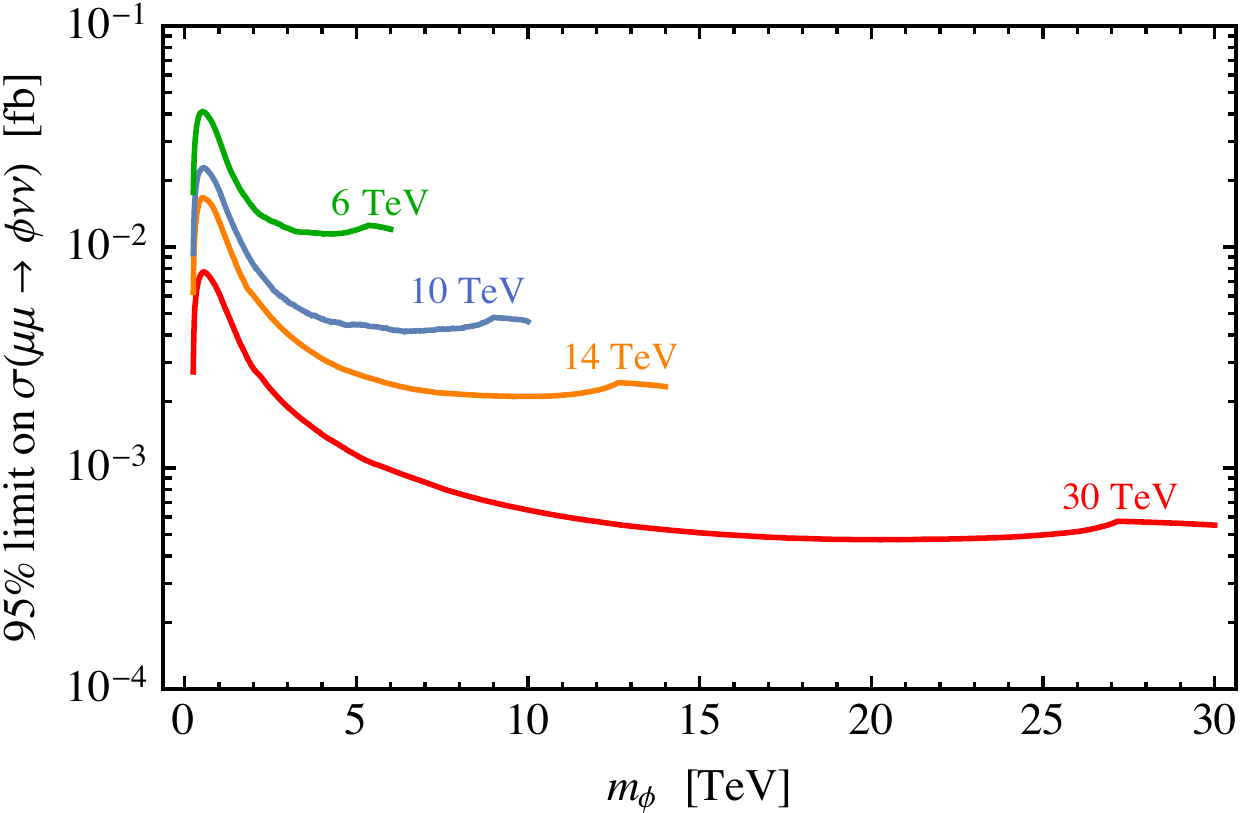}
\caption{Left: Distribution of double-Higgs events in the $hh$ invariant mass at a muon collider with energy $E_{\rm cm} = 30$\,TeV. The SM background $\mu^+\mu^- \to hh\nu\bar\nu$ is shown in blue, and the resonant $\phi\to hh$ production (with $s_\gamma = 10^{-3}$) is superimposed for $m_\phi = 5$\,TeV (orange) and $m_\phi = 20$\,TeV (red). Right: Expected 95\% C.L.\ exclusion on the $\mu^+\mu^- \to \phi\nu\bar\nu$ fiducial cross section at the various muon collider benchmarks of Table~\ref{tab:energylumi}.\label{fig:limits}}
\end{figure}

The invariant mass distribution of the SM events is shown in blue in Fig.~\ref{fig:limits} (left), where two examples of signal are also shown.
The limit on the cross section is obtained performing a cut-and-count experiment around the resonance peak, by requiring the di-Higgs invariant mass $M_{hh}$ to lie within $\pm 15\%$ of the resonance $m_\phi$. The reach on the cross section is obtained imposing
\begin{equation}
\frac{S}{\sqrt{S+B + \alpha^2 B^2}} = 2\,,
\end{equation}
where the factor proportional to $\alpha = 3\%$ takes into account possible systematic uncertainties. The results for the various collider benchmarks are shown in Fig.~\ref{fig:limits} (right).

\begin{figure}
\centering
\includegraphics[width=0.6\textwidth]{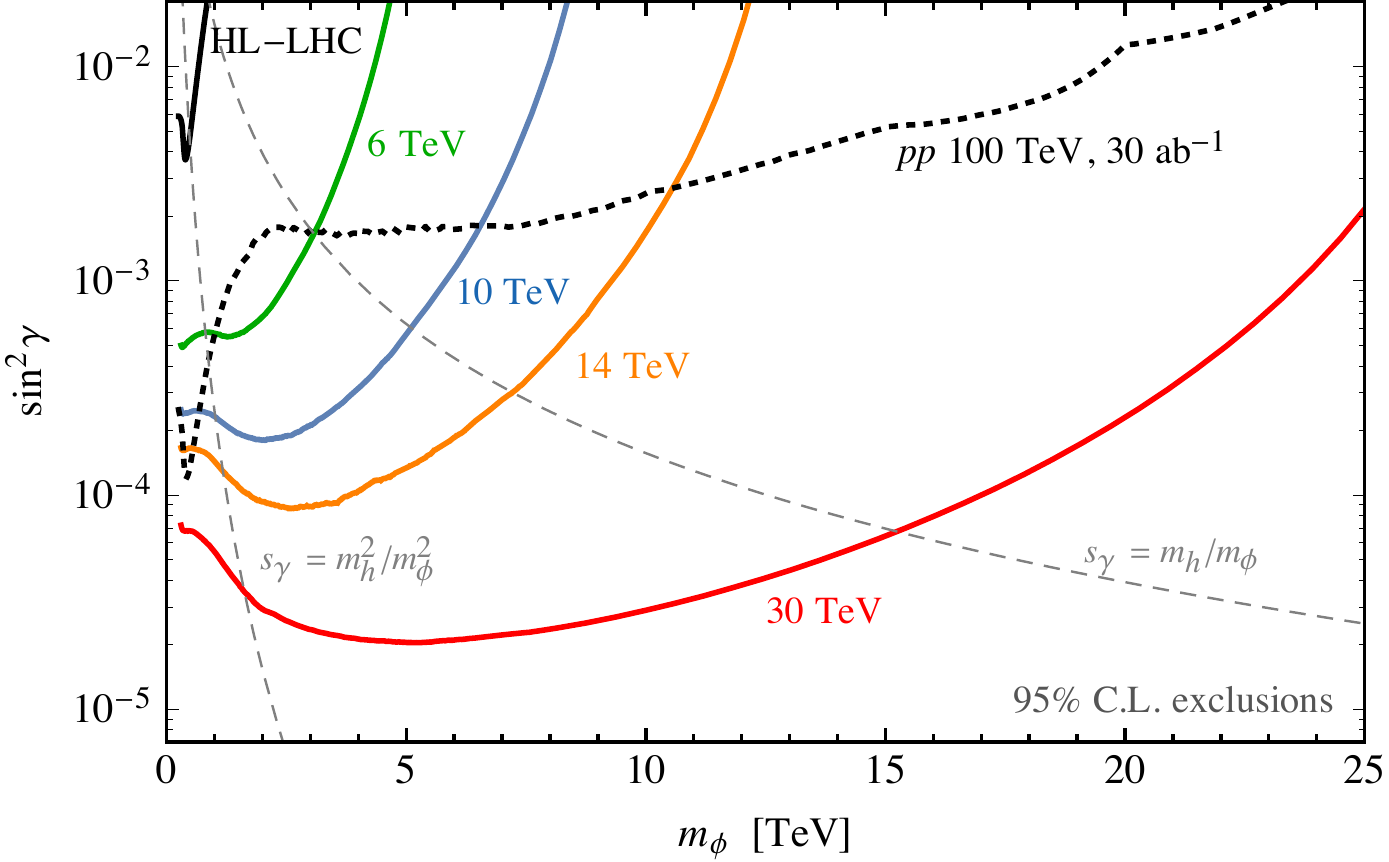}
\caption{Exclusions on the mixing angle of a generic scalar singlet, $\sin^2\gamma = \kappa_V - 1$, as a function of the singlet mass $m_\phi$ for the various collider benchmarks (colored lines). The expected limits at HL-LHC (solid) and a FCC-hh (dashed) are shown as black lines for comparison. The thin dashed lines indicate the two possible scalings of the mixing angle with $m_\phi$ in realistic models with fixed coupling. \label{fig:results}}
\end{figure}

The limits on the cross sections are then translated into a reach on the mixing angle $\sin\gamma$ by comparing it with the signal cross section. The differential cross section as a function of the Higgs scattering angle $\theta_h$ for $\ell^+\ell^-\to \phi\nu\nu \to hh\nu\nu$ can be computed analytically in the EVA (see~\cref{sec:EVA}), and reads
\begin{align}
\frac{\dd\sigma_{\ell^+ \ell^- \to hh\nu\nu}}{\dd \cos\theta_h} &= \frac{m_W^4}{16\pi^3 v^6}\frac{1}{\sin^2\!\theta_h}\Bigg[\!\frac{m_{\phi}^2}{E_{\rm cm}^2} - 1+ \frac{2 m_{\phi}}{E_{\rm cm}\sin\theta_h}\Bigg(\!\!\arctan\frac{E_{\rm cm}\!\cot\!\frac{\theta_h}{2}}{m_{\phi}} - \arctan\frac{m_{\phi}\!\cot\!\frac{\theta_h}{2}}{E_{\rm cm}}\!\Bigg)\!\Bigg].\notag\\[3pt]
\label{differential}
\end{align}
The angular distribution of $hh$ events is plotted in the right panel of \cref{fig:xsec}, for different values of the singlet mass.
This expression is then integrated over the phase-space region defined by the acceptance cuts to find the number of signal events.\footnote{Integrating Eq.~\eqref{differential} over $0<\theta_h<\pi$ gives Eq.~\eqref{singlet_xsec} in the limit $m_\phi \ll E_{\rm cm}$.} The result has been cross-checked by generating the signal events with {\sc MadGraph}, after implementing the singlet model in {\sc FeynRules}, finding perfect agreement. The reach in $\sin^2\gamma$ is shown in \cref{fig:results} (left) as a function of the singlet mass. In the same figure we also show the reach of HL-LHC and FCC-hh for comparison. Notice that a muon collider with c.m.~energy in the tens of TeV range could reach a sensitivity $s_\gamma^2 \lesssim 10^{-4}$ or lower, corresponding to deviations in Higgs couplings of $\mathcal{O}(10^{-4})$, which are beyond the capability of any present or other proposed future collider. This example highlights the great qualitative advantage of a high-energy muon collider in probing the Higgs sector: the longitudinal enhancement of VBF and relatively modest backgrounds give exquisite sensitivity to any additional scalars participating in electroweak symmetry breaking.

\subsection{Dark matter}
\label{sec:dm}
The predominance of apparently non-baryonic matter in the universe remains one of the few unambiguous indicators of physics beyond the Standard Model, and identifying the microscopic properties of dark matter is a central goal of multiple fields. Among the many candidates for particle dark matter, the Weakly Interacting Massive Particle (WIMP) paradigm has long been one of the most compelling. Within this paradigm, dark matter candidates arising as the lightest member of an electroweak (EW) multiplet form a particularly simple class of models \cite{Cirelli:2005uq,Cirelli:2009uv,DiLuzio:2018jwd}. The thermal relic abundance of such ``minimal'' dark matter is fixed strictly in terms of the quantum numbers of the electroweak multiplet in question, picking out a high mass scale between 1 - 23 TeV for $SU(2)_L$ representations ranging from doublets to septuplets. This makes minimal dark matter a motivated but difficult scenario for colliders in light of the high mass scale.  Additionally,  it is challenging from the detector point of view, because the typically small splittings of the EW multiplets suppress the amount of visible energy (and hence, missing momentum)  in a typical event. Nevertheless, the abundant electroweak cross sections and relatively low irreducible backgrounds at a muon collider make it well positioned to search for minimal dark matter, to the point where a muon collider of sufficient energy could potentially render a decisive verdict on the scenario. In this section, we summarize the studies performed in Ref.~\cite{Han:2020uak}, adapting their projections to the optimistic and conservative luminosity targets presented here.

\begin{table}[t]
  \centering
  \renewcommand{\arraystretch}{1.5}
\setlength{\arrayrulewidth}{.3mm}
\setlength{\tabcolsep}{0.8 em}
    \begin{tabular}{c|c|r|c|c|c|c}
    \multicolumn{2}{c|}{\multirow{2}[-3]{*}{Model}} & \multicolumn{1}{c|}{\multirow{2}[-3]{*}{Thermal}} & \multicolumn{4}{c}{5$\sigma$ discovery coverage (TeV)}  \\ \cline{4-7}
    \multicolumn{2}{c|}{$({\rm color}, n, Y)$} & target \hspace{3pt}  & \multicolumn{1}{c|}{mono-$\gamma$} & \multicolumn{1}{c|}{mono-$\mu$} & \multicolumn{1}{c|}{di-$\mu$'s} & \multicolumn{1}{c}{disp. tracks} \\  \hline
    (1,2,\sfrac{1}{2})$$ & Dirac & 1.1 TeV & ---  & 2.8  & ---  & $1.8\text{ - }3.7$  \\ \hline  \hline
    (1,3,0)$$ & Majorana & 2.8 TeV & ---   & 3.7  & ---  & $13\text{ - }14$  \\ \hline
    (1,3,$\epsilon$)$$ & Dirac & 2.0 TeV & 0.9  & 4.6  & ---  & $13\text{ - }14$  \\ \hline  \hline
    (1,5,0)$$ & Majorana & 14 TeV & 3.1  & 7.0  & 3.1  & $10\text{ - }14$ \\ \hline
    (1,5,$\epsilon$)$$ & Dirac & 6.6 TeV &  6.9 & 7.8 & 4.2  & $11\text{ - }14$   \\ \hline  \hline
    (1,7,0)$$ & Majorana & 23 TeV  & 14   & 8.6   & 6.1  & $8.1\text{ - }12$ \\ \hline
    (1,7,$\epsilon$)$$ & Dirac & 16 TeV & 13  & 9.2 & 7.4 & $8.6\text{ - }13$  
    \end{tabular}%
    \caption{
    The fermionic minimal dark matter multiplets considered in this paper (and Ref.~\cite{Han:2020uak}), the mass target set by thermal abundance, and a brief summary of the $5\sigma$ discovery reach at a 30 TeV muon collider under the optimistic luminosity scaling in the four individual search channels described in the text. The 5$\sigma$ discovery reach for muon colliders at $\sqrt s= 3, 6, 10, 14, 30, 100 \text{ TeV}$ under both conservative and optimistic luminosity assumptions is provided in the summary plots in \cref{fig:summary}. More details can be found in Ref.~\cite{Han:2020uak}. 
    \label{tab:WIMP}
    }
\end{table}%

Perhaps the best-known examples of minimal dark matter are the $SU(2)_L$ doublet and triplet, which can be mapped onto the higgsino and wino in supersymmetric theories. However, it is also interesting to consider multiplets with quantum numbers $(1, n, Y)$ under the SM gauge group $SU(3)_{C} \times SU(2)_{L} \times U(1)_{Y}$. For definiteness, we restrict our attention to fermions, whose only renormalizable SM interactions are with electroweak gauge bosons and whose mass arises from a vector-like mass parameter. The resulting mass degeneracy among members of the multiplet is split by EW loop corrections \cite{Thomas:1998wy,Buckley:2009kv,Cirelli:2005uq,Cirelli:2009uv,Ibe:2012sx}. A number of considerations shape the motivated values of $n$ and $Y$. For $n > 7$, the large electroweak charge induces a Landau pole in the Standard Model gauge couplings about one to two orders of magnitude above the mass of the EW multiplet \cite{DiLuzio:2015oha}. As such, we will restrict our attention to $n\leq 7$. For a given $n$, a specific value of $Y$ ensures that the lightest eigenstate of the EW multiplet is neutral and hence a suitable dark matter candidate. 

For odd-dimensional multiplets $(1, n=2T+1, Y)$ with $T \in \mathbb{Z}^+$, $Y=0$ ensures the electrically neutral member $\chi$ is always the lightest mass eigenstate in the multiplet, and fermions in these multiplets may be either Majorana or Dirac. Beyond the renormalizable level, irrelevant operators could allow the dark matter to decay. Such operators could be forbidden by introducing a small hypercharge $Y = \epsilon$ \cite{DelNobile:2015bqo} such that the dark matter acquires an electric charge $Q = \epsilon$. However, other alternatives are possible, e.g., gauged discrete symmetries, which avoid the existence of global symmetries and ensure that the neutral particle is a good dark matter candidate. We will remain agnostic as to the particular mechanism ensuring the stability of the dark matter candidate, since it is largely irrelevant for the signals at a muon collider. In what follows, the notation $(1, n=2T+1, \epsilon)$ denotes a Dirac multiplet stabilized by additional considerations, while $(1, n=2T+1, 0)$ denotes a Majorana multiplet. 

For even-dimensional multiplets, $Y=(n-1)/2$ ensures the lightest mass eigenstate of the multiplet will be electrically neutral. Limits from direct detection already exclude all minimal cases with $Y \neq 0$, but even-dimensional multiplets may be rendered viable by introducing another state that mixes with the minimal multiplet after electroweak symmetry breaking. This generates a small Majorana mass splitting between the neutral Dirac fermion pair \cite{Cirelli:2009uv}, evading bounds from direct detection. For simplicity, in this case, we assume EW loop corrections still dominate the mass splitting between the neutral and charged members of the multiplet. This assumption is compatible with a relatively high scale of additional physics. For instance, when the additional state of mass $M$ is heavy enough to be integrated out, the leading mass splitting is typically due to a dimension-5 operator which generates $\Delta m \propto v^2/M$. Requiring this splitting to be large enough to evade direct detection bounds but small enough to avoid altering the mass ordering from electroweak loop corrections implies $M \sim (10$ - $1000)$ TeV. While this additional physics may itself be probed at a muon collider, this is more model-dependent and will not be further studied here. In what follows, we will focus on the case of the electroweak doublet ($n=2$) as representative of the class of even-dimensional multiplets, with the above assumptions to reconcile limits from direct detection.

Of course, minimal dark matter candidates may also arise from real or complex scalar multiplets carrying electroweak quantum numbers. Scalars admit more renormalizable couplings to the Standard Model, most notably through Higgs portal operators of the schematic form $\chi \chi^\dagger H H^{\dagger}$. These couplings can induce significant tree-level mass splittings after electroweak symmetry is broken, introducing a high degree of model dependence. We leave this case for future study.

As noted above, the leading interactions between a minimal dark matter multiplet and the Standard Model are strictly controlled by the multiplet's EW quantum numbers. These interactions control the thermal relic abundance of cold dark matter resulting from freeze-out. Assuming that this is the sole source of the dark matter's abundance, matching current observations~\cite{Aghanim:2018eyx} determines the dark matter's ``thermal target'' mass. The various fermionic dark matter candidates and their corresponding thermal target masses are enumerated in \cref{tab:WIMP}. It bears emphasizing that the perturbative calculation of the thermal target mass is subject to large corrections from both Sommerfeld enhancement~\cite{Belotsky:2005dk,Hisano:2006nn,Cirelli:2007xd} and bound state effects~\cite{An:2016gad,Mitridate:2017izz}. For the purposes of this discussion, we mainly use the thermal targets presented in Ref.~\cite{DiLuzio:2018jwd}, which themselves are primarily obtained by including Sommerfeld corrections to results in Ref.~\cite{DelNobile:2015bqo}. A notable exception is the quintuplet Majorana fermion, for which bound state effects are significant; these lift the thermal target from 9 TeV to 14 TeV~\cite{Mitridate:2017izz}. This is in contrast to the triplet Majorana fermion, for which bound state effects do not shift the thermal target relative to the Sommerfeld calculation~\cite{Mitridate:2017izz}. For the septuplet Majorana fermion, we obtain an approximate target by using the fact that the degrees of freedom decrease by a factor of two relative to the Dirac fermion, pushing the thermal target higher by a factor of $\sqrt{2}$ relative to the Dirac case \cite{DelNobile:2015bqo}. Needless to say, all thermal targets quoted here are subject to residual theoretical uncertainties. Experimental coverage of these targets is a compelling goal for a future collider program. 

High-energy muon colliders are exceptionally well-positioned in this regard. There are a number of promising channels in which to search for minimal dark matter, including mono-photon, mono-muon, and VBF di-muon final states with an inclusive missing mass signature~\cite{Han:2020uak}. Alternately, the production of charged particles in the multiplet followed by decay into dark matter (and soft tracks) gives rise to a promising disappearing track signature, where the small mass splitting due to EW corrections translates into a macroscopic distance traversed by the charged particles before they decay. This channel's performance is subject to considerable uncertainties owing to the currently-unknown beam-induced backgrounds at a muon collider, but may significantly enhance the reach~\cite{Capdevilla:2021fmj}. Here we summarize the performance of each channel, following~\cite{Han:2020uak}.

\begin{itemize}
\item \textbf{Mono-muon:} In this channel, a charged particle of the EW multiplet is produced in association with a neutral one, leading to a single muon in the final state.  This is a unique signal for a muon collider, with considerable discovery potential on account of a high signal-to-background ratio.  Signal production is predominantly from the VBF processes $ZW, \gamma W \to \chi \chi$, although the process $\gamma Z \to \chi\chi$ also contributes to this channel when one of the initial state muons escapes detection following collinear emission of a $\gamma$ or $Z$. Given the central role of VBF production for the signal, this channel is especially promising for $m_\chi \ll \sqrt{s}/2$ and lower-dimensional EW multiplets, where the reach can exceed the mono-photon channel for $n \leq 3$. Higher-dimensional representations enjoy correspondingly higher rates and reach, but this channel cannot quite cover the (much higher) thermal target in these cases. The signal decreases with the dark matter mass as $1/m_\chi^4$, so that the reach in this channel does not extend all the way to the kinematic limit set by $\sqrt{s}/2$. The reach in mass scales approximately linearly with energy under our optimistic luminosity scaling, $\mathcal{L} \propto s$.
\item \textbf{Mono-photon:} In this channel, the particles in the EW multiplet are pair produced in association with a photon from either initial or final state radiation. This channel is particularly effective for higher-dimensional multiplets due to the corresponding coupling enhancement, which scales as $n^2$, and the reach exceeds that of the mono-muon channel for $n \geq 5$. It again does not quite reach the kinematic limit of $\sim \sqrt{s}/2$, despite coming close for $n=7$. The primary challenge for the reach in this channel is the sizable irreducible mono-photon background, leading to a signal-to-background ratio on the order of of $S/B < 10^{-2}$. 
\item \textbf{Di-muon:} In this channel, the particles in the EW multiple are pair-produced in neutral VBF from fusion of $ZZ$, $Z\gamma$, and $\gamma\gamma$, where the two final state muons are tagged.  The acceptance of this channel is sensitive to the assumed muon angular acceptance; here, we have taken the muon acceptance to extend out to $|\eta_\mu|<2.5$, leading to a significant reduction of the signal rate on account of the low likelihood of tagging two muons in the forward and backward regions. Even so, for higher-dimensional EW multiplets such as $n=7$, this channel provides coverage competitive with the mono-photon channel while being more robust against systematics. Improving the reach in this channel motivates advanced detector design that would cover more of the forward regime, for instance, covering $2.5<|\eta_\mu|<4.0$ (or potentially even out to $|\eta_\mu| < 8$ \cite{delphesTalk}).
\item \textbf{Disappearing tracks:} In this channel, the charge $\pm 1$ particle in the EW multiplet is pair produced and decays into the dark matter plus soft particles with a long lifetime due to the small radiative mass splitting. For the cases considered here, $c \tau$ ranges from 0.37~cm to 5.6~cm. If the charged particle hits several layers of the tracker before decaying, this results in the unique signature of a disappearing track, a potentially low-background process. However, making accurate projections for the reach of this channel at a muon collider is hampered by our current ignorance of tracker design and beam-induced backgrounds. Here we present an estimate based on the combination of a singlet displaced track plus another tagging object such as a photon (with the expectation that this will be required to suppress backgrounds), requiring tens of signal events for discovery. Requiring two displaced tracks would necessarily provide further background suppression, albeit with a significant loss of rate; this channel would be useful to study further in the event that backgrounds for the single-track channel prove to be prohibitive. Focusing on the single-track final state with an additional tagging object, the mono-photon channel with one disappearing track will have the largest signal rate, significantly extending the reach for all odd-dimensional cases. However, this channel fails to reach the kinematic threshold owing to the boost required for the charged particle to leave enough hits in the tracker before decaying.\footnote{In this case, further sensitivity may be obtained from using timing information~\cite{Liu:2018wte}. However, the large out-of-time contribution from beam-induced backgrounds requires more detailed studies.}
The triplet enjoys the greatest increase in sensitivity from this channel, coming close to the kinematic threshold, while for the doublet, this channel is stronger than the mono-muon channel. 
\end{itemize}

\begin{figure}[t]
\centering
\includegraphics[width = 0.49 \textwidth]{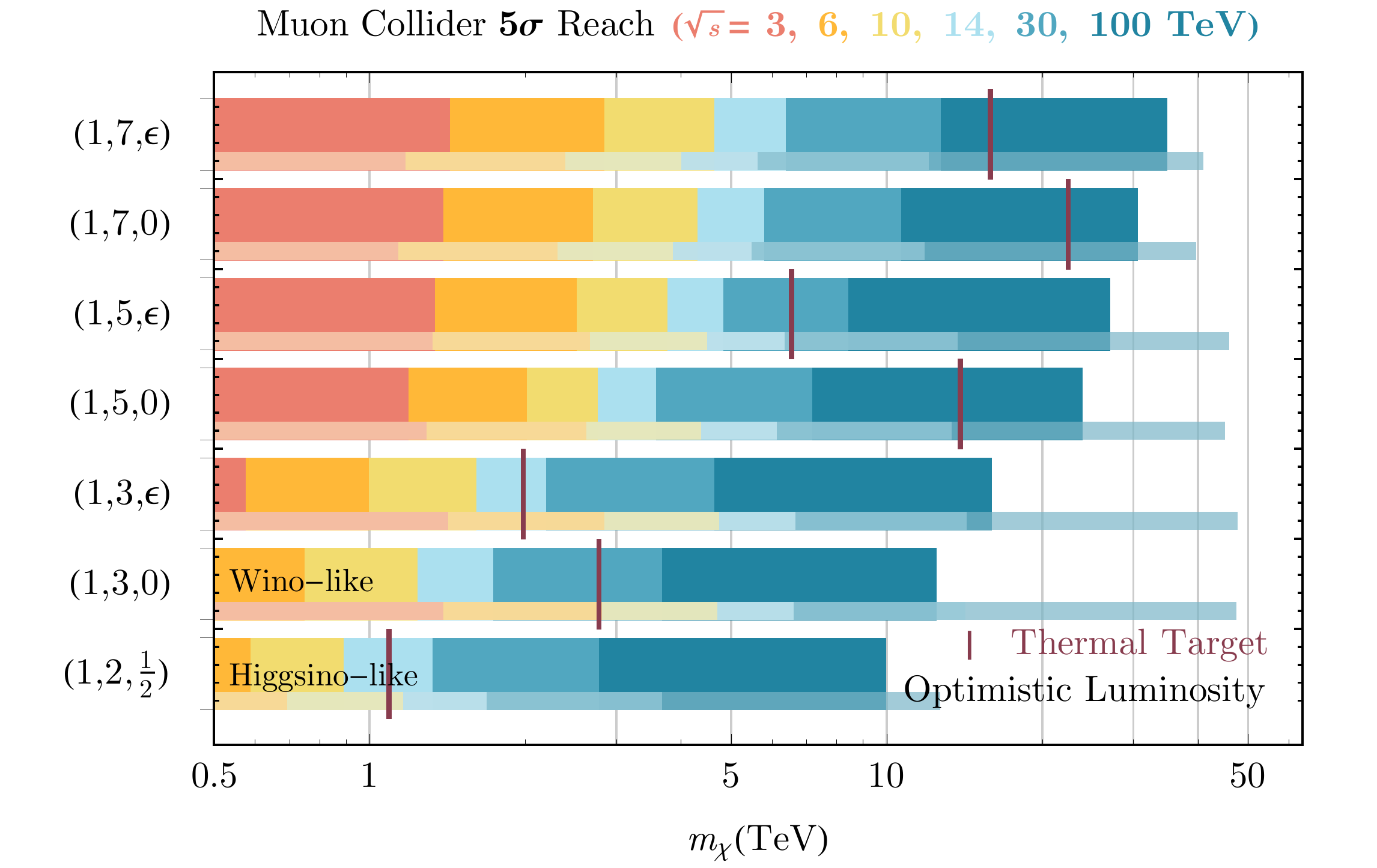}
\includegraphics[width = 0.49 \textwidth]{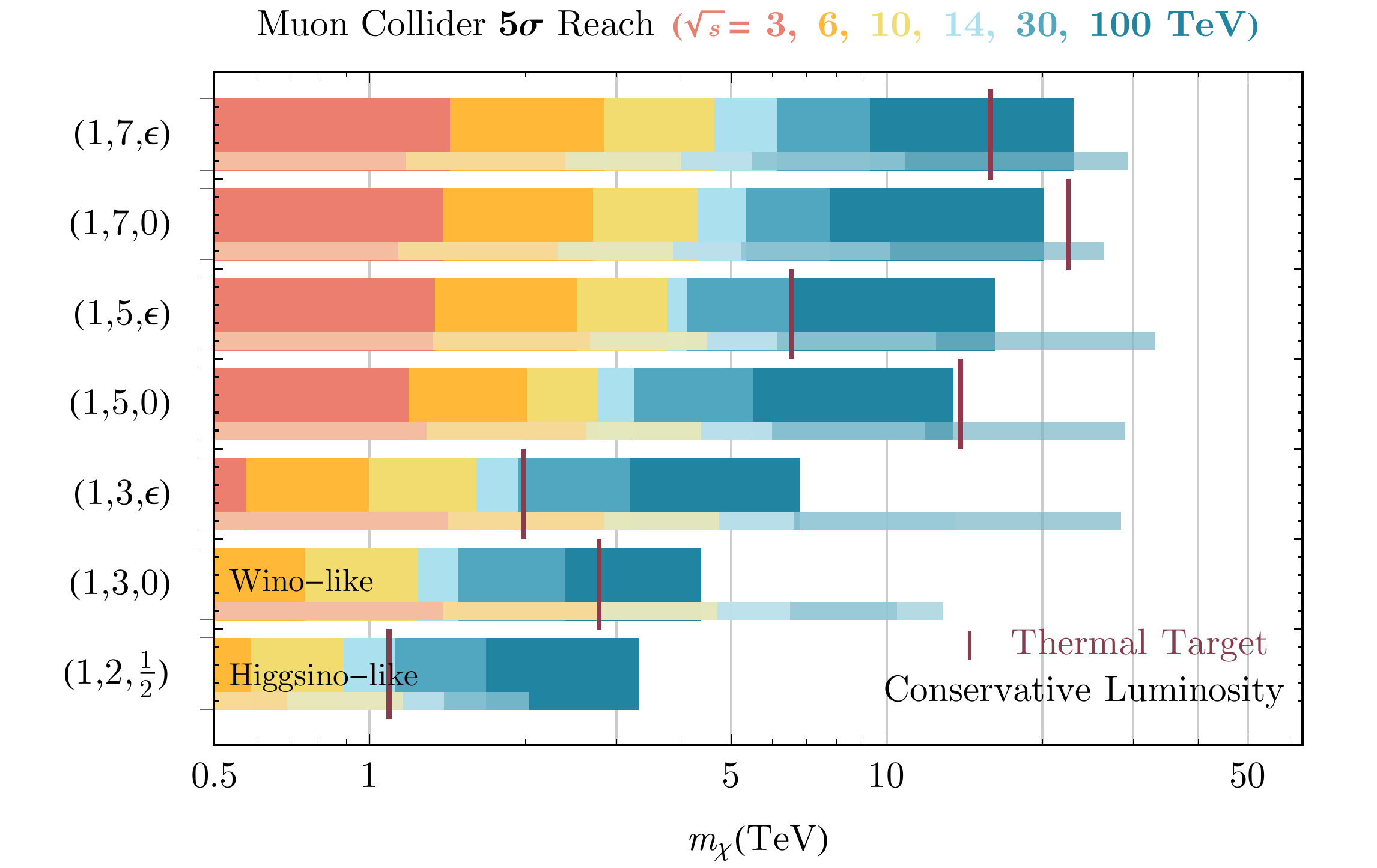}
\caption{Summary of the $5\sigma$ discovery reach for minimal dark matter of muon colliders operating at various center of mass energies with the optimistic luminosity scaling (left) and conservative luminosity scaling (right). The wider bars indicate the combined reach from missing mass searches in the mono-photon, mono-muon, and VBF di-muon channels. The thinner bars indicate the estimated reach from the mono-photon plus one disappearing track search. The maroon vertical bars denote the thermal target for a given minimal DM multiplet. More details, including the detailed reaches for each channel and collider energies, can be found in Ref.~\cite{Han:2020uak}. We also note a recent dedicated study of disappearing tracks which projects a 10 TeV muon collider can cover 2.75 TeV (4.55 TeV) higgsinos (winos)~\cite{Capdevilla:2021fmj}.
}
\label{fig:summary}
\end{figure}

The $5\sigma$ discovery reach of muon colliders operating at various center of mass energies is summarized in \cref{fig:summary} for the optimistic (conservative) integrated luminosity scaling scenarios defined in \cref{tab:energylumi}. The sensitivity obtained by the combination of missing mass searches in the mono-photon, mono-muon, and VBF di-muon channels is shown separately from the sensitivity of the displaced track search. When combining the missing mass channels in the optimistic luminosity scaling scenario (left panel), the overall reach does not extend to the kinematic limit $m_\chi \sim \sqrt{s}/2$ (most notably for multiplets with $n \leq 3$) due to the low signal-to-background ratio. It is possible to cover (with $2 \sigma$ confidence) the thermal targets of the doublet and Dirac triplet with a 10 TeV muon collider, while a 30 TeV option would suffice for the Majorana triplet. The thermal targets of Dirac (Majorana) quintuplet would be covered by muon colliders operating at 30 (100) TeV, while a 100 TeV collider would also cover the thermal target for the septuplet. 

Rather than considering the reach of the benchmark collider energies, it is also interesting to note the minimum collider energy that would cover a given multiplet, assuming integrated luminosity scales with $s$. From this perspective, a Majorana triplet can be reached by a 20 TeV muon collider (still assuming integrated luminosity scales with $s$). A Majorana quintuplet can be covered by a 50 TeV muon collider, while a septuplet can be covered by a 70 TeV muon collider. The thermal targets of all the minimal multiplets considered here could be discovered at the $5 \sigma$ level by a muon collider operating at 75 TeV.

Finally, we emphasize that the disappearing track signal has excellent potential, bringing the reach close to the kinematic threshold $m_\chi \sim \sqrt{s}/2$ on the basis of the current study \cite{Han:2020uak}. For instance, a 10 TeV muon collider alone could reach the thermal target of both doublet and triplet cases with a disappearing track search, motivating further studies and careful consideration for detector design.


\subsection{Naturalness}
\label{sec:nat}
The hierarchy problem is a prime motivator that new physics should be accessible at colliders, as it strongly correlates the mass scale of additional degrees of freedom with those of the Standard Model. Precisely what degrees of freedom appear at scales indicated by the hierarchy problem is much less definite; in recent years it has become increasingly clear that there exists a plethora of solutions with wildly varying signatures. Nonetheless, the spectrum of solutions can be usefully divided into two categories: solutions of the ``big'' hierarchy problem, namely those reaching from the weak scale all the way to the putative scale of quantum gravity, and solutions of the ``little'' hierarchy problem, extending from the weak scale to the highest scales directly probed (thus far) by experiments. 

There are two known solutions to the ``big'' hierarchy problem: compositeness and supersymmetry. The lack of weak scale evidence for either solution suggests the existence of a mass gap between the Higgs and whatever physics resolves the hierarchy problem. Such a mass gap implies a significant degree of fine tuning, somewhere between the percent and per mille level depending on the details of the UV completion. While it is entirely possible that the weak scale is finely tuned -- after all, we do not actually know how Nature computes fine tuning -- a robust commitment to naturalness could suggest the existence of additional physics that bridges the gap between the weak scale and the appearance of supersymmetry or compositeness. Resolutions to this ``little'' hierarchy problem need only span an order of magnitude in energy in order to reconcile the paucity of new physics at the weak scale with the expectations of naturalness. In contrast to the sparsity of qualitative solutions to the big hierarchy problem, there are innumerable solutions to the little hierarchy problem consistent with current data, ranging from dynamical mechanisms that relax the Higgs mass \cite{Graham:2015cka} to symmetry-based mechanisms that reside in ``dark'' hidden sectors \cite{Chacko:2005pe, Craig:2015pha}. Their signatures are equally diverse, often falling outside the scope of conventional collider signals. 

Solutions to the hierarchy problem reduce the UV sensitivity of the Higgs mass parameter, making it possible to understand what sets the value of the weak scale and, ultimately, why electroweak symmetry is broken in the first place. To systematically test the naturalness of the weak scale, one would ideally like to pursue two lines of experimental inquiry: leveraging precision to directly test solutions to the ``little'' hierarchy problem at or around the weak scale, and leveraging energy to reach the scale at which states associated with resolution of the ``big'' hierarchy problem begin to appear. The great advantage of a high-energy muon collider is that, provided sufficient energy and luminosity, it may achieve both goals. On the one hand, the relatively low background rate and clean environment make it a promising tool for discovering new light states with weak (or no) Standard Model quantum numbers. On the other hand, the high c.m.~energy gives it the reach to discover states well above the weak scale. In what follows, we illustrate this potential by considering aspects of muon collider sensitivity to representative solutions of the ``big'' and ``little'' hierarchy problems.

\subsubsection{The ``big'' hierarchy problem: supersymmetry}

The many superpartners predicted by supersymmetry give rise to a host of experimental signatures; see e.g.~\cite{Craig:2013cxa} for an overview. Here we will focus on the three states most closely tied to the naturalness of the Higgs potential: the higgsino, stop, and gluino, which respectively contribute to the Higgs potential at tree level, one loop, and two loops; these are the calling cards of ``natural supersymmetry'' \cite{Dimopoulos:1995mi, Pomarol:1995xc, Cohen:1996vb, Brust:2011tb, Papucci:2011wy}. In addition, we will explore a unique opportunity available to high-energy lepton colliders: probing low-scale supersymmetry breaking sectors through a direct search for the gravitino.

\paragraph{Higgsinos, stops, and gluinos} \leavevmode\\ 

The specific collider reach for a given superpartner requires detailed simulation as well as knowledge of beam and detector effects that is not available at this stage. Nonetheless, we can make some meaningful estimates that should hold up to $\mathcal{O}(1)$ factors. In general, the mass reach for superpartners with electroweak quantum numbers is closely correlated with the c.m.~energy of the collider. Although the $s$-channel production for these states falls as $1/s$, it remains $\gtrsim \mathcal{O}({\rm few\, tens})$ of attobarns at $\sqrt{s} = 30$ TeV and $\gtrsim \mathcal{O}({\rm few})$ attobarns at $\sqrt{s} = 100$ TeV independent of the mass $\widetilde{m}$ of the superpartner, provided $\widetilde{m} \ll \sqrt{s} /2$. As $\widetilde{m} \rightarrow \sqrt{s} / 2$, the mass dependence of $s$-channel production typically enters with the velocity scaling $\propto \beta$ for fermionic superpartners and $\propto \beta^3$ for scalar superpartners. Even in the latter case, the cross section only falls by an order of magnitude once $\widetilde{m} \sim 0.9 \times \sqrt{s}/2$. As long as the integrated luminosity at a collider with c.m.~energy $\sqrt{s} \gtrsim 30$ TeV is at the level of inverse attobarns, there will be enough signal events to discover superpartners up to $\widetilde{m} \lesssim \sqrt{s}/2$ provided backgrounds can be eliminated while retaining high signal efficiency.

The mass reach then becomes a question of the distinctiveness of the final state. For decays with large available phase space, given the relatively low irreducible backgrounds compared to hadron colliders, the final state can be made essentially background-free with high signal efficiency. Consider, for example, the decay $\tilde t \rightarrow t + \chi^0_1$ with $m_{\tilde t} \gg m_t + m_{\chi}$. The cross section for $s$-channel $t \bar t$ production is always within an order of magnitude of the $\tilde t\, \tilde t^*$ cross section until velocity suppression becomes significant. The cross section for the VBF mode $\mu^+ \mu^- \rightarrow t \bar t + \nu \bar \nu$ does not meaningfully exceed the $s$-channel $t \bar t$ cross section until $\sqrt{s} \sim 5$ TeV, at which point it grows relative to the $s$-channel cross section as $\sim s \log^3(s)$, exceeding the signal cross section by 3-4 orders of magnitude by $\sqrt{s} = 100$ TeV. Even at such high energies, this is far less daunting than the $t \bar t$ background to the same final state at the LHC, and sufficiently strong missing energy requirements and cuts on the distribution of visible particles in the final state should substantially reduce background with high signal efficiency. Thus, to first approximation, for distinctive final states we might assume the mass reach will be of order $\widetilde{m} \sim 0.9\times \sqrt{s} /2$ up to $\sqrt{s} = 100$ TeV.

Not all supersymmetric final states are so distinctive, however. For the higgsino, if Standard Model radiative corrections are the only source of mass splitting, then there is little phase space for missing energy without additional initial-state radiation. The mass reach then becomes more sensitive to backgrounds. Of course, if there is additional splitting in the higgsino multiplet, e.g., due to mixing with a partially decoupled bino or wino, then the final state rapidly becomes more distinctive and the estimate of the reach follows the same logic as above.

With this in mind, we estimate the reach for various superpartners, beginning with the higgsino. The mass of the higgsino is the most immediate measure of fine tuning in the Higgs potential, since supersymmetry relates the masses of the higgsino and Higgs doublets at tree level. In general the contribution of the higgsino to the tuning of the weak scale\footnote{Quantified here by the Barbieri-Giudice measure \cite{Barbieri:1987fn} $\Delta_{M^2} = |\partial \ln m_h^2 / \partial \log M^2|$ in terms of $m_h^2$. As always, we emphasize that individual fine-tuning measures should be taken with a grain of salt given our ignorance of nature's prescription, and is only used here to provide a qualitative guide.} is parametrically of the form
\begin{equation}
\Delta_{\tilde h} \simeq \frac{2 m_{\tilde h}^2}{m_h^2}\,.
\end{equation}
From this we can conclude that 10\% tuning ($\Delta = 10$) corresponds to $m_{\tilde h} \sim 300$ GeV, while the percent and per mille levels are reached by $m_{\tilde h} \sim 900$ GeV and $m_{\tilde h} \sim 2.8$ TeV, respectively.

For a pure higgsino multiplet with only Standard Model radiative splittings, the final state is indistinct and a detailed study is required to forecast reach as a function of $\sqrt{s}$. Here we may rely on the dark matter analysis presented in Section \ref{sec:dm} for a pair of $SU(2)_L$ doublets with hypercharge $\pm 1/2$, which details the potential reach under both optimistic and conservative luminosity assumptions. Under the optimistic (conservative) luminosity assumptions, this analysis suggests that percent-level tuning can be probed at the $2 \sigma$ level by a muon collider operating at $\sqrt{s} = 6$ TeV (10 TeV), while per mille-level tuning is accessible by a $\sqrt{s} = 30$ TeV (100 TeV) collider. The ultimate exclusion limit of a $\sqrt{s} = 100$ TeV collider under optimistic luminosity scaling approaches $m_{\tilde h} \sim 15$ TeV, corresponding to tuning at more than the {\it per myriad} level. But this is ultimately the worst-case scenario; even a modest splitting in the higgsino multiplet would make it possible to significantly suppress backgrounds beyond those considered here, and once sufficient phase space becomes available for on-shell decays within the higgsino multiplet, bounds should approach the $m_{\tilde h} \sim 0.9 \sqrt{s} /2$ reach corresponding to distinctive final states. In this limit, percent tuning becomes accessible at $\sqrt{s} \gtrsim 2$ TeV, per mille tuning at $\sqrt{s} \gtrsim 6$ TeV, and per myriad tuning at $\sqrt{s} \gtrsim 20$ TeV. 

Now we turn to the stop, whose contribution to the tuning of the weak scale at leading-logarithmic order is
\begin{equation}
\Delta_{\tilde t} \simeq \frac{3y_t^2}{4 \pi^2}  \frac{m_{\tilde t}^2}{m_h^2} \log\frac{\Lambda}{m_{\tilde t}}\,,
\end{equation}
where $\Lambda$ is a UV scale at which SUSY breaking is communicated; here we conservatively take $\Lambda = 10 \times m_{\tilde t}$. Here 10\% tuning corresponds to $m_{\tilde t} \sim 950$ GeV, percent tuning to $m_{\tilde t} \sim 3$ TeV, and per mille tuning to $m_{\tilde t} \sim 9.5$ TeV. 

\begin{figure}[t!] 
   \centering
   \includegraphics[trim = 20 0 0 0, width = 0.95\textwidth]{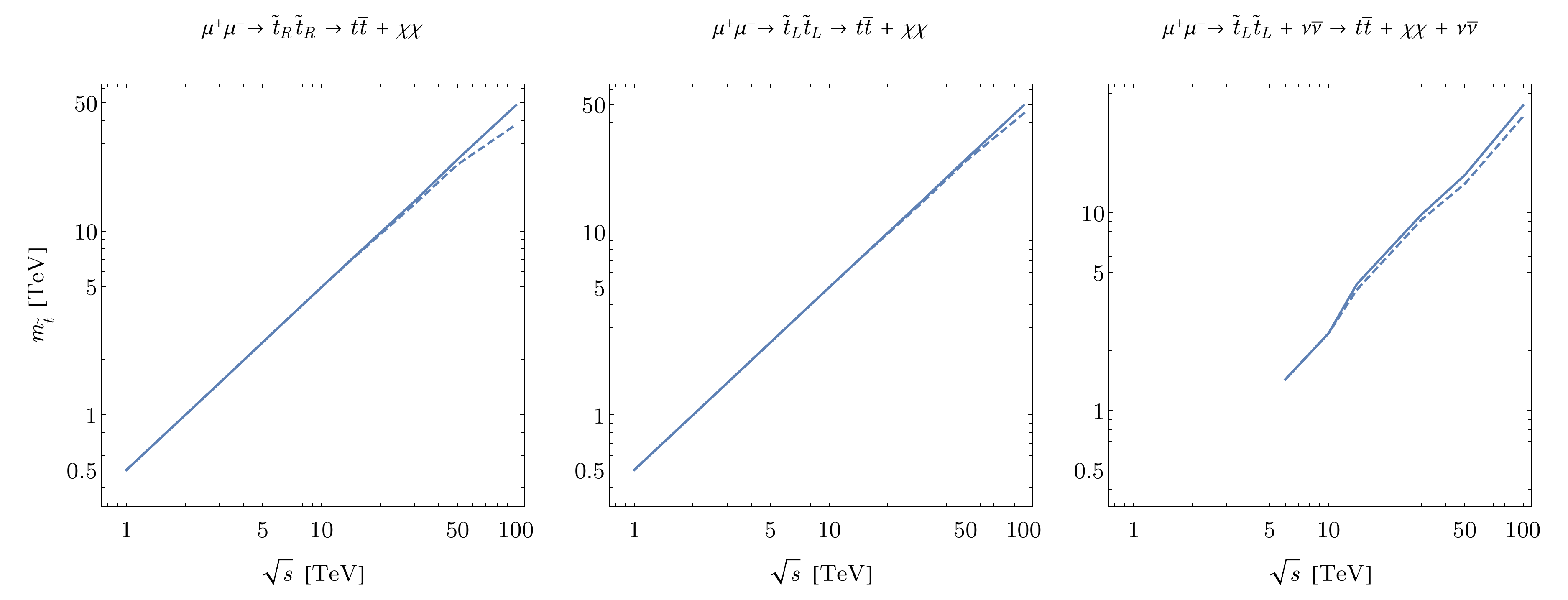} 
   \caption{Exclusion reach on a stop-top-neutralino simplified model as a function of $\sqrt{s}$ and $m_{\tilde t}$ with $m_\chi = 1$ GeV, assuming ``optimistic'' (solid) and ``conservative'' (dashed) integrated luminosity scaling as detailed in the body of the text. Left: Limit on pair production of an RH stop in muon annihilation. Center: Limit on pair production of an LH stop in muon annihilation. Right: Limit on pair production of an LH stop in vector boson fusion. }
   \label{fig:stopex}
\end{figure}

Barring a high degree of degeneracy between the lightest stop and its decay products, the final state is quite distinctive and we can anticipate reach scaling as $m_{\tilde t} \sim 0.9 \sqrt{s} /2$. To validate this expectation, we perform a parton-level analysis for two simplified stop-top-neutralino models: one in which the stop is an $SU(2)_L$ singlet denoted $\tilde t_R$, the other in which it is part of an $SU(2)_L$ doublet denoted $\tilde t_L$. In both models the neutralino $\chi$ is taken to be bino-like. To ensure the relevant production processes are not contaminated by contributions from heavy superpartners, we construct both models in {\sc FeynRules} using the $t$-channel dark matter framework \cite{Arina:2020udz} with couplings set to their supersymmetric values. We then simulate pair production of the stop in both muon annihilation and VBF in {\sc Whizard}, followed by the decay of the stop to a top quark and neutralino. We generate samples for a range of stop masses up to $\sqrt{s}/2$, keeping the neutralino effectively massless with $m_\chi = 1 \text{ GeV}$. We consider the leading background to be VBF $t \bar t$ production, for which the $W^+W^-$ fusion contribution yields a $t \bar t$ + missing energy final state closely mimicking the signal. For simplicity, we consider perfect reconstruction of the top quarks in the final state. We then separately consider two possible cuts to separate signal from background: (1) a cut on the $M_C$ variable \cite{Tovey:2008ui} constructed from the four-momenta of the $t$ and $\bar t$,
\begin{align}
M_C = \sqrt{2 \big(E_1 E_2 + \vec p_1 \cdot \vec p_2 + m_t^2\big)}\,,
\end{align}
for which the signal has an endpoint at $M_C^{\rm max} = m_{\tilde t}$ assuming a massless neutralino, and (2) a cut on the missing transverse momentum $\slashed{p}_T$. For each signal benchmark, we determine the $M_C$ and $\slashed{p}_T$ cuts that maximize $S/\sqrt{B}$, translating it into the exclusion reach shown in Fig.~\ref{fig:stopex} corresponding to $S/\sqrt{B} = 2$ under either the optimistic or the conservative luminosity scaling. For the right-handed stop, the cross section for VBF pair production is generally too small to be a meaningful contribution to the total signal rate, while for the left-handed stop both muon annihilation and VBF production play an interesting role. In each case, the $\slashed{p}_T$ cut alone is sufficient to provide sensitivity all the way up to $m_{\tilde t} \sim \sqrt{s} /2$, save for the largest values of $\sqrt{s}$ where some degradation of the limit is observed under conservative luminosity assumptions due to the significant size of the $t \bar t + \nu \bar \nu$ background at these energies. Of course, this highly simplified analysis neglects a host of relevant effects associated with the decay and reconstruction of top quarks, energy resolution, forward reconstruction, and beam-induced backgrounds, but it supports the supposition that significant kinematic separation of signal and background is possible, and similar performance is likely achievable with an optimized selection using all kinematic information.

As such, we consider the stop mass reach of muon colliders operating at various values of $\sqrt{s}$ assuming that the reach scales as $m_{\tilde t} \sim 0.9 \times \sqrt{s} /2$. As a starting point, the anticipated $2 \sigma$ exclusion reach at the HL-LHC is $m_{\tilde t} \sim 1.7$ TeV, which suggests that a muon collider operating at $\sqrt{s} \gtrsim 4$~TeV is required to compete with LHC limits. Reaching percent-level tuning requires $\sqrt{s} \gtrsim 6$~TeV, while per mille requires $\sqrt{s} \gtrsim 20$~TeV and per myriad $\sqrt{s} \gtrsim 66$~TeV. Ultimately, the fine-tuning reach as a function of $\sqrt{s}$ is comparable to the {\it worst-case} higgsino scenario with only Standard Model radiative splittings, the one loop suppression of the stop contribution to fine tuning competing with the more challenging backgrounds faced by the higgsino. 

Apart from considerations of fine tuning, the stop mass is also central to the supersymmetric prediction of the observed Higgs mass. Famously, accommodating $m_h \sim 125$ GeV in the minimal supersymmetric extension of the Standard Model without significant mixing between the stop gauge eigenstates suggests that the stops lie around or above $\sim 5\text{ - }10$ TeV; see, e.g., \cite{Draper:2016pys} for a review. A high-energy muon collider operating at $\sqrt{s} = 30$ TeV could cover the typical scale of the stop mass suggested by the observed Higgs mass at large values of $\tan \beta$, subject to further dependence on mixing angles and the remaining sparticle spectrum. At moderate values of $\tan \beta$, the stops can be heavier, $\sim 100 \text{ - } 1000$ TeV in well-motivated models. However, in such scenarios the electroweakinos may be accessible (e.g., via the searches discussed in \cref{sec:dm}), and the lightest sfermions, such as the right-handed stau, may be an order of magnitude lighter than the stops and could be directly accessible. Thus, the full range of searches for superparticles could cover a substantial portion of the parameter space motivated by the measured Higgs mass.

Finally, we turn to the gluino, which evades our reach estimates because it does not carry electroweak quantum numbers. Nonetheless, it may still be produced by a variety of higher-order processes, including gluon splitting from $q \bar q$ production. The gluino contribution to the tuning of the weak scale at leading logarithmic order scales as
\begin{equation}
\Delta_{\tilde g} \simeq \frac{\alpha_s y_t^2}{\pi^3} \frac{m_{\tilde g}^2}{m_h^2} \log^2 \frac{\Lambda}{m_{\tilde g}}
\end{equation}
Again taking $\Lambda = 10 \times m_{\tilde g}$, 10\% tuning corresponds to $m_{\tilde g} \sim 3$ TeV, percent tuning to $m_{\tilde g} \sim 9.5$ TeV, and per mille tuning to $m_{\tilde g} \sim 30$ TeV. 

\begin{figure}[t!] 
   \centering
 \includegraphics[trim = 20 0 0 0, width = 0.4\textwidth]{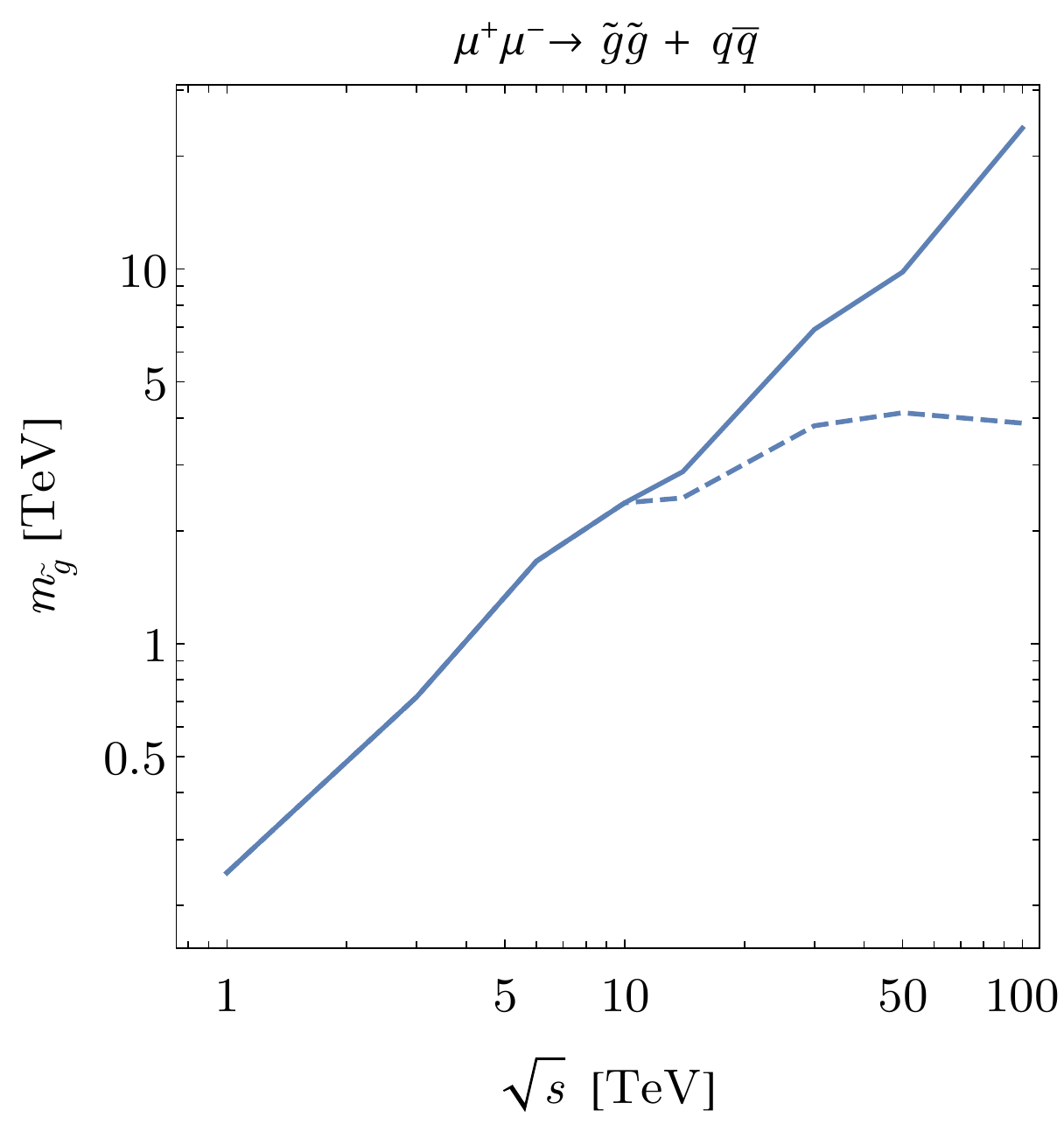} 
   \caption{Gluino discovery reach from $\mu^+ \mu^- \rightarrow \tilde g \tilde g + q \bar q$ as a function of $\sqrt{s}$ , assuming ``optimistic'' (solid) and ``conservative'' (dashed) integrated luminosity scaling as detailed in the body of the text.}
   \label{fig:gluinodisc}
\end{figure}

To estimate the production rate, we consider the process $\mu^+ \mu^- \rightarrow \tilde g \tilde g + q \bar q$ with modest phase space cuts, as detailed in the Appendix. Although the final state is quite distinctive -- presumably zero background is achievable with high signal efficiency -- the ultimate reach of a muon collider is quite sensitive to the integrated luminosity due to the relatively small cross section. In Fig.~\ref{fig:gluinodisc}, we plot the discovery reach (assuming zero background, i.e., 5 signal events) for both optimistic and conservative luminosity assumptions. The optimistic luminosity assumption leads to a mass reach scaling with collider energy, while the conservative luminosity assumption leads to a relatively constant reach for $\sqrt{s} \gtrsim 10$ TeV. 

As a starting point, the anticipated $2 \sigma$ exclusion reach at the HL-LHC is $m_{\tilde g} \sim 3.2$ TeV, which suggests that the discovery reach of a muon collider could just reach the edge of LHC limits for $\sqrt{s} \gtrsim 10$ TeV and 10 ab$^{-1}$. The optimistic luminosity scaling paints a rosier picture, providing sensitivity to $m_{\tilde g} \sim 20$ TeV by $\sqrt{s} = 100$ TeV. Ultimately, the gluino reach lags behind the fine-tuning sensitivity of higgsino and stop reach, but nonetheless a muon collider is comparable to or better than the LHC depending on the c.m.~energy and integrated luminosity.

Taken together, these projections suggest that a high-energy muon collider operating at $\sqrt{s} \sim 20\text{ - }30$ TeV is capable of probing a natural supersymmetric explanation of the weak scale beyond the per mille level, with meaningful sensitivity to both of the states most important for understanding the scale and origin of electroweak symmetry breaking: the higgsinos and stops. Setting aside fine tuning, which may or may not be a sharp guide to the scale of new physics, a collider operating at $\sqrt{s} \sim 30$ TeV could reach the scale suggested by the Higgs mass prediction of the MSSM, regardless of the the stop sector mixing.

\paragraph{The gravitino} \leavevmode\\ 

Although impressive in reach, the potential for a high-energy muon collider to probe Standard Model superpartners represents a continuation of the already-vast search program currently under way at the LHC. But a muon collider offers more than incremental progress in the search for supersymmetry; it would be perhaps the first collider with the potential to directly discover supersymmetry through its universal feature, the goldstone fermion of spontaneous supersymmetry breaking.

A universal prediction of spontaneous breaking of supersymmetry in the rigid limit is the presence of a massless Majorana fermion, the goldstino. When gravity effects are taken into account, the goldstino is eaten through super-Higgs mechanism by the spin 3/2 gravitino~\cite{Deser:1977uq}, which obtains a mass
\begin{equation}
m_{3/2}=\frac{F}{\sqrt{3}M_{\text{Pl}}}\, ,
\end{equation}
where $\sqrt{F}$ is the scale of SUSY breaking and $M_{\text{Pl}}=2.4\times 10^{18}\text{ GeV}$. The scale of the superpartners is set schematically by the supersymmetry-breaking contributions   
\begin{equation}
m_{\text{soft}}\sim \frac{ g_{\text{eff}}^2 }{16\pi^2} \sqrt{F}+ m_{3/2}\, ,\label{eq:soft}
\end{equation}
The first term in \eqref{eq:soft} is the gauge mediation contribution, with $g_{\text{eff}}$ an effective coupling encoding suppressions or enhancements in the soft masses which are model dependent.
For instance for the gaugino masses one typically has 
$g_{\text{eff}}^2 \sim g_{\text{SM}}^2 \, s_g \, N_{\text{mess}} \, \frac{\sqrt{F}}{M_{\text{mess}}} $, where 
$s_{g} \lesssim 1$ parametrizes possible gaugino screening effects.
The second term in \eqref{eq:soft} is the gravity mediation contribution. We have assumed that a single supersymmetry-breaking scale controls both. Low-energy supersymmetry breaking models are those in which the gauge mediation contribution dominates, such that the gravitino is the lightest supersymmetric particle (LSP) in the spectrum and hence cosmologically stable unless $R$-parity is broken. 

In low-energy supersymmetry breaking scenarios, the stable gravitino is dominantly produced in the early universe through gluon-gluino scattering processes as computed in Refs.~\cite{Bolz:2000fu,Pradler:2006qh,Pradler:2006hh,Rychkov:2007uq}. The gravitino yield can be written as
\begin{equation}
Y_{3/2}^{\text{UV}}=C_{\text{UV}}\frac{m_{\tilde{g}}^2T_{\text{r.h.}}}{m_{3/2}^2 M_{\text{Pl}}}\,, \qquad \text{with} \qquad C_\text{UV}=\frac{45 \sqrt{5}f_3}{8\pi^{13/2}g_*^{3/2}}\simeq 4\times 10^{-5}\, ,
\end{equation}
where $m_{\tilde{g}}$ is the gluino soft mass and $T_{\text{r.h.}}$ is the reheating temperature. Requiring that the gravitino yield does not overclose the universe typically results in strong constraints on the maximal reheating temperature in these scenarios~\cite{Moroi:1993mb,Kawasaki:1994af,Moroi:1995fs,Hall:2013uga}.  

An interesting allowed region is the so called \emph{ultralight gravitino window} where $T_{\text{r.h.}}$ is required to be high enough to make the gravitino thermalize in the early universe. Requiring $Y_{3/2}^{\text{UV}}> Y_{\text{eq}}$ with $Y_{\text{eq}}=1.8\times 10^{-3}$ gives
\begin{equation}
T_{\text{r.h.}}>\frac{45 m_{3/2}^2 M_{\text{Pl}}}{m_{\tilde{g}}^2}\simeq 9\text{ MeV}\left(\frac{m_{3/2}}{16\text{ eV}}\right)^2\left(\frac{2\text{ TeV}}{m_{\tilde{g}}}\right)^2\, .
\end{equation} 
Since the gravitino freezes out when it is still relativistic, the matter power spectrum is going to be damped at small scales~\cite{Pierpaoli:1997im}. In Ref.~\cite{Viel:2005qj} a combination of CMB data from WMAP and Lyman-$\alpha$ forest data was used to set an upper bound on the gravitino mass (and the SUSY-breaking scale) 
\begin{equation}
m_{3/2}<16\text{ eV}\quad \Longleftrightarrow\quad \sqrt{F}<260\text{ TeV}\, . 
\end{equation}
This bound will presumably be improved with current Planck data and even further with future cosmological surveys. 

Given the upper bound on the SUSY-breaking scale, the gluino can be made to lie above the 2 TeV reach of the LHC provided
\begin{equation}
g_{\text{eff}}^2> 1.2\times\left( \frac{m_{\tilde{g}}}{2\text{ TeV}}\right)\left(\frac{260\text{ TeV}}{\sqrt{F}}\right)\, .
\end{equation}
following the parametrics in Eq.~\eqref{eq:soft}. We refer to~\cite{Hook:2015tra,Hook:2018sai} for explicit models which realize a heavy superpartner spectrum with such a low SUSY-breaking scale. If we are agnostic about UV completions, the perturbativity of $g_{\text{eff}}$ gives a lower bound on $\sqrt{F}$ which is a loop factor below 260 TeV. The gravitino window can then be defined as 
\begin{equation}
10^{-3}\text{ eV}<m_{3/2}<16\text{ eV}\quad \Longleftrightarrow\quad 2.5\text{ TeV}<\sqrt{F}<260\text{ TeV}\, ,\label{eq:gravitinowindow}
\end{equation}
where the lower bound is given by the current LHC bound on the gluinos plus perturbativity of the SM couplings, while the upper bound is from cosmological constraints on warm dark matter.

Within the light gravitino window defined in \cref{eq:gravitinowindow}, we can safely assume that the superpartner spectrum is decoupled and the gravitino interactions with the SM are described by a universal EFT with couplings controlled by the supersymmetry breaking scale only~\cite{Brignole:1997pe,Brignole:1997sk,Brignole:1998me}. Collider searches for direct gravitino production can hence provide a robust limit on the supersymmetry breaking scale, independent of the specific details of the superpartner spectrum. Traditionally, this has not been an emphasis of the supersymmetry search program at hadron colliders because direct searches for colored superpartners always exceed the sensitivity of direct gravitino searches. In contrast, at high-energy lepton colliders the reduced background and rapid growth of signal cross sections with $\sqrt{s}$ makes this a competitive channel for the discovery of supersymmetry.

\begin{figure}[t!]
  \centering
 \includegraphics[trim = 20 0 0 0, width = 0.6\textwidth]{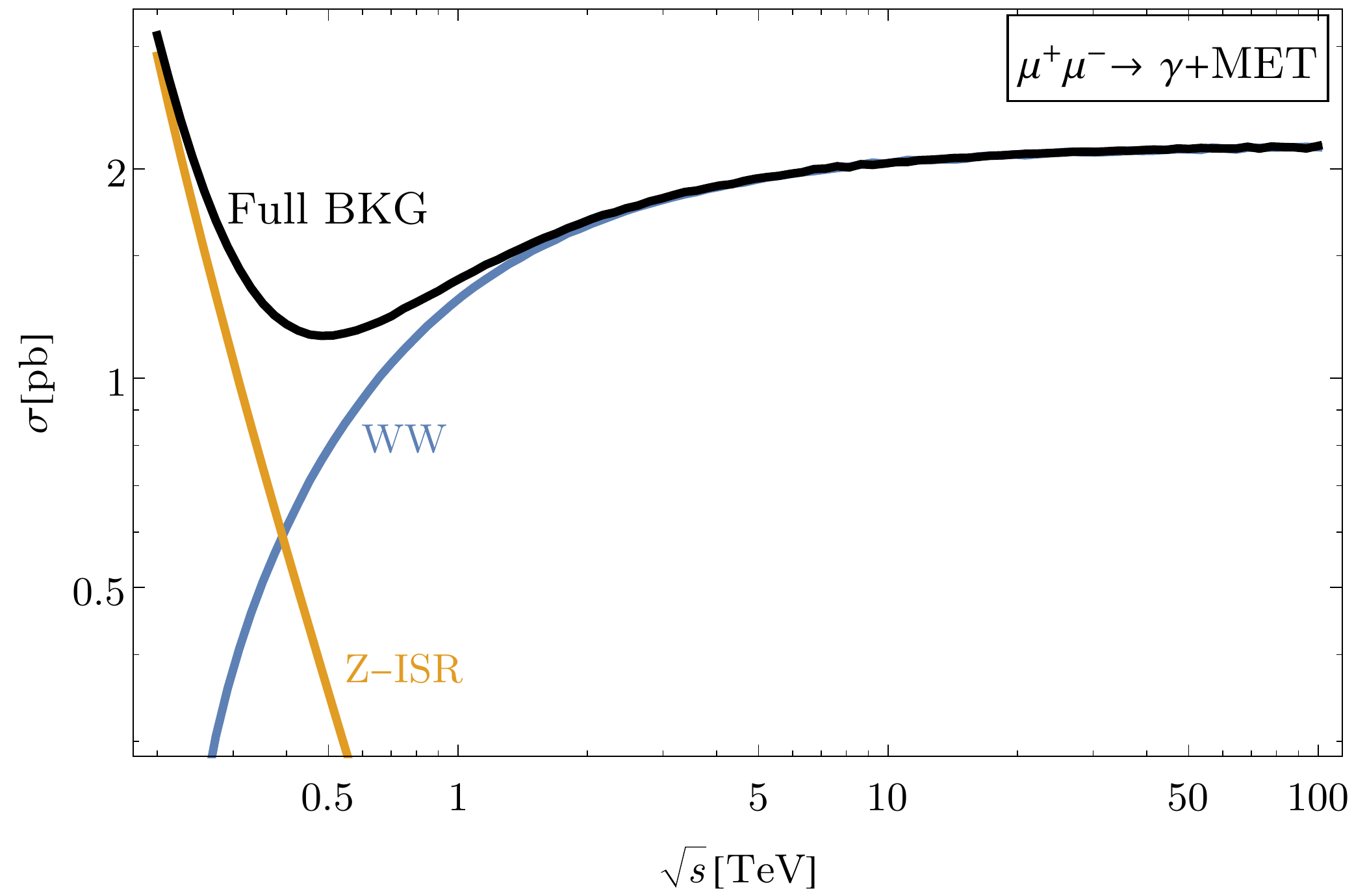}
 \caption{Background cross section for the $\gamma+\text{MET}$ final state as a function of the c.m.~energy $\sqrt{s}$. The black line is the total background cross section in pb while the red and blue lines show the separate contributions of $Z\gamma$ and $W^+W^-$ respectively.\label{fig:SM_monophoton}}
\end{figure}

Here, we explore the sensitivity of a muon collider to gravitino pair production in the mono-photon final state via the production mode
$\mu^+ \mu^- \to \tilde G \tilde G \gamma$.
The analytic cross section for this process in the gravitino EFT has been computed in Ref.~\cite{Brignole:1997sk}; it includes ISR photon topologies from contact interactions between leptons and gravitinos, as well as contact interactions involving leptons, gravitinos and a photon.
In the limit of soft and collinear photon emission the cross section reads
\be
\label{eq:monophoton_simp}
\sigma ( \mu^+ \mu^- \to \tilde G \tilde G \gamma) = \frac{\alpha s^3}{160 \pi^2 F^4} \left [\frac{247}{60}+2 \log \left(\frac{2 E_{\text{min}}^{\gamma}}{\sqrt{s}} \right) \right] \log \left(\frac{1-\cos \theta_{\text{min}}}{1+\cos \theta_{\text{min}}} \right)\, ,
\ee
where $\sqrt{s}$ is the c.m.~energy, $E^{\gamma}_{\text{min}}$ is the minimal energy of the photon, and $\theta_{\text{min}}$ is the minimal 
photon angle with respect to the beam direction.
In the following numerical analysis we will employ the complete analytic formula, but we can use the above equation to get some reasonable estimates for the reach in terms of $\sqrt{F}$.

\begin{figure}[t!]
  \centering
 \includegraphics[trim = 20 0 0 0, width = 0.4\textwidth]{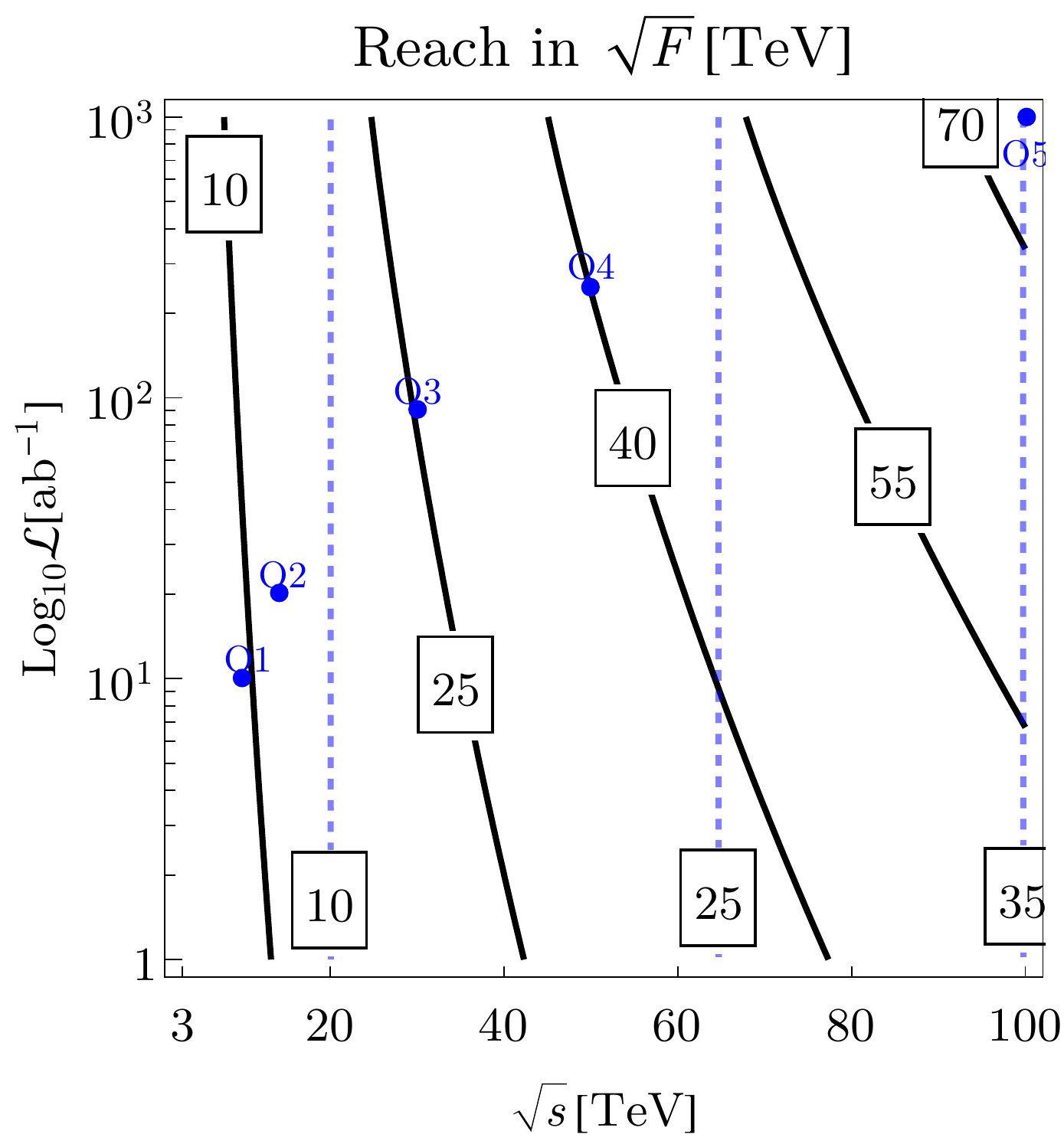}
 \caption{\label{fig:sqrts_reach}
Estimated reach of the gravitino search interpreted as a constraint on the supersymmetry breaking scale $\sqrt{F}$.  The
solid black lines denote limit contours assuming only statistical errors: $S/\sqrt{S+B}=1.96$. The dashed blue lines denote limit contours that include a systematic error: $S/\sqrt{S+B+\epsilon^2(S^2+B^2)}=1.96$ with $\epsilon=1\%$. 
The points O1$\dots$O5 indicate the optimistic benchmarks of Table \ref{tab:energylumi}, starting from $\sqrt{s} = 10 \text{ TeV}$ and with increasing center-of-mass energies. 
}
\end{figure}

We assume minimal cuts on the photon ($E^{\gamma}_{\text{min}}>50$ GeV, $|\eta_{\gamma}| < 2.4$). 
The SM background at high $\sqrt{s}$ is dominated by $W^+W^-$ fusion as can be seen in \cref{fig:SM_monophoton}, and it is roughly constant and equal to $2$ pb for
$\sqrt{s} \gtrsim 3 $ TeV.
We further define the signal region by requiring the photon energy not to exceed the endpoint of the signal process, $E^{\gamma}_{\rm max} = \sqrt{s}/2$. With these cuts, the $2\sigma$ sensitivity to $\sqrt{F}$ using \cref{eq:monophoton_simp} is
\be
\sqrt{F} \lesssim 61.7 \text{ TeV} \left(\frac{\mathcal{L}}{1000 \text{ ab}^{-1}} \right)^{1/16} \left( \frac{\sqrt{s}}{100 \text{ TeV}}\right)^{3/4} \left(4.8 + \log\left[\frac{\sqrt{s}}{100 \text{ TeV}} \right] \right)^{1/8}\, ,
\ee
where we have only included statistical errors. The corresponding numerical result is displayed in \cref{fig:sqrts_reach}.
In the same figure, we show the decrease in the reach assuming a $1\%$ systematic error on both signal and background.
Given that the signal cross section grows as the 6$^\text{th}$ power of the c.m.~energy, the energy increase is the most beneficial in increasing the sensitivity
to high values of $\sqrt{F}$.

We conclude that a future high energy muon collider can almost certainly push up the lower bound of the ultralight gravitino window by one order of magnitude. An improvement of the cosmological bounds with respect to the ones derived from WMAP data could allow the light gravitino window to be completely closed in the future. More broadly, this exemplifies the ability of a high-energy muon collider to directly probe the mechanism of supersymmetry breaking.

\subsubsection{The ``big'' hierarchy problem: compositeness}

We next turn our attention to the competing solution to the ``big'' hierarchy problem, compositeness. Though compositeness may take many guises, its realizations all share a set of common features given the discovery of a light Standard Model-like Higgs with an apparent mass gap.  The expectation is that fermionic top partners should exist at a scale $m_T \sim y_t f$ parametrically below the scale of compositeness $m_* = g_* f$ at which a host of resonances appear. See e.g.~\cite{Panico:2015jxa} for an excellent overview.

Perhaps the strongest tests of compositeness come from its distinctive imprints on the electroweak sector of the SM, including modification of Higgs couplings \cite{Giudice:2007fh}. To the extent that the tuning associated with vacuum misalignment is generally on the order of $\Delta \sim f^2 / v^2$ and the leading deviations from Standard Model predictions can be captured by irrelevant operators suppressed by $f^2$ or $(g_*/g)^2 f^2$, tests of such deviations typically provide the most powerful constraints. Here, high-energy muon colliders enjoy a particular advantage, leveraging the energy growth of these deviations without the considerable backgrounds of $pp$ counterparts. As demonstrated in \cite{Buttazzo:2020uzc}, this allows a high-energy muon collider to access Higgs compositeness scales well above any other future collider project. For example, sensitivity to operators such as
\begin{align}
\mathcal{O}_H = \frac{1}{2} \left( \partial_\mu |H|^2 \right)^2 \qquad \text{and} \qquad \mathcal{O}_W = \frac{ig}{2} \left( H^\dag \sigma^a \dfbd H \right) D^\nu W_{\mu \nu}^a\,,
\label{eq:OHandOW}
\end{align}
allows a muon collider operating at $\sqrt{s} = 10$ TeV to probe compositeness scales as high as $m_* \sim 45$ TeV with 10 ab$^{-1}$, while a collider operating at $\sqrt{s} = 30$ TeV can probe compositeness up to $m_* \sim 140 \text{ TeV}$ with 90 ab$^{-1}$ \cite{Buttazzo:2020uzc}. The sensitivity at these two energies covers fine tuning associated with vacuum misalignment from the percent level to the per mille level.
 
To the extent that it is possible to decrease the tuning associated with vacuum alignment, sensitivity to fermionic top partners provides a complementary probe. Here the test of fine tuning is expected to be qualitatively similar to that of supersymmetric tops, with two relative advantages: for fermions the $s$-channel production cross section is only linearly suppressed by velocity near threshold, and in the absence of additional structure they typically decay directly to visible and highly distinct final states. Following the fine-tuning estimates in the supersymmetric case, reaching percent-level tuning requires $\sqrt{s} \gtrsim 6$ TeV, while per mille requires $\sqrt{s} \gtrsim 20$ TeV and per myriad $\sqrt{s} \gtrsim 66$ TeV. As in the case of supersymmetry, a high-energy muon collider operating at $\sqrt{s} \sim 20\text{ - }30$ TeV would decisively test compositeness as a natural explanation for the weak scale.

\subsubsection{The ``little'' hierarchy problem}

As we have seen, a high-energy muon collider operating at tens of TeV could provide satisfying coverage of known solutions to the ``big'' hierarchy problem. It is no less suited to constraining or discovering solutions to the ``little'' hierarchy problem, whose subtle signatures will remain largely untouched by the LHC. There are a plethora of such solutions which operate by relying on novel field theoretical~\cite{ArkaniHamed:2001nc, ArkaniHamed:2002pa, ArkaniHamed:2002qx, ArkaniHamed:2002qy, Fox:2002bu, Chacko:2005pe, Burdman:2006tz, Craig:2014aea, Craig:2014roa, Cohen:2018mgv, Cheung:2018xnu, Cohen:2020ohi} or cosmological~\cite{Graham:2015cka, Arkani-Hamed:2016rle, Hook:2016mqo, Geller:2018xvz, Csaki:2020zqz} ingredients.
Here we will focus on solutions involving ``neutral naturalness'' such as the Twin Higgs \cite{Chacko:2005pe} or Hyperbolic Higgs \cite{Cohen:2018mgv}, in which the partner particles are entirely neutral under the Standard Model. Such models feature (at least) four possible avenues to discovery at colliders: Higgs coupling deviations from mixing between scalars; direct production of SM singlet partner particles; displaced vertices from exotic Higgs decays; and direct production of the ``radial mode'' associated with spontaneous breaking of the discrete symmetry. 

In general, the Higgs coupling deviations are a subset of those expected in composite models, including most notably the oblique operator $\mathcal{O}_H$ defined in \cref{eq:OHandOW}. As with composite models, constraints on $\mathcal{O}_H$ translate directly into bounds on $f^2/v^2$, a typical measure of the fine tuning. The results of \cite{Buttazzo:2020uzc} suggest sensitivity to $f \sim 16 v$ at $\sqrt{s} = 10$ TeV and $f \sim 46 v$ at $\sqrt{s} = 30$ TeV, probing the tuning of the weak scale between the percent and per mille levels within the framework of the Twin Higgs. The prospects for discovery via direct production of partner particles are considerably weaker. Although the precise reach depends on whether these partner particles are fermions (as in the Twin Higgs) or bosons (as in the Hyperbolic Higgs), the qualitative sensitivity can be inferred from the reach for the ``nightmare scenario'' of a $\mathbb{Z}_2$-symmetric singlet scalar summarized in \cref{sec:EWSB} and \cref{app:simplified}, extending at most to the hundreds of GeV. 

A more exotic collider signature of models of neutral naturalness is the prediction of displaced decays of dark particles into Standard Model states \cite{Craig:2015pha}. The scalar mixing produces a non-zero branching ratio of Higgses into neutral partner states, which gives a portal for energy to be transferred to the partner sector at a collider. Some of the produced partner sector states are unstable to decaying through an off-shell Higgs into Standard Model states, bearing out ``Hidden Valley'' type phenomenology \cite{Strassler:2006im,Han:2007ae}. This produces spectacular signatures -- for example vertices displaced from the beamline from which Standard Model jets appear -- and this allows search strategies with very low background \cite{Liu:2016zki,Liu:2018wte}. The prospects for probing these signatures in neutral naturalness models at future lepton colliders have been studied in the context of Higgs factories \cite{Alipour-Fard:2018lsf,Wang:2019xvx}, and a branching ratio reach is projected which is competitive with or better than the LHC forecasts. Dedicated study for a higher-energy muon collider has yet to be performed.   

The prospects for discovering the ``radial mode'' are much stronger than for other partner particles. These can be obtained by interpreting the model independent results of Fig.~\ref{fig:results} in the Twin Higgs, where the mixing angle scales as $\sin\gamma\sim g_* v/m_\phi$. Concretely, the Higgs potential in the simplest realization of the Twin Higgs can be written as 
\be
V= \lambda_*\bigg(|H_A|^2+|H_B|^2 -\frac{f_0^2}{2}\bigg)^2 +  \kappa \Big(|H_A|^4+|H_B|^4\Big) + \sigma_{\rm soft} f^2\, |H_A|^2 \,.
\label{eq:V_TH}
\ee
The SM Higgs sector is extended by the addition of the twin Higgs $H_B$, a singlet under the SM and doublet under a mirror electroweak gauge group $SU(2)_B$. The twin Higgs is coupled to the SM Higgs $H_A$ via a portal coupling $\lambda_*$. With only this quartic included, the potential linearly realizes an $SO(8)$ symmetry, spontaneously broken to $SO(7)$ at the scale $f$. The radiative stability of the construction is ensured by a $\mathbb{Z}_2$ symmetry between the SM and the mirror sector, which is softly broken by $\sigma_{\rm soft}$ to allow for $f > v$ and a viable phenomenology.\footnote{Here for simplicity, we set the $\mathbb{Z}_2$-breaking quartic to zero. Having it non-zero will only  impact the phenomenology at small $m_\phi$, allowing the Higgs mass constraint to be satisfied at large $f$ and small $m_\phi$. See \cite{Katz:2016wtw,Buttazzo:2018qqp} for a discussion.}  The $SO(8)$-breaking quartic $\kappa$ receives IR contributions from (mirror) top loops $\kappa\simeq (3y_t^4/8\pi^2)\log m_*/m_t$. Finally, $f_0\simeq f$ up to corrections of order $\kappa/\lambda_*$.

After the spontaneous breaking of the $SO(8)$ symmetry, we are left with two real scalars in the spectrum: the SM-like Higgs $h$ and the radial mode~$\phi$ with  masses $m_\phi^2 \simeq 2 \lambda_* f^2$ and $m_h^2\simeq 4 \kappa v^2$ in the limit $\lambda_\ast\gg \kappa\, ,\sigma_{\rm soft}$. This shows that the Higgs mass is already of the correct size given the typical size of the irreducible contributions to $\kappa$. The requirement to reproduce the electroweak scale $v$ and the Higgs mass $m_h$ fixes 2 out of the 4 free parameters in Eq.~\eqref{eq:V_TH}. The remaining two are chosen to be  ($f$, $m_\phi$) while the mixing between the twin Higgs and the SM Higgs is predicted, and it scales as $\sin\gamma \simeq v/f$ in the $m_\phi\gg m_h$ limit.  

\begin{figure}[t!]
\centering
\includegraphics[width=0.6\textwidth]{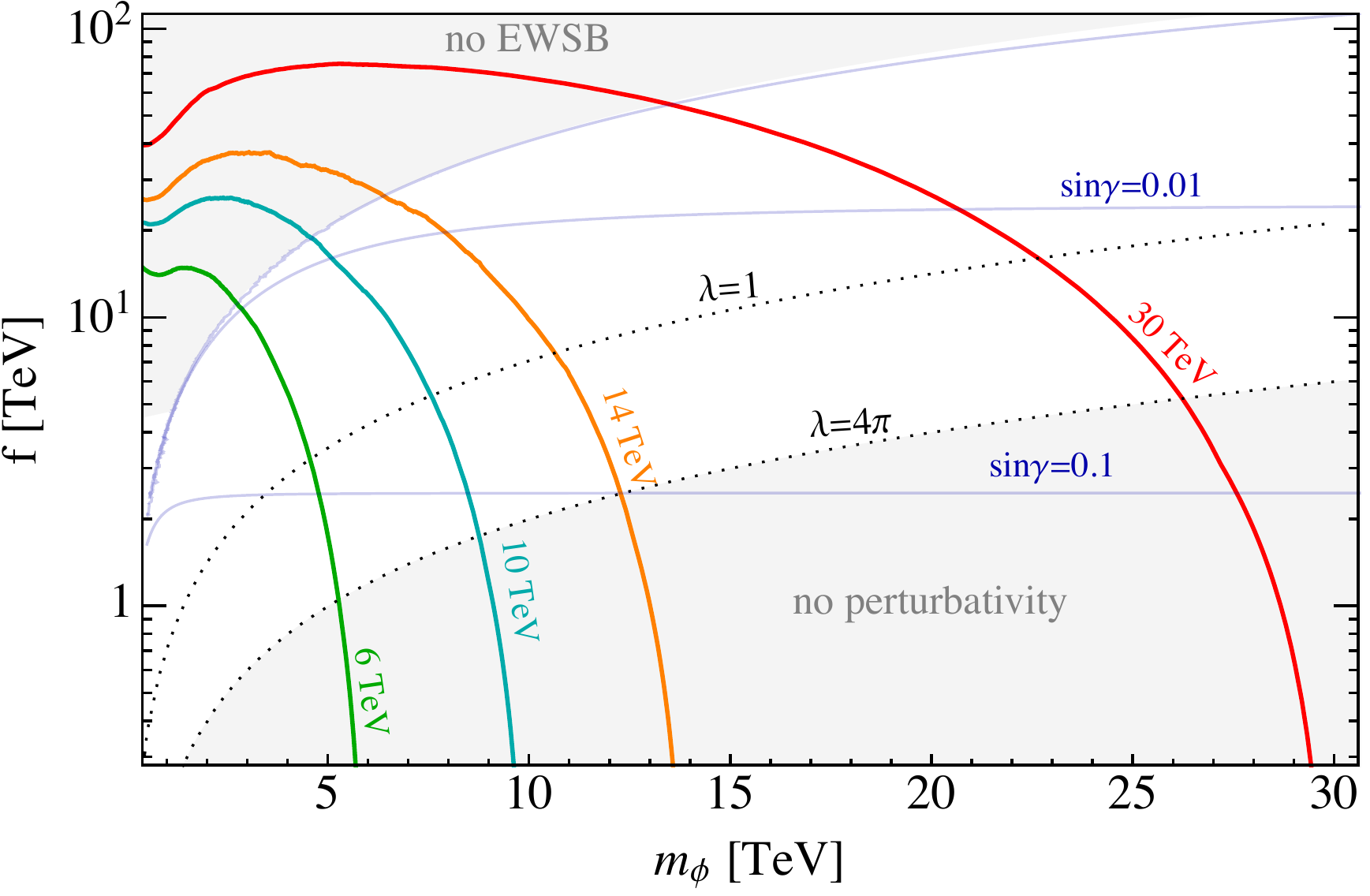}
\caption{The same limits on a singlet scalar Higgs as shown in \cref{fig:results}, interpreted in terms of the reach on the sigma-model scale $f$ in the context of a Twin Higgs model. \label{fig:twinresults}}
\end{figure}

In Twin Higgs models, the fine tuning of the electroweak scale generically scales as
\begin{equation}
\frac{\delta m_h^2}{m_h^2}\simeq \frac{3y_t^4}{2\pi^2\lambda_*}\left(\frac{m_*}{m_h}\right)^2\log\frac{m_*}{m_t}\ ,\label{eq:FT}
\end{equation}
where $m_*$ should be interpreted as the cutoff of the potential in Eq.~\eqref{eq:V_TH}, where colored states are expected to be necessary to stabilize the scale $f$ from radiative corrections. Fixing the scale of the colored states $m_*$, the fine tuning of the electroweak scale in Twin Higgs models is parametrically reduced with respect to that of conventional supersymmetry or composite scenarios by $y_t^2/\lambda_\ast$; see, e.g.,~\cite{Katz:2016wtw,Contino:2017moj}. The gain in fine tuning is limited by the perturbativity requirement on the linear potential in Eq.~\eqref{eq:V_TH}, which roughly requires $\lambda_*\lesssim 4\pi$, allowing colored states as heavy as 5 TeV with fine tuning on the order of 1\%. The implications of a muon collider's sensitivity to the radial mode are shown in \cref{fig:twinresults}, reinterpreting the reach illustrated in \cref{fig:results} in terms of the Twin Higgs parameter space. Keeping the scale of the colored states fixed, the fine tuning of the electroweak scale  goes like $1/\lambda_*$, making the sliver of parameter space with largest $\lambda_*$ (or equivalently with largest $m_\phi/f$) the most appealing.

Apart from fine tuning considerations, \cref{fig:twinresults} illustrates the correlation between direct searches and Higgs coupling deviations in the Twin Higgs parameter space. Interestingly, a high energy muon collider of $\sqrt{s}=14\text{ TeV}$ will be able to fully probe the parameter space corresponding to a \% deviation in Higgs couplings ($\sin\gamma \simeq 0.1$) by means of direct searches for the extra scalar. This may be taken as further motivation to reach such a high c.m.~energy at a future collider. Conversely, moving into the region with small $\lambda_*$, the deviations in the Higgs couplings will be difficult to observe, but direct production at a muon collider could still cover a large portion of this parameter space.


\section{Complementarity}
\label{sec:comp}
Given the time required to achieve first collisions at a future collider, a number of planned and proposed experiments capable of extending our sensitivity to indirect signs of new physics could lead to orders of magnitude improvement of current limits. 
Among others, experiments searching for electric dipole moments, anomalous flavor violation beyond the Standard Model, and stochastic gravitational wave backgrounds will probe scales ranging from tens to hundreds of TeV on relevant timescales.\footnote{There are already hints that something interesting might be going with the muon, which could have exciting potential implications for a future muon collider, e.g., see recent studies regarding the muon $g-2$ \cite{Capdevilla:2020qel, Buttazzo:2020eyl, Capdevilla:2021rwo,Chen:2021rnl,Yin:2020afe}, $B$ meson~\cite{Huang:2021nkl}, and $K$ meson~\cite{Huang:2021biu} anomalies.}  Ultimately, a signature at any of these experiments would provide an indication of new physics at a scale amenable to further exploration. This motivates asking what energies and luminosities would be required for a muon collider to directly test the origin of indirect signals, providing another set of sharp goalposts.

\subsection{EDMs}
Electric dipole moments of elementary particles offer a nearly background-free probe of new physics beyond the Standard Model, since all CP-violating SM effects are accompanied by flavor-violating spurions and give rise to extremely small EDMs. Recently, substantial progress has been made in experimental searches for EDMs using paramagnetic molecules.
For instance, by studying the polar ThO molecule, the ACME collaboration has set a bound on the electron EDM  of $|d_e| \leq 1.1 \times 10^{-29}\,e\,\mathrm{cm}$ at 90\% confidence \cite{Andreev:2018ayy}. Rapid experimental progress towards the use of atoms and molecules, including novel approaches using polyatomic molecules, is likely to improve the electron EDM sensitivity to at least $10^{-32}\,e\,\mathrm{cm}$ in the coming decade (see Refs.~\cite{Cairncross:2019air, Hutzler:2020pze} and references therein). These approaches will also offer novel opportunities to probe hadronic EDMs.

\begin{figure}[t!]
\centering
      \includegraphics[width=0.38\textwidth]{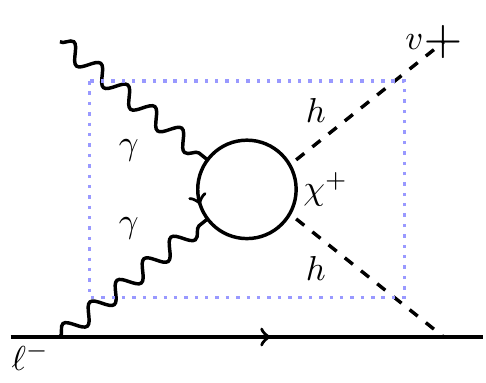}
\qquad \qquad
      \includegraphics[width=0.38\textwidth]{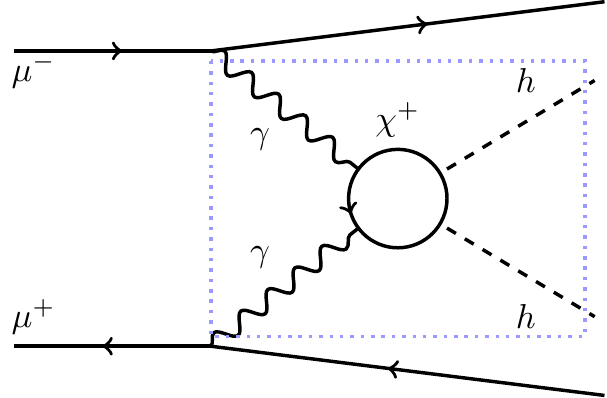}
\caption{One possibility for testing the physics associated with a Barr-Zee type contribution to a lepton EDM at a future muon collider. Left: The two-loop Barr-Zee contribution to a lepton EDM. Right: A $\gamma \gamma \to h h$ process at a muon collider, sensitive to loops of charged particles that couple to the Higgs.  The dotted blue box shows that both processes probe the same underlying physics.} 
\label{fig:BarrZeeComp}
\end{figure} 

The discovery of an electric dipole moment could definitively prove the existence of new physics beyond the SM. In a typical model, weakly-coupled $k$-loop physics, at a mass scale $M$ with CP-violating phase $\delta_\text{CP}$, could generate an electron EDM of the size
\begin{equation}
d_e \sim \sin(\delta_\text{CP}) \frac{e\, m_e}{M^2} \left(\frac{\alpha}{4\pi}\right)^k  \simeq 10^{-32}\,e\,{\rm cm} \sin(\delta_\text{CP})  \times
\begin{cases} 
\left(1~{\rm PeV}/M\right)^2 & \text{for } k = 1 \\[5pt]  
\left(20~{\rm TeV}/M\right)^2 & \text{for } k = 2  
\end{cases}\,\,.
\end{equation}
This is only a rough, order-of-magnitude guide, but it shows that an EDM experiment could provide the first discovery of the effects of new physics beyond the reach of current colliders. A one-loop EDM of this size could arise, for example, from sleptons with masses of order PeV and order one CP-violating phases~\cite{Altmannshofer:2013lfa, McKeen:2013dma}, or from lighter sleptons with smaller CP-violating phases. A two-loop EDM could arise, for instance, from chargino loops in split SUSY \cite{ArkaniHamed:2004yi, Giudice:2005rz}. More  generally, {\em any} new particles interacting with electroweak gauge bosons and  the Higgs  can produce a 2-loop EDM through Barr-Zee diagrams \cite{Barr:1990vd}, see the left panel of \cref{fig:BarrZeeComp}.

An EDM discovery would tell us that new physics exists, but would tell us very little about the nature of the new physics. Colliders will have a crucial role to play, by providing more insight on the new particles responsible for the EDM and allowing us to determine which extension of the SM explains the effect. A one-loop EDM would be associated with new particles carrying electric charge which could be directly pair-produced at a muon collider if they are kinematically within reach, e.g., for sleptons we would search for $\mu^+ \mu^- \to {\tilde \ell}^+ {\tilde \ell}^-$. The range of possibilities at two loops is broader (e.g., \cite{Cesarotti:2018huy, Panico:2018hal}), but the case of the Barr-Zee diagram offers a particularly appealing target. It involves new electroweak particles which could be pair produced directly. However, it {\em also} implies that these particles alter the interactions between Higgs and gauge bosons, as illustrated in \cref{fig:BarrZeeComp}. Since high-energy lepton colliders are electroweak gauge boson colliders, they  offer a unique prospect to directly probe the same underlying electroweak interactions that generate the EDM, via precision studies of processes like $\gamma \gamma \to h h$.  We leave a detailed assessment of the prospects for EDM/collider complementarity along these lines for future work.

\subsection{Flavor}
\newcommand{\sh}[1]{[\textcolor{blue}{{\bf SH: }#1]}}

One of the biggest puzzles in the SM is the pattern of fermion masses and mixings. Both the quark and lepton sectors have significant mass hierarchies, whereas the mixing matrices take a very different form in the two sectors. We expect that at high energies, where the flavor pattern of the SM is established, there may be much larger rates of flavor-changing processes than the SM predicts. This is a strong motivation for searching for flavor-violating processes at high-energy colliders.

Conversely, some of our most stringent bounds on physics beyond the SM come from low- or  medium-energy precision tests of flavor-changing processes, e.g., $K-{\bar K}$  or $D-{\bar  D}$ mixing~\cite{Isidori:2010kg}. Of particular interest, in the context of a muon collider, are precision tests of charged lepton flavor violation processes like $\mu \to e  \gamma$,  $\mu \to  3e$, $\tau \to 3 \mu$, or $\mu$-to-$e$  conversion within atomic nuclei. As with EDMs, these processes are expected to be much more strongly constrained  in the coming decades than they are at present, due to a number of currently operating or planned future experiments~\cite{Baldini:2018uhj}. These experiments can indirectly probe physics at energies of 10s of TeV or even higher. A high-energy muon collider can probe the same physics (e.g., through direct searches for flavor-changing processes like $\mu^+ \mu^- \to \tau^\pm \mu^\mp$), or can help to elucidate the underlying mechanism of flavor violation by directly producing new particles with flavor-violating interactions, such as the mixed slepton production process $\mu^+ \mu^- \to {\tilde \ell}_i^+ {\tilde \ell}_j^-$. Below, we will give first estimates of the physics reach for both of these scenarios. We will see that there is a powerful complementarity between precision lepton flavor experiments and high-energy muon collider searches.

\subsubsection{Lepton-flavor violating contact interactions}

Belle has set a limit of $\textrm{Br}(\tau \to 3\mu) < 2.1\times 10^{-8}$~\cite{Hayasaka:2010np}, and this bound will be improved to $3.5\times 10^{-10}$ by Belle~II~\cite{Kou:2018nap}. 
These bounds can be contrasted with constraints on the $\mu \to 3e$ branching ratio, which is currently constrained to be $1.0\times 10^{-12}$ by SINDRUM~\cite{Bellgardt:1987du}, but will be dramatically improved, eventually to a sensitivity $\sim 1\times 10^{-16}$, by the Mu3e experiment~\cite{Baldini:2018uhj}.

We parametrize the four-fermion operators relevant for the $\tau \to 3\mu$ decay via
\begin{equation}
\mathcal{L} \supset 
V_{LL}^{\tau 3\mu}\big( \bar{\mu} \gamma^{\mu} P_L \mu \big) \big(\bar{\tau} \gamma_{\mu} P_L \mu\big)
+ V_{LR}^{\tau 3\mu} \big( \bar{\mu} \gamma^{\mu} P_L \mu\big) \big(\bar{\tau} \gamma_{\mu} P_R \mu\big) + \big( L \leftrightarrow R \big) + \textrm{h.c.}\,,
\end{equation}
with an equivalent set for the $\mu \to 3e$ decay. In what follows, we will assume all the $\tau 3\mu$ coefficients are equal:
\begin{equation}
V_{LL}^{\tau 3\mu}  = V_{LR}^{\tau 3\mu} = V_{RL}^{\tau 3\mu} = V_{RR}^{\tau 3\mu} = \frac{c^{\tau 3\mu}}{ \Lambda^2 }
\end{equation}
where $c^{\tau 3\mu}$ is a dimensionless coefficient and $\Lambda$ is to be interpreted as the scale of new physics.
Setting the Wilson coefficients to unity, we see that the current Belle (future Belle~II) constraint above translates into a bound of $\Lambda > 14.7\text{ TeV}$ ($\Lambda > 40.9\text{ TeV}$), far below the corresponding constraints from SINDRUM (with $\tau 3 \mu$ replaced  by $\mu 3 e$), which gives $\Lambda > 273 \text{ TeV}$.
Generically, though, one might expect new physics violating lepton flavor to manifest with hierarchies similar to the pattern of hierarchies in the SM lepton Yukawas, in which case the bounds from $\tau$ and $\mu$ might be more comparable. We will discuss this more in what follows.

At a muon collider, these same operators can be probed directly via $\mu^+\mu^- \to \ell^+_i \ell^-_j$. We will focus on $\mu\tau$ production, since it can be compared directly with the sensitivity from tau decays. Our analysis closely follows an analogous study at an $e^+ e^-$ collider in Ref.~\cite{Murakami:2014tna}.
With only the four-fermion operator insertion, the rate for $\mu^+\mu^- \to \mu\,\tau$ grows as $s$. Taking $c^{\tau 3\mu} / \Lambda^2 = 1 / (50\text{ TeV})^2$, the cross section grows from $0.059\text{ fb}$ at a $3\text{ TeV}$ muon collider to $66.1\text{ fb}$ at $\sqrt{s} = 100\text{ TeV}$.

Ignoring detector effects, there are two primary backgrounds for the flavor-violating process in the SM: $\tau^+\tau^-$ production, where (at least) one of the taus decays to a muon and a neutrino and $\mu^+ \mu^- \to \mu\,{\nu}_{\mu}\,\tau\,{\nu}_{\tau}$ production via intermediate $W$ bosons.
These backgrounds can be significantly suppressed with simple cuts on the muon energy and the missing 3-momentum. In particular, we demand that the most energetic muon in the event has at least $90\%$ of the beam energy and that the direction of the total missing 3-momentum vector be at least $170^{\circ}$ from the 3-momentum of the most energetic muon. These cuts suppress the $\tau^+\tau^-$ background by a factor $\sim 300$, and the $W^+W^-$ background by a factor $\sim 10$, depending slightly on the c.m.~energy.

\begin{table}[htbp]
   \centering
     \renewcommand{\arraystretch}{1.5}
\setlength{\arrayrulewidth}{.3mm}
\setlength{\tabcolsep}{0.8 em}
   \begin{tabular}{c|c|c|c} 
 & \multicolumn{3}{c}{$N_{\text{post-cuts}}$}\\
\hline
$\sqrt{s}$ (TeV)&\quad$\mu^+\mu^-\to\mu\,\tau$ &\quad$\mu^+\mu^-\to\mu\,\tau\, \nu_{\mu}\,\nu_{\tau}$ & \quad$\mu^+\mu^-\to\tau^+\tau^-$\\
\hline\hline
$.125$ &  0.0948  & 30.8 & $3.42\times 10^{4}$\\ \hline
$ 3 $ & $53.3$ & $6.32\times 10^{3}$  & 40.4\\ \hline
$ 6 $ & $212$ & $3.26\times 10^{3}$ & 9.52\\ \hline
$14$ & $1.14\times 10^{3}$ & $1.14\times 10^{3}$  & 0.138\\ \hline
$100$ & $5.73\times 10^{4}$ & 60.9  & 0.0312\\
   \end{tabular}
   \caption{Number of signal and background events after kinematic cuts and estimating the loss of signal efficiency due to initial state radiation for $1\text{ ab}^{-1}$ of data and $c^{\tau 3\mu} / \Lambda^2 = 1 / (50\text{ TeV})^2$.}
   \label{tab:4fermi.nEvents}
\end{table}

\begin{figure}[t!]
\centering
\includegraphics[width=0.8\linewidth]{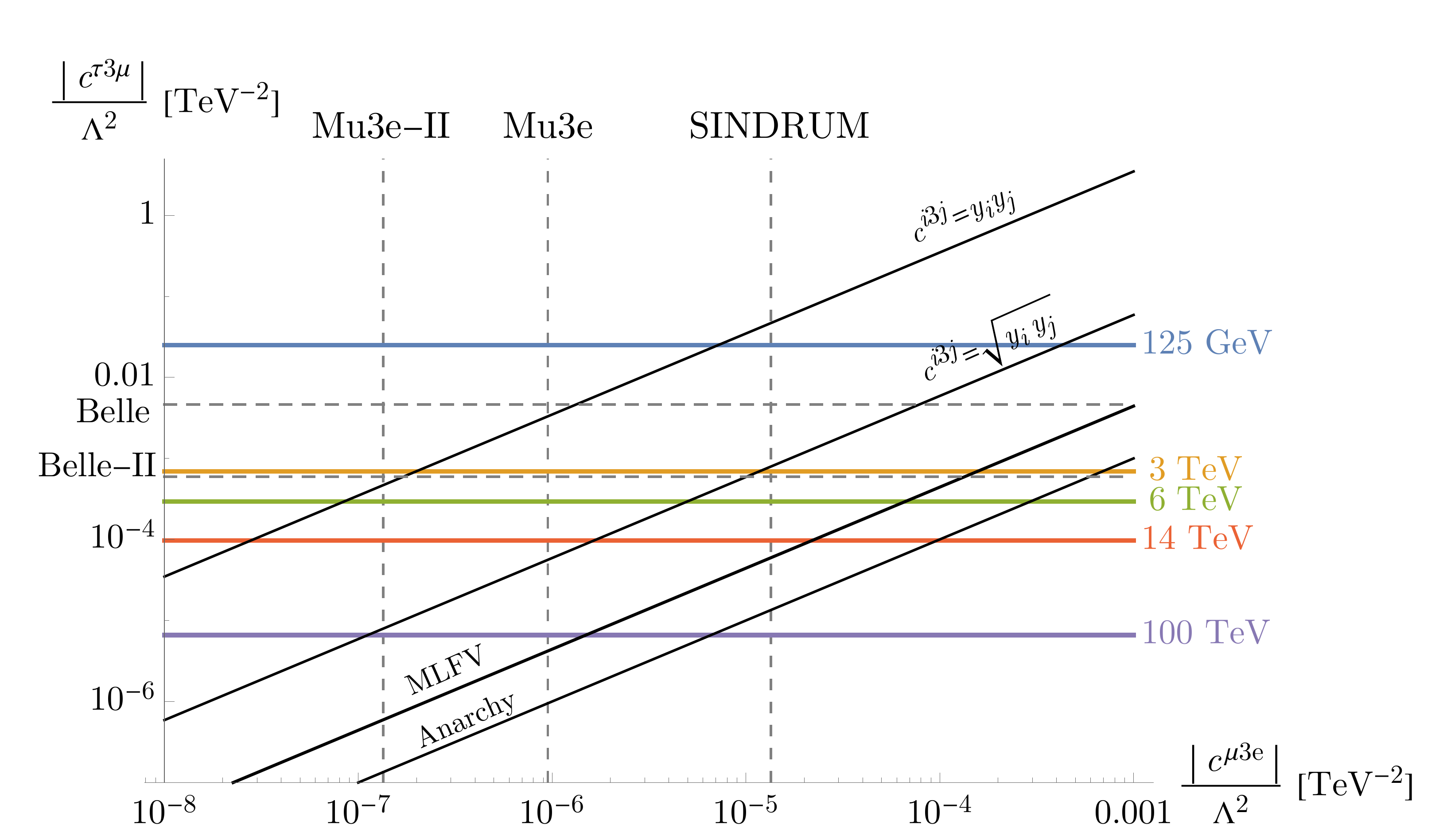}
\caption{Summary of muon collider and precision constraints on flavor-violating 3-body decays. The colored horizontal lines show the sensitivity to the $\tau 3\mu$ operator at various energies, all assuming $1\text{ ab}^{-1}$ of data. The dashed horizontal (vertical) lines show the current or expected sensitivity from $\tau \to 3\mu$ ($\mu \to 3e$) decays for comparison. The diagonal black lines show the expected relationship between different Wilson coefficients with various ansatz for the scaling of the flavor-violating operators (e.g., ``Anarchy'' assumes that all Wilson coefficients are $\mathcal{O}(1)$).}
\label{fig:flavor.4fermi}
\end{figure}

The signal process, on the other hand, passes these cuts with near perfect efficiency. Kinematically, there is no loss of signal events, and the only degradation is due to initial state radiation. This leads to a $\sim 10\%$ reduction at $\sqrt{s} = 3\text{ TeV}$, slowly increasing to a $13\%$ reduction at $\sqrt{s} = 100\text{ TeV}$. 
The number of signal and background events after applying the kinematic cuts and initial state radiation degradation for $1\text{ ab}^{-1}$ of data with $c^{\tau 3\mu} / \Lambda^2 = 1/(50\text{ TeV})^2$  are shown in Table \ref{tab:4fermi.nEvents}. We find a signal-to-background ratio $\sim 1$ to be achievable at a $14\text{ TeV}$ machine, and this dramatically rises to $\sim 10^3$ at a $100\text{ TeV}$ machine.
The resulting bounds, assuming integrated luminosities of $1\text{ ab}^{-1}$ at $0.125, 3, 6, 14$, and $100\text{ TeV}$ are shown in \cref{fig:flavor.4fermi}.
It is clear that even a $3\text{ TeV}$ machine would be able to set a direct bound at the same level as the future Belle~II sensitivity, and this constraint can be improved by up to $\sim 2$ orders of magnitude at a higher energy machine.

In Fig.~\ref{fig:flavor.4fermi}, these results are compared to the constraints on the analogous 4-fermion operator in the $\mu \to 3 e$ decay with various ansatz regarding flavor violation. The diagonal lines show the expected relationship between the two Wilson coefficients assuming (i) flavor anarchy (all coefficients $\sim 1$), (ii) Minimal Leptonic Flavor Violation (MLFV)~\cite{Cirigliano:2005ck}, (iii) the Wilson coefficients scale like the square root of the Yukawa couplings of the leptons involved in the flavor violation, and (iv) the Wilson coefficients scale like the product of the same Yukawa couplings.\footnote{
One should be cautious that if a flavorful ansatz is used, the inferred scale from the $100\text{ TeV}$ bounds may be low enough that an effective field theory description is no longer valid.}

While the muon decay sets the strongest limits assuming anarchical coefficients, we see that a $14\text{ TeV}$ muon collider could set a bound comparable to the current SINDRUM limit in the case of MLFV, and would be comparable to the Stage-I Mu3e sensitivity if the coefficients scale like the square root of the Yukawa couplings.
In the extreme case, where the Wilson coefficients behave like the product of the two Yukawas, even a $3\text{ TeV}$ muon collider would provide a bound complementary to the final Mu3e sensitivity, with higher energy machines improving this bound by orders of magnitude.

In addition to the $\tau 3\mu$ operators considered here, we expect roughly similar sensitivity to the $\mu^+ \mu^- \to \mu^{\pm} e^{\mp}$ process, as well as to the processes such as $\mu^+ \mu^- \to \tau^{\pm} e^{\mp}$ that violate lepton flavor by two units.
Overall, we see that a muon collider would be capable of directly probing flavor-violating interactions that are quite complementary to future precision constraints.

\subsubsection{Direct probes of lepton-flavor violation in the MSSM}

Charged lepton flavor violation in the MSSM arises as a result of the soft-breaking terms in the slepton mass matrix having non-diagonal entries in the basis where the SM lepton Yukawas are diagonal. In this case, the physical sleptons will be mixtures of different flavors, and their interactions with leptons and neutralinos/charginos will be flavor-violating. These flavor violating interactions lead to processes such as rare muon decays or muon-to-electron conversion at loop level, and thus, low energy experiments can indirectly probe these interactions with sensitivities extending beyond the TeV scale, depending on the flavor structure of the theory~\cite{Altmannshofer:2013lfa, Ellis:2016yje}.
A high-energy muon collider, on the other hand, would not only be capable of producing superpartners at high masses, but would also provide direct measurements of the lepton-flavor violating processes that would complement these low-energy probes and provide detailed insight into the mechanism of supersymmetry breaking.

For simplicity, we will consider a simplified scenario where the effects of all scalar superpartners except for $\tilde{e}_R$ and $\tilde{\mu}_R$ decouple. In this case, the slepton mixing reduces to a $2 \times 2$ problem with slepton-mass squared matrix
\begin{equation}
\mathcal{M}^2_{\tilde{\ell},RR} = 
\begin{pmatrix} \Delta_{RR,11} & \widetilde{m}^2_{E,12} \\ \widetilde{m}^2_{E,12} & \Delta_{RR,22} \end{pmatrix} \, ,
\end{equation}
where the diagonal terms are the sum of both soft-SUSY-breaking scalar masses ($\widetilde{m}^2_E$) and $D$-terms as well as terms dictated by supersymmetry, and we have assumed the off-diagonal soft-breaking terms are CP conserving. This mass matrix can be diagonalized via a unitary matrix $U_R$ to yield mass eigenstates $m^2_{\tilde{e}_1}, m^2_{\tilde{e}_2}$ with the mixing angle given by
\begin{equation}
\frac{1}{2} \sin(2\theta_R) = \frac{\widetilde{m}^2_{E,12}}{m^2_{\tilde{e}_1} - m^2_{\tilde{e}_2}} \, .
\end{equation}
We will further consider the situation where the lightest supersymmetric particle is a pure bino with mass $M_1$ and assume the other neutralinos can be ignored. With $m_{\tilde{\ell}} > M_1$, the sleptons decay directly to a lepton and bino, and the latter will appear as missing momentum.

\paragraph{Nearly-degenerate sleptons:}

As a first benchmark scenario we consider the situation where the selectron and smuon are nearly degenerate in mass.
This situation is well-motivated from gauge-mediated supersymmetry breaking scenarios, and also leads to a strong suppression of the lepton-flavor violation via a ``super-GIM'' mechanism, allowing the superpartners to be relatively light.
Such a scenario was previously studied in the context of $e^+e^-$ collisions in Ref.~\cite{ArkaniHamed:1996au}, but for relatively light superpartners. A high-energy muon collider would allow similar tests with a substantially more impressive mass reach.

In the limit of a small mass splitting, the parameter governing the amount of flavor violation is given by
\begin{equation}
(\delta_{RR})^{12} \equiv \frac{\widetilde{m}^2_{E,12} }{ \sqrt{ \Delta_{RR,11} \Delta_{RR,22} } } \simeq \frac{1}{2} \frac{\Delta m^2}{\bar{m}^2} \sin(2\theta_R)\,,
\end{equation}
where we have introduced the average mass squared, $\bar{m}^2$ and mass-splitting $\Delta m^2 = m^2_{\tilde{e}_1} - m^2_{\tilde{e}_2}$.

\begin{figure}[htbp]
\centering
\includegraphics[width=0.45\linewidth]{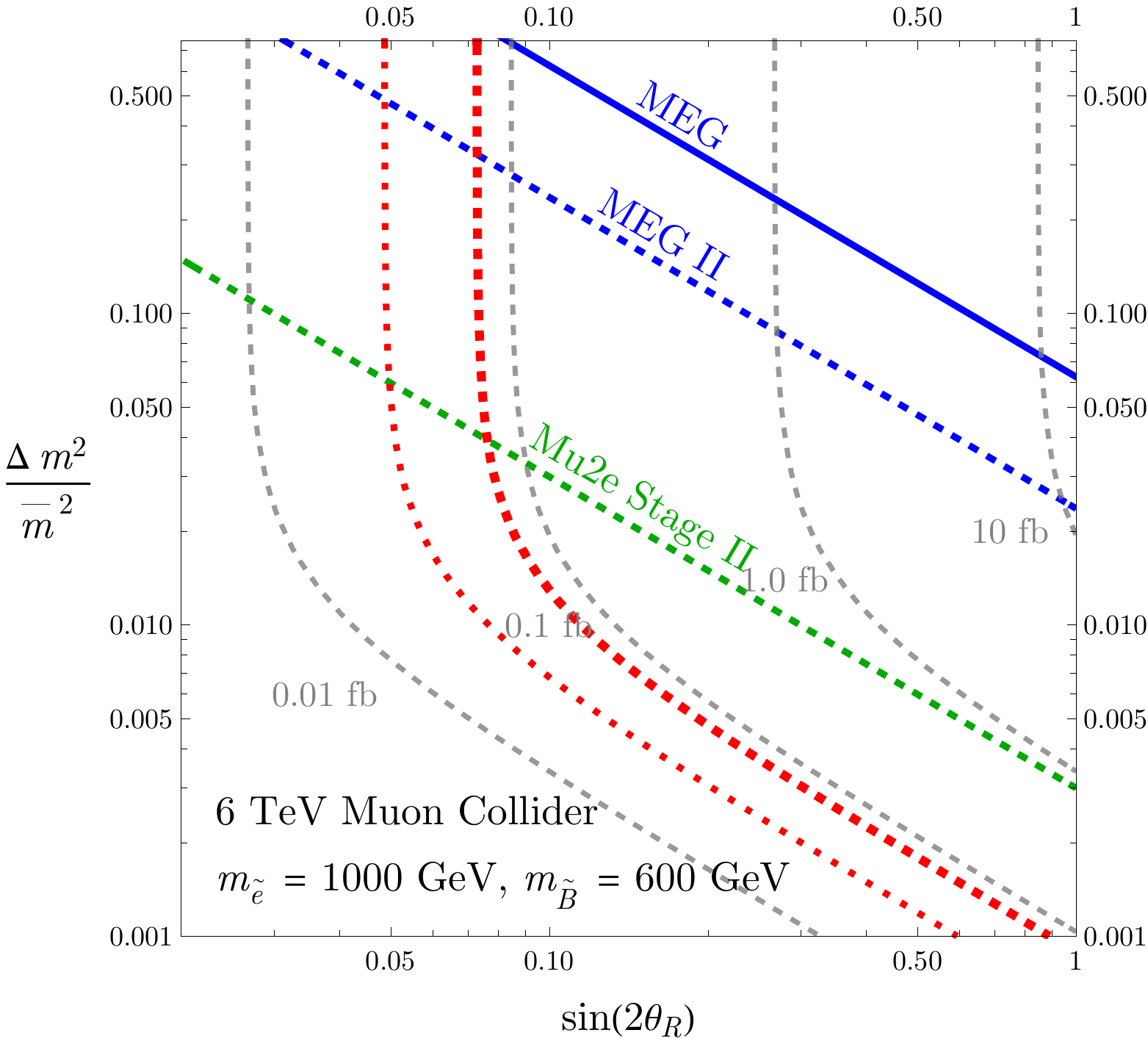}
~~~
\includegraphics[width=0.45\linewidth]{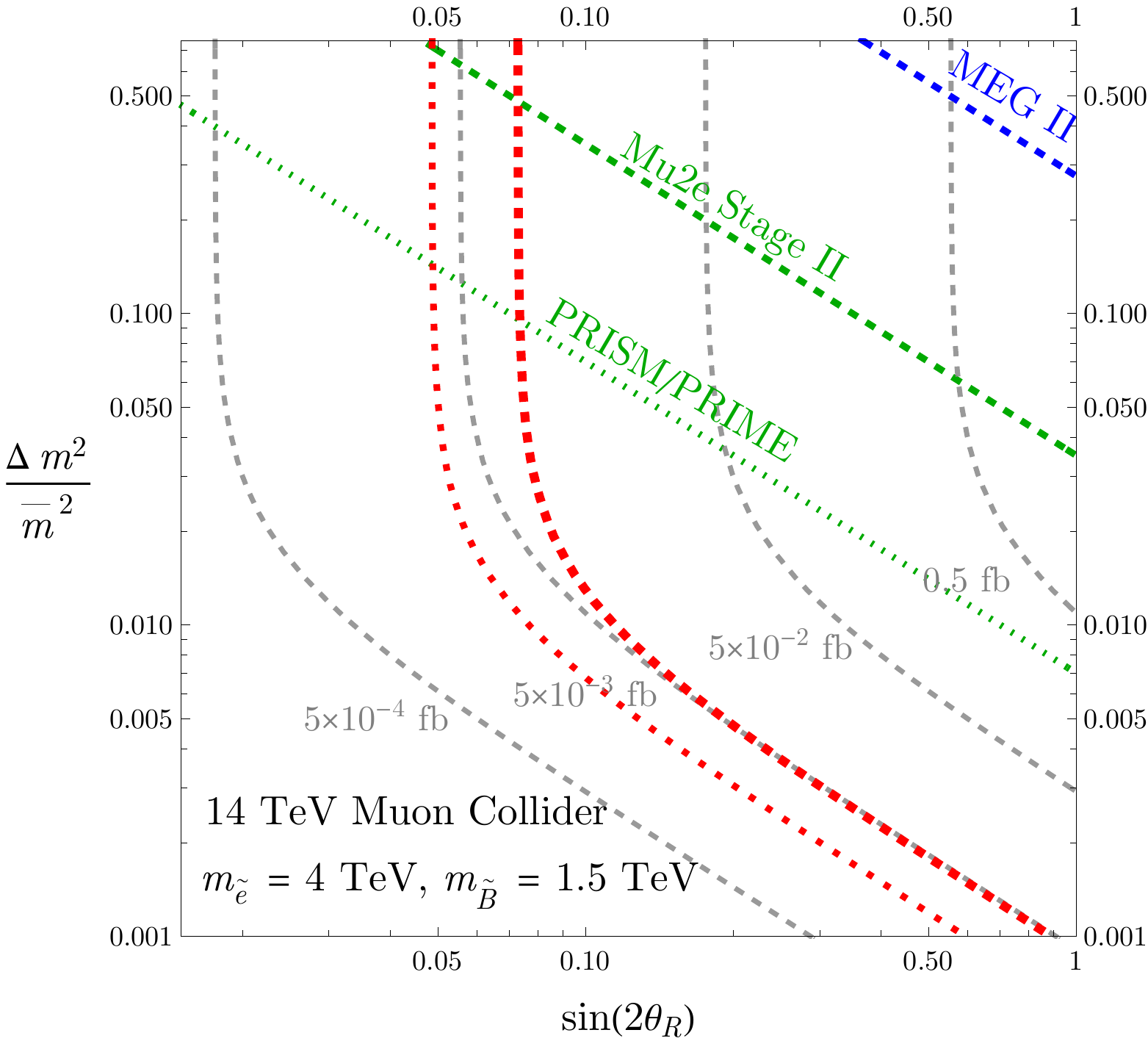}\\
\includegraphics[width=0.45\linewidth]{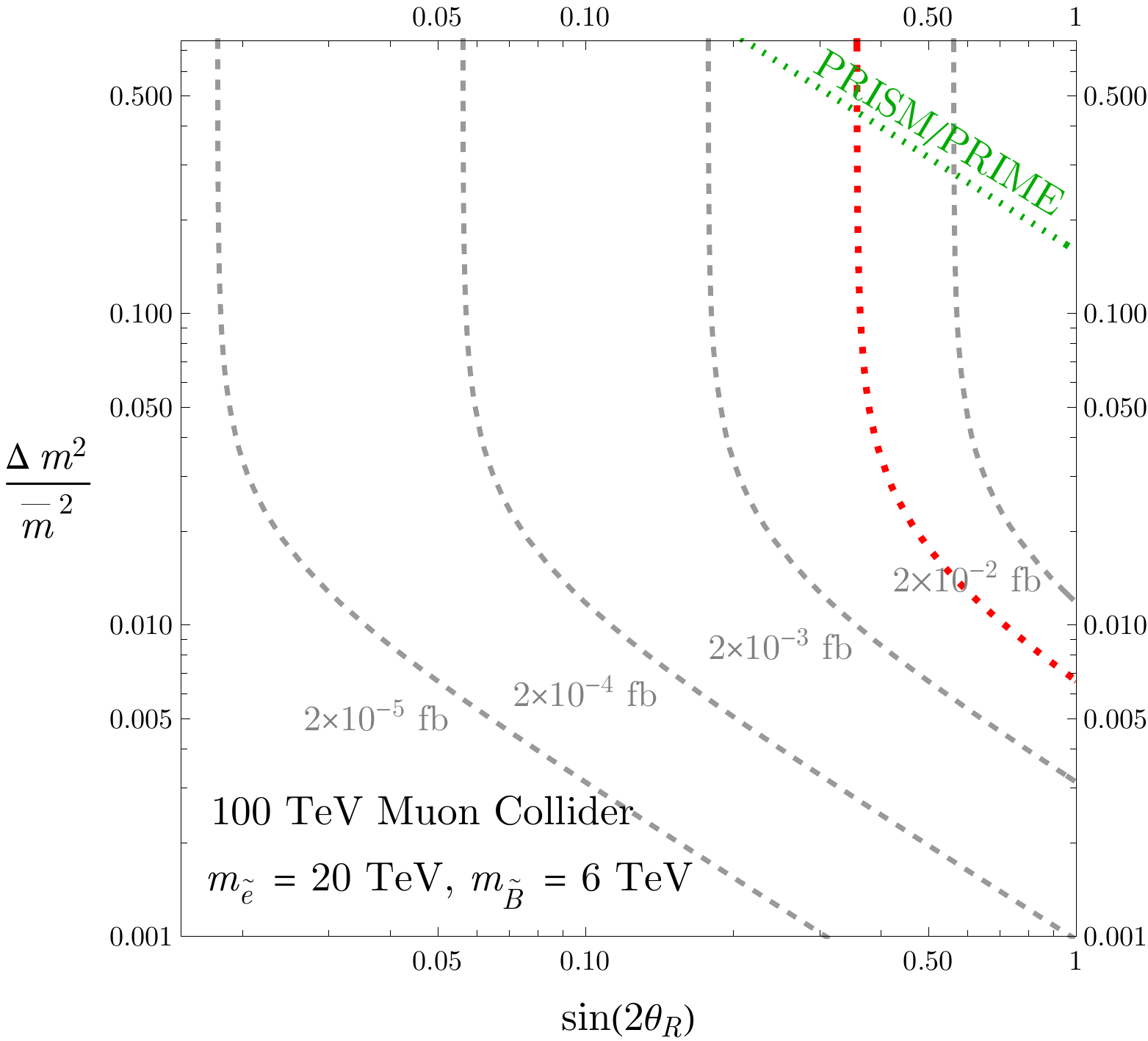}
\caption{Contours (in dashed gray) of the cross section in the $\delta m^2 / \bar{m}^2$ vs. $\sin(2\theta_R)$ plane for the flavor violating process $\mu^+ \mu^- \to \mu^{\pm}\, e^{\pm}\, \chi^0_1\, \chi^0_1$ for the nearly-degenerate slepton scenario described in the text. The bounds from current and future precision searches are overlaid in blue and green, along with red dashed and dotted curves showing the expected reach with 1 and $5\text{ ab}^{-1}$ of data.  The different panels correspond to benchmark choices of c.m.~energy, and superpartner masses.}\label{fig:flavor.degenerate}
\end{figure}

Following \cite{ArkaniHamed:1996au}, we can compute the probability that a gauge eigenstate $\tilde{\mu}$ decays into a final state with an electron -- including the interference effects when $\bar{m}\Gamma$ and $\Delta m^2$ are of similar size -- and thus find the cross section for the flavor violating process $\mu^+\mu^- \to \tilde{e}^+_{1,2}\, \tilde{e}^-_{1,2} \to \mu^{\pm}\,e^{\mp}\, \chi_1^0\, \chi_1^0$ as a function of the mass splitting and mixing angle.
The results are shown in \cref{fig:flavor.degenerate} for several choices of slepton and bino mass and at several different several of mass energies alongside the current bounds from $\mu \to e\gamma$ from MEG~\cite{TheMEG:2016wtm} and the expected sensitivities from several future experiments~\cite{Baldini:2018nnn, Bartoszek:2014mya, Abusalma:2018xem}.

The primary background for these flavor-violating processes at a lepton collider is production of the different flavor final states and missing energy via intermediate $W$ bosons.
The total cross section for this background, including branching ratios, is $52\text{ fb}$ at a $6\text{ TeV}$ collider ($15 / 0.6 \text{ fb}$ at a $14 / 100\text{ TeV}$ machine), but the kinematics of this process are quite different from the slepton-pair production signal of interest.
Moreover, for the flavor-violating scenarios at hand, it is likely that the relevant slepton and neutralino masses will have already been measured from the corresponding flavor-conserving processes (for details on how this can be done, see e.g., Refs.~\cite{AguilarSaavedra:2001rg, Freitas:2004re}).
With the slepton masses known, it is then possible to fully reconstruct the two final state neutralino momenta by requiring that the neutralino and lepton momenta satisfy the slepton mass-shell constraint, along with conservation of energy and momentum. In general, these conditions will be impossible to satisfy for the background events, and indeed, we find in simulation that only $\sim 1 / 500$ background events can reconstruct the neutralino momenta while satisfying conservation of energy, while $\sim 98\%$ of the signal events reconstruct the momenta successfully.

In \cref{fig:flavor.degenerate}, we show the $5\sigma$ discovery reach for the three benchmark scenarios assuming $1\text{ ab}^{-1}$ ($5\text{ ab}^{-1}$) of data in red dashed (dotted) lines, assuming the efficiencies just quoted apply at all masses and energies. We see that a $6\text{ TeV}$ muon collider would probe a great deal of parameter space complementary to the Stage-II Mu2e bounds in the case of $1\text{ TeV}$ nearly-degenerate sleptons. More energetic colliders would be able to measure parameters that are beyond the reach of even the PRISM/PRIME future sensitivity.

\paragraph{A single light slepton:}

\begin{figure}[htbp]
\centering
\includegraphics[width=0.45\linewidth]{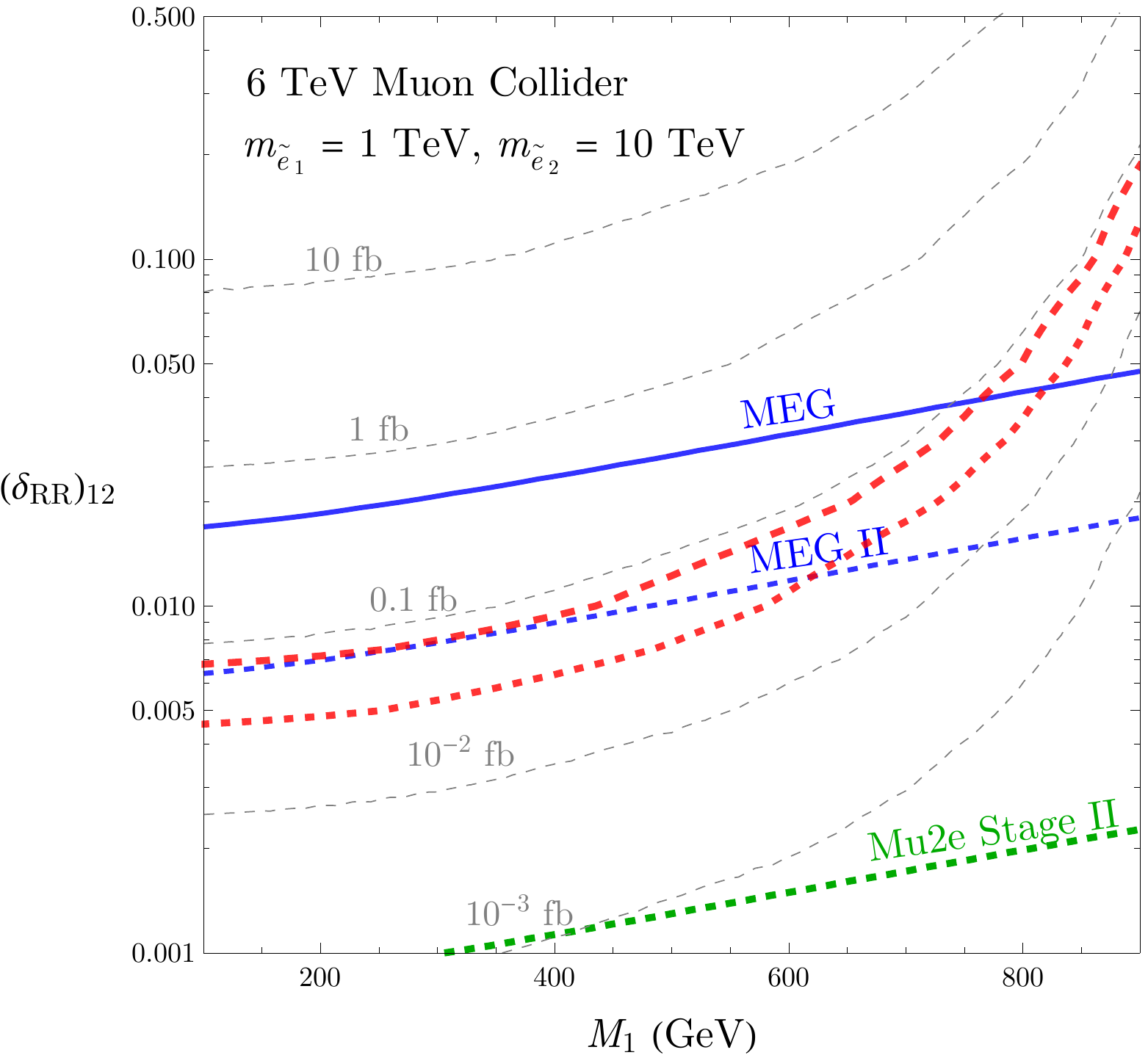}
~~~
\includegraphics[width=0.45\linewidth]{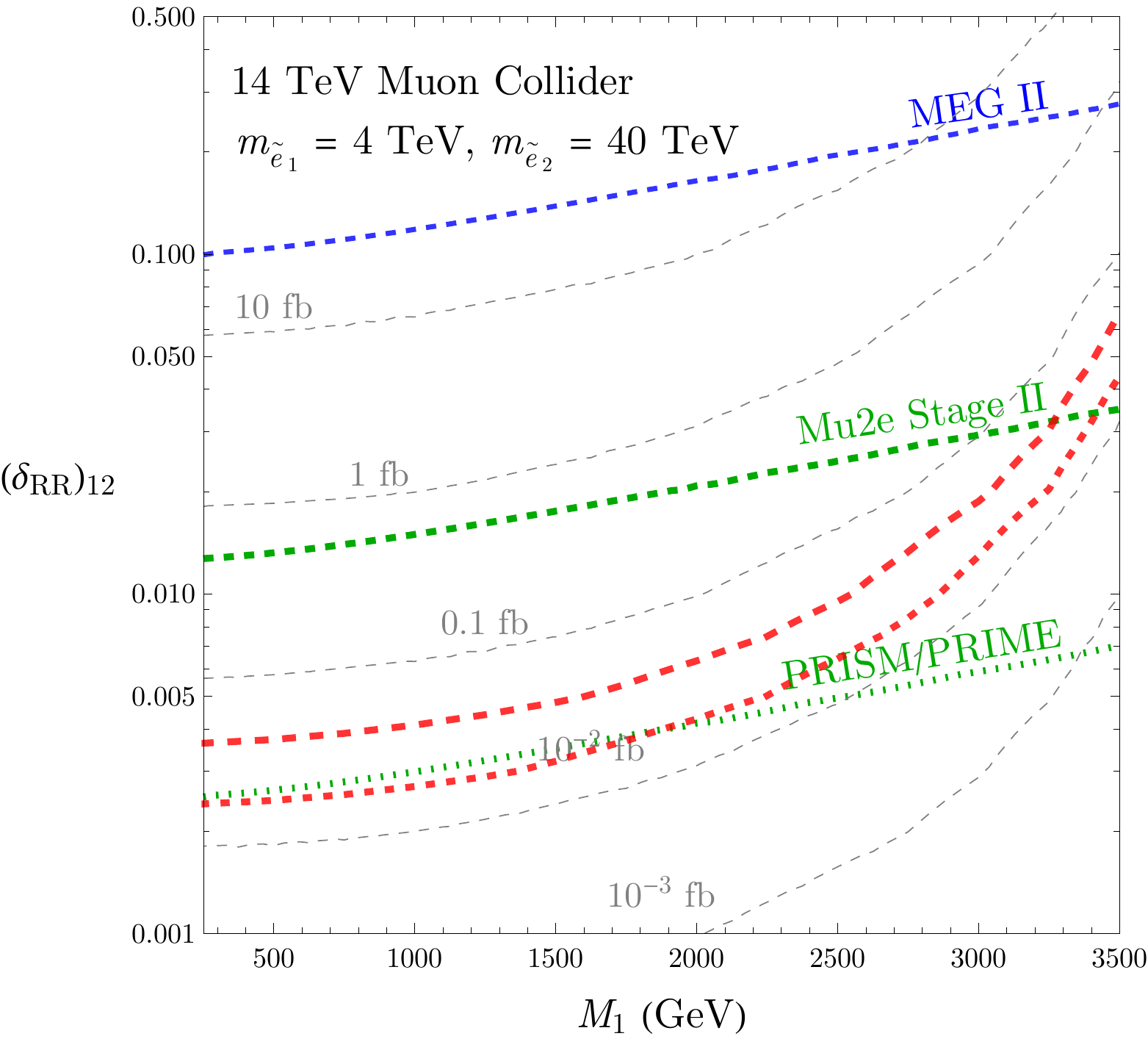}\\
\includegraphics[width=0.45\linewidth]{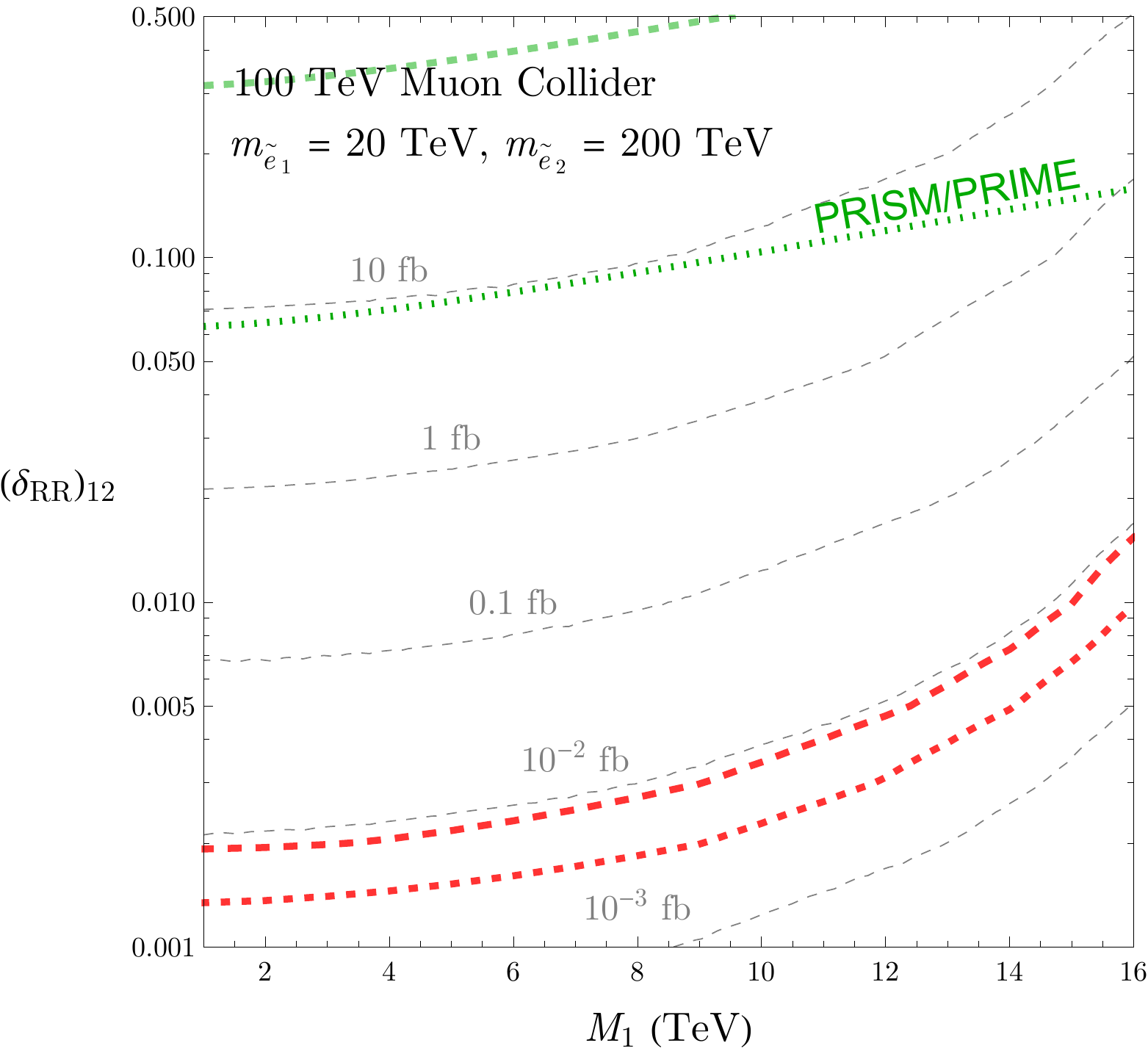}
\caption{Similar to \cref{fig:flavor.degenerate}, but in the $(\delta_{RR})_{12}$ vs. $M_1$ plane for the single light slepton scenario.}\label{fig:flavor.vary_bino}
\end{figure}

As an alternative scenario, we consider a situation with only one slepton within reach of the collider that is a nearly pure right-handed selectron, but with some small mixing with a heavier right-handed smuon, i.e., $\delta_{RR}^{12} > 0$. We can then examine the cross sections for the flavor violating process as a function of $\delta_{RR}^{12}$ and $M_1$. The results are shown in \cref{fig:flavor.vary_bino}, where we again superimpose the current and future sensitivities from the low-energy experiments.

The backgrounds for this scenario are the same as in the nearly-degenerate case already described, and we assume that the same efficiencies for removing the background described above can be achieved for all neutralino masses and collider energies here as well. The resulting $1$ and $5\text{ ab}^{-1}$ discovery reaches are again shown as dashed and dotted red lines, respectively. In this situation, a $6\text{ TeV}$ collider would cover much of the parameter space probed by MEG II, except when the neutralino is close to the light slepton mass, but would fall short of the eventual Mu2e Stage II sensitivity. A higher energy muon collider, however, would surpass the Mu2e sensitivity, and could measure flavor violating insertions as small as a few $\times~10^{-3}$ for light sleptons at 4 TeV and 20 TeV with a muon collider of 14 and 100 TeV respectively. 
Overall, these benchmarks illustrate that muon colliders would have impressive capabilities for not only discovering superpartners, but measuring their flavor structure in detail.

\subsection{Gravitational waves}
The central importance and capabilities of a future collider must be seen within the context of the broader experimental efforts in 
particle physics. We have already discussed the connections between possible discoveries in precision flavor and CP violation 
experiments,  direct and indirect dark matter detection, as well as searches for new light fields and dark forces. On the cosmological front,
we are entering an era in which dramatic new forms of ``fossil'' evidence for BSM physics may be found,  within stochastic gravitational wave backgrounds (SGWB) \cite{Caprini:2018mtu, Christensen:2018iqi} and within primordial non-Gaussianities in Large Scale Structure (LSS) and high-redshift 21-cm 3D ``maps'' \cite{Meerburg:2019qqi}.   While discoveries in any of these non-collider experiments would be spectacular, powerful new colliders would provide the ``gold standard'' laboratory conditions to corroborate, connect, extend, and analyze their full significance. 

First order cosmological phase transitions that could occur due to extensions of 
the SM or within dark sectors could be powerful sources of SGWB, while possibly providing the non-equilibrium conditions needed for generating matter asymmetries. 
The peak SGWB frequency, after redshifting from the time of production  in the very early universe, is given by 
\begin{equation}
\omega \simeq 0.03\, {\rm mHz}  \frac{\beta}{H_\textrm{PT}} \frac{T_\textrm{PT}}{\rm TeV}\,,
\end{equation}
where $T_\textrm{PT}$ is the temperature immediately after the phase transition,  $1/\beta$ is essentially its duration, and $H_\textrm{PT}$ is the Hubble expansion rate during this era \cite{Kamionkowski:1993fg, Huber:2008hg}. Typically, one expects
$\beta \sim {\cal O}(10 \text{ - } 100) H_\textrm{PT}$. Fortuitously then, for $T_\textrm{PT}$ in the BSM-motivated range, TeV - $100$ TeV, we can expect SGWB in roughly the mHz - Hz range accessible to proposed gravitational wave detectors such as LISA, BBO, and DECIGO. If a SGWB from a phase transition is detected, it would be critical to piece together the information in its frequency spectrum with the 
complementary microphysics accessible within collider experiments to whatever extent possible. Quite plausibly, these elements relate to extensions of the SM Higgs sector. A high energy muon collider provides a balance of potential to probe the Higgs sector, to create very massive BSM states related to the phase transition, and given its clean environment to possibly produce and diagnose a small number of events resulting from the presence of a dark sector in case the SGWB originated there. 

Let us illustrate the interplay between gravitational wave detection and a muon collider  by one of the central 
questions of particle physics: Is the Higgs boson truly elementary, or a composite of new strongly-coupled confined constituents? 
In the latter case, the compositeness would greatly mitigate the electroweak hierarchy problem (see \cite{Bellazzini:2014yua,Panico:2015jxa} for review). 
We would then expect to see excited composites at high-energy colliders as well as a deconfinement-to-confinement 
phase transition at gravitational wave detectors. This physics is usually discussed in its more 
theoretically tractable AdS/CFT dual formulation as a 5D Randall-Sundrum I (RS1) model \cite{Randall:1999ee}, 
translating to the collider phenomenology of Kaluza-Klein excitations and the transition between the black-brane horizon and IR-brane phases of RS1~\cite{Creminelli:2001th, Randall:2006py, Nardini:2007me, Konstandin:2010cd, Konstandin:2011dr}. For a very recent work refining phase transition dynamics and for references, see \cite{Agashe:2020lfz}.
For example, if LISA were to detect a SGWB with peak frequency $0.5$ mHZ and gravitational power spectrum of $3 \times 10^{-8}$ at the peak, 
in the RS1 framework this would correspond to a supercooled phase transition with $T_\textrm{PT} \simeq 1.3$ TeV, with Kaluza-Klein resonances in the (multi-)TeV regime accessible to muon colliders.\footnote{Given a high degree of supercooling, say by a factor of $10^3$, we are assuming that bubble collisions dominantly source the SWGB signal and that these are captured by the ``envelope approximation''. This approximation can be invalid in some regimes, but its replacements would not invalidate the general complementarity of the collider and gravitational wave experiments. See the discussion reviewed in \cite{Caprini:2019egz}.} If say a $2$ TeV ``radion'' spin-$0$ excitation were discovered at a muon collider via vector boson fusion, its mass and cross-section would independently determine  $T_\textrm{PT}$, giving a valuable cross-check of the underlying physics. Further, it would bound the heavy Kaluza-Klein spectrum to begin below $40$ TeV.
If say a spin-$2$ Kaluza-Klein graviton with mass $20$ TeV is eventually discovered in associated production with the radion, the cross-section for this process (along with cross-sections for radion processes) together with the frequency and power spectrum from the SGWB data would allow us to extract the small ``critical exponent'' responsible for generating the large electroweak hierarchy, in this example $\epsilon=0.1$.  We emphasize that this central parameter of the composite dynamics would be difficult to extract experimentally without combining collider and gravitational wave data. All the relevant measurements would be challenging, but worthy ambitions for future experiments.

Upcoming precision LSS and 21-cm surveys offer the potential to detect heavy particle production and propagation during inflation, imprinted on the non-Gaussian bispectrum in distinctive non-local effects (non-analytic in co-moving momenta). This field of ``cosmological collider physics'' is sensitive to particle masses of order the inflationary Hubble scale or even somewhat higher \cite{Chen:2009zp, Arkani-Hamed:2015bza}. (For very recent work and references, see \cite{Bodas:2020yho}). We do not as yet know the scales of inflation. If new particles are discovered in cosmological non-Gaussianities, they may lie far above the reach of terrestrial colliders, in which case they would give complementary information to what we learn from even a powerful muon collider. But there are two scenarios in which they could give us a (pre)view of collider-accessible physics: (i) if the inflationary Hubble scale is of order $100$ TeV or less, then obviously ``cosmological collider physics'' may directly be within reach of future terrestrial colliders; (ii) even if the inflationary Hubble scale is orders of magnitude above the TeV scale, there is a ``heavy-lifting'' mechanism \cite{Kumar:2017ecc} whereby the particles seen in non-Gaussianities were given inflationary scale masses through strong curvature effects, but such effects are negligible today so that the particles may now be within terrestrial collider reach.


\section{Summary and Future Directions}
\label{sec:conc}
The goal of this work is to paint the physics case of a high-energy muon collider with a broad brush, emphasizing the sense in which such a collider is positioned to answer the many questions posed or sharpened by the discovery of the Higgs. 
The broad outlines of this case are drawn by the physics of both muon annihilation and vector boson fusion, which in tandem provide compelling rates for Standard Model and beyond-the-Standard Model processes across a range of energies. Relative to recent work highlighting the significance of vector boson fusion, we have emphasized the value of muon annihilation as a discovery mode for sufficiently distinctive new physics.
To characterize the rich physics of the initial state, we surveyed descriptions of the virtual electroweak gauge boson content of high-energy muons ranging from the Effective Vector Approximation to electroweak PDFs. As a practical matter, we emphasized the sense in which the simple ``leading log PDF'' is sufficient to capture most of the qualitative physics, and briefly explored the impact of finite $W$ and $Z$ masses on electroweak PDFs.

Turning to the physics case itself, we highlighted the potential of a muon collider to illuminate various aspects of electroweak symmetry breaking, dark matter, and naturalness. The measurement of the top Yukawa coupling provided a particularly sharp test case for probing the physics of electroweak symmetry breaking, in which deviations from the Standard Model prediction lead to significant changes in the rates for longitudinal vector boson fusion into $t \bar t$ pairs. This is sufficient for a muon collider at or above $\sqrt{s} = 14$ TeV to test the top Yukawa coupling beyond the expected sensitivity of the LHC. The increase in reach provided by polarized beams exemplifies their potential value.  As a second test case relevant to the Higgs potential and the electroweak phase transition, we proposed two avenues for constraining the so-called ``nightmare scenario'' in which the electroweak phase transition is strengthened by a light, $\mathbb{Z}_2$-symmetric singlet scalar that only couples to the Higgs. In addition to the canonical search for pair production of the scalar in missing energy final states, we noted the possibility of again using longitudinal vector boson fusion into $t \bar t$ pairs, this time looking for a feature in the invariant mass distribution associated with the scalar threshold. In both channels, a collider operating above $\sqrt{s} = 10$ TeV can cover much of the motivated parameter space. We further considered direct searches for additional Higgs bosons, focusing on the case of a singlet scalar mixing with the Higgs. Here the abundant production of such scalars via vector boson fusion would allow a collider operating above $\sqrt{s} = 10$ TeV to probe mixing angles whose indirect imprints (in the form of Higgs coupling deviations) lie beyond the reach of any proposed collider, and a collider operating above $\sqrt{s} = 30$ TeV to far exceed the mass reach of a 100 TeV $pp$ collider.  
 
 The implications of a muon collider for dark matter are exemplified by its coverage of ``minimal dark matter'' models, in which the dark matter particle resides in an electroweak multiplet whose interactions with SM gauge bosons can generate the observed abundance. We highlighted two classes of search strategies for these multiplets, using either a missing mass or a disappearing track signature. Depending on the integrated luminosity, a high-energy muon collider operating at $\sqrt{s} = 10$ or 14 TeV can {\it discover} smaller electroweak representations (such as an $SU(2)$ doublet or triplet) at their thermal targets, while a collider operating at $\sqrt{s} = 30$ - $100$ TeV can cover the thermal targets for higher electroweak representations as well. 
 
 We illustrated the potential of a muon collider to discover or constrain natural explanations of the weak scale by considering the reach for both solutions to the ``big'' hierarchy problem (supersymmetry or compositeness) and the ``little'' hierarchy problem (represented here by the Twin Higgs). Within the context of supersymmetry, searches for the higgsino and the stop would allow a muon collider operating at $\sqrt{s} \gtrsim 20$ TeV to comprehensively cover tuning beyond the per mille level, and collisions at $\sqrt{s} = 30$ TeV would reach stop masses consistent with the observed Higgs mass even in the absence of significant stop mixing. Despite its lack of electroweak quantum numbers, even the gluino could be probed beyond LHC limits leveraging radiation off of $q \bar q$ pairs at $\sqrt{s} \gtrsim$ 10 TeV. In addition to these conventional channels, a muon collider can provide the first direct collider test of the supersymmetry-breaking sector itself, achieving sensitivity to direct production of the gravitino consistent with LHC bounds on Standard Model superpartners, provided a low scale of supersymmetry breaking. The coverage of compositeness is comparable, with the addition of powerful indirect tests coming from the energy growth of irrelevant operators parameterizing mixing of the Higgs and composite states. Even solutions to the ``little'' hierarchy problem without conventional LHC signatures, such as the Twin Higgs, lie within reach of a muon collider. In addition to its considerable sensitivity to indirect effects, a high-energy muon collider would be able to directly access the radial mode of these models and decisively confirm or falsify them as a viable explanation of the weak scale.
 
 A host of other experiments will indirectly probe physics as high as the PeV scale in the years preceding the first beams at a high-energy muon collider. To this end, we have considered the complementarity between a muon collider and potential signals of new physics in electric dipole moments, flavor violation, and gravitational waves. Of particular interest are precision tests of charged lepton flavor violation in processes such as $\mu \rightarrow e \gamma, \mu \rightarrow 3 e, \tau \rightarrow 3 \mu$ and $\mu$-to-$e$ conversion. Here we have explored the detailed reach of a muon collider for both indirect sources of lepton flavor violation (such as flavor-violating four-fermion operators) and direct sources (such as flavor-violating slepton interactions in the MSSM), in both cases finding that muon colliders below $\sqrt{s} \sim 10$ TeV provide complementary sensitivity to experiments such as Mu2e and Mu3e, while more energetic colliders are capable of probing parameter space beyond the reach of future proposals.

Broadly speaking, we find that a muon collider operating at tens of TeV and tens of ab$^{-1}$ is capable of surpassing the indirect reach of proposed $e^+ e^-$ Higgs factories and the direct reach of proposed 100 TeV $pp$ colliders, covering a broad swathe of motivated physics beyond the Standard Model and probing explanations for potential signals in experiments across the many frontiers of particle physics. Perhaps the most compelling gains stand to be made by colliders operating between $\sqrt{s} = 14$ - $30$ TeV, for which most of the questions posed by the discovery of the Higgs boson may be decisively answered. 

Needless to say, many of the projections made in this work are naive in light of the significant uncertainties and many unresolved challenges facing both accelerators and detectors. Nonetheless, we hope that they provide a qualitative guide to the energies and luminosities that would position a future muon collider as a comprehensive successor to the LHC, pinpointing a variety of directions that merit more careful study. Beyond characterizing the reach in conventional benchmarks for a future collider program, we have identified a number of opportunities uniquely suited to a muon collider, including direct tests of low-energy supersymmetry breaking. Key outstanding questions include the performance and prospects of searches involving missing energy, which are central to the coverage of dark matter, the electroweak phase transition, and supersymmetry breaking; the invariant mass resolution in heavy Standard Model final states, essential to making the most of the abundant opportunities provided by the fusion of longitudinal vector bosons; and the feasibility of fully instrumenting the forward region, which will shape the set of available observables and the composition of signal processes. 

Much remains before us. But the muons are calling, and we must go.


\section*{Acknowledgments}

We are grateful to Ken Van Tilburg for collaboration during the early stages of this work. We thank Kaustubh Agashe, Tao Han, Sergo Jindariani, Fabio Maltoni, and Dave Soper for useful conversations.

The work of I.~Banta, T.~Cai, N.~Craig, G.~Koszegi, A.~McCune, and T.~Trott is supported by the U.S.~Department of Energy under the grant DE-SC0011702. The work of D.~Buttazzo and A.~Tesi is supported by MIUR under grant number 2017L5W2PT.
The work of T.~Cohen is supported by the U.S.~Department of Energy under grant number DE-SC0011640.
The work of J.~Fan is supported by the U.S.~Department of Energy under grant number DE-SC-0010010. 
The work of I.~Garcia~Garcia is funded by the Gordon and Betty Moore Foundation through Grant GBMF7392, and in part by the National Science Foundation under Grant No.~NSF PHY-1748958.
The work of S.~Homiller, Q.~Lu, and M.~Reece was supported in part by the DOE Grant DE-SC0013607 and by the Alfred P.~Sloan Foundation Grant No.~G-2019-12504.
The work of P.~Meade is supported in part by the National Science Foundation under Grant No.~NSF PHY-1915093.
The work of A.~Mariotti is supported in part by the SRP HEP-VUB and by the EOS-be.h n.30820817.
D.~Sutherland has received funding from the European Union's Horizon 2020 research and innovation programme under the Marie Skłodowska-Curie grant agreement No.~754496.
The work of L.-T. Wang is supported by the DOE grant DE-SC0013642. The work of S.~Koren is funded by a Mafalda and Reinhard Oehme Postdoctoral Fellowship.
The work of M.~Ekhterachian and R.~Sundrum is supported by the NSF grant PHY-1914731.

\appendix
\addcontentsline{toc}{section}{Appendix} 
\section{Simplified Models}
\label{app:simplified}
\newcommand{\pr}[1]{\ensuremath{\mathtt{\color{red}#1}}}
\newcommand{\prt}[1]{\ensuremath{\mathtt{#1}}}

In order to benchmark the physics potential of muon colliders over a range of energies, we compute the rates for various processes using representative simplified models. Simplified models are defined in {\sc FeynRules} using a combination of public and custom model files. Wherever possible, we simulate processes in both {\sc Whizard}~\cite{Moretti:2001zz, Kilian:2007gr, Christensen:2010wz} and {\sc Madgraph5}~\cite{Alwall:2014hca}. We group the majority of processes into three categories: ``annihilation'' when states are produced directly in $\mu^+ \mu^-$ collisions; ``neutral VBF'' when states are produced via vector boson fusion from combinations of electroweak vectors carrying zero total charge; and ``charged VBF'' when states are produced via vector boson fusion from combinations of electroweak vectors carrying nonzero total charge. For the most part, we present results in annihilation and neutral VBF channels. For both charged and neutral VBF processes, we exclude contributions from on-shell $W$ and $Z$ bosons by imposing appropriate invariant mass cuts on the final state, as in \cite{Costantini:2020stv}. In what follows, we typically do not display simulated VBF cross sections for mass points close to threshold given the strong sensitivity to phase space cuts; as noted in the text, annihilation production dominates near threshold. Where relevant, we also consider QCD ``bremsstrahlung'' processes, in which a gluon is radiated off a hard final state quark and then ``splits'' into the strongly-interacting final state of interest; this is particularly relevant for gluino production in supersymmetric models. 

To validate the results of our simulations, we compute a selection of representative cross sections analytically using a combination of {\sc FeynArts}, {\sc FormCalc}, and {\sc FeynCalc}. We compute select annihilation cross sections explicitly, while for select VBF cross sections we compute the ``partonic'' cross sections explicitly and obtain total inclusive cross sections via numerical convolution with the LL luminosity functions detailed in \cref{sec:PDF}. 

\subsection{Standard Model}
We begin by exploring some SM rates as a function of $\sqrt{s}$.  This provides some interesting benchmarks on its own, and also allows one to get a sense of the raw rate for backgrounds that are relevant when estimating the new physics potential of a muon collider.

\begin{figure}[h!] 
   \centering
   \includegraphics[width=3in]{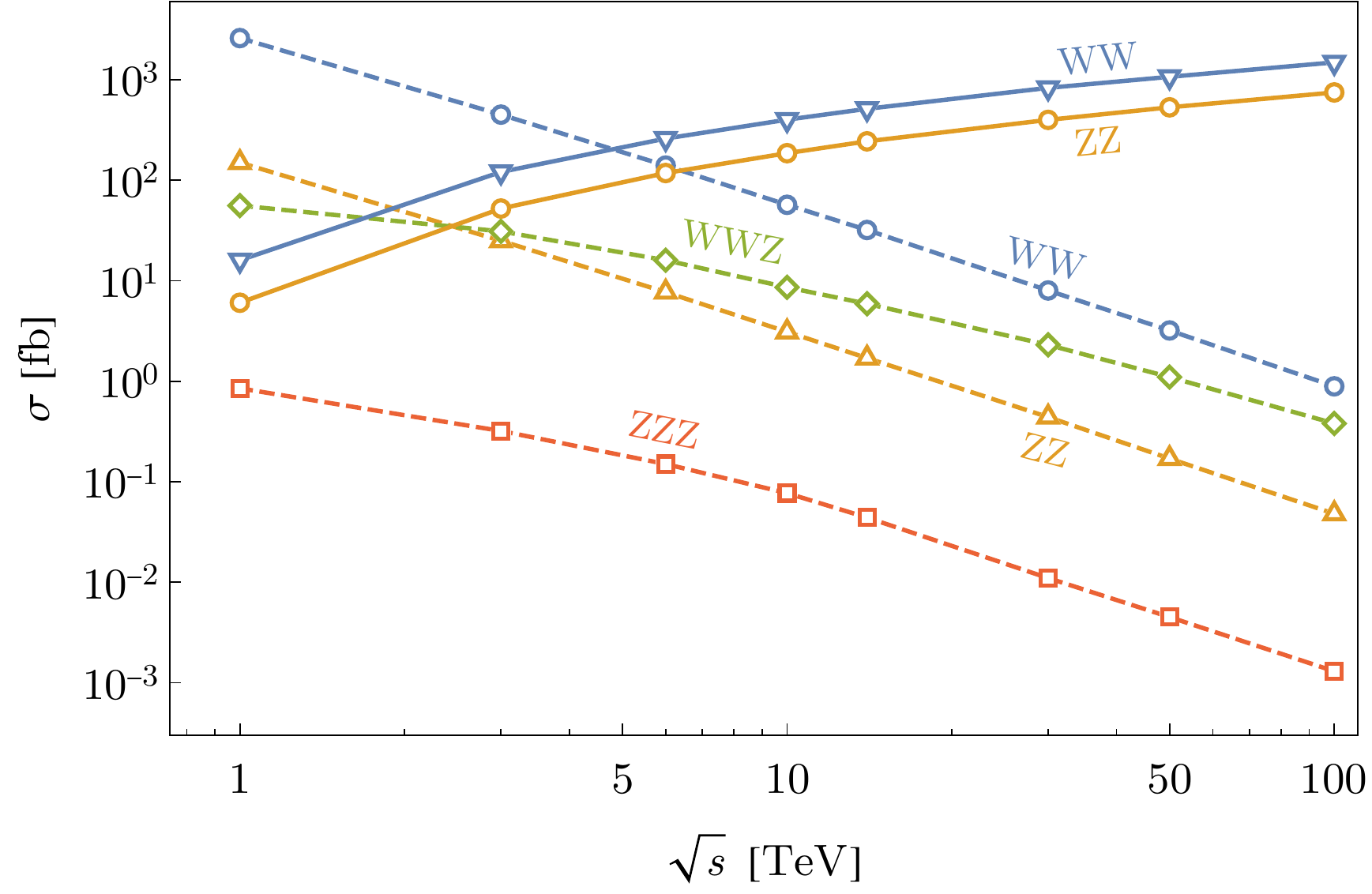} 
      \includegraphics[width=3in]{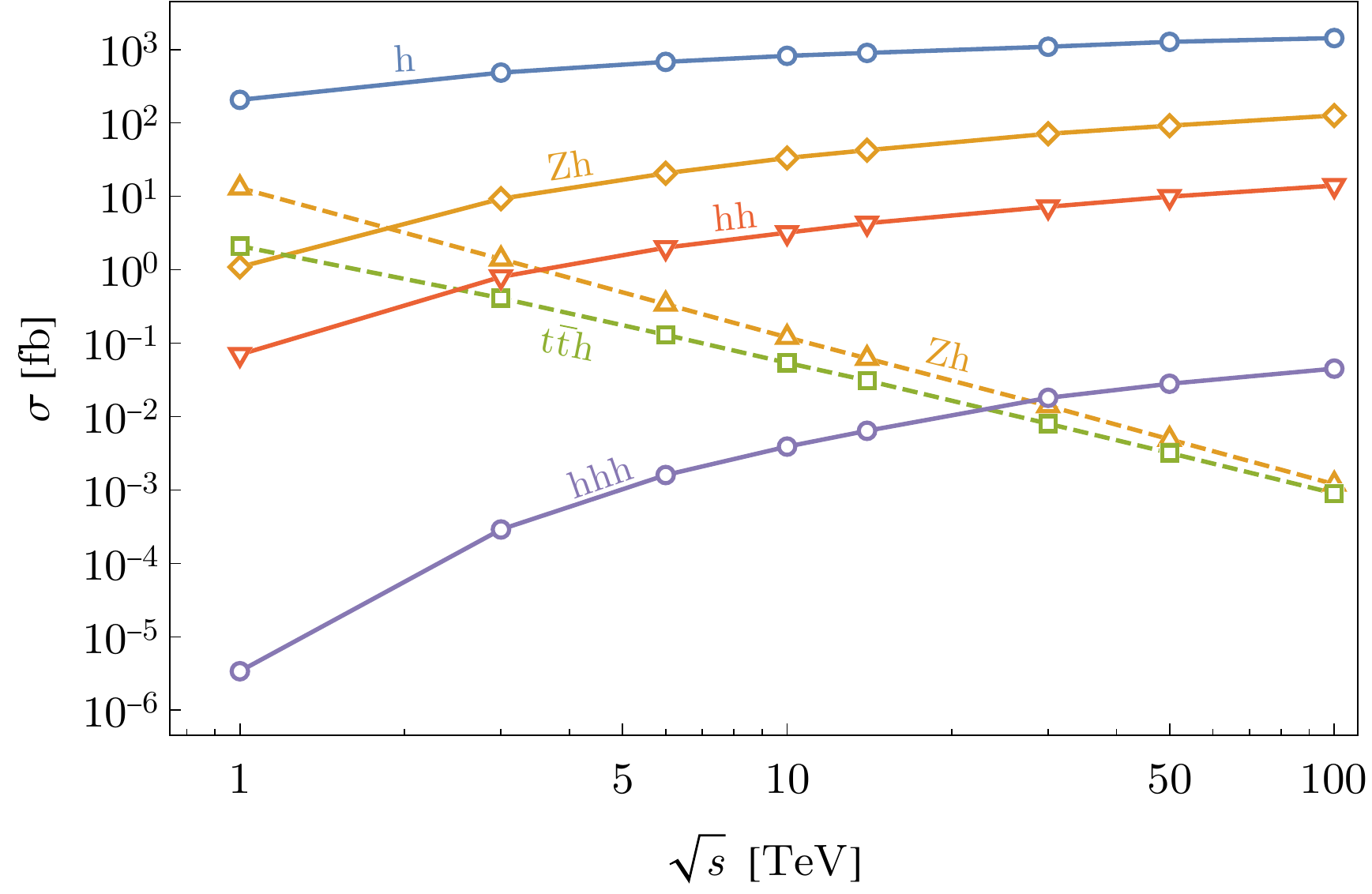} \\[15pt]
         \includegraphics[width=3in]{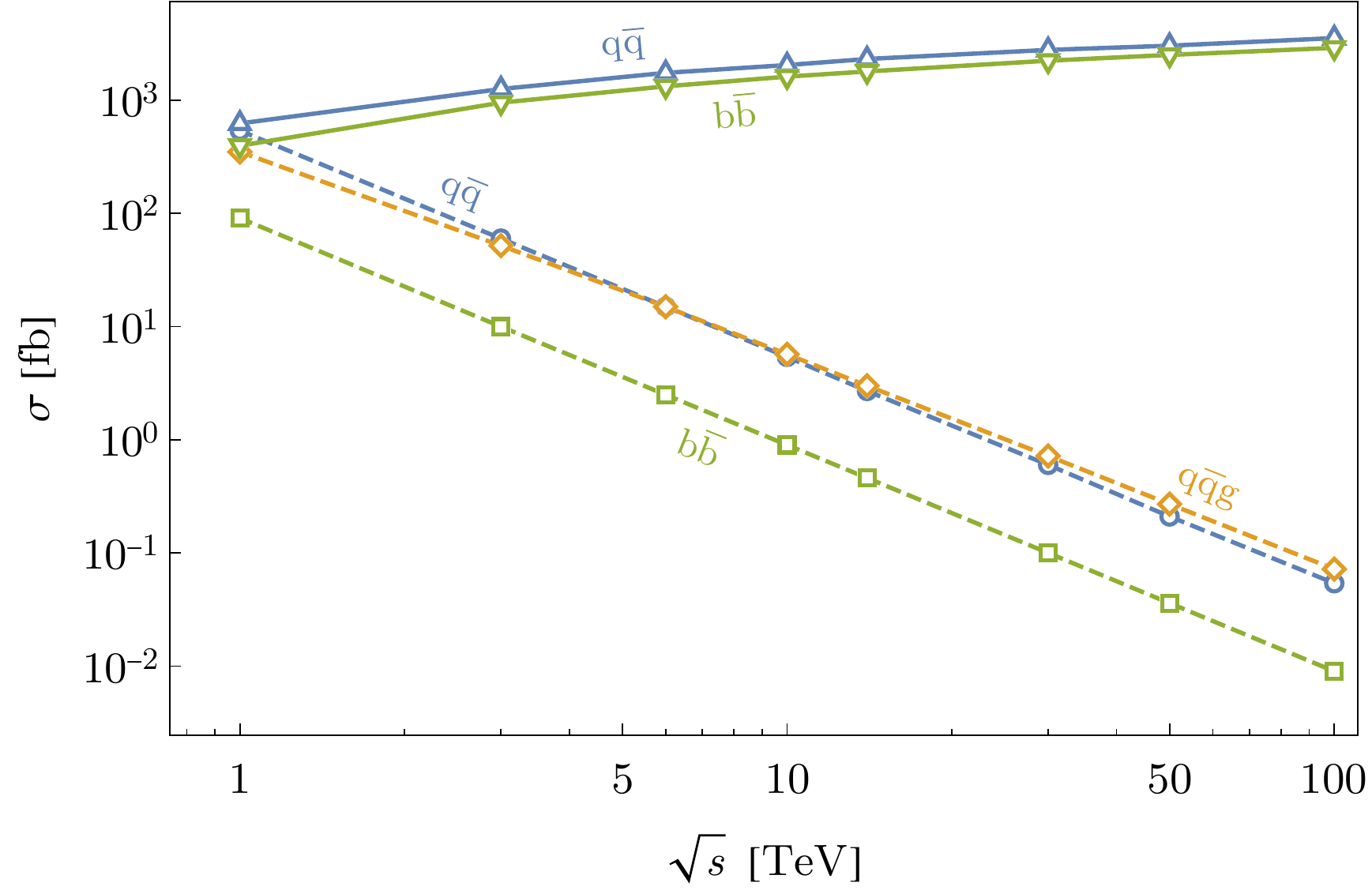} 
            \includegraphics[width=3in]{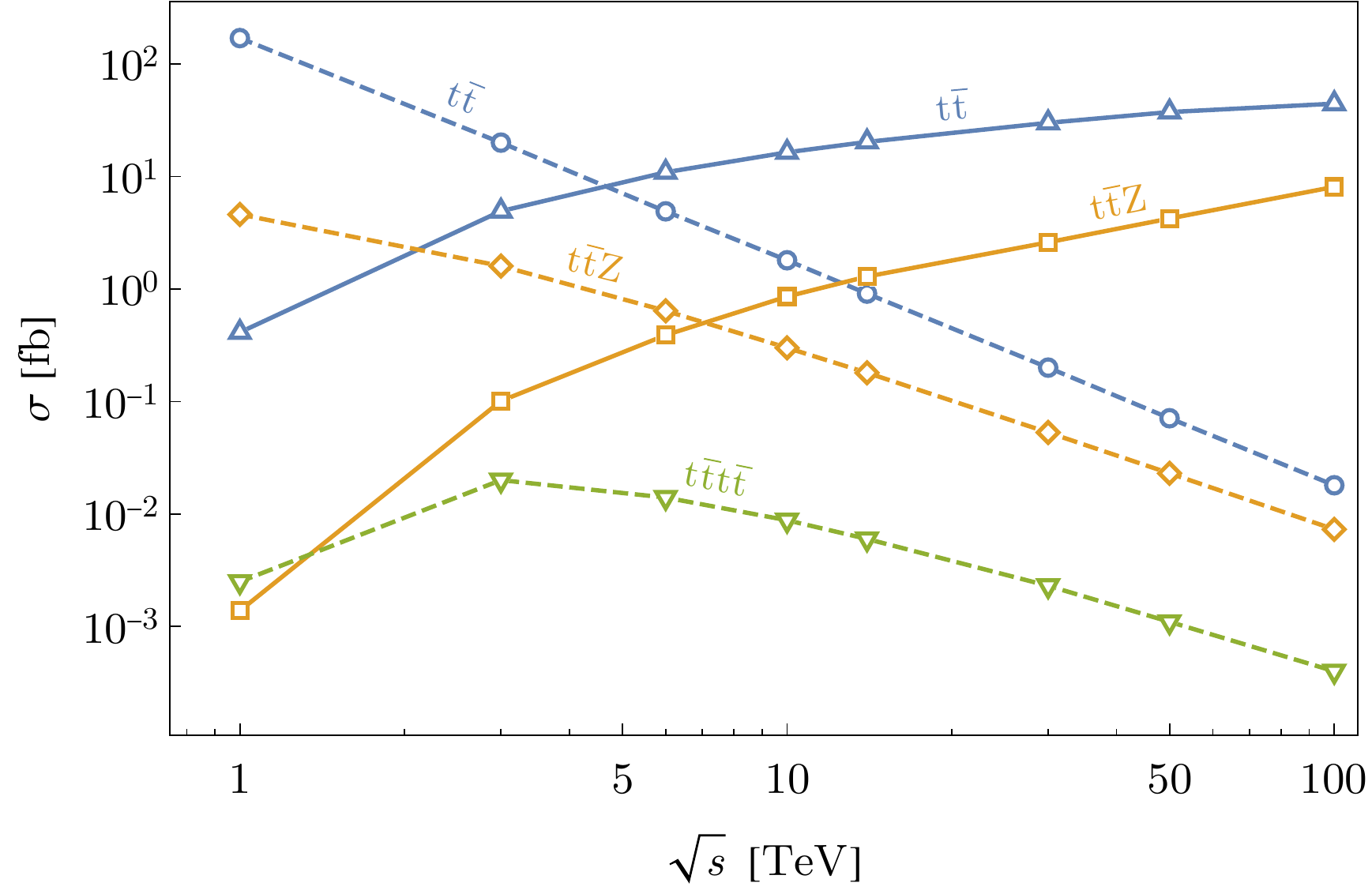} 
   \caption{Cross sections for representative annihilation (dashed) and VBF (solid) SM processes as a function of $\sqrt{s}$.}
   \label{fig:sm}
\end{figure}

For SM interactions, we treat the first and second generation fermions as massless and we use a diagonal CKM matrix for simplicity. We choose to work in terms of the measured value $\alpha(m_Z)^{-1}=127.9$, at the cost of having $m_W=79.8$ GeV, which is calculated using the tree-level relations in terms of $G_F,\alpha$. We take $m_t = 173$ GeV and $m_h = 125$ GeV. Cross sections for representative processes computed using {\sc Whizard} are illustrated in \cref{fig:sm}, and we find excellent agreement with {\sc Madgraph5}. Unsurprisingly, as $\sqrt{s}$ is increased the $s$-channel rates decrease as a power law, while the VBF processes grow logarithmically.

\newpage
\subsection{Supersymmetry}
\begin{figure}[htbp] 
   \centering
   \includegraphics[width=2.5in]{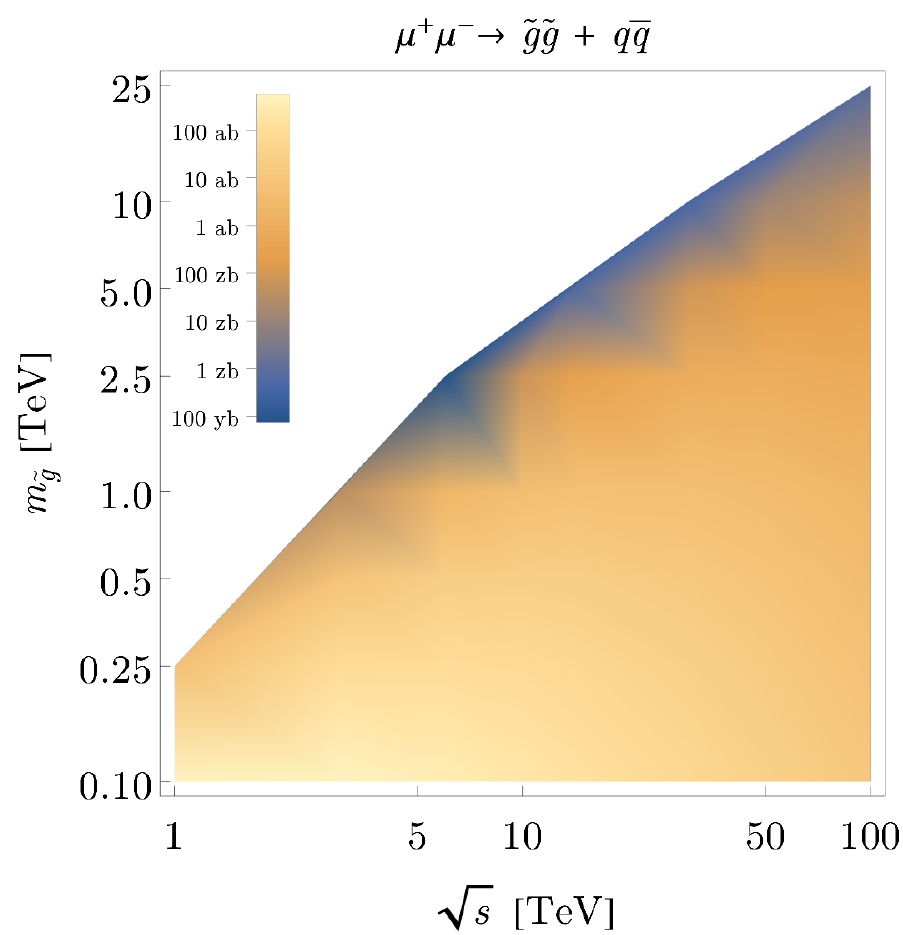} 
   \caption{Cross section for $\tilde g \tilde g$ production resulting from the ``bremsstrahlung'' of a gluon off of a $q \bar {q}$ pair as a function of $m_{\tilde g}$ and $\sqrt{s}$, computed using {\sc Whizard}.}
   \label{fig:SUSYGG}
\end{figure}

\begin{figure}[htbp] 
   \centering
   \includegraphics[width=2.5in]{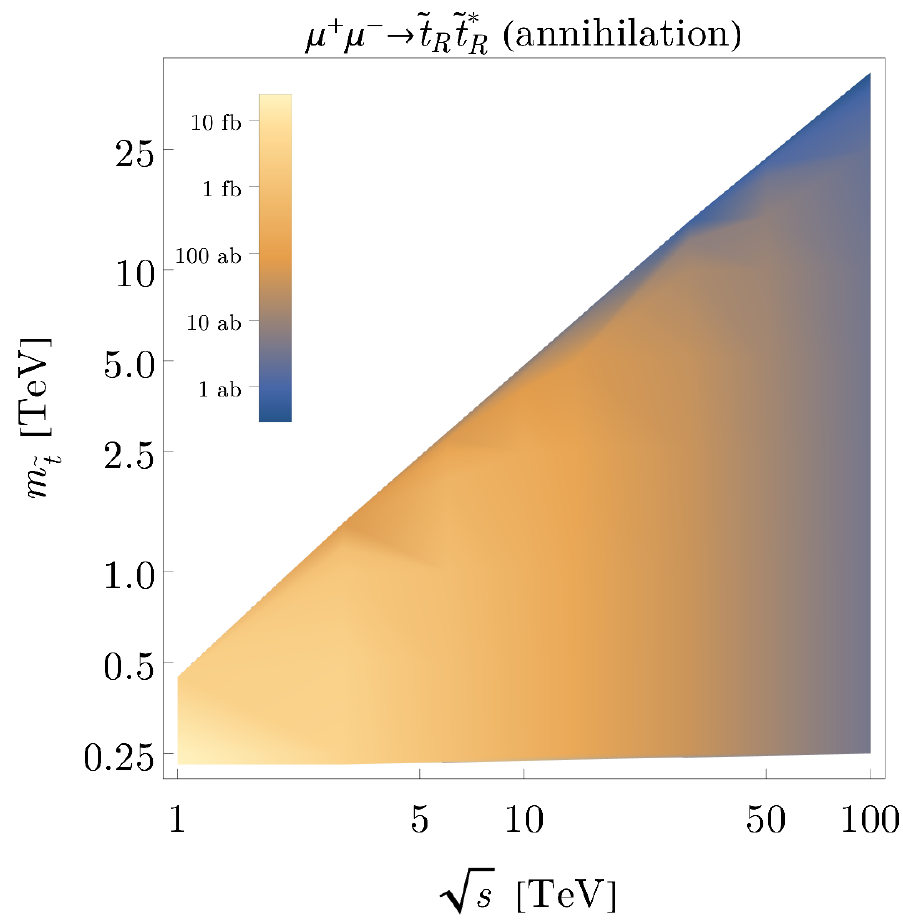} 
      \includegraphics[width=2.5in]{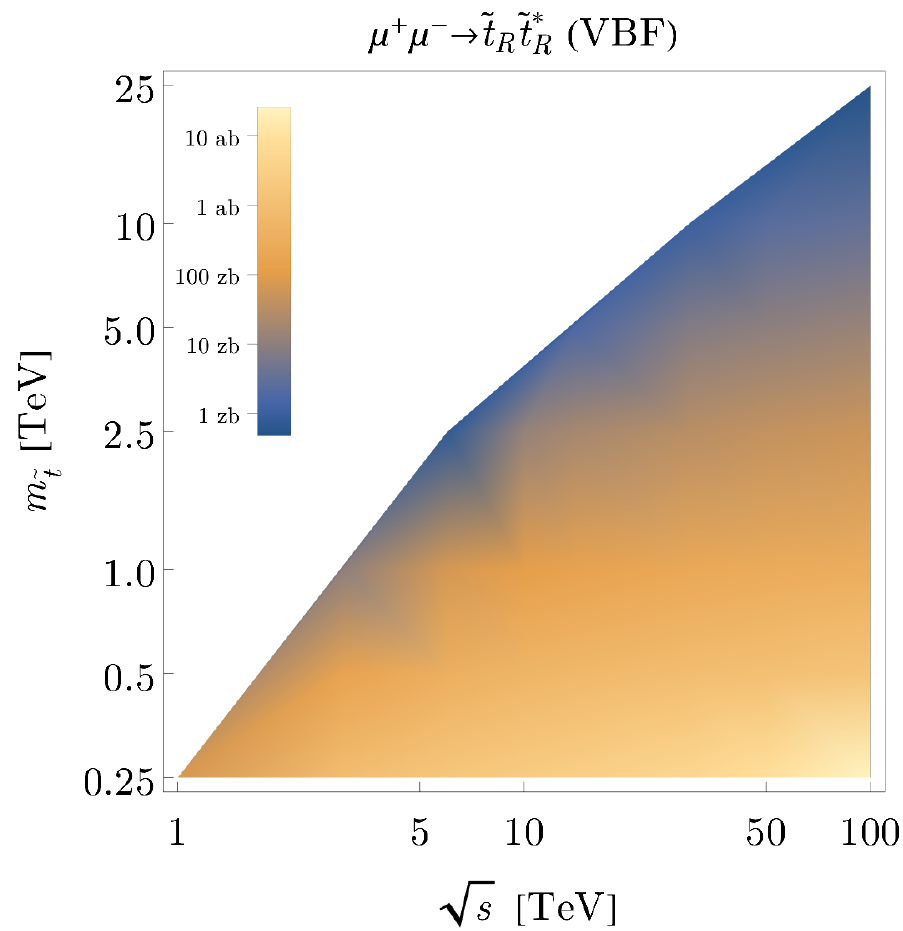} 
   \caption{Cross section for $\tilde t_R \tilde t_R^*$ production via annihilation (left) and VBF (right) as a function of $m_{\tilde t}$ and $\sqrt{s}$, computed using {\sc Whizard}.}
   \label{fig:SUSYUU}
\end{figure}

\begin{figure}[htbp] 
   \centering
   \includegraphics[width=2.5in]{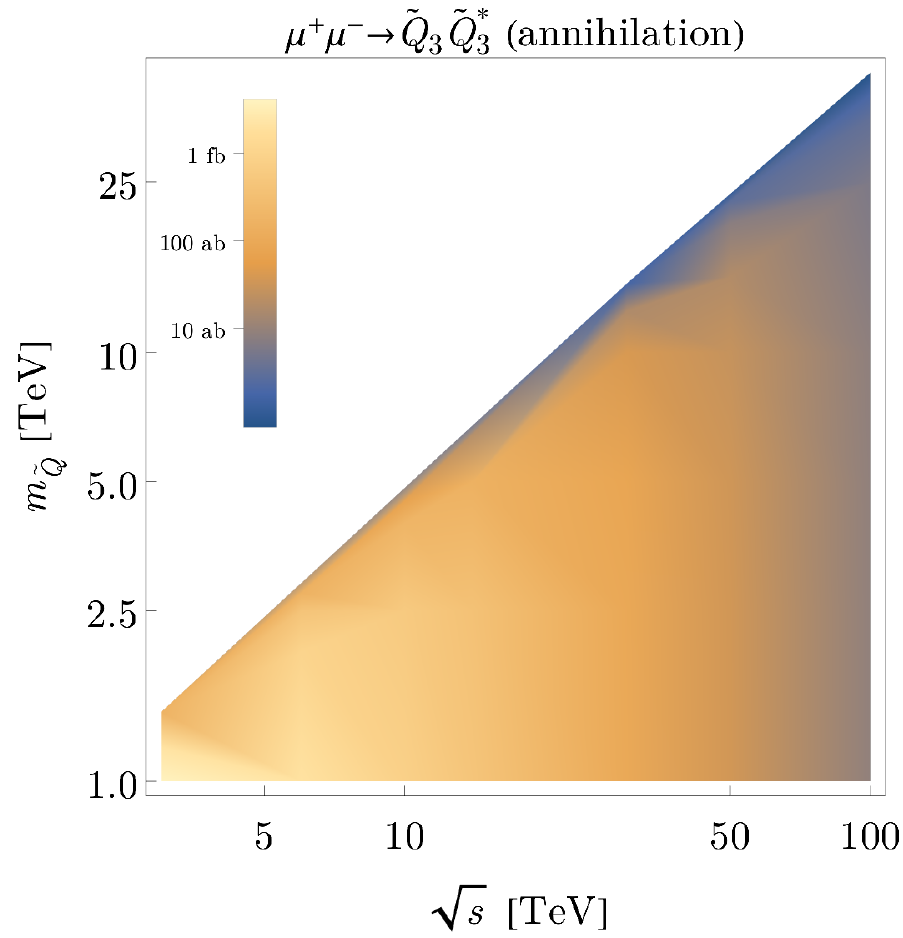} 
      \includegraphics[width=2.5in]{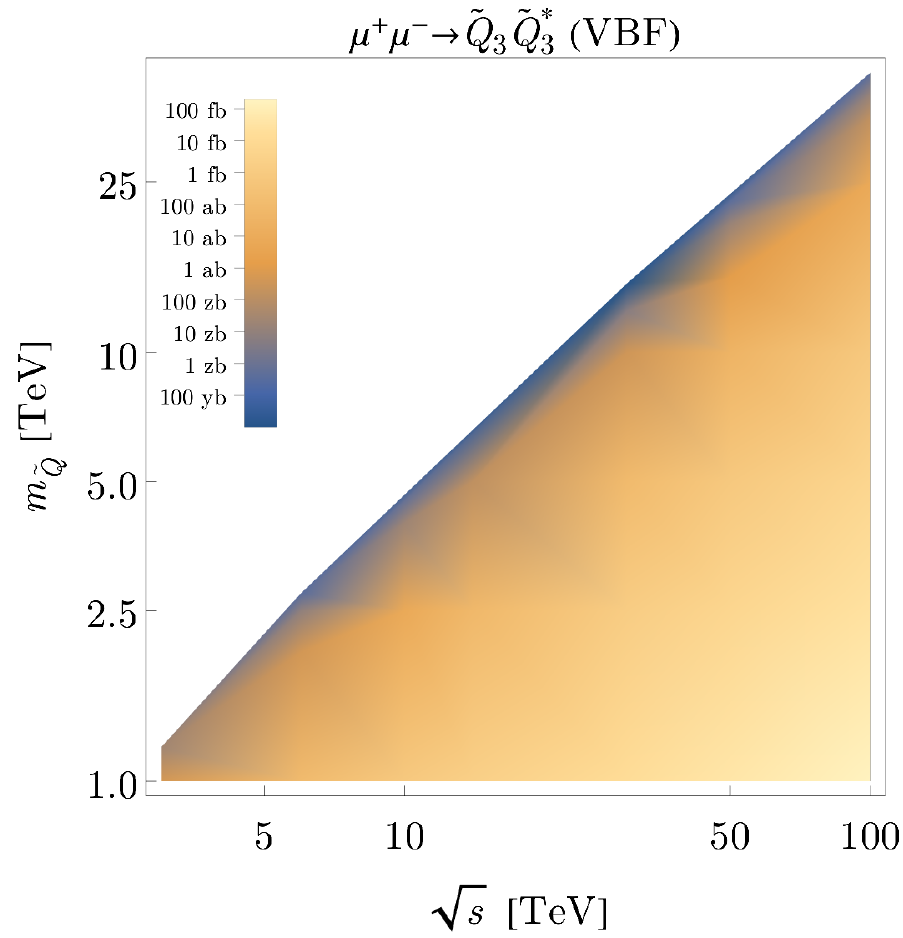} 
   \caption{Cross section for $\tilde Q_3 \tilde Q^*_3$ production via annihilation (left) and VBF (right) as a function of $m_{\tilde Q}$ and $\sqrt{s}$, computed using {\sc Whizard}.}
   \label{fig:SUSYQQ}
\end{figure}

\begin{figure}[htbp] 
   \centering
   \includegraphics[width=2.5in]{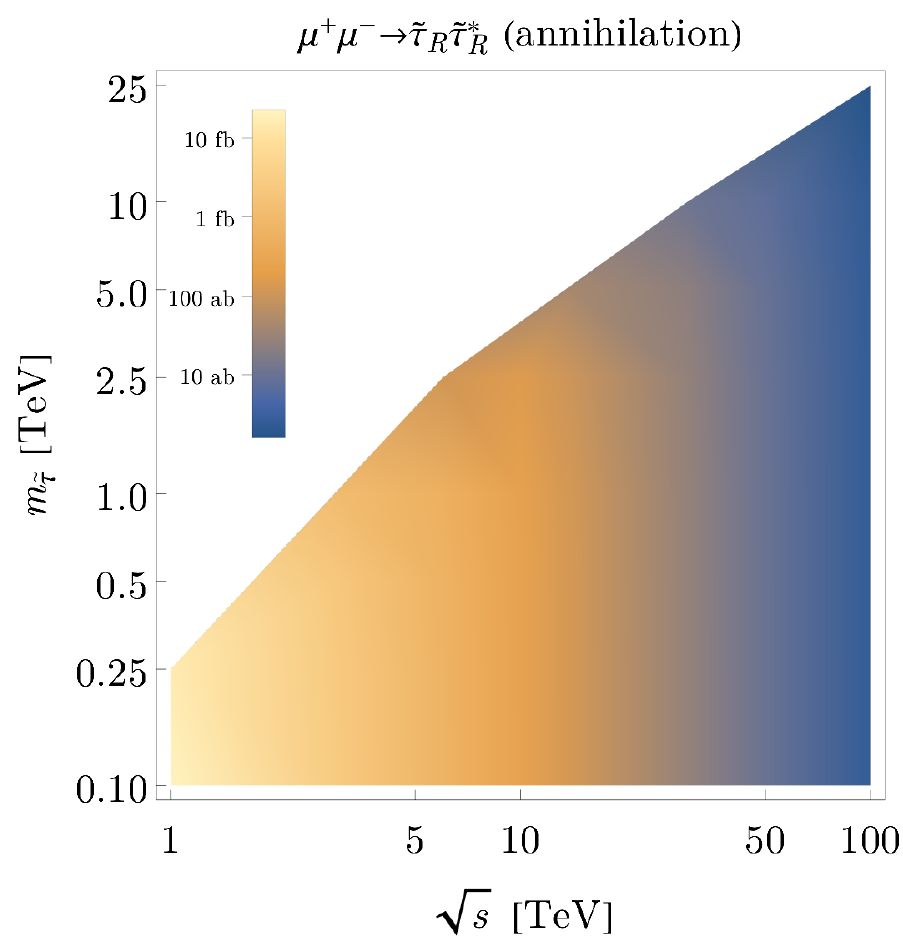} 
      \includegraphics[width=2.5in]{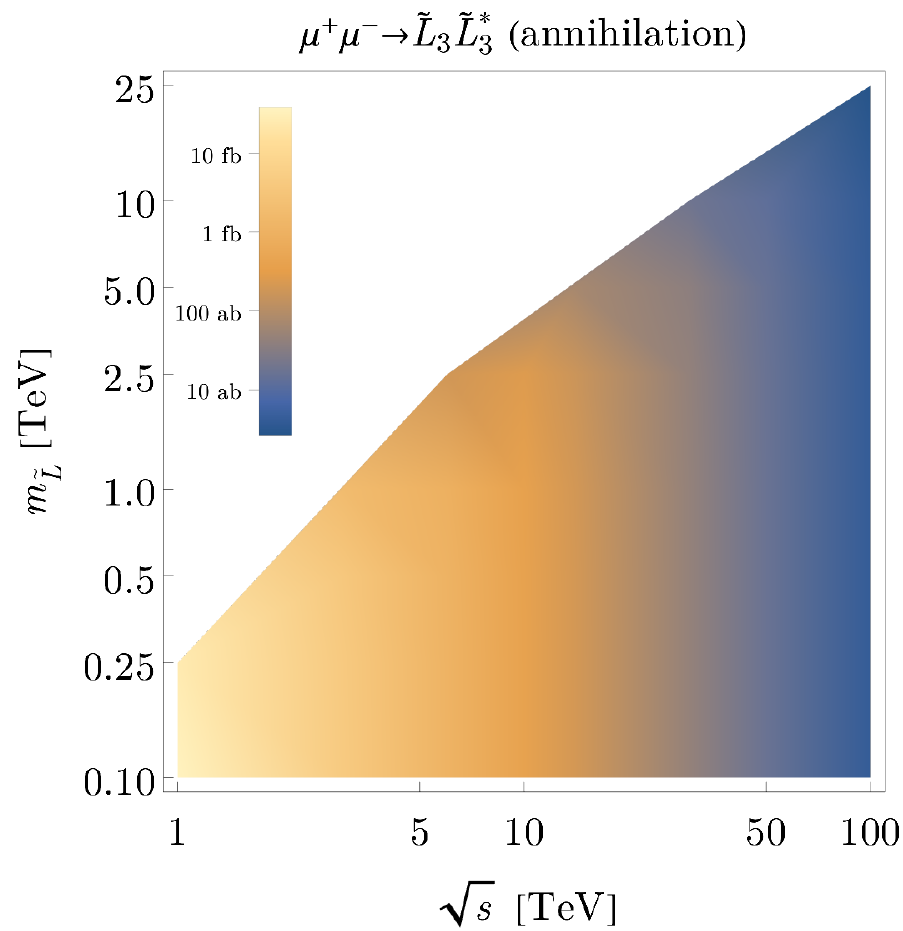} 
   \caption{Cross section for $\tilde \tau_R \tilde \tau_R^*$ production (left) and $\tilde L_3 \tilde L^*_3$ production (right) via annihilation as a function of $m_{\tilde t}$ and $\sqrt{s}$, computed using {\sc Whizard}.}
   \label{fig:SUSYSL}
\end{figure}

\begin{figure}[htbp] 
   \centering
   \includegraphics[width=2.5in]{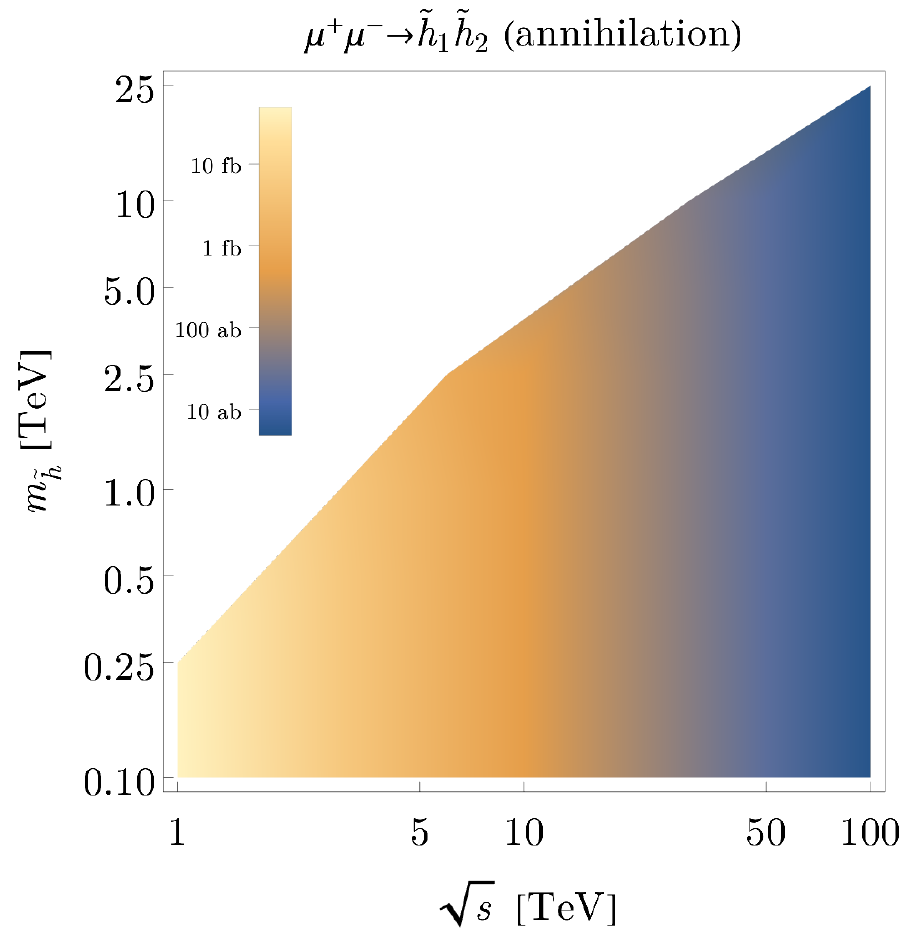} 
          \includegraphics[width=2.5in]{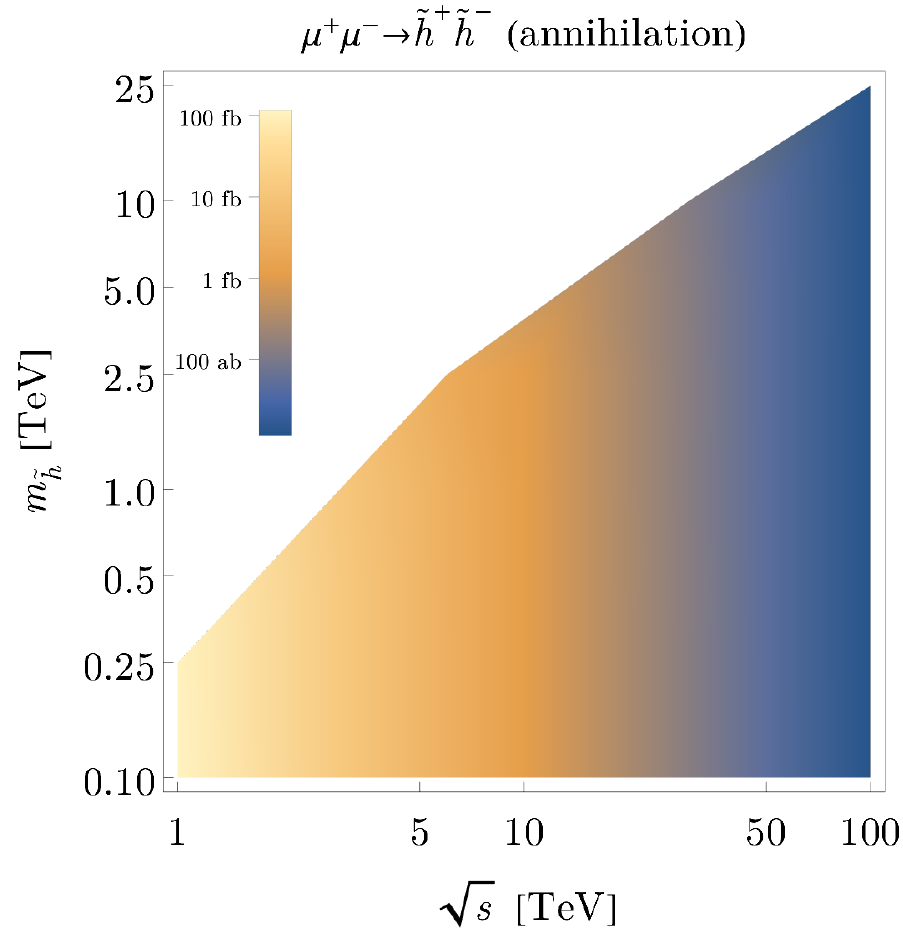} \\[15pt]
      \includegraphics[width=2.5in]{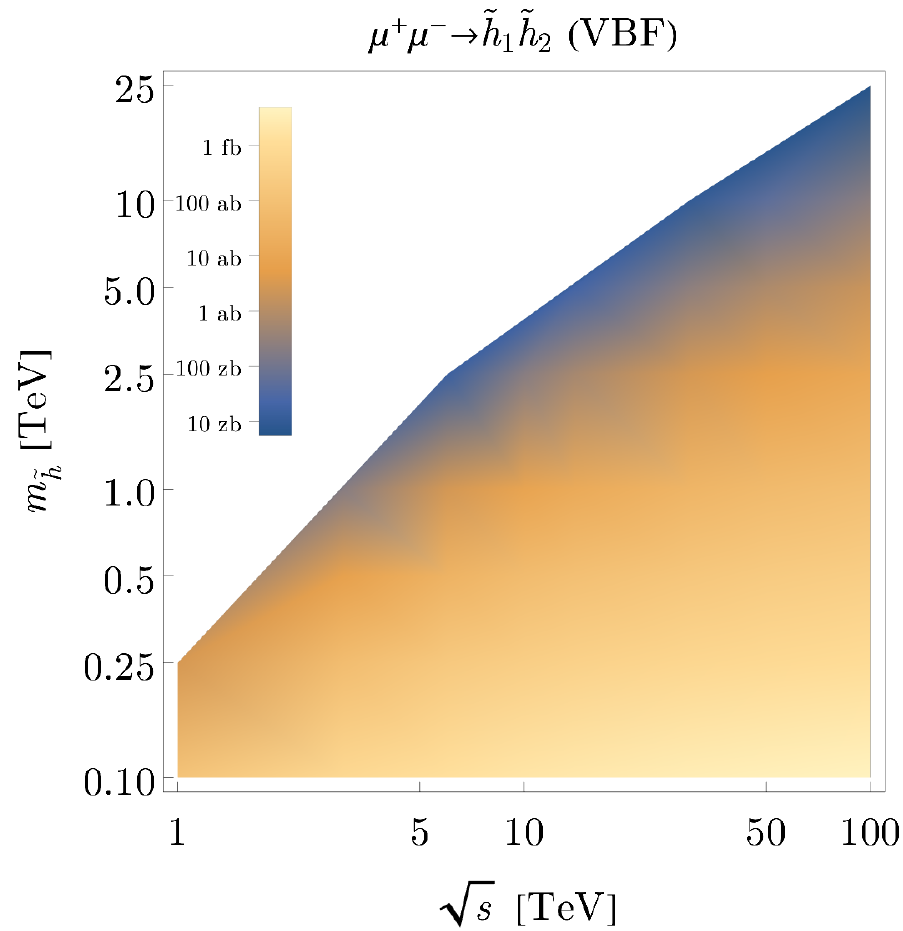} 
      \includegraphics[width=2.5in]{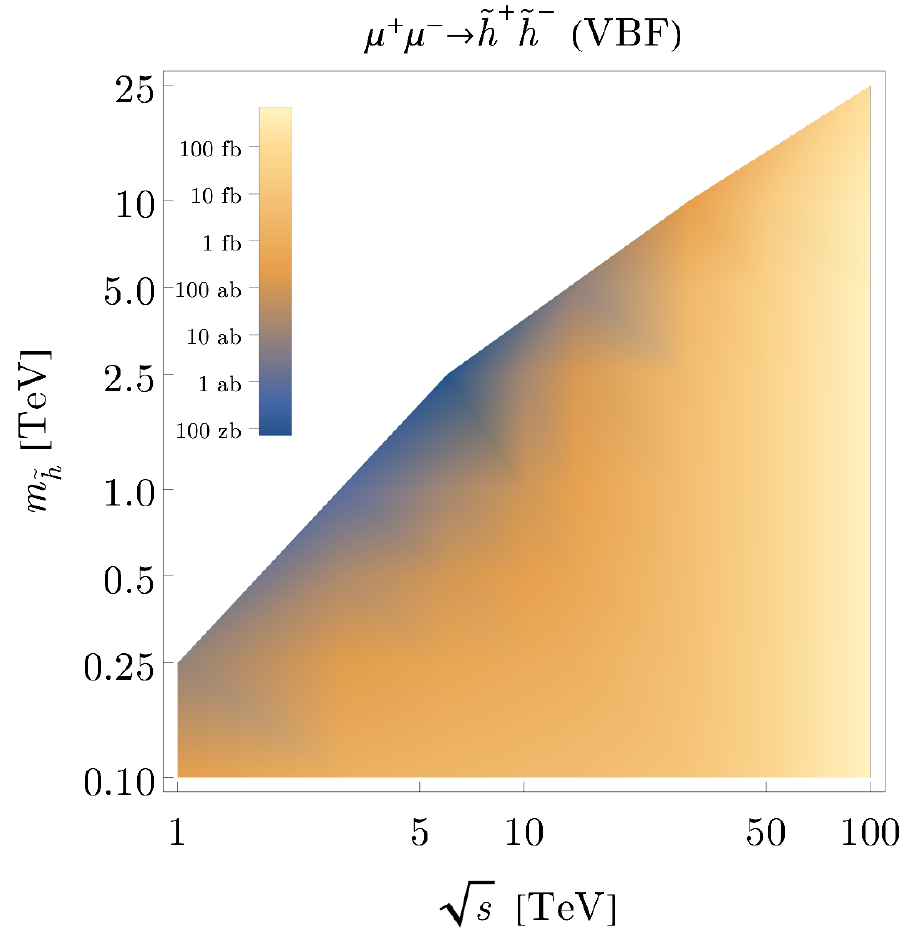} 
   \caption{Cross section for $\tilde h_1^0 \tilde h_2^0$ and $\tilde h^+ \tilde h^-$  production as a function of $m_{\tilde h}$ and $\sqrt{s}$ via annihilation (top) and VBF (bottom), computed using {\sc Whizard}. }
   \label{fig:SUSYH}
\end{figure}

\begin{figure}[htbp] 
   \centering
   \includegraphics[width=2.5in]{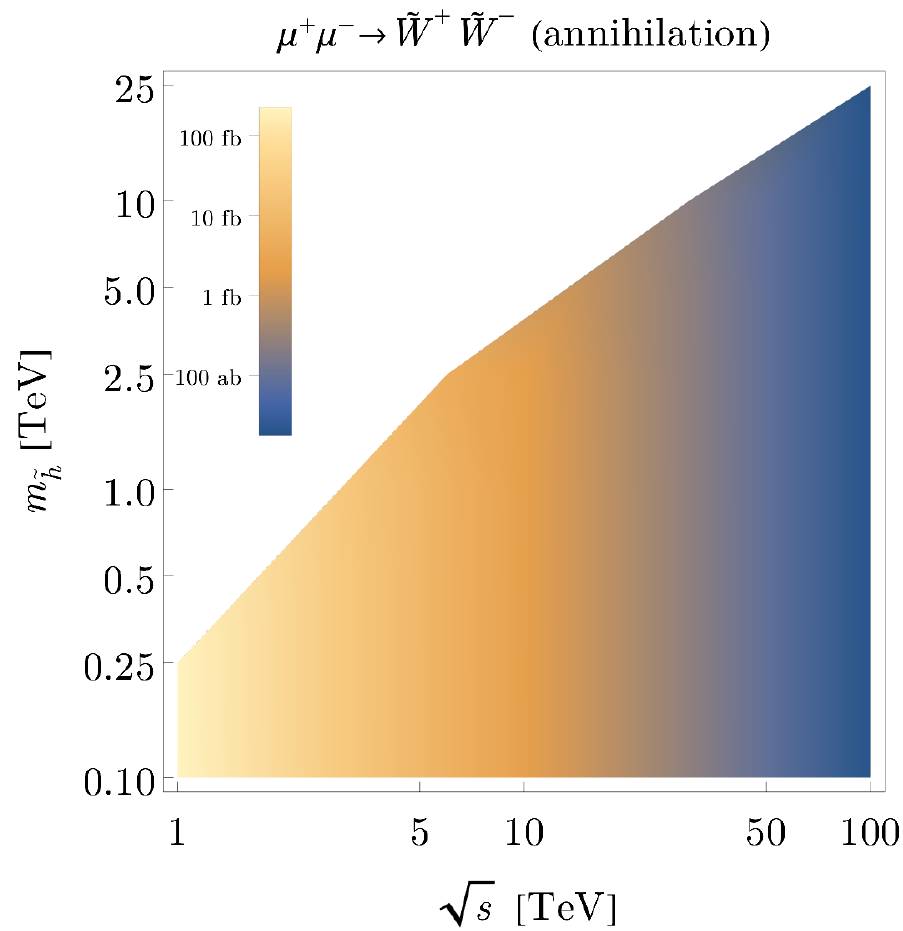} 
          \includegraphics[width=2.5in]{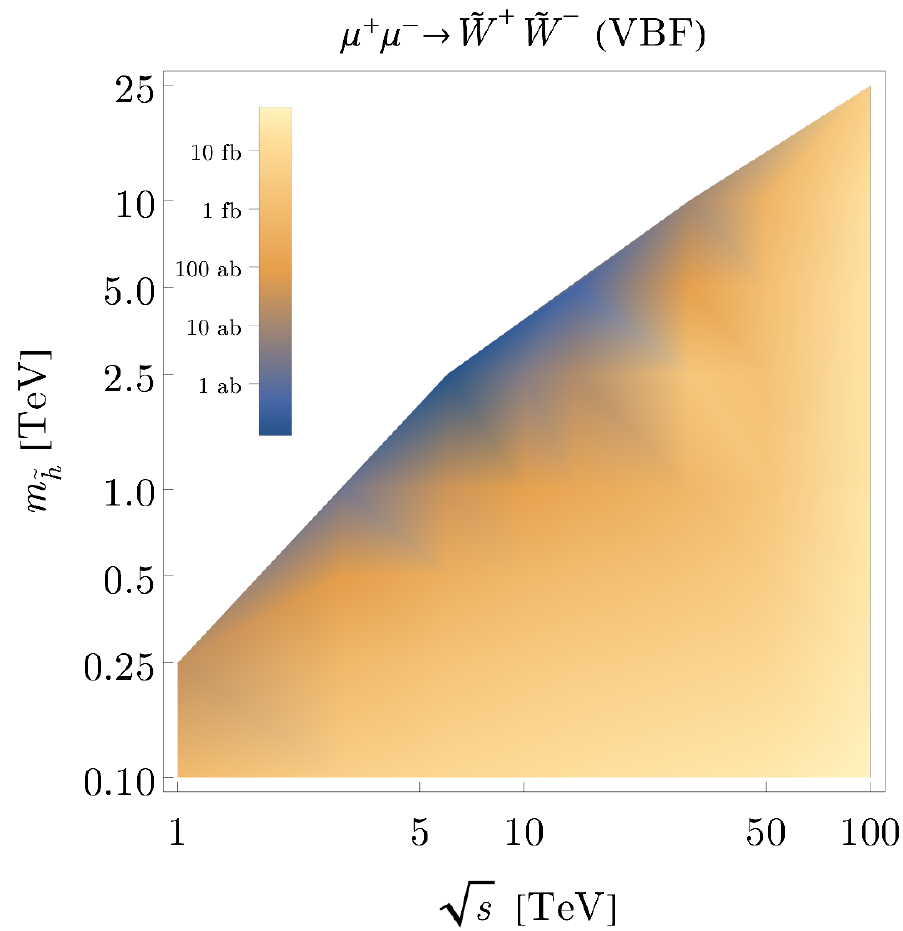} \\[15pt]
      \includegraphics[width=2.5in]{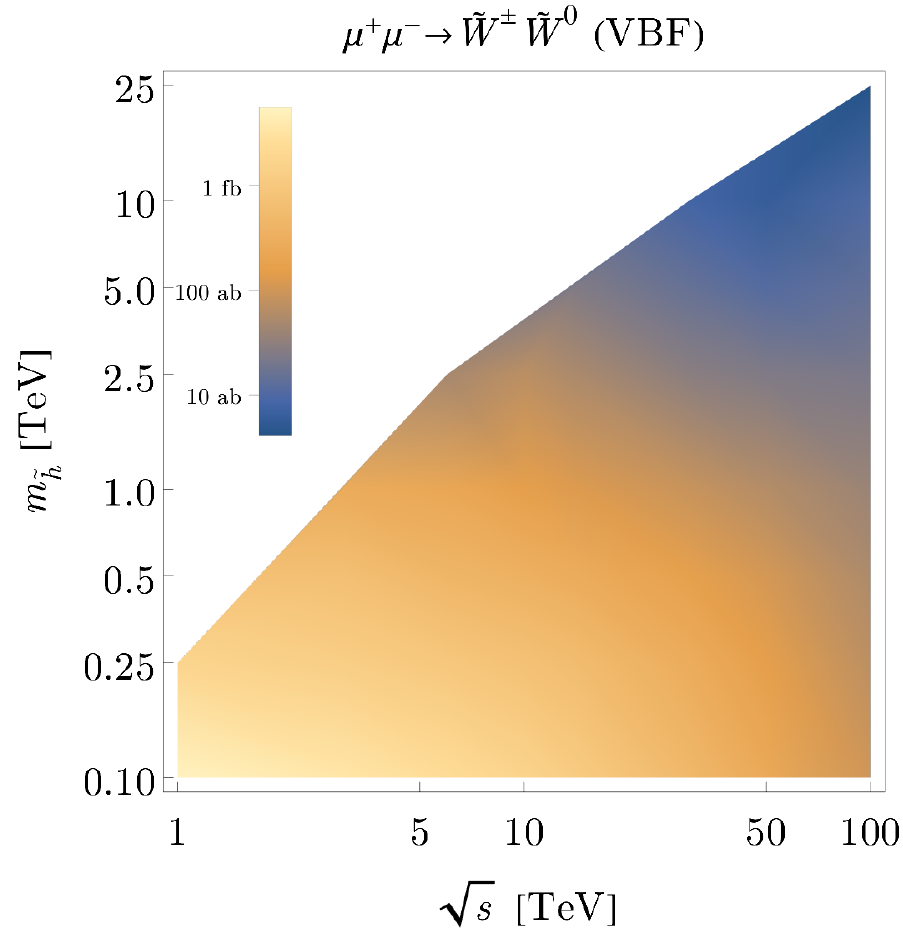} 
   \caption{Cross section for  $\tilde W^+ \tilde W^-$ production via annihilation (top left) and VBF (top right), $\tilde W^\pm \tilde W^0$ production via VBF (bottom) as a function of $m_{\tilde W}$ and $\sqrt{s}$, computed using {\sc MadGraph5}. }
   \label{fig:SUSYW}
\end{figure}

Perhaps the best studied simplified models are motivated by the Minimal Supersymmetric Standard Model (MSSM).  For supersymmetric processes, we use a variety of {\sc FeynRules} models. For gluino pair production we use the default {\sc MSSM} model in {\sc FeynRules}, keeping the gluino light while decoupling all other sparticles. The cross section for $\tilde g \tilde g$ production resulting from the ``bremsstrahlung'' of a gluon off of a $q \bar {q}$ pair is shown in \cref{fig:SUSYGG}. For stop pair production we use the $t$-channel dark matter {\sc FeynRules} model \cite{Arina:2020udz} with couplings set to their supersymmetric values.  Cross sections for the production of the $SU(2)$ singlet $\tilde t_R$ and $SU(2)$ doublet $\tilde Q_3$ via  annihilation and VBF are shown in \cref{fig:SUSYUU,fig:SUSYQQ}. For slepton pair production, we use the default {\sc MSSM} model in {\sc FeynRules}, alternately keeping either $\tilde \tau_R$ or $\tilde L_3$ light while decoupling all other sparticles. Cross sections for the production of the $SU(2)$ singlet $\tilde \tau_R$ and $SU(2)$ doublet $\tilde L_3$ via annihilation are shown in \cref{fig:SUSYSL}. For higgsino pair production, we use the default {\sc MSSM} model in {\sc FeynRules}, keeping the components $\chi_1^\pm$ and $\chi_{1,2}^0$ light while decoupling all other sparticles and taking care to keep the neutralino and chargino mixing matrices appropriately aligned. Cross sections for both chargino and neutralino pair production via annihilation and VBF are shown in \cref{fig:SUSYH}. For wino pair production, we use the default {\sc MSSM} model in {\sc FeynRules}, keeping the components $\chi_1^\pm$ and $\chi_{1}^0$ light while decoupling all other sparticles and taking care to keep the neutralino and chargino mixing matrices appropriately aligned. Cross sections for both chargino and neutralino pair production via annihilation and VBF are shown in \cref{fig:SUSYW}. Across all processes, we see the same trend as with the Standard Model processes: when both $s$-channel and VBF production is possible, the former dominates at low $\sqrt{s}$ while the latter takes over for high $\sqrt{s}$.  Another interesting point to note is that the squark production rate is significantly larger than the gluino rate, which implies that gluino production could be dramatically impacted when considering a more complete model that includes both gluinos and squarks.  

\newpage

\subsection{Vector-like quarks}

\begin{figure}[htbp] 
   \centering
   \includegraphics[width=2.5in]{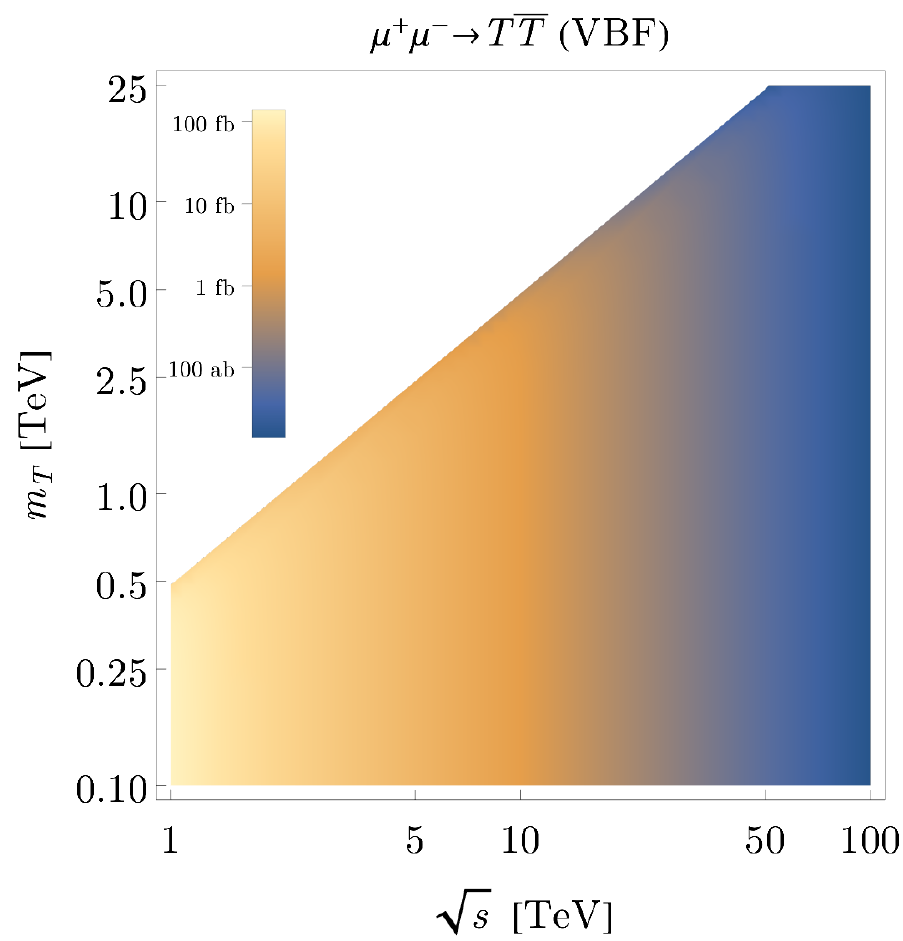} 
      \includegraphics[width=2.5in]{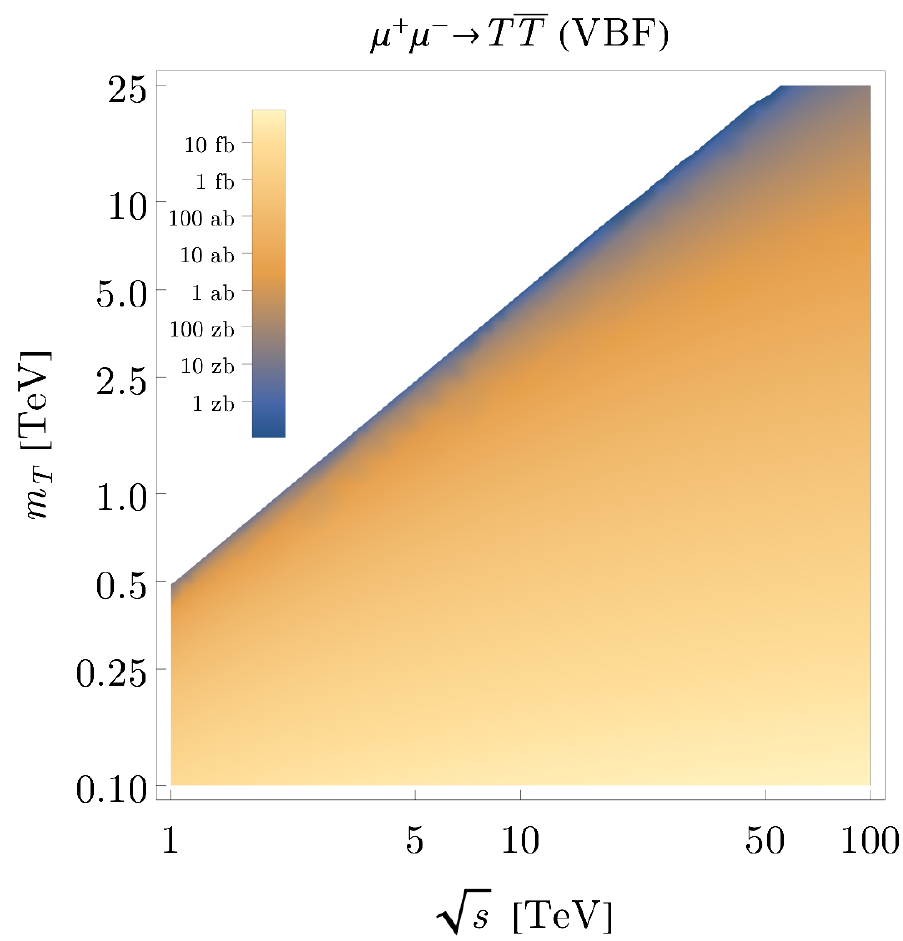} 
   \caption{Cross section for $T \overline T$ production via annihilation (left) and VBF (right) as a function of $m_T$ and $\sqrt{s}$, computed by convolving partonic amplitudes obtained via {\sc FeynArts} and {\sc FormCalc} with the LL luminosity functions detailed in \cref{sec:PDF}.}
   \label{fig:VLQ}
\end{figure}

Next we turn to vector-like quarks, an important example that arises in scenarios ranging from global symmetry approaches to the hierarchy problem to parity solutions to the strong CP problem. We restrict our considerations to the gauge interactions of an $SU(3)$ triplet, $SU(2)$ singlet, hypercharge $+\frac23$ Dirac fermion $T$, which we implement in {\sc FeynRules}. Cross sections for $T \overline T$ production via annihilation and VBF are shown in \cref{fig:VLQ}, computed by convolving partonic amplitudes obtained via {\sc FeynArts} and {\sc FormCalc} with the LL luminosity functions detailed in \cref{sec:PDF}.

\newpage
\subsection{Higgs portal}
Next, we turn to one of the models that has implications for baryogenesis, dark matter, and models of neutral naturalness.
Furthermore, this model is simultaneously very simple to state while being very difficult to discover.
Specifically, we extend the SM using a $\mathbb{Z}_2$ symmetric singlet $S$ that couples via the Higgs portal:
\begin{equation}
  \mathcal{L} = \frac12 (\partial S)^2 - \frac12 M_S^2 S^2 - \lambda_{HS} \, S^2 |H|^2\,.
 \label{eq:lag_singlet}
\end{equation}
We omit a possible singlet quartic, which does not influence the phenomenology, and assume $M_S, \lambda_{HS}$ are such that $S$ is stabilized at the origin and the $\mathbb{Z}_2$ symmetry is unbroken. The physical mass of the singlet in the broken phase is 
\begin{align}
m_S^2 = M_S^2 + \lambda_{HS} v^2\,.
\end{align}
As written, the singlet is absolutely stable and may constitute a dark matter candidate. It also may be rendered unstable by explicit $\mathbb{Z}_2$ breaking, leading to prompt or long-lived signatures.

\begin{figure}[h!] 
   \centering
   \includegraphics[width=2.5in]{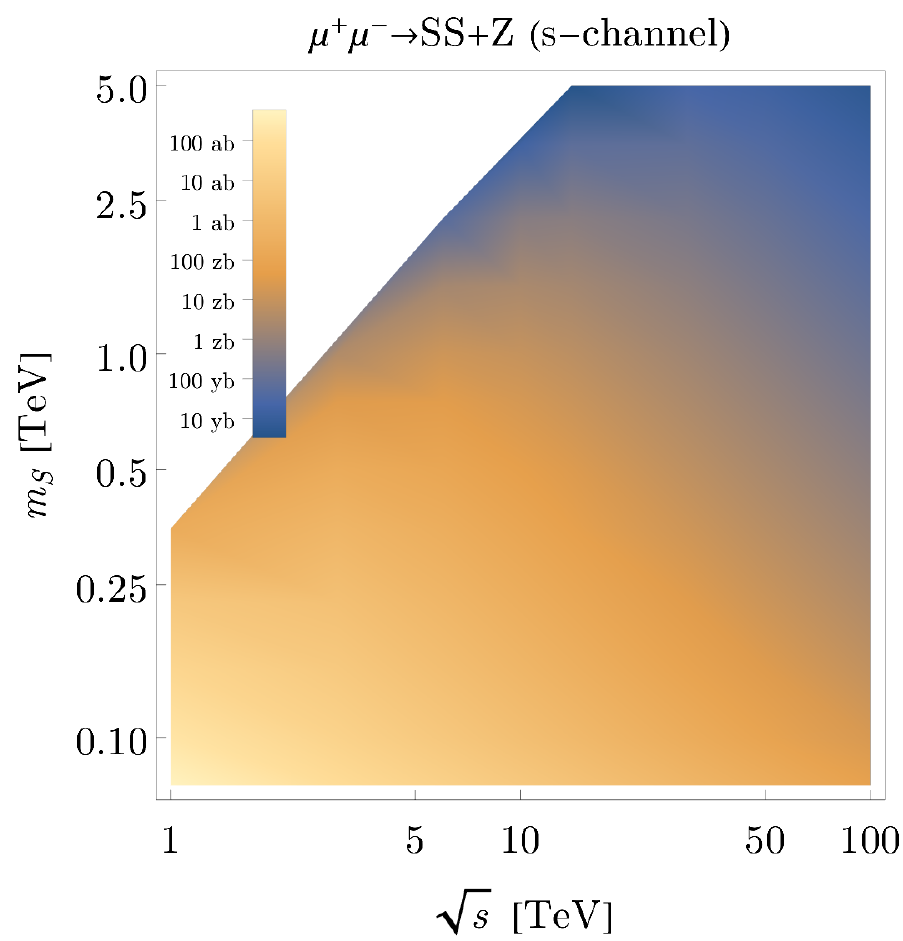} 
          \includegraphics[width=2.5in]{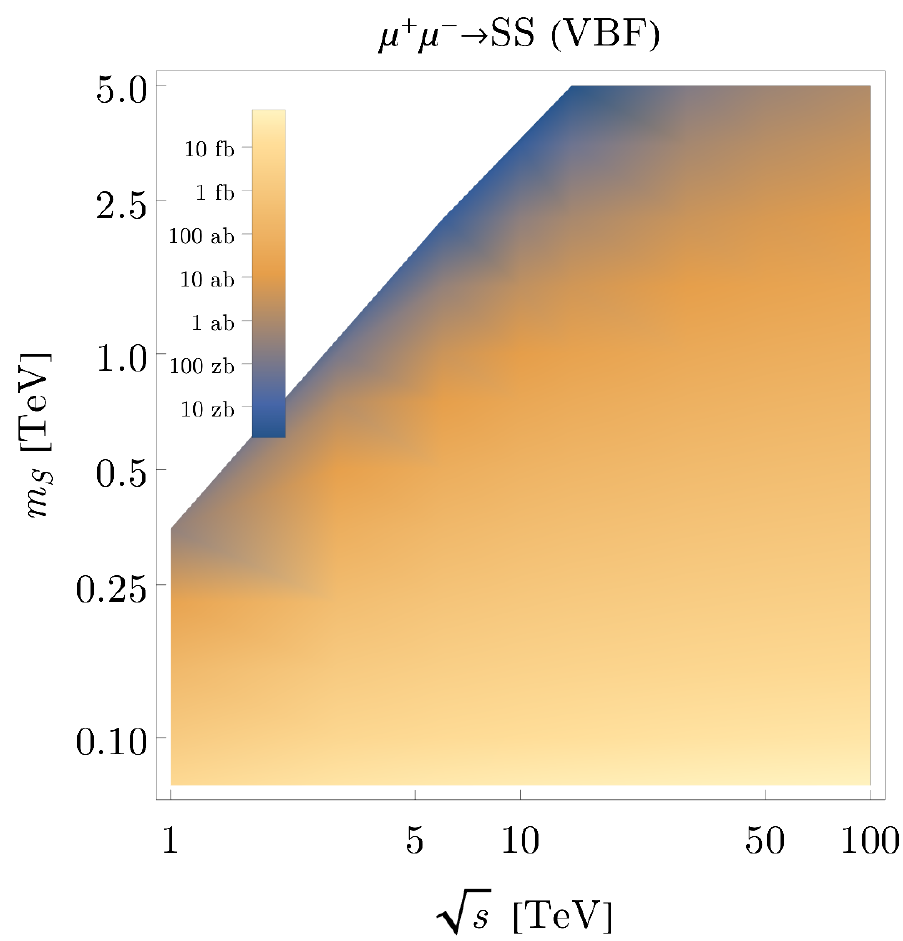} \\[15pt]
      \includegraphics[width=2.5in]{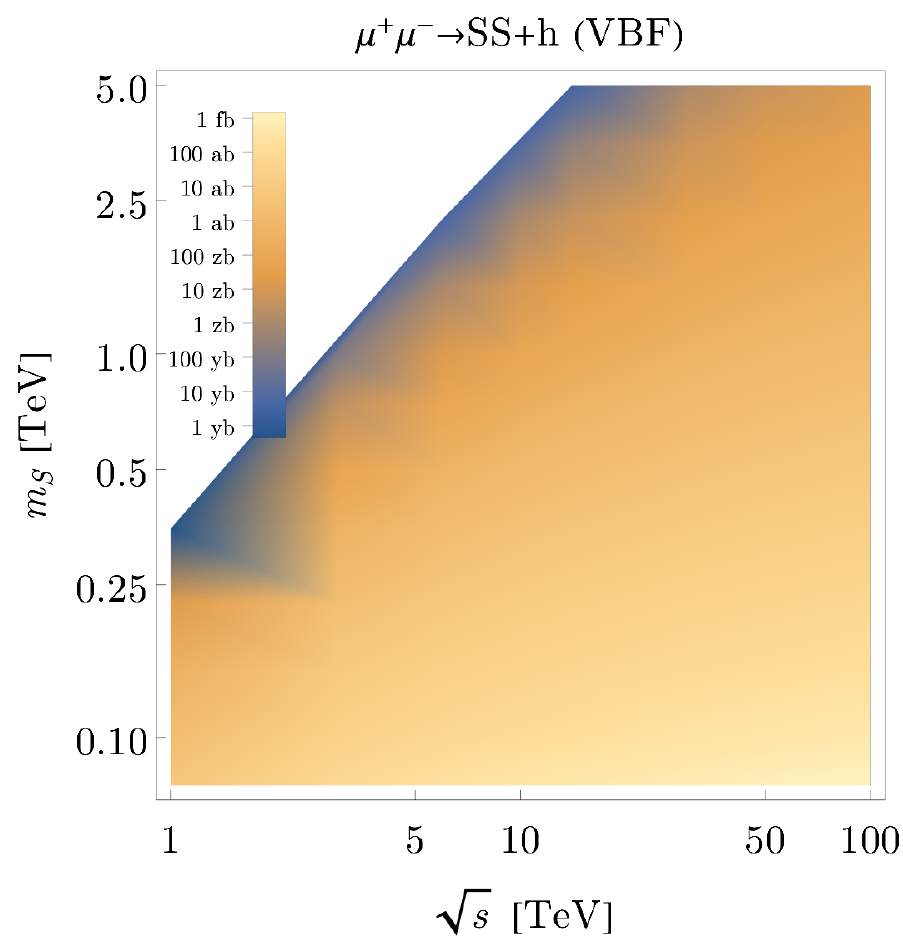} 
       \includegraphics[width=2.5in]{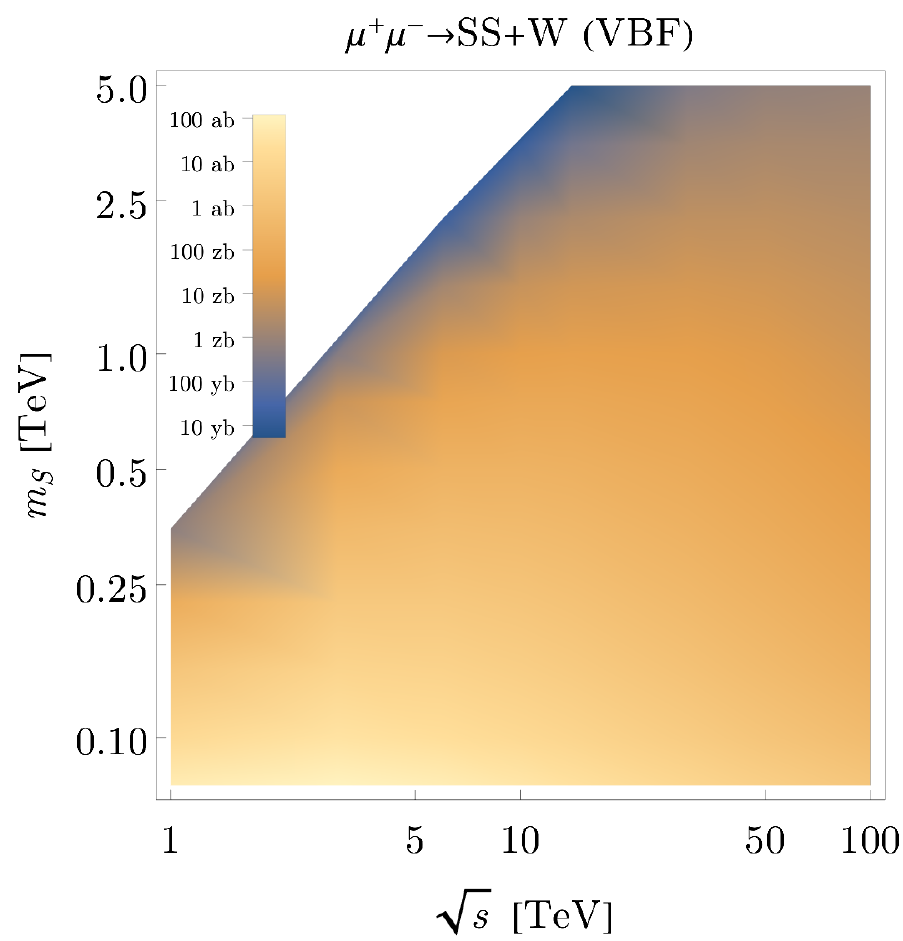} 
   \caption{Cross section for various Higgs portal production modes as a function of $m_S$ and $\sqrt{s}$ computed using {\sc Whizard}, including $s$-channel $SS+Z$ production (upper left), VBF $SS$ production (upper right), VBF $SS+h$ production (lower left), and VBF $SS+W$ production (lower right). As a benchmark, we take $\lambda_{HS} = 1$ here. }
   \label{fig:HP}
\end{figure}

The differential cross section for $s$-channel production of $SS$ in association with a $Z$ boson is \cite{Burgess:2000yq}
\begin{align} \nonumber
\hspace{-5pt}\frac{\dd \sigma}{\dd t\, \dd u} (\mu^+ \mu^- \rightarrow Z SS) &= \frac{\big[\big(-\frac{1}{2} + s_W^2\big)^2 + s_W^4\big] \lambda_{HS} v^2 m_Z^2}{16 (2 \pi)^3 s} \bigg(\!\frac{e}{s_W c_W}\! \bigg)^4 \bigg[\frac{s + (t-m_Z^2)(u-m_Z^2)/m_Z^2}{(s-m_Z^2)^2+ \Gamma_Z^2 m_Z^2} \bigg]   \\[6pt]
&\hspace{50pt} \times \left[ \frac{1}{s+t+u-m_Z^2 - m_h^2 } \right]^2 \left[1 - \frac{4 m_S^2}{s+t+u-m_Z^2} \right]^{1/2}\,,
\end{align}
where the Mandelstam variables are defined in the standard way for $2\to2$ processes such that $s+t+u=m_Z^2 + Q^2$, where $Q^2$ is the square of the invariant mass of the two singlets in the final state. At large $s$, this leads to a cross section falling as $1/s^2$. This motivates considering a variety of VBF-like processes, including $W^+W^-$ fusion into $SS$ and $HSS$ final states, as well as $WZ/W\gamma$ fusion into the $WSS$ final state. We implement the model in {\sc FeynRules} and compute cross sections primarily in {\sc Whizard}, finding good agreement with both {\sc MadGraph5} and the result of convolving the partonic cross sections with our LL PDFs. Cross sections for all four processes as a function of $m_S$ and $\sqrt{s}$ are shown in \cref{fig:HP}.

\newpage

\subsubsection{Missing mass analysis}
\label{sec:singlet_missing}

Given the central importance of the singlet scalar Higgs portal in BSM scenarios ranging from electroweak baryogenesis to neutral naturalness, we carry out a series of simplified analyses to benchmark the reach of dedicated searches for $S$, assuming that $S$ is detector stable. These analyses cover three different production modes: $W^+W^-$ fusion into an $SS$ pair in association with an ISR photon; $W^+W^-$ fusion into an $SS$ pair in association with a Higgs boson; and $s$-channel production of an $SS$ pair in association with a $Z$ boson. In each case and for each benchmark value of $\sqrt{s}$, we simulate 100k background events and 10k signal events for each of $m_S = 75, 100, 150, 225, 350, 500, 750, 1000 \text{ GeV}$ in {\sc Whizard}. 

\newpage
\paragraph{\textbf{\textit{SS}}$\bm{\; + \; \gamma}$ analysis}

For this signal, we consider the production of $SS$ via $W^+W^-$ fusion accompanied by an ISR photon, $\mu^+ \mu^- \rightarrow S S + \nu \bar \nu + \gamma$, where $S$ is treated as invisible. The primary SM background for this final state is $\mu^+ \mu^- \rightarrow \nu \bar{\nu} + \gamma$, i.e., production of a single photon via $W^+W^-$ fusion. We require that the ISR photon falls within the detector acceptance, here defined as $10^\circ < \theta < 170^\circ$, and employ a ``missing mass'' strategy in which the photon energy is required to lie between 
\begin{equation}
50 \, {\rm GeV}  < E_\gamma < \frac{s - 4 m_S^2}{2 \sqrt{s}}\,.
\end{equation}
The background spectrum is harder than the signal spectrum, so the lower bound on the photon energy serves primarily to avoid singular regions of phase space. The upper bound on the photon energy reflects the lower bound on the missing mass distribution.

\begin{figure}[h!] 
   \centering
   \includegraphics[trim = 20 0 0 0, width = 0.95\textwidth]{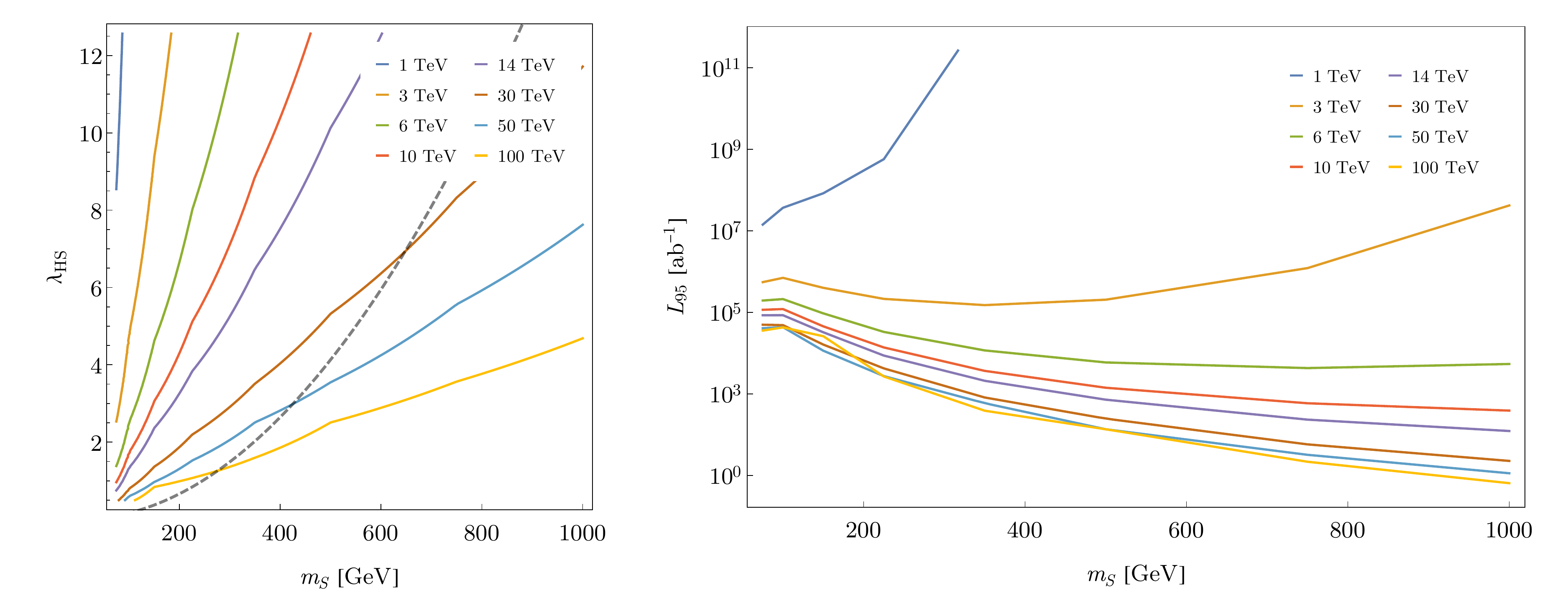}
   \caption{Left: Projected 95\% CL exclusion limits as a function of $m_S$ and $\lambda_{HS}$ from the $SS + \gamma$ final state for various energy and luminosity benchmarks described in the text. The dashed line corresponds to a singlet scalar acquiring mass entirely from electroweak symmetry breaking. Right: The integrated luminosity $\mathcal{L}$ in units of ab$^{-1}$ required to exclude a singlet scalar whose mass $m_S$ is due entirely to electroweak symmetry breaking at 95\% CL in the $SS + \gamma$ final state.}
   \label{fig:salex}
\end{figure}

The integrated luminosity required to exclude a singlet scalar at 95\% CL using this analysis is shown in Fig.~\ref{fig:salex}.

\newpage
\paragraph{\textbf{\textit{SS}}$\bm{\; + \;}$\textbf{\textit{h}} analysis}

For this signal, we consider the production of $hSS$ via $W^+W^-$ fusion, $\mu^+ \mu^- \rightarrow S S + h + \nu \bar \nu$, where $S$ is treated as invisible. The primary SM background for this final state is $\mu^+ \mu^- \rightarrow \nu \bar{\nu} + h$, i.e.,  production of a single Higgs via $W^+W^-$ fusion. We require the Higgs boson falls within the detector acceptance, here defined as $10^\circ < \theta < 170^\circ$, and employ a ``missing mass'' strategy in which the Higgs energy is required to lie between 
\begin{equation}
E_{\rm min}  < E_h < \frac{s - 4 m_S^2 + m_h^2}{2 \sqrt{s}}\,.
\end{equation}
The signal spectrum is harder than the background spectrum, so the lower bound $E_{\rm min}$ on the Higgs energy is chosen for each value of $m_S$ to maximize $S / \sqrt{B}$. The upper bound on the Higgs energy reflects the lower bound on the missing mass distribution.

\begin{figure}[h!] 
   \centering
 \includegraphics[trim = 20 0 0 0, width = 0.95\textwidth]{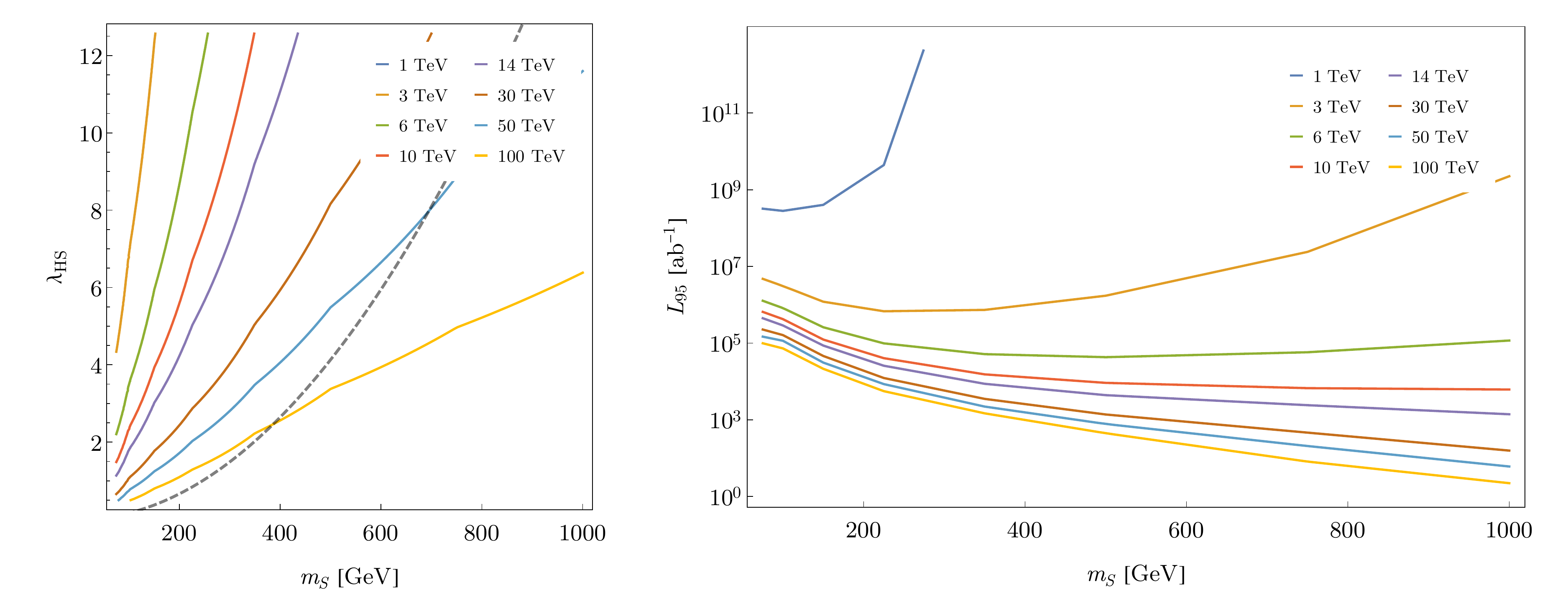}
   \caption{Left: Projected 95\% CL exclusion limits as a function of $m_S$ and $\lambda_{HS}$ from the $SS + h$ final state for various energy and luminosity benchmarks described in the text. The dashed line corresponds to a singlet scalar acquiring mass entirely from electroweak symmetry breaking. Right: The integrated luminosity $\mathcal{L}$ in units of ab$^{-1}$ required to exclude a singlet scalar whose mass $m_S$ is due entirely to electroweak symmetry breaking at 95\% CL in the $SS +h$ final state.}
   \label{fig:shlex}
\end{figure}

The integrated luminosity required to exclude a singlet scalar at 95\% CL using this analysis is shown in \cref{fig:shlex}.

\newpage

\paragraph{\textbf{\textit{SS}}$\bm{\; + \;}$\textbf{\textit{Z}} analysis}

For this signal, we consider $s$-channel production of $ZSS$ $\mu^+ \mu^- \rightarrow S S + Z$, where $S$ is treated as invisible. The primary SM background for this final state is $\mu^+ \mu^- \rightarrow Z + \nu \bar{\nu}$, which has contributions from both $W^+W^-$ fusion and $\mu^+ \mu^-$ annihilation. We require that the $Z$ boson falls within the detector acceptance, here defined as $10^\circ < \theta < 170^\circ$, and employ a ``missing mass'' strategy in which the $Z$ energy is required to lie between 
\begin{equation}
E_{\rm min}  < E_Z < \frac{s - 4 m_S^2 + m_Z^2}{2 \sqrt{s}}\,.
\end{equation}
The signal spectrum is harder than the background spectrum, so the lower bound $E_{\rm min}$ on the $Z$ energy is chosen for each value of $m_S$ to maximize $S / \sqrt{B}$. The upper bound on the $Z$ energy reflects the lower bound on the missing mass distribution. At higher energies, the optimal value of $E_{\rm min}$ leads to background acceptance of less than $10^{-4}$ for lower values of $m_S$, in which case we generate additional exclusive background samples to improve Monte Carlo statistical error. 

\begin{figure}[h!] 
   \centering
 \includegraphics[trim = 20 0 0 0, width = 0.95\textwidth]{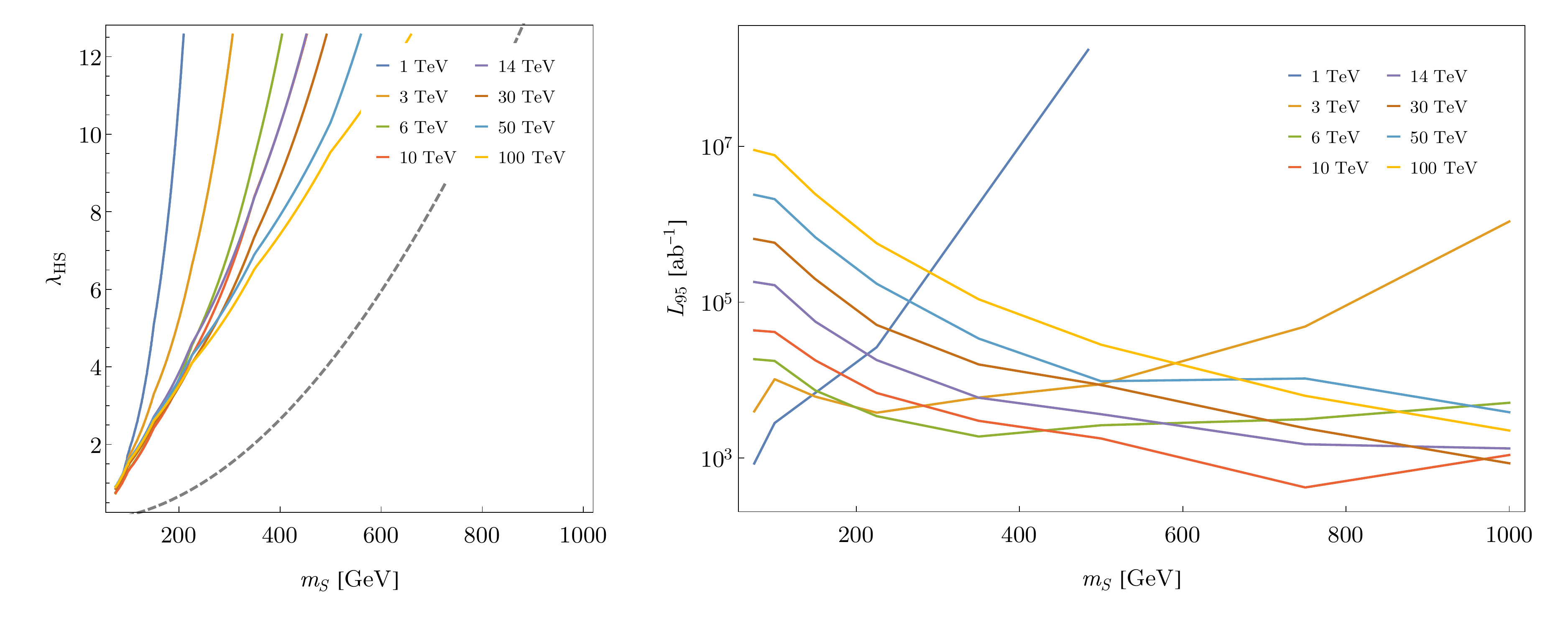}
   \caption{Left: Projected 95\% CL exclusion limits as a function of $m_S$ and $\lambda_{HS}$ from the $SS + Z$ final state for various energy and luminosity benchmarks described in the text. The dashed line corresponds to a singlet scalar acquiring mass entirely from electroweak symmetry breaking. Right: The integrated luminosity $\mathcal{L}$ in units of ab$^{-1}$ required to exclude a singlet scalar whose mass $m_S$ is due entirely to electroweak symmetry breaking at 95\% CL in the $SS +Z$ final state.}
   \label{fig:szlex}
\end{figure}

The integrated luminosity required to exclude a singlet scalar at 95\% CL using this analysis is shown in \cref{fig:szlex}.

\newpage
\paragraph{Missing mass combination}

\begin{figure}[h!] 
   \centering
 \includegraphics[trim = 20 0 0 0, width = 0.95\textwidth]{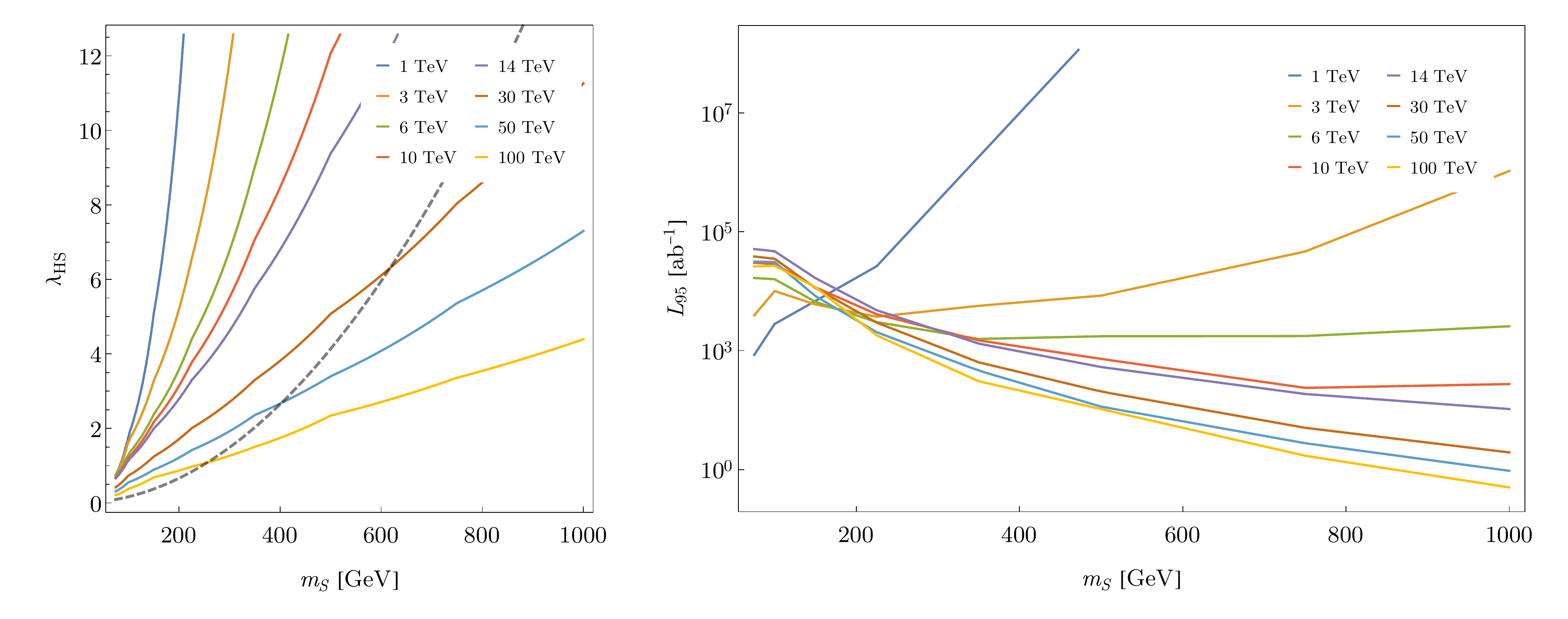}
   \caption{Left: Projected 95\% CL exclusion limits as a function of $m_S$ and $\lambda_{HS}$ from the naive combination of $SS + \gamma/h/Z$ final states for various energy and luminosity benchmarks described in the text. The dashed line corresponds to a singlet scalar acquiring mass entirely from electroweak symmetry breaking. Right: The integrated luminosity $\mathcal{L}$ in units of ab$^{-1}$ required to exclude a singlet scalar whose mass $m_S$ is due entirely to electroweak symmetry breaking at 95\% CL from the naive combination of $SS + \gamma/h/Z$ final states.}
   \label{fig:scolex}
\end{figure}

To benchmark the combined sensitivity of these three searches, we consider the naive combination that results from adding their significance in quadrature. The integrated luminosity required to exclude a singlet scalar at 95\% CL using the combination of analyses is shown in \cref{fig:scolex}.

\subsubsection{Invariant mass analysis}
\label{sec:singlet_invt}
An alternative strategy to probe the Higgs portal singlet scenario is to search for the virtual effects of $S$ through its influence on the Higgs sector, rather than looking for this state on-shell as described in \cref{sec:singlet_missing}. At 1-loop order, the simplified model of \cref{eq:lag_singlet} includes modifications in both the Higgs 2-point function, as well as the couplings between the Higgs and SM degrees of freedom. As already noted in \cref{sec:EWSB}, this kind of tampering in the Higgs sector will affect the behavior of scattering amplitudes involving longitudinally polarized gauge bosons, given the role of the Higgs in maintaining perturbative unitarity of the SM. In this section, we illustrate how the resulting features in the differential cross section for the process $\mu^+ \mu^- \rightarrow t \bar t +X$ may be leveraged to probe this model.

The 1-loop correction to the Higgs 2-point function in the model of \cref{eq:lag_singlet} includes both mass and wave-function renormalization effects, and it is given by
\begin{equation}
	\Sigma_2 (p^2) = \frac{(\lambda_{HS} v)^2}{16 \pi^2} \int_0^1 \dd x\, \log \left( \frac{m_S^2 - p^2 x (1-x)}{m_S^2 - m_h^2 x (1-x)} \right) - (p^2 - m_h^2) \delta Z \ ,
\label{eq:sigma2_singlet}
\end{equation}
where
\begin{equation}
	\delta Z = - \frac{(\lambda_{HS} v)^2}{16 \pi^2} \int_0^1 \dd x\, \frac{x (1-x)}{m_S^2 - m_h^2 x (1-x)} \, ,
\label{eq:deltaZ_singlet}
\end{equation}
and we have imposed renormalization conditions such that the Higgs propagator has a single pole at $p^2 = m_h^2$ with residue $i$. Wavefunction renormalization further affects SM Higgs couplings, which are modified by a universal factor of $(1 + \delta_*)$, with $\delta_* \simeq \delta Z / 2$.

\begin{figure}[t!] 
   \centering
   \includegraphics[scale=0.38]{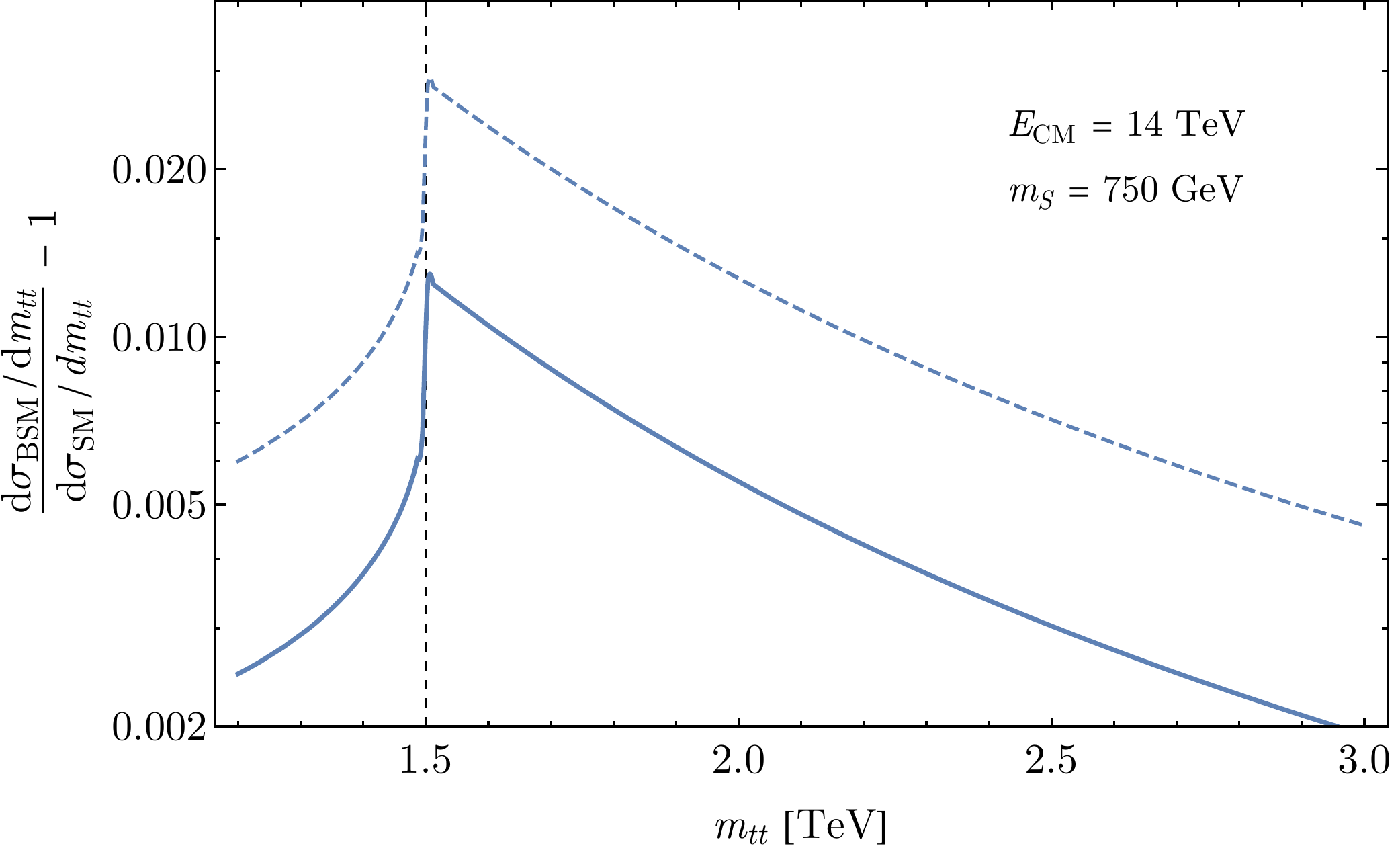} 
   \caption{Fractional deviation in the differential cross section for the process $\mu^+ \mu^- \rightarrow t \bar t +X$ as a result of the radiative corrections present in the model of \cref{eq:lag_singlet} (see Eqs.~(\ref{eq:sigma2_singlet})-(\ref{eq:deltaZ_singlet}) and surrounding discussion). The center-of-mass-energy of the colliding muons is taken to be $14$~TeV, and the mass of the singlet $m_S = 750$~GeV, all of which arises from electroweak symmetry breaking. The solid line includes both charged and neutral intermediate VBF processes, whereas the dashed line includes only contributions from $W^+W^-$ fusion, as would be appropriate if background events could be eliminated by identifying the outgoing muons. The vertical dashed line corresponds to $m_{t \bar t} = 2 m_s = 1.5$ TeV, at which the peak of the kinematic feature in the differential cross section takes place. }
   \label{fig:dsigmadm}
\end{figure}

At the partonic level, the processes most sensitive to these modifications in the Higgs sector are those involving longitudinally polarized $W$ and $Z$ bosons in the initial state, with $W^+_L W^-_L \rightarrow t \bar t$ being the most relevant given the enhanced $W$ content of the muons. This results in a distinctive kinematic feature in the differential cross section for the process $\mu^+ \mu^- \rightarrow t \bar t +X$ that peaks at a scale $m_{t \bar t} \simeq 2 m_S$, where $m_{t \bar t}$ is the invariant mass of the $t \bar t$ pair. For illustration, \cref{fig:dsigmadm} shows the fractional deviation in $\dd \sigma / \dd m_{t \bar t}$ with respect to the SM for a collision center-of-mass energy $E_{\rm CM} = 14$~TeV, and a singlet mass $m_S = 750$~GeV that results entirely from electroweak symmetry breaking. The maximum size of the deviation can be of order $\sim 1 \%$. Our calculation includes the 1-loop effects described in the previous paragraph, but is otherwise performed at tree-level. This approximation is justified as long as $\lambda_{HS}$ is larger than any of the couplings of the SM, which is always case when the singlet mass arises entirely from its coupling to the Higgs and provided $m_S > m_t$.

\begin{figure}[t!] 
   \centering
   \includegraphics[scale=0.38]{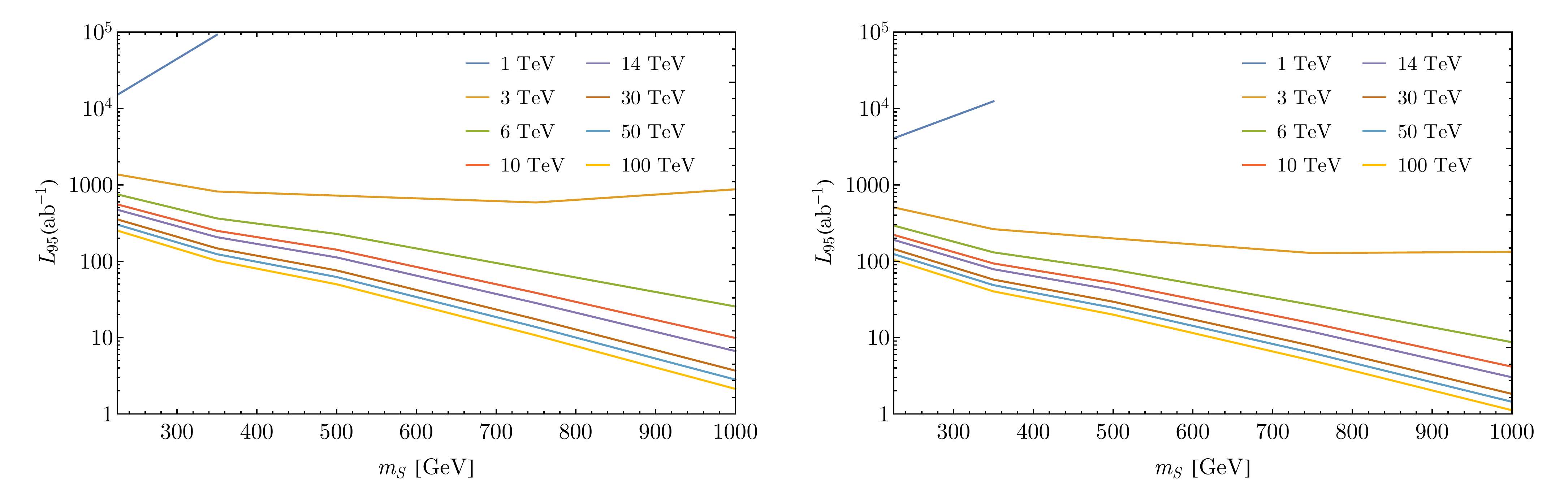} 
   \caption{Estimate of the integrated luminosity required to exclude a Higgs portal singlet $S$ whose mass arises entirely from electroweak symmetry breaking, for various center-of-mass energies. Left: $t \bar t$ production through intermediate neutral and charged VBF is treated as background. Right: Only $t \bar t$ production through charged VBF is treated as background, under the assumption that the outgoing muons accompanying neutral VBF can be identified.}
   \label{fig:singlet_mtt_lum}
\end{figure}

An estimate of the integrated luminosity required to exclude a singlet whose mass is due entirely to electroweak symmetry breaking at 95\% CL using this analysis is shown in \cref{fig:singlet_mtt_lum}. The number of signal and background events have been computed by integrating the differential cross section over a mass window spanning the range between $90\%$ and $150\%$ of $2 m_S$. The corresponding $S/ \sqrt{B}$ ratio has been computed for $m_S = 225, 350, 500, 750$, and $1000$~GeV and linearly interpolated for intermediate masses.\footnote{For a center-of-mass energy of the incoming muons $E_{\rm CM} = 1$ TeV, we only estimate the required luminosity for $m_S = 225$ and $350$ GeV, and linearly interpolate between the two. For the larger values of $m_S$, it is not possible for the deviation in the differential cross section to reach its peak at $m_{t \bar t} = 2 m_S$, as the singlet never becomes on-shell, and therefore the method discussed here is not applicable.}

Although a more sophisticated analysis would be required to draw robust conclusions, the results in \cref{fig:singlet_mtt_lum} suggest that this search strategy could improve on the individual channels described in \cref{sec:singlet_missing}, and may be competitive with their combination. The analysis could be further improved by considerations related to the angular distribution of the outgoing $t \bar t$ pair, or through the use of polarized beams. More importantly, our discussion illustrates how muon colliders offer qualitatively new possibilities to search for new physics, by taking advantage of both their high energy reach as well as their underlying identity as gauge boson colliders. 

\newpage
\subsection{Hidden valleys}

To benchmark the sensitivity of muon colliders to Hidden Valley scenarios \cite{Strassler:2006im, Han:2007ae}, we consider a particularly minimal realization in which the vector current of a SM-neutral Dirac fermion $\chi$ is coupled to the muon vector current of the Standard Model via a dimension-6 contact term:
\begin{equation}
  \mathcal{L} = i \bar \chi \slashed{D} \chi - m_\chi \bar \chi \chi + \frac{c}{\Lambda^2} (\bar\mu \gamma^\mu \mu) (\bar\chi \gamma_\mu \chi)\,.
\end{equation}

\begin{figure}[h!] 
   \centering
   \includegraphics[width=2.5in]{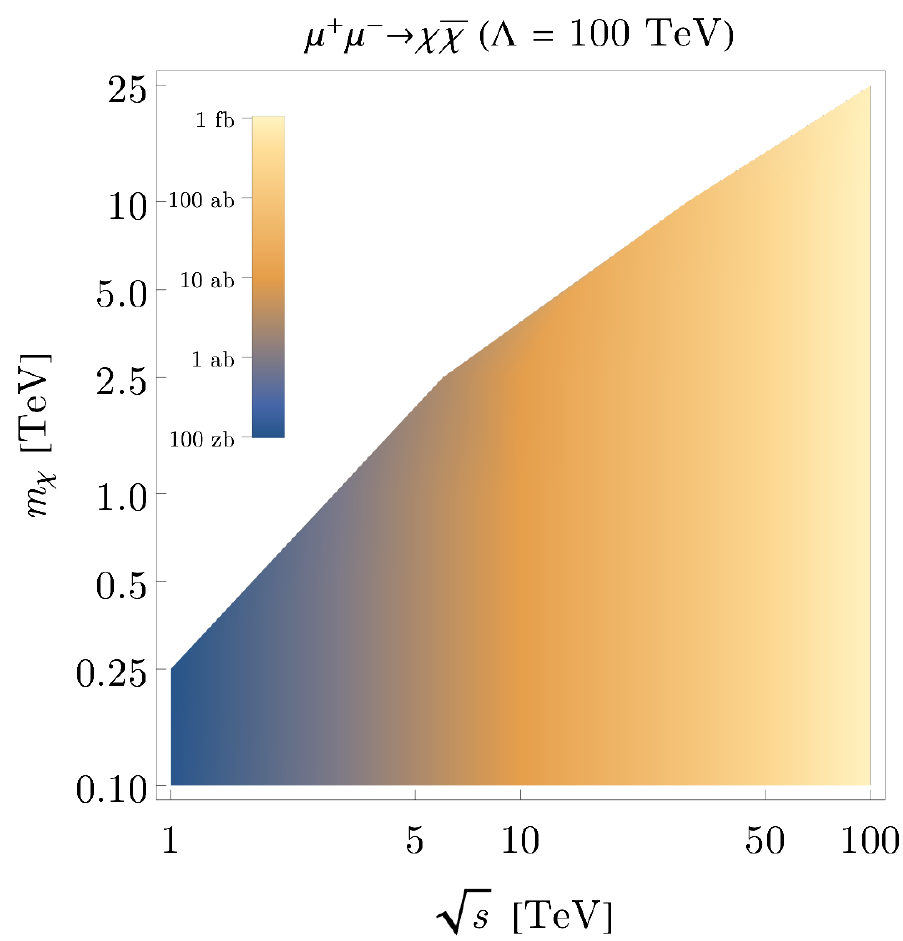} 
   \caption{Cross section for $\chi \bar \chi$ production in the hidden valley simplified model as a function of $m_\chi$ and $\sqrt{s}$ for $\Lambda = 100$ TeV and $c= 1$, computed using {\sc MadGraph5}.}
   \label{fig:HV}
\end{figure}

We implement this model in {\sc FeynRules} and compute the cross section using {\sc MadGraph5}. The cross section for $\chi \bar \chi$ production as a function of $m_\chi$ and $\sqrt{s}$ is shown in \cref{fig:HV} for $\Lambda = 100$ TeV and $c = 1$.

\newpage
\subsection{Axion-like particles}
As our final example, we consider a simplified model in which an axion-like particle $a$ couples to electroweak field strengths. We use the {\sc ALPsEFT} {\sc FeynRules} model file documented in \cite{Brivio:2017ije}, taking the Lagrangian to be
\begin{equation}
  \mathcal{L} \supset \frac12 (\partial a)^2 - \frac12  m_a^2 a^2 
  - \frac{c_{\tilde B}}{f_a} a B_{\mu\nu} \tilde B^{\mu\nu}
  - \frac{c_{\tilde W}}{f_a} a W^a_{\mu\nu} \tilde W^{a,\mu\nu}\,.
\end{equation}

\begin{figure}[h!] 
   \centering
   \includegraphics[width=2.5in]{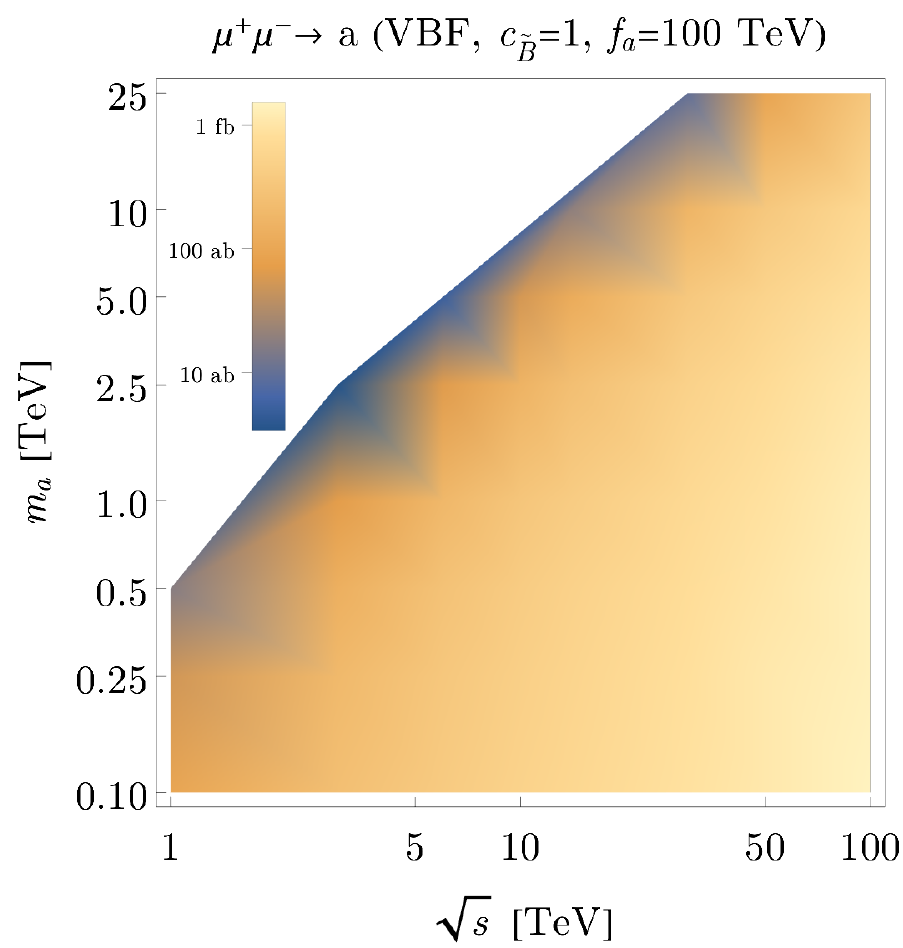} 
      \includegraphics[width=2.5in]{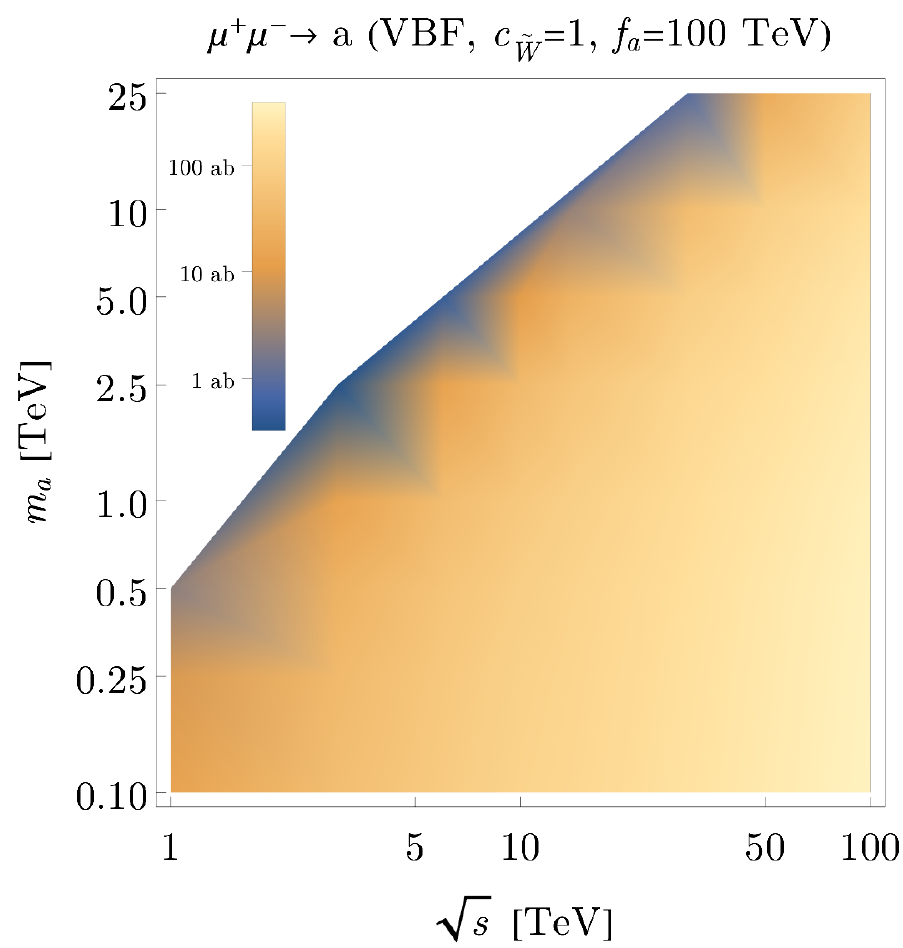} 
   \caption{Cross section for VBF production of $a$ in the axion-like particle simplified model as a function of $m_a$ and $\sqrt{s}$ for $f_a = 100$ TeV, computed using {\sc MadGraph5}. \hspace{100pt} Left: $c_{\tilde B}= 1, c_{\tilde W} = 0$.  Right: $c_{\tilde B} = 0, c_{\tilde W} = 1$.}
   \label{fig:ALP}
\end{figure}

Cross sections for VBF production of $a$ as a function of $m_a$ and $\sqrt{s}$ are shown in \cref{fig:ALP} for $f_a = 100$ TeV and $(c_{\tilde B}, c_{\tilde W}) = (1,0)$ [left] and $(0,1)$ [right].

\clearpage


\end{spacing}

\addcontentsline{toc}{section}{References} 

\bibliographystyle{utphys}
\bibliography{muonbib}

\end{document}